\documentclass[12pt]{article}
\usepackage[english]{babel}
\usepackage{hyperref}
\usepackage{amsmath}
\usepackage{amsfonts}
\usepackage{amssymb}
\renewcommand{\theequation}{\thesection.\arabic{equation}}
\begin{document}

\title{
%\hfill{\normalsize{}hep-th/1012.0000}\\[5mm]
{\bf{} General Lagrangian Formulation for Higher Spin Fields with
Arbitrary Index Symmetry.  2. Fermionic fields}}

\author{\sc
 A. Reshetnyak${}^{a,b}$\thanks{reshet@ispms.tsc.ru}
\\[0.5cm]
\it  ${}^{a}$Laboratory of Computer-Aided Design of Materials, Institute of
\\ Strength Physics and Materials Science, 634021 Tomsk, Russia,\\
${}^{b}$Tomsk State Pedagogical University, 634041 Tomsk, Russia}
\date{}
\maketitle \thispagestyle{empty}

\begin{abstract}
We continue the construction of a Lagrangian description of
irreducible half-integer higher-spin representations of the
Poincare group with an arbitrary Young tableaux having $k$ rows,
on a basis of the BRST--BFV approach suggested for bosonic fields
in our first article (Nucl. Phys. B862  (2012) 270,
[arXiv:1110.5044[hep-th]). Starting from a description of
fermionic mixed-symmetry higher-spin fields in a flat space of any
dimension in terms of an auxiliary Fock space associated with a
special Poincare module, we realize a conversion of the initial
operator constraint system (constructed with respect to the
relations extracting irreducible Poincare-group representations)
into a system of first-class constraints. To do this, we find,
 in first time, by means of generalized Verma module the auxiliary representations of
 the constraint
 subsuperalgebra, to be isomorphic due to Howe duality to $osp(1|2k)$
 superalgebra, and
containing the subsystem of second-class constraints  in terms of
new  oscillator variables. We suggest a
universal procedure of finding unconstrained gauge-invariant
Lagrangians with reducible gauge symmetries, describing the
dynamics of both massless and massive fermionic fields of any
spin. It is shown that the space of BRST cohomologies with a
vanishing ghost number is determined only by constraints
corresponding to an irreducible Poincare-group representation. As
examples of the general approach, we propose a method of
Lagrangian construction for fermionic fields subject to an
arbitrary Young tableaux having $3$ rows, and obtain a
gauge-invariant Lagrangian for a new model of a massless rank-3
spin-tensor field of spin $(5/2,3/2)$ with first-stage reducible
gauge symmetries and a non-gauge Lagrangian for a massive rank-3
spin-tensor field of spin $(5/2,3/2)$.
\end{abstract}


\section{Introduction}

Higher-spin field theory, in its various aspects, has been under a
long and intense study, in the hopes to re-examine the problems of
a unified description for the variety of elementary particles, and
to discover new approaches to a unification of the known
fundamental interactions.  Higher-spin field theory is in close relation
to superstring theory, which operates an infinite set of bosonic
and fermionic fields of various spins, providing the consideration
of higher-spin theory as a tool of investigating the structure of
superstring theory. For the current progress in higher-spin field
theory, see the reviews \cite{reviews}, whereas some recent
directions in higher-spin theory, starting from the pioneering
papers \cite{flatin}, \cite{Singh}, \cite{Fronsdal}, are examined
in \cite{Heslop}--\cite{Boulanger}.

The dynamics of totally symmetric free higher-spin fields is
currently the most well-developed area in the variety of unitary
representations of the Poincare and AdS algebras \cite{Singh},
\cite{Fronsdal}, \cite{massless AdS}, \cite{massive AdS}. This
situation is due to the fact that a $4d$ space-time does not admit
any mixed-symmetry irreducible representations, except for dual
theories. It is well-known that in higher space-time dimensions
there arise mixed-symmetry representations, determined by
spin-like parameters being more than one in number
\cite{Labastida}, \cite{Vasilievmix}, \cite{metsaevmixirrep},
whereas their field-\-theoretic description is not so
well-developed as for totally symmetric representations. While the
simplest mixed-symmetric HS bosonic fields were examined in
\cite{Curtright}, attempts to construct Lagrangian descriptions of
free and interacting higher-spin field theories have met with
consistency problems, which have not yet been completely solved.
Unconstrained Lagrangians for half-integer HS fields with higher
derivatives in the ``metric-like'' formulation in Minkowski
space-time for massless irreducible Poincare group
representations, and without higher derivatives in the case of
reducible ones, have been derived on a basis of the Bianchi
identities resolution in \cite{franciamixfermi}, whereas in the
case of arbitrary irreducible Poincare group representations with
a half-integer spin the resulting unconstrained action (given by
Eq. (6.31) in \cite{franciamixfermi}) contains some special
projection operators that have not yet been found explicitly, thus
making the Lagrangian formulation unclosed, and therefore requires
some additional efforts to find them\footnote{In \cite{BRmixbos},
we analyze in Footnote 2 the same problem of an unconstrained
Lagrangian description for bosonic fields in a flat space of any
dimension, subject to an arbitrary Young tableaux in
\cite{franciamixbos}, where the projectors $\Pi^{ijk}_{klm}$ in
the action given by Eq. (5.25) have a determined status only in
the case of totally symmetric fields, see Eq.(5.28).}. The main
result in the task of a constrained (with off-shell
gamma-traceless algebraic constraints) Lagrangian construction for
arbitrary massless mixed-symmetry fermionic HS fields in a
$d$-dimensional Minkowski space-time has been recently obtained in
\cite{framefermimix} within the ``frame-like'' formulation.
Meanwhile, in the (anti-)de Sitter case, the same results for
massless and massive mixed-symmetry fermionic HS fields in the
``frame-like''formulation with off-shell gamma-traceless
constraints are known in the case of a Young tableaux with two
rows \cite{Zinovievfermi}.

In this article, which continues our investigation started in
\cite{BRmixbos} for tensor fields (see also Ref.
\cite{mixboseResh}), we construct a gauge-invariant Lagrangian
description in the ``metric-like'' formalism for both massless and
massive mixed-symmetry spin-tensor fields of Lorentz rank $n_1 +
n_2 + ... + n_k$ and spin $\mathbf{s} = (n_1 +1/2, n_2+1/2,
 ... , n_k+1/2)$, with any integer numbers $n_1 \geq
n_2 \geq ... \geq n_k \geq 1$ for $k \leq [(d-1)/2]$ in a
$d$-dimensional Minkowski space, corresponding to a unitary
irreducible Poincare-group $ISO(1,d-1)$ representation with a
Young tableaux having $k$ rows. In the case of Minkowski space,
several approaches have been suggested to study mixed-symmetry
higher-spin fields \cite{Labastida}, \cite{Pashnev1},
\cite{Bekaert}, \cite{Medeiros}. Our approach is based on the
BFV--BRST construction \cite{BFV} (see also the reviews \cite{bf},
\cite{Henneaux}), which was initially developed for a Hamiltonian
quantization of dynamical systems subject to first-class
constraints. We recall that the application of the BRST
construction to higher-spin field theory in constant-curvature
spaces consists of four steps, being reduced to three steps in the
case of flat spaces. At the first stage, the conditions
determining representations with a given spin and mass are
regarded as a system of mixed-class operator constraints in an
auxiliary Fock space. Second, the system of initial constraints is
converted, with a preservation of the initial algebraic structure,
into a system of first-class constraints only, acting in an
enlarged Fock space, with respect to which one constructs a BRST
operator (being the (nontrivial) third step in the case of fields
in AdS spaces). Finally, a Lagrangian for a higher-spin field is
constructed in terms of the BRST operator, in such a way that the
corresponding equations of motion reproduce the initial
constraints. It should be emphasized that this approach
automatically implies a gauge-invariant Lagrangian description,
reflecting the general fact of BV--BFV duality \cite{AKSZ},
\cite{BV-BFV}, \cite{GMR}, realized in order to reproduce a
Lagrangian action by means of a Hamiltonian object.

The development of flat dynamics for mixed-symmetry gauge fields
has been examined in \cite{Francia}, \cite{Zinoviev_m},
\cite{Ouvry}, \cite{Metsaev-1}, \cite{Mets-amb} for massless
bosonic higher-spin fields with two rows of the Young tableaux
\cite{BurdikPashnev}, and recently also for interacting bosonic
higher-spin fields \cite{Metsaev-0}, \cite{Tsulaiaint},
\cite{Manvelyan}. Lagrangian descriptions of massless
mixed-symmetry fermionic and bosonic higher-spin fields in (A)dS
spaces have been suggested within a ``frame-like'' approach in
\cite{AlkalaevVasiliev}, whereas for massive fields of lower
superspins in flat and (A)dS spaces they have been examined in
\cite{Zinoviev}. In turn, the aspects of interacting higher-spin fields and its relation to string theory spectrum was considered in \cite{Tsulaia}. For the sake of completeness, we emphasize that
in the case of free totally symmetric higher-spin fields of
half-integer spin the BRST approach has been used to obtain
Lagrangians in the flat space \cite{symferm-flat} and in the (A)dS
space \cite{symferm-ads}, whereas for totally symmetric bosonic HS
tensors in AdS spaces the same has been done in \cite{symint-ads},
and for mixed-symmetric ones, subject to a Young tableaux with two
rows, in \cite{BRmixads}.

The paper is organized as follows. In Section~\ref{Symmalgebra},
we present a closed superalgebra of operators (using the Howe
duality), based on constraints in an auxiliary Fock space with a
symmetric basis, that determines a massless irreducible
representation of the Poincare group in $\mathbb{R}^{1,d-1}$ with
a generalized spin $\mathbf{s} = (n_1+\frac{1}{2}, ...,
n_k+\frac{1}{2})$. In Section~\ref{Vermamodule}, we construct an
auxiliary representation for the
rank-$\left([\frac{(d-1)}{2}],[\frac{(d-1)}{2}]\right)$
orthosymplectic $osp(1|2k)$ subsuperalgebra of the superalgebra of
initial constraints, corresponding to the subsystem of
second-class constraints in terms of new (additional) creation and
annihilation operators in Fock space\footnote{Notice that a
similar construction for fermionic HS fields subject to a Young
tableaux with 2 rows in a flat space has been presented in
\cite{mixfermiflat}.}. As a result, the initial system of first-
and second-class odd and even constraints is converted into a
system of first-class constraints in the space being the tensor
product of the initial and new Fock spaces. Next, we construct the
standard BRST operator for the converted constraint superalgebra
in Section~\ref{BRSToperator}. The construction of an action and
of a sequence of reducible gauge transformations describing the
propagation of a mixed-symmetry fermionic field of an arbitrary
spin is realized in Section~\ref{LagrFormulation}. We show, after
applying dimensional reduction to the massless half-integer
mixed-symmetry HS field in a $(d+1)$-dimensional flat space, that
the Lagrangian description for a massive HS field in a
$d$-dimensional Minkowski space of the same type is deduced. In
Section~\ref{examples}, we demonstrate that the general procedure
contains, first, a previously known algorithm of Lagrangian
construction for fermionic fields subject to a Young tableaux with
two rows, and, second, a new algorithm for spin-tensor fields with
three rows in the corresponding Young tableaux. In
Subsections~\ref{ex5232},~\ref{ex5232m}, we construct new
unconstrained Lagrangian descriptions for both third-rank massless
and massive spin-tensor fields with spin $(5/2,3/2)$, which have
not been obtained before. In Conclusion, we summarize the results
of this work and outline some open problems. Finally, in
Appendix~\ref{addalgebra} we construct an auxiliary representation
for the $osp(1|2k)$ algebra on a basis of the (generalized) Verma
module, shortly described in Appendix~\ref{defVerma}.
Appendix~\ref{oscrealsp2kdet} is devoted to finding a polynomial
representation of the operator superalgebra given by
Tables~\ref{table in},{}~\ref{table inodd} in terms of creation
and annihilation operators. In Appendix~\ref{reductionC}, we prove
the fact that the constructed general Lagrangian actually
reproduces correct conditions for the field that determine an
irreducible representation of the Poincare group; we also suggest
a new form of gauge-fixing procedure. In
Appendix~\ref{example5232}, the expressions for the field and all
the Fock-space gauge vectors are presented in the powers of ghost
creation operators, to be applied to a Lagrangian construction for
the third-rank Dirac spin-tensor.

In addition to the conventions of \cite{BRmixbos},
\cite{BurdikPashnev}, \cite{symferm-flat}, \cite{0001195}, we use,
first, the mostly minus signature for the metric tensor
$\eta_{\mu\nu} = diag (+, -,...,-)$, with Lorentz indices $\mu,
\nu = 0,1,...,d-1$, second, the relations $\{\gamma^{\mu},
\gamma^{\nu}\} = 2\eta^{\mu\nu}$ for the Dirac matrices
$\gamma^{\mu}$, third, the  notation $\varepsilon(A)$, $gh(A)$ for
the respective values of Grassmann parity and ghost number of a
quantity $A$, and denote by $[A,\,B\}$ the supercommutator of
quantities $A, B$, which, in case they possess definite values of
Grassmann parity, is given by $[A\,,B\}$ = $AB -
(-1)^{\varepsilon(A)\varepsilon(B)}BA$.

\section{Derivation of Half-Integer HS
Symmetry Superalgebra in $\mathbb{R}^{1,d-1}$}\label{Symmalgebra}

Let us study a massless half-integer irreducible representation of
the Poincare group in a $d$-dimensional Minkowski space, which is
to be described by a spin-tensor field:
$\Psi_{(\mu^1)_{n_1},(\mu^2)_{n_2},...,(\mu^k)_{n_k}}
\hspace{-0.2em}$ $\equiv \hspace{-0.2em}
\Psi_{\mu^1_1\ldots\mu^1_{n_1},\mu^2_1\ldots\mu^2_{n_2},...,}$
${}_{ \mu^k_1\ldots \mu^k_{n_k}{}A}(x)$, with the Dirac index $A$
(being suppressed in what follows) of rank $\sum_{i\geq 1}^k n_i$
and the generalized spin
 $\mathbf{s} = (n_1 +1/2, n_2+1/2,
 ... , n_k+1/2)$ ($n_1 \geq n_2\geq ... \geq n_k>0, k \leq [(d-1)/2])$,
 which corresponds to a Young tableaux with $k$
rows of length $n_1, n_2,  ..., n_k$, respectively,
\begin{equation}\label{Young k}
\Psi_{(\mu^1)_{n_1},(\mu^2)_{n_2},...,(\mu^k)_{n_k}}
\hspace{-0.3em}\longleftrightarrow \hspace{-0.1em}
\begin{array}{|c|c|c c c|c|c|c|c|c| c|}\hline%\vphantom{\biggm|}
  \!\mu^1_1 \!&\! \mu^1_2\! & \cdot \ & \cdot \ & \cdot \ & \cdot\  & \cdot\  & \cdot\ &
  \cdot\    &\!\! \mu^1_{n_1}\!\! \\
   \hline%\vphantom{\biggm|}
    \! \mu^2_1\! &\! \mu^2_2\! & \cdot\
   & \cdot\ & \cdot  & \cdot &  \cdot & \!\!\mu^2_{n_2}\!\!   \\
 \cline{1-8} \cdots & \cdot &\cdot &\cdot & \cdot & \cdot &\cdot   \\
   \cline{1-7}
     \! \mu^k_1\! &\! \mu^k_2\! & \cdot\
   & \cdot\ & \cdot  & \cdot &   \!\!\mu^k_{n_k}\!\!   \\
   \cline{1-7}%\vphantom{\biggm|}
\end{array}\ ,
\end{equation}
The field is symmetric with respect to permutations of each type
of Lorentz indices
 $\mu^i$,
  and
obeys the Dirac equations (\ref{Eq-0}), as well as those of
gamma-tracelessness (\ref{Eq-1}) and mixed-symmetry (\ref{Eq-2})
[for $i,j=1,...,k;\, l_i,m_i=1,...,n_i$],
\begin{eqnarray}
\label{Eq-0} &&
\imath\gamma^{\mu}\partial_{\mu}\Psi_{(\mu^1)_{n_1},(\mu^2)_{n_2},...,(\mu^k)_{n_k}}
=0\,,
\\
&& \gamma^{\mu^i_{l_i}}\Psi_{
(\mu^1)_{n_1},(\mu^2)_{n_2},...,(\mu^k)_{n_k}} =0\,,
 \label{Eq-1}\\
&& \Psi_{
(\mu^1)_{n_1},...,\{(\mu^i)_{n_i}\underbrace{,...,\mu^j_{1}...}\mu^j_{l_j}\}...\mu^j_{n_j},...(\mu^k)_{n_k}}=0,\quad
i<j,\ 1\leq l_j\leq n_j,
 \label{Eq-2}
\end{eqnarray}
where the figure bracket below denotes that the indices included
in it do not take part in symmetrization, which thus concerns only
the indices $(\mu^i)_{n_i}, \mu^j_{l_j} $ in
$\{(\mu^i)_{n_i}\underbrace{,...,\mu^j_{1}...}\mu^j_{l_j}\}$.

All irreducible representations can be described simultaneously if
we make one of the two choices of introducing an auxiliary Fock
space $\mathcal{H}$. We consider, as usual, the Fock space
$\mathcal{H}$ generated by the bosonic (the case of a symmetric
basis) creation and annihilation operators $a^{i+}_{\mu^i},
a^{j}_{\nu^j}$ with additional internal indices, $i,j =1,...,k,
\mu^i,\nu^j =0,1...,d-1$\footnote{There exists another realization
of all irreps be means of a different Fock space
$\mathcal{H}^{as}$, generated by fermionic oscillators
(antisymmetric basis) $\hat{a}^m_{\mu^m}(x)$,
$\hat{a}^{\hat{n}+}_{\nu^n}(x)$ with the anticommutation relations
$\{\hat{a}^m_{\mu^m},
\hat{a}_{\nu^n}^{n+}\}=-\eta_{\mu^m\nu^n}\delta^{mn}$, for $m, n =
1,..., n_1$, and one can complete the below procedure, which
follows the prescription of \cite{brst1} for totally antisymmetric
spin-tensors for $n_1=n_2=...=n_k$.}:
\begin{eqnarray}\label{comrels}
[a^i_{\mu^i}, a_{\nu^j}^{j+}]=-\eta_{\mu^i\nu^j}\delta^{ij}\,,
\qquad \delta^{ij} = diag(1,1,\ldots 1)\,,
\end{eqnarray}

The general state of the Fock space is a Dirac-like spinor, having
the form
\begin{eqnarray}
\label{PhysState} |\Psi\rangle &=&
\sum_{n_1=0}^{\infty}\sum_{n_2=0}^{n_1}\cdots\sum_{n_k=0}^{n_{k-1}}
\Psi_{(\mu^1)_{n_1},(\mu^2)_{n_2},...,(\mu^k)_{n_k}}\,
\prod_{i=1}^k\prod_{l_i=1}^{n_i} a^{+\mu^i_{l_i}}_i|0\rangle,
\end{eqnarray}
which provides the symmetry property of
$\Psi_{(\mu^1)_{n_1},(\mu^2)_{n_2},...,(\mu^k)_{n_k}}$ with
respect to permutations of indices of the same type. Using the
accepted terminology, we refer to the vector (\ref{PhysState}) as
the basic vector\footnote{One
 may regard the set of
 all finite Dirac-like vectors with finite upper limits for $n_1$
 and different choices of spin $\mathbf{s}$ as a vector space
 of polynomials $P_k^d(a^{+})$, being Dirac-like spinors
 in powers of $a^{+\mu^i}_i$. The Lorentz algebra is realized
 by the action on $P_k^d(a^{+})$
 of Lorentz transformations, $M^{\mu\nu}=  \sum_{i\geq 1}^k a^{+[\mu}_i a^{
 \nu]i}+ \frac{1}{2}\gamma_{\mu\nu}$, with the standard rule
 $A^{[\mu}B^{\nu]}\equiv A^{\mu}B^{\nu}-A^{\nu}B^{\mu}$ and $\gamma_{\mu\nu}=
 \frac{1}{2}\gamma_{[\mu}\gamma_{\nu]}$.}.

Because of the property of translational invariance of the vacuum,
$\partial_\mu |0\rangle = 0$, the conditions
(\ref{Eq-0})--(\ref{Eq-2}) can be equivalently expressed in terms
of the bosonic operators
\begin{eqnarray}
  && \tilde{t}_0= -i\gamma^{\mu}\partial_\mu\,, \qquad\qquad \tilde{t}^i= \gamma^{\mu}a^i_\mu\,,
  \label{tildet0} \\
   &&  t^{i_1j_1} = a^{i_1+}_\mu
a^{j_1\mu}, \quad {i_1 < j_1} \label{tij}
\end{eqnarray}
as follows:
\begin{equation}\label{t0t1t}
    \tilde{t}_0|\Psi\rangle =
\tilde{t}^i|\Psi\rangle = t^{i_1j_1}|\Psi\rangle =  0.
\end{equation}
Thus, the set of $(\frac{1}{2}k(k+1)+1)$ primary constraints
(\ref{t0t1t}) with $\{o_\alpha\}$ = $\bigl\{{\tilde{t}}_0,
\tilde{t}^i,  t^{i_1j_1} \bigr\}$ for each component
$\Psi_{(\mu^1)_{n_1},(\mu^2)_{n_2},...,(\mu^k)_{n_k}}$ of the
vector (\ref{PhysState}) is equivalent to Eqs.
(\ref{Eq-0})--(\ref{Eq-2}) for all values of spin subject to the
condition $n_1 \geq n_2\geq ... \geq n_k>0$.
 In turn, if we impose,
in addition to Eqs.(\ref{t0t1t}), the constraints
\begin{equation}\label{gocond}
g_0^i|\Psi\rangle =(n_i+\frac{d}{2}) |\Psi\rangle\,,
\end{equation}
 then these
combined conditions are in one-to-one correspondence with Eqs.
(\ref{Eq-0})--(\ref{Eq-2}) for the spin-tensor
$\Psi_{(\mu^1)_{n_1},(\mu^2)_{n_2},...,(\mu^k)_{n_k}}$ with a
given value of spin $\mathbf{s} = (n_1+\frac{1}{2},
n_2+\frac{1}{2}, ... , n_k+\frac{1}{2})$.

Because of the fermionic nature of equations
(\ref{Eq-0})--(\ref{Eq-2}) with respect to the standard
Lorentz-like Grassmann parity, and due to the bosonic nature of
the primary constraint operators $\tilde{t}_0, \tilde{t}^i$,
$\varepsilon(\tilde{t}_0) = \varepsilon(\tilde{t}^i)= 0$, in order
to equivalently transform these operators into fermionic
ones\footnote{Indeed, the relations $\tilde{t}^i\tilde{t}^i =
\frac{1}{2}{\gamma}^\mu{\gamma}^\nu(a^i_\mu a^i_\nu+a^i_\nu
a^i_\mu)= \frac{1}{2}\{{\gamma}^\mu,{\gamma}^\nu\}a^i_\mu a^i_\nu
= 2 l^{ii}$, see Eqs. (\ref{lilijpr}), imply the validity of the
anticommutator representation $\{\tilde{t}^i,
\tilde{t}^i\}=4l^{ii} $, however, in the case of a bosonic
$\tilde{t}^i$ this is contradictory from the viewpoint of the
spin-statistic theorem.}, we follow Refs. \cite{symferm-flat},
\cite{symferm-ads} and introduce a set of $d+1$ Grassmann-odd
gamma-matrix-like objects $\tilde{\gamma}^\mu$, $\tilde{\gamma}$,
subject to the conditions
\begin{eqnarray}
\{\tilde{\gamma}^\mu,\tilde{\gamma}^\nu\} = 2\eta^{\mu\nu}, \qquad
\{\tilde{\gamma}^\mu,\tilde{\gamma}\}=0, \qquad
\tilde{\gamma}^2=-1, \label{tgammas}
\end{eqnarray}
and related to the conventional gamma-matrices as follows:
\begin{eqnarray}\label{gammas}
\gamma^{\mu} = \tilde{\gamma}^{\mu} \tilde{\gamma}.
\end{eqnarray}

We can now define Grassmann-odd constraints,
\begin{eqnarray}
\label{t0ti} {t}_0 = -\imath\tilde{\gamma}^\mu \partial_\mu\,,
\qquad\qquad {t}^i =  \tilde{\gamma}^\mu a^i_\mu ,
\end{eqnarray}
% in case of massless Y(2) fields and AdS Y(1) case the mistake was made in sign below
%
related to the operators (\ref{tildet0}) as follows:
\begin{eqnarray}
\left(t_0, t^i\right) = \tilde{\gamma}\left(\tilde{t}_0,
\tilde{t}^i\right) .
\end{eqnarray}

An essential part of the procedure of Lagrangian description
concerns the property of BFV--BRST operator $Q$, $Q = C^\alpha
o_\alpha + more$, to be Hermitian, which is equivalent to the
requirements $\{o_\alpha\}^+ = \{o_\alpha\}$ and closedness for
$\{o_\alpha\}$ with respect to the supercommutator multiplication
$[\ ,\ \}$. It is evident that the set of $\{o_\alpha\}$ violates
the above conditions. To provide them, we follow the case of
totally-symmetric \cite{symferm-flat} and mixed-symmetry
spin-tensors with $Y(n_1,n_2)$ \cite{mixfermiflat}, and define an
odd scalar product in $\mathcal{H}$,
\begin{eqnarray}
\label{sproduct} \langle\tilde{\Phi}|\Psi\rangle & =  & \int
d^dx\sum_{n_1=0}^{\infty}\sum_{n_2=0}^{n_1}\cdots\sum_{n_k=0}^{n_{k-1}}
         \sum_{p_1=0}^{\infty}\sum_{p_2=0}^{p_1}\cdots\sum_{p_k=0}^{p_{k-1}}
\langle 0|\prod_{j=1}^k\prod_{m_j=1}^{p_j}
a^{\nu^j_{m_j}}_j\Phi^+_{(\nu^1)_{p_1},(\nu^2)_{p_2},...,(\nu^k)_{p_k}}(x)\nonumber\\
&& \times \tilde{\gamma}_0
\Psi_{(\mu^1)_{n_1},(\mu^2)_{n_2},...,(\mu^k)_{n_k}}(x)\,
\prod_{i=1}^k\prod_{l_i=1}^{n_i} a^{+\mu^i_{l_i}}_i|0\rangle ,
\end{eqnarray}
for nonnegative integers $n_i$, $p_j$. As a result, the set of
primary constraints $\{o_\alpha\}$, being extended, first, due to
the closedness condition by means of the D'Alembert operator, and
the divergentless and traceless operators
\begin{equation} \label{lilijpr}
l_0= \partial^\mu\partial_\mu \,,\qquad l^i=- ia^i_\mu
\partial^\mu \,,\qquad l^{ij}={\textstyle\frac{1}{2}}\,a^{i{}\mu}
a^j_{\mu}\,,
\end{equation}
second, due to the Hermitian conjugation properties by the
operators
\begin{eqnarray} \label{lilijt+} \hspace{-2ex} && \hspace{-2ex}
\bigl(t^{i+},\ {l}^{i+},\ l^{ij+},\ t^{i_1j_1+} \bigr) =
\bigl(\tilde{\gamma}^\mu a^{i+}_\mu,\ -i a^{i+}_\mu
\partial^\mu,\ \textstyle\frac{1}{2}a^{i+}_\mu a^{j\mu+},\
a^{i_1}_\mu a^{j_1\mu+}\bigr) ,\ i\leq j;\ i_1 < j_1,
\end{eqnarray}
satisfies both requirements if the particle number operators
$g_0^i$ are included in the set of all constraints $o_I$, thus
having the structure
\begin{eqnarray}
\{o_I\} = \{o_\alpha, o_\alpha^+;\ g_0^i\}\equiv \{o_a, o_a^+ ;\
{t}_0, l_0,\ l^i,\ l^{i+};\ g_0^i\}. \label{inconstraints}
\end{eqnarray}
Being combined, the set $\{o_a, o_a^+\}$ in Eq.
(\ref{inconstraints}), for $\{o_a\} = \{t^i, l^{ij},
t^{i_1j_1}\}$, and the one $\{o_A\}= \{{t}_0, l_0,\ l^i,\
l^{i+}\}$, may be interpreted in the Hamiltonian analysis of
dynamical systems as respective operator-valued $2k^2$ bosonic and
$2k$ fermionic second-class, as well as $(2k+1)$ bosonic and $1$
fermionic first-class constraint subsystems among $\{o_I\}$ for a
topological gauge system (one with the zero Hamiltonian), since
\begin{eqnarray}
[o_a,\; o_b^+\} = f^c_{ab} o_c +\Delta_{ab}(g_0^i),\ [o_A,\;o_B\}
= f^C_{AB}o_C, \  [o_a,\; o_B\} = f^C_{aB}o_C .
\label{inconstraintsd}
\end{eqnarray}
Here, the constants $f^c_{ab}, f^C_{AB}, f^C_{aB}$ possess the
generalized antisymmetry property with respect to permutations of
lower indices, whereas the quantities $\Delta_{ab}(g_0^i)$ form a
non-degenerate $(k\times k; k^2\times k^2)$ supermatrix,
$\|\Delta_{ab}\|$, in the Fock space $\mathcal{H}$ on the surface
$\Sigma \subset \mathcal{H}$: $\|\Delta_{ab}\|_{|\Sigma} \ne 0 $,
which is determined by the equations $(o_a, t_0,\ l_0,\
l^i)|\Psi\rangle = 0$. The set of $o_I$ contains the operators
$g_0^i$, which are not constraints in $\mathcal{H}$, due to Eqs.
(\ref{gocond}).

Explicitly, the operators $o_I$ obey a Lie superalgebra with the
commutation relations
\begin{equation}\label{geninalg}
    [o_I,\ o_J\}= f^K_{IJ}o_K, \  f^K_{IJ}= - (-1)^{\varepsilon(o_I)\varepsilon(o_J)}f^K_{JI},
\end{equation}
where the structure constants $f^K_{IJ}$ used in
Eq.(\ref{inconstraintsd}) include the constants $f^{[g_0^i]}_{ab}:
f^{[g_0^i]}_{ab}g_0^i \equiv \Delta_{ab}^{[g_0^i]}(g_0^i)$ and are
determined by Multiplication Table~\ref{table in}, with
commutators only, and Multiplication Table~\ref{table inodd},
composed from anticommutators of fermionic constraints only.
\hspace{-1ex}{\begin{table}[t] {{\footnotesize
\begin{center}
\begin{tabular}{||c||c|c|c|c|c|c|c||c||}\hline\hline
$\hspace{-0.2em}[\; \downarrow, \rightarrow
]\hspace{-0.5em}$\hspace{-0.7em}&
 $t^{i_1j_1}$ & $t^+_{i_1j_1}$ &
$l_0$ & $l^i$ &$l^{i{}+}$ & $l^{i_1j_1}$ &$l^{i_1j_1{}+}$ &
$g^i_0$ \\ \hline\hline $t_0$
    & $0$ & $0$
   & $0$&\hspace{-0.3em}
    $0$\hspace{-0.5em} &
    \hspace{-0.3em}
    $0$\hspace{-0.3em}
    &\hspace{-0.7em} $0$ \hspace{-1.2em}& \hspace{-1.2em}$
    0$\hspace{-1.2em}& $0$ \\
\hline $t^{i_2}$
    & $-t^{j_1}\delta^{i_2i_1}$ & $-t_{i_1}\delta^{i_2}{}_{j_1}$
   & $0$&\hspace{-0.3em}
    $\hspace{-0.2em}0$\hspace{-0.5em} &
    \hspace{-0.3em}
    $-t_0\delta^{i_2i}$\hspace{-0.3em}
    &\hspace{-0.7em} $0$ \hspace{-1.2em}& \hspace{-1.2em}$
    -\frac{1}{2}t^{\{i_1+}\delta^{j_1\}i_2}\hspace{-0.9em}$\hspace{-1.2em}& $t^{i_2}\delta^{i_2i}$ \\
\hline$t^{i_2+}$
    & $t^{i_1+}\delta^{i_2j_1}$ & $t^+_{j_1}\delta_{i_1}{}^{i_2}$
   & $0$&\hspace{-0.3em}
    $\hspace{-0.2em}t_0\delta^{i_2i}$\hspace{-0.5em} &
    \hspace{-0.3em}
    $0$\hspace{-0.3em}
    &\hspace{-0.7em} $\hspace{-0.7em}\frac{1}{2}t^{\{i_1}\delta^{j_1\}i_2}
    \hspace{-0.9em}$ \hspace{-1.2em}& \hspace{-1.2em}$0\hspace{-0.9em}$\hspace{-1.2em}& $-t^{i_2+}\delta^{i_2i}$ \\
\hline\hline $t^{i_2j_2}$
    & $A^{i_2j_2, i_1j_1}$ & $B^{i_2j_2}{}_{i_1j_1}$
   & $0$&\hspace{-0.3em}
    $\hspace{-0.2em}l^{j_2}\delta^{i_2i}$\hspace{-0.5em} &
    \hspace{-0.3em}
    $-l^{i_2+}\delta^{j_2 i}$\hspace{-0.3em}
    &\hspace{-0.7em} $\hspace{-0.7em}l^{\{j_1j_2}\delta^{i_1\}i_2}
    \hspace{-0.9em}$ \hspace{-1.2em}& \hspace{-1.2em}$
    -l^{i_2\{i_1+}\delta^{j_1\}j_2}\hspace{-0.9em}$\hspace{-1.2em}& $F^{i_2j_2,i}$ \\
\hline $t^+_{i_2j_2}$
    & $-B^{i_1j_1}{}_{i_2j_2}$ & $A^+_{i_1j_1, i_2j_2}$
&$0$   & \hspace{-0.3em}
    $\hspace{-0.2em} l_{i_2}\delta^{i}_{j_2}$\hspace{-0.5em} &
    \hspace{-0.3em}
    $-l^+_{j_2}\delta^{i}_{i_2}$\hspace{-0.3em}
    & $l_{i_2}{}^{\{j_1}\delta^{i_1\}}_{j_2}$ & $-l_{j_2}{}^{\{j_1+}
    \delta^{i_1\}}_{i_2}$ & $-F_{i_2j_2}{}^{i+}$\\
\hline $l_0$
    & $0$ & $0$
& $0$   &
    $0$\hspace{-0.5em} & \hspace{-0.3em}
    $0$\hspace{-0.3em}
    & $0$ & $0$ & $0$ \\
\hline $l^j$
   & \hspace{-0.5em}$- l^{j_1}\delta^{i_1j}$ \hspace{-0.5em} &
   \hspace{-0.5em}$
   -l_{i_1}\delta_{j_1}^{j}$ \hspace{-0.9em}  & \hspace{-0.3em}$0$ \hspace{-0.3em} & $0$&
   \hspace{-0.3em}
   $l_0\delta^{ji}$\hspace{-0.3em}
    & $0$ & \hspace{-0.5em}$- \textstyle\frac{1}{2}l^{\{i_1+}\delta^{j_1\}j}$
    \hspace{-0.9em}&$l^j\delta^{ij}$  \\
\hline $l^{j+}$ & \hspace{-0.5em}$l^{i_1+}
   \delta^{j_1j}$\hspace{-0.7em} & \hspace{-0.7em}
   $l_{j_1}^+\delta_{i_1}^{j}$ \hspace{-1.0em} &
   $0$&\hspace{-0.3em}
      \hspace{-0.3em}
   $-l_0\delta^{ji}$\hspace{-0.3em}
    \hspace{-0.3em}
   &\hspace{-0.5em} $0$\hspace{-0.5em}
    &\hspace{-0.7em} $ \textstyle\frac{1}{2}l^{\{i_1}\delta^{j_1\}j}
    $\hspace{-0.7em} & $0$ &$-l^{j+}\delta^{ij}$  \\
\hline $l^{i_2j_2}$
    & \hspace{-0.3em}$\hspace{-0.4em}-l^{j_1\{j_2}\delta^{i_2\}i_1}\hspace{-0.5em}$
    \hspace{-0.5em} &\hspace{-0.5em} $\hspace{-0.4em}
    -l_{i_1}{}^{\{i_2+}\delta^{j_2\}}_{j_1}\hspace{-0.3em}$\hspace{-0.3em}
   & $0$&\hspace{-0.3em}
    $0$\hspace{-0.5em} & \hspace{-0.3em}
    $ \hspace{-0.7em}-\textstyle\frac{1}{2}l^{\{i_2}\delta^{j_2\}i}
    \hspace{-0.5em}$\hspace{-0.3em}
    & $0$ & \hspace{-0.7em}$\hspace{-0.3em}
L^{i_2j_2,i_1j_1}\hspace{-0.3em}$\hspace{-0.7em}& $\hspace{-0.7em}  l^{i\{i_2}\delta^{j_2\}i}\hspace{-0.7em}$\hspace{-0.7em} \\
\hline $l^{i_2j_2+}$
    & $ l^{i_1 \{i_2+}\delta^{j_2\}j_1}$ & $ l_{j_1}{}^{\{j_2+}
    \delta^{i_2\}}_{i_1}$
   & $0$&\hspace{-0.3em}
    $\hspace{-0.2em} \textstyle\frac{1}{2}l^{\{i_2+}\delta^{ij_2\}}$\hspace{-0.5em} & \hspace{-0.3em}
    $0$\hspace{-0.3em}
    & $-L^{i_1j_1,i_2j_2}$ & $0$ &$\hspace{-0.5em}  -l^{i\{i_2+}\delta^{j_2\}i}\hspace{-0.3em}$\hspace{-0.2em} \\
\hline\hline $g^j_0$
    & $-F^{i_1j_1,j}$ & $F_{i_1j_1}{}^{j+}$
   &$0$& \hspace{-0.3em}
    $\hspace{-0.2em}-l^i\delta^{ij}$\hspace{-0.5em} & \hspace{-0.3em}
    $l^{i+}\delta^{ij}$\hspace{-0.3em}
    & \hspace{-0.7em}$\hspace{-0.7em}  -l^{j\{i_1}\delta^{j_1\}j}\hspace{-0.7em}$\hspace{-0.7em} & $ l^{j\{i_1+}\delta^{j_1\}j}$&$0$ \\
   \hline\hline
\end{tabular}
\end{center}}} \vspace{-2ex}\caption{even-even and odd-even parts of HS symmetry  superalgebra  $\mathcal{A}^f(Y(k),
\mathbb{R}^{1,d-1})$.\label{table in} }\end{table} {\begin{table}[b]
{{\footnotesize
\begin{center}
\begin{tabular}{||c||c|c|c||}\hline\hline
$\hspace{-0.2em}[\; \downarrow, \rightarrow
\}\hspace{-0.5em}$\hspace{-0.7em}&
 $t_0$ & $t_{i_1}$ &
$t^{i_1{}+}$  \\
\hline\hline $t_0$
    & $-2l_0$ & $2l_{i_1}$
   & $2l^{i_1{}+}$\\
   \hline $t_{i_2}$
    & $2l_{i_2}$ & $4l_{i_2i_1}$
   & $C_{i_2}{}^{i_1}$
   \\
   \hline
   $t^{i_2+}$
    & $2l^{i_2{}+}$ & $(C_{i_1}{}^{i_2})^+$
   & $4l^{i_2i_1+}$
   \\\hline\hline
\end{tabular}
\end{center}}} \vspace{-2ex}\caption{odd-odd part of HS symmetry  superalgebra  $\mathcal{A}^f(Y(k),
\mathbb{R}^{1,d-1})$.\label{table inodd} }\end{table}

First, note that in Table~\ref{table in}, except for the first
three rows with the fermionic constraints $t_0, t_{i_2},
t^+_{i_2}$, described in the case of an integer HS symmetry
algebra $\mathcal{A}(Y(k), \mathbb{R}^{1,d-1})$ in
\cite{BRmixbos}, the operators $t^{i_2j_2}, t^+_{i_2j_2}$ obey (by
definition) the properties
\begin{equation} \label{thetasymb} (t^{i_2j_2}, t^+_{i_2j_2}) \equiv
(t^{i_2j_2},t^+_{i_2j_2})\theta^{j_2i_2}, \ \theta^{j_2i_2} = 1(0)\texttt{ for }j_2>i_2(j_2 \leq i_2)
\end{equation}
with the Heaviside $\theta$-symbol $\theta^{ji}$ and without
summation with respect to the indices $i_2, j_2$. The figure
brackets for the indices $i_1$, $i_2$ in the quantity
$A^{\{i_1}B^{i_2\}i_3}\theta^{i_3i_2\}}$ imply the symmetrization
 $A^{\{i_1}B^{i_2\}i_3}\theta^{i_3i_2\}}$ =
$A^{i_1}B^{i_2i_3}\theta^{i_3i_2}+
A^{i_2}B^{i_1i_3}\theta^{i_3i_1}$; these indices are raised and
lowered by means of the Euclidian metric tensors $\delta^{ij}$,
$\delta_{ij}$, $\delta^{i}_{j}$. Second, the products
$B^{i_2j_2}_{i_1j_1}, A^{i_2j_2, i_1j_1}, F^{i_1j_1,i},
L^{i_2j_2,i_1j_1}$ are determined by explicit relations (see, Ref.
\cite{BRmixbos} for details),
%\noindent
\begin{eqnarray}
  {}B^{i_2j_2}{}_{i_1j_1} &=&
  (g_0^{i_2}-g_0^{j_2})\delta^{i_2}_{i_1}\delta^{j_2}_{j_1} +
  (t_{j_1}{}^{j_2}\theta^{j_2j_1} + t^{j_2}{}^+_{j_1}\theta^{j_1j_2})\delta^{i_2}_{i_1}
  -(t^+_{i_1}{}^{i_2}\theta^{i_2i_1} + t^{i_2}{}_{i_1}\theta^{i_1i_2})
  \delta^{j_2}_{j_1}
\,,\label{Bijkl}
\\
   A^{i_2j_2, i_1j_1} &=&  t^{i_1j_2}\delta^{i_2j_1}-
  t^{i_2j_1}\delta^{i_1j_2}  ,   \label{Aijkl}\\
   {} F^{i_2j_2,i} &=&
   t^{i_2j_2}(\delta^{j_2i}-\delta^{i_2i}),\label{Fijk} \\
  L^{i_2j_2,i_1j_1} &=&   \textstyle\frac{1}{4}\Bigl\{\delta^{i_2i_1}
\delta^{j_2j_1}\Bigl[2g_0^{i_2}\delta^{i_2j_2} + g_0^{i_2} +
g_0^{j_2}\Bigr]  - \delta^{j_2\{i_1}\Bigl[t^{j_1\}i_2}\theta^{i_2j_1\}} +t^{i_2j_1\}+}\theta^{j_1\}i_2}\Bigr] \nonumber \\
&& - \delta^{i_2\{i_1}\Bigl[t^{j_1\}j_2}\theta^{j_2j_1\}}
+t^{j_2j_1\}+}\theta^{j_1\}j_2}\Bigr] \Bigr\}
 \,.\label{Lklij}
\end{eqnarray}
Third, the bosonic operator-valued quantities $C_{i_2}{}^{i_1}$ in
Table~\ref{table inodd} have the following definition and
Hermitian conjugation properties:
\begin{equation}\label{Cij}
   C_{i_2}{}^{i_1} = 2\Bigl(-g_0^{i_1}\delta_{i_2}{}^{i_1} +
   t^{i_1}{}_{i_2}\theta^{i_2i_1} +
   t_{i_2}{}^{i_1+}\theta^{i_1i_2} \Bigr)
  ,\qquad (C_{i_2}{}^{i_1})^+ = C_{i_2}{}^{i_1}.
\end{equation}
For completeness, we  list below the obvious additional properties
of antisymmetry and Hermitian conjugation for the operators in
Eqs.(\ref{Bijkl})--(\ref{Lklij}),
\begin{align}
  & A^{i_2j_2,  i_1j_1} = -A^{ i_1j_1, i_2j_2} &&
  {} A^+_{i_1j_1,  i_2j_2}=(A_{i_1j_1,  i_2j_2})^+ = t^+_{i_2j_1}\delta^{j_2i_1}
   -   t^+_{i_1j_2}\delta^{i_2j_1},\\
  & ({L^{i_2j_2,i_1j_1}})^+ =  L^{i_1j_1, i_2j_2} && {F^{i_2j_2,i}}^+
  =(F^{i_2j_2,i})^+= t^{i_2j_2+}(\delta^{j_2i}-\delta^{i_2i})
\end{align}
\vspace{-5ex}
\begin{equation}\label{Bijk+}
   {B^{i_2j_2}_{i_1j_1}}^+ = (g_0^{i_2}-g_0^{j_2})\delta^{i_2i_1}\delta^{j_2j_1} +
  (t_{j_1j_2}\theta^{j_2j_1} + t^+_{j_2j_1}\theta^{j_1j_2})\delta^{i_2i_1}
  -(t^+_{i_1i_2}\theta^{i_2i_1} +
  t_{i_2i_1}\theta^{i_1i_2})\delta^{j_2j_1}.
\end{equation}

We call the algebra of operators $o_I$ (\ref{inconstraints}) a
\emph{half-integer higher-spin symmetry algebra in Minkowski space
with a Young tableaux having $k$ rows}\footnote{As in the case of
bosonic fields \cite{BRmixbos}, one should not identify the term
``\emph{higher-spin symmetry superalgebra}'' used here for a free
HS formulation starting from the paper \cite{symferm-flat} with
the algebraic structure known as ``\emph{higher-spin
superalgebra}'' (see, for instance, Ref.\cite{Vasiliev_inter})
used to describe HS interactions.} (or, simply, \emph{half-integer
HS symmetry superalgebra in Minkowski space}) and denote it as
$\mathcal{A}^f(Y(k), \mathbb{R}^{1,d-1})$.

From Table~\ref{table in}, it is obvious that the D'Alembertian
$l_0$, being a Casimir element of the Poincare algebra
$iso(1,d-1)$, belongs to the center of the superalgebra
$\mathcal{A}^f(Y(k), \mathcal{R}^{1,d-1})$ as well.

Now, we are able to describe shortly the structure of the
Lorentz-module $P^d_k(a^+)$ of all finite string-like Dirac
vectors of the form given by the Eq. (\ref{PhysState})  (see
footnote~5) on a base of generalization of Howe duality
\cite{Howe1} on a case of half-integer spin representations of
Lorentz group $SO(1,d-1)$. The Howe dual superalgebra to
$so(1,d-1)$ is $osp(1|2k)$ if $k=\left[\frac{d-1}{2}\right]$
with the following basis elements \cite{Howe1} for arbitrary $i,j
= 1,...,k$,
\begin{equation}\label{basissp2n}
   \hat{t}_{i} = \tilde{\gamma}^{\mu} a_{i{}\mu}^+,\qquad \hat{t}^{i} = \tilde{\gamma}^{\mu} a^i_{\mu},\qquad  \hat{l}_{ij} = a_{i{}\mu}^+a^{\mu+}_j,\qquad  \hat{t}_{i}{}^j = \frac{1}{2}\{a_{i{}\mu}^+,\;a^{j{}\mu }\},\qquad \hat{l}^{ij} =
    a_{{}\mu}^ia^{j\mu},
\end{equation}
which is distinguished from the elements of $\mathcal{A}^f(Y(k),
\mathcal{R}^{1,d-1})$ by the sign "hat". Their non-vanishing
supercommutator's relations have the form
\begin{align}
 & \{\hat{t}_{i},\; \hat{t}_{j}\} \ =\  2\hat{l}_{ij} ,&&
 \{\hat{t}^{i},\; \hat{t}^{j}\} \ =\  2\hat{l}^{ij},  \nonumber \\
& [\hat{t}_{i},\; \hat{t}^{j}\} \ =\  2\hat{t}_{i}{}^j ,&&
 [\hat{t}^{i},\; \hat{t}_{i_1}{}^{j_1}] \ =\ -
 \hat{t}^{j_1}\delta^i_{i_1},  \nonumber \\
& [\hat{t}_{i},\;\hat{t}_{i_1}{}^{j_1}] \ =\
 \hat{t}_{i_1}\delta_i^{j_1} ,&&
 [\hat{t}^{i},\; \hat{l}_{i_1{}j_1}] \ =\  -\hat{t}_{\{i_1}\delta^i_{j_1\}},  \nonumber \\
& [\hat{t}_{i},\; \hat{l}^{i_1{}j_1}] \ =\
\hat{t}^{\{i_1}\delta_i^{j_1\}} ,&&
   \label{oddsp} \\
 & [\hat{t}_{i_1}{}^{j_1},\; \hat{t}_{i_2}{}^{j_2}] \ =\  \hat{t}_{i_1}{}^{j_2}
  \delta_{i_2}^{j_1} - \hat{t}_{i_2}{}^{j_1}
  \delta_{i_1}^{j_2},&&  [\hat{l}^{i_2{}j_2},\; \hat{l}_{i_1{}j_1}]\ =\ \delta^{\{i_2}_{\{i_1}\hat{t}_{j_1\}}{}^{j_2\}},  \nonumber \\
 & [\hat{t}_{i_1}{}^{j_1},\; \hat{l}_{i_2j_2}]  =  \hat{l}_{i_1\{j_2}\delta_{i_2\}}^{j_1} ,
 &&  [\hat{t}_{i_1}{}^{j_1},\; \hat{l}^{i_2j_2}]\ =\
 -\hat{l}^{j_1\{j_2}\delta^{i_2\}}_{i_1} .
  \label{comrelsp}
\end{align}
The elements $t^{i}, t^{i+}$, $l^{ij}, l^{ij+}, t^{i_1j_1},
t^+_{i_1j_1}, g_0^i$  from  HS symmetry superalgebra
$\mathcal{A}^f(Y(k), \mathbb{R}^{1,d-1})$ are derived from the
basis elements of $osp(1|2k)$ by the rules (for $sp(2k)$ case see
Ref. \cite{BRmixbos}),
\begin{equation}\label{osp2nhssa}
    t^{i} = \hat{t}^{i},\quad t^{i+}= \hat{t}_{i};\quad l_{ij}^+ = \frac{1}{2}\hat{l}_{ij}, \quad {l}^{ij} =
    \frac{1}{2}\hat{l}^{ij},\quad
     {t}_{i}{}^j = \hat{t}_{i}{}^j\theta^{ji},\quad  {{t}^{j}{}_{i}}^+{} = \hat{t}_{i}{}^j\theta^{ij},\quad
     g_0^i=-
     \hat{t}_{i}{}^i.
\end{equation}
The rest elements $\{l^i, l^{i+}, t_0, l_0\}$ of the superalgebra
$\mathcal{A}^f(Y(k), \mathbb{R}^{1,d-1})$  forms the
subsuperalgebra which describes the isometries of Minkowski space
$\mathbb{R}^{1,d-1}$. It may be realized as direct sum of
$k$-dimensional commutative algebra $T^k = \{l_i\}$ and its dual
$T^{k*}=\{l^{i+}\}$,
\begin{equation}\label{TkTk}
    \{l^i, l^{i+}, t_0,  l_0\} = (T^k \oplus T^{k*}\oplus [T^k,
    T^{k*}]),\quad [T^k,
    T^{k*}] \sim l_0 = - t_0^2,
\end{equation}
so that half-integer HS symmetry  algebra $\mathcal{A}^f(Y(k),
\mathbb{R}^{1,d-1})$ represents the semidirect sum of the
orthosymplectic superalgebra $osp(1|2k)$ [as an algebra of
internal derivations of $(T^k \oplus T^{k*})]$ with $(T^k \oplus
T^{k*}\oplus [T^k,
    T^{k*}])$\footnote{The construction of algebra  $\mathcal{A}^f(Y(k), \mathbb{R}^{1,d-1})$ in the
    Eq. (\ref{identalg}) is similar to the realization of
    the Poincare algebra
    $iso(1,d-1)$ by means of Lorentz algebra and Abelian subalgebra $T(1,d-1)$ of
    space-time translations which looks as follows, $iso(1,d-1) = T(1,d-1)+ \hspace{-1em} \supset  so(1,d-1)$.},
\begin{equation}\label{identalg}
    \mathcal{A}^f(Y(k), \mathbb{R}^{1,d-1}) = \left(T^k \oplus T^{k*}\oplus [T^k,
    T^{k*}]\right) + \hspace{-1em} \supset  osp(1|2k).
\end{equation}
Note, the elements $g_0^i$, form a basis in the Cartan subalgebra
whereas $t^i, l^{ij}, {t}_{i}{}^j$ are the basis of low-triangular
subsuperalgebra in $osp(1|2k)$.

Having the identification (\ref{identalg}), we are able to
conclude that half-integer-spin finite-dimen\-sio\-nal irreducible
representations of the Lorentz algebra $so(1,d-1)$, subject to the
Young tableaux $YT(k)$, and realized on spin-tensor fields
(\ref{Young k}), are equivalently extracted by the annihilation of
all elements from the $so(1,d-1)$-module $P^d_k(a^+)$ by the
low-triangular subalgebra of $osp(1|2k)$, along with the weight
conditions, given by Eqs. (\ref{gocond}), with respect to its
Cartan subalgebra, which has the form, in terms of independent
relations,
\begin{equation}\label{Cartcond}
   t^{i}|\Psi\rangle =0, \qquad
   {t}_{i}{}^j|\Psi\rangle=0,\qquad \hat{t}_i{}^i |\Psi\rangle \equiv - g_0^i|\Psi\rangle = -
    \left(s_i+\textstyle\frac{d}{2}\right)|\Psi\rangle.
\end{equation}
Half-integer spin finite-dimensional irreducible representations
of the Poincare algebra $iso(1$, $d-1)$ can be easily obtained
from those for the Lorentz algebra, by adding the only independent
condition, given by the Dirac operator:
\begin{equation}\label{addcond}
      t_0|\Psi\rangle = 0,
\end{equation}
which lifts the set $P^d_k(a^+)$ to the Poin\-care module  (for
another realization of the bosonic Poincare module from the
Lorentz module, see Ref.\cite{mg}).

The derivation of the HS symmetry algebra has not yet provided a
construction of the BRST operator $Q$ with respect to the elements
$o_I$ from  $\mathcal{A}^f(Y(k), \mathcal{R}^{1,d-1})$, due to the
presence of non-degenerate (in the Fock space $\mathcal{H}$)
operators $g_0^i$, determining, according to Eqs.
(\ref{inconstraints}), the system of $o_I$ as that with a
second-class constraint system. Because of the general property
\cite{BFV,Henneaux} of the BFV-method, such a BRST operator $Q$
would not reproduce a correct set of initial constraints
(\ref{t0t1t}), (\ref{lilijpr}) in the zero ghost $Q$-cohomology
subspace of the total Hilbert space, $\mathcal{H}_{tot}$
($\mathcal{H} \subset \mathcal{H}_{tot}$). To solve this problem,
we consider a procedure of converting the set of $o_I$ into that
of $O_I$, which would be of first-class constraints only, in
subspaces controlled by extended particle-number operators
$G_0^I$.

%%%%%%%%%%%%%%%%%%%%%%%%%%%%%%%%%%%%%%%%%%%%%%%%
\section{Converted HS Symmetry Superalgebras for YT with $k$ Rows}\label{Vermamodule}
\setcounter{equation}{0}

\setcounter{equation}{0}

To convert the set of $o_I$ operators, we shall present the
construction of an auxiliary representation for the
orthosymplectic superalgebra $osp(1|2k)$ with second-class
constraints only, in terms of oscillator operators in an auxiliary
Fock space over an appropriate Heisenberg--Weyl superalgebra, and
then extend the latter to the case of massive half-integer HS
fields subject to the same Young tableaux $Y(s_1,...,s_k)$.

\subsection{Auxiliary Representation for $osp(1|2k)$
Superalgebra}\label{oscrealsp2k}

Since the $osp(1|2k)$ generators alone are the second-class
constraints in $\mathcal{A}^f(Y(k),\mathbb{R}^{1,d-1})$, which are
to be converted, then, instead of the additional parts $o'_I$ in
the representation of converted constraints $O_I = o_I + o'_I$
within an additive conversion procedure of the BRST approach (see,
for details, e.g., \cite{BurdikPashnev},  \cite{mixfermiflat},
\cite{0001195}, and for the general concept, \cite{conversion}),
it is sufficient to use only some of them, namely $\{o'_a,
{o'}^+_a\}$. These additional parts $o'_I$ act in a new Fock space
$\mathcal{H}'$, subject to the standard relation
$\mathcal{H}'\bigcap \mathcal{H} = \emptyset$. Algebraic
structures of the sets $o'_I$ and $O_I$ are determined by the
requirement of supercommutativity, $[\ o_I, \ o'_J\} = 0$. This
provides the fact that these sets have the same multiplication
laws as those for the $osp(1|2k)$ superalgebra and for the
superalgebra $\mathcal{A}^f_c(Y(k),\mathbb{R}^{1,d-1})$,
respectively, for $o'_I$ and $O_I$.

Therefore, one has to obtain a new operator realization for the
$osp(1|2k)$ algebra $o'_I$. An effective solution of this problem
can be provided by a special procedure, known in the mathematical
literature as the generalized Verma module construction
\cite{Dixmier, genVM}, applied to the latter superalgebra. This
procedure is presented explicitly in Appendix~\ref{addalgebra}.

\subsection{Scalar Oscillator Realization of
Additional Parts to Constraints}\label{oscrealosp2k}

Prior to an explicit oscillator realization for the additional
parts $o'_I$, it should be noted that in the case of the
superalgebra of half-integer HS mixed-symmetric fields
$\mathcal{A}^f(Y(2),\mathbb{R}^{1,d-1})$ an auxiliary
representation of its converted subsuperalgebra $osp(1|4)$ of
second-class constraints was constructed in \cite{mixfermiflat}.
For a more general case of fermionic HS fields characterized by a
Young tableaux with $k\geq 2$ rows in a symmetric basis, we
enlarge the results of \cite{Burdik} from the case of a Lie
algebra to case of the orthosymplectic superalgebra $osp(1|2k)$,
so as to transform the generalized Verma module (a special
representation, whose detailed construction for $osp(1|2k)$ is
described in Appendix~\ref{addalgebra}) to an oscillator form,
being suitable to obtain the BRST operator. As a result, we
present here the oscillator representation (found on a basis of
calculations made in Appendix~\ref{oscrealsp2k}) for the operators
$o'_I$, first, for the  operators with the Hermitian conjugation
sign, "+", $t^{\prime+}_i, l^{\prime+}_{ij}, t^{\prime+}_{rs}$,
\begin{eqnarray}
\label{t'+iFf}
 t^{\prime  +}_i & = & f^+_i + 2b_{ii}^+f_i
 +4\sum_{l=1}^{i-1}b_{li}^+f_l
  \,,
 \qquad\qquad
 l^{\prime+}_{ij} \ = \ b_{ij}^+\,,
\\
 t^{\prime+}_{rs}   & = & d^+_{rs} - \sum_{n=1}^{r-1}d_{nr}d^+_{ns}
  - \sum_{n=1}^{k}(1+\delta_{nr})b^+_{ns}b_{rn} -\bigl[4\sum_{n=r+1}^{s-1}b^+_{ns}f_n +({f}^+_{s}+2b^+_{ss}f_s)\bigr]f_{r}
 \,,
 \label{t'+lmf}
 \end{eqnarray}
second, for the particle number operators, $g_0^{\prime i}$,
\begin{eqnarray}
   g_0^{\prime i}& = & f_i^+f_i + \sum_{l\leq m}
 b_{lm}^+b_{lm}(\delta^{il}+\delta^{im}) + \sum_{r< s}d^+_{rs}d_{rs}(\delta^{is}-
 \delta^{ir}) +h^i
 \,.\label{g'0iFf}
  \end{eqnarray}
The quantities $h^i$, $i=1,\ldots,k$ in (\ref{g'0iFf}) and below
are arbitrary dimensionless constants, introduced in
Appendix~\ref{addalgebra}, whose specific values
are determined in Section~\ref{LagrFormulation}
and based on a solution of a special spectral problem.

Third, for the Grassmann-odd ``gamma-traceless'' elements  $t^{\prime}_i$
we have (for $k_{-1}\equiv 1$),
\begin{eqnarray}
\label{t'iFf}% \nonumber to remove numbering (before each equation)
  t^{\prime }_i &=&
2\sum_{n=1}^{i-1}\Bigl\{
\sum_{p=0}^{i-n-1} \bigg[\sum_{k_1=n+1}^{i-1}\ldots \sum_{k_p=n+p}^{i-1}\Big\{
 C^{k_{p}i}(d^+,d)- \sum_{n'=k_{p-1}}^{k_p-1}d^+_{n'k_p}d_{n'i} \Big\} \prod_{j=1}^pd_{k_{j-1}k_{j}}\bigg]\\
 &&  -\sum_{m=1}^{k}(1+\delta_{mi})
b^+_{mn}b_{mi}  + \bigl[4\sum_{m=n+1}^{i-1} b^+_{nm} f_m -f^+_n
\bigr]f_i
  \Bigr\}f_n\nonumber\\
  &&
  +2\sum_{n=i+1}^{k}\Bigl\{
d^+_{in} - \sum_{m=1}^{i-1}d^+_{mn} d_{mi}  - \sum_{m
=1}^{k}(1+\delta_{mi})b^+_{nm}b_{im}
  \Bigr\}f_n \nonumber\\
&& -2\left(\sum_{l=1}^k(1+\delta_{il}) b^+_{il}b_{il}  -
\sum_{s>i}d^+_{is}d_{is}+\sum_{r<i}d^+_{ri}d_{ri} + h^{i}\right)f_i \nonumber \\
 && + \textstyle \sum\limits_{n=1}^k
(1+\delta_{ni}) \Bigl\{2\sum\limits_{m=n+1}^{k} b^+_{nm}f_m
-\frac{1}{2}\bigl(f_n^+ - 2b_{nn}^+f_n\bigr)
  \Bigr\}b_{ni}.  \nonumber
 \end{eqnarray}
Next, for the ``traceless''  elements $l^{\prime}_{lm}$,
separately for $l=m$ and for $l<m$, corresponding
to the secondary constraints, we obtain
\begin{eqnarray}
\label{l'llf}
l^{\prime}_{ll} &=&
 \hspace{-0.25em}
  -\Biggl[2\sum_{n=l+1}^{k}\Bigl\{
d^+_{ln} -
\sum_{n'=1}^{l-1}d^+_{n'n}d_{n'l}-\sum_{n'=1}^k(1+\delta_{n'l})b^+_{n'n}b_{n'l}\Bigr\}f_n
\\
&&-  \sum\limits_{n=1}^k(1+\delta_{nl})
 \Bigl\{-2\sum\limits_{m=n+1}^{k}b^+_{nm}f_{m}
 +\frac{1}{2}\bigl[f^+_n -(1-\delta_{nl})2b^+_{nn}f_n\bigr]  \Bigr\}b_{ln}\Biggr]f_{l}+ l^{\prime b}_{ll},
 \nonumber
 \end{eqnarray}
\vspace{-3ex}
\begin{eqnarray}\label{l'lmf}
  l^{\prime }_{lm} &=&
\hspace{-0.3em}-\Biggl[ \hspace{-0.1em}\sum_{n=l+1}^{m-1}\Bigl\{
     \sum_{p=0}^{m-n-1}\bigg[\sum_{k_1=n+1}^{m-1}\ldots \sum_{k_p=n+p}^{m-1}
\Big\{C^{k_{p}m}(d^+,d)- \sum_{n'=k_{p-1}}^{k_p-1}d^+_{n'k_p}d_{n'm} \Big\} \prod_{j=1}^pd_{k_{j-1}k_{j}}\bigg]\\
 && -\sum_{n'=1}^{k}(1+\delta_{n'm})b^+_{n'n}b_{n'm}
 + \Bigl[4\sum_{n'=n+1}^{m-1}b^+_{n'n}f_{n'}
  - f^+_{n}\Bigr]f_{m} \Bigr\}f_{n}\nonumber\\
 &&
  +\sum_{n=m+1}^{k}\Bigl\{d^+_{mn}  -
\sum_{n'=1}^{m-1}d^+_{n'n}d_{n'm} -
\sum_{n'=1}^{k}(1+\delta_{n'm})b^+_{n'n}b_{mn'} \Bigr\}f_n \nonumber \\
&& -\Bigl(\sum_{n=1}^k (1+\delta_{nm})b^+_{mn}b_{mn}  -
\sum_{s>m}d^+_{ms}d_{ms}+\sum_{r<m}d^+_{rm}d_{rm} + h^{m}\Bigr)f_m
 \nonumber \\
 &&
+ \frac{1}{2}\textstyle \sum\limits_{n=1}^k(1+\delta_{nm})
 \Bigl\{2\sum\limits_{n'=n+1}^{k}b^+_{nn'}f_{n'}
 -\frac{1}{2}\bigl[f^+_{n} -(1-\delta_{nl})2b^+_{nn}f_{n}\bigr]  \Bigr\}b_{nm}\Biggr]f_{l}
 \nonumber\\
&& -\Biggl[ \sum_{n=m+1}^{k}\Bigl\{d^+_{ln} -
\sum_{n'=1}^{l-1}d^+_{n'n}d_{n'l}  -
\sum_{n'=1}^{k}(1+\delta_{n'l})b^+_{nn'}b_{ln'}
  \Bigr\}f_{n} \nonumber \\
&& + \frac{1}{2}\textstyle \sum\limits_{n=1}^k(1+\delta_{nl})
 \Bigl\{2\sum\limits_{n'=n+1}^{k}b^+_{nn'}f_{n'}
 - \frac{1}{2}\bigl[f^+_{n} -(1-\delta_{nm})2b^+_{nn}f_{n}\bigr]  \Bigr\}b_{nl}
 \Biggr]f_{m} + l^{\prime b}_{lm},  \nonumber
 \end{eqnarray}
 with the use of the $(f_l,f^+_l)$-independent bosonic operators $l^{\prime b}_{lm}$,
 first obtained for the symplectic
 $sp(2k)$ algebra in \cite{BRmixbos}, for $l=m$, and, for $l>m$,
 respectively,
 \begin{eqnarray}
 l^{\prime b}_{ll}
  &=& \frac{1}{4}\sum_{n=1,n\neq l}^{k}b^+_{nn}{b}^2_{ln}
 + \frac{1}{2}\sum_{n=1}^{l-1}\bigg(
\sum_{n'=n+1}^{k}(1+\delta_{n'l})b^+_{nn'}b_{n'l}
\label{l'llbosef}\\
  && -
\sum_{p=0}^{l-n-1}\bigg[\sum_{k_1=n+1}^{l-1}\ldots \sum_{k_p=n+p}^{l-1}
\Big\{ C^{k_{p}l}(d^+,d)- \sum_{n'=k_{p-1}}^{k_p-1}d^+_{n'k_p}d_{n'l} \Big\}\prod_{j=1}^pd_{k_{j-1}k_{j}} \bigg]\bigg){b}_{nl}
 \nonumber\\
 && + \left(\sum_{n= l}^k b_{nl}  - \sum_{s>l}d^+_{ls}d_{ls}+\sum_{r<l}d^+_{rl}
 d_{rl} + h^{l}\right)b_{ll}\nonumber\\
 && - \frac{1}{2}\sum_{n=l+1}^{k}\Bigl[d^+_{ln} -
\sum\limits_{n'=1}^{l-1}d^+_{n'n}d_{n'l} -
\sum_{n'=n+1}^{k}(1+\delta_{n'l})b^+_{n'n}b_{n'l}
 \Bigr]{b}_{ln}\,, \nonumber\\
   l^{\prime b}_{lm}&=&
 -
\frac{1}{4}\sum\limits_{n=1}^{m-1}(1+\delta_{nl}) \bigg(
- \sum\limits_{n'=1}^{n-1}d^+_{n'n}d_{n'm}
-\sum_{n'=n}^{k}(1+\delta_{n'm})
 b^+_{n'n}b_{n'm}\label{l'lmbosef}\\
 && +
\sum_{p=0}^{m-n-1}\bigg[\sum_{k_1=n+1}^{m-1}\ldots \sum_{k_p=n+p}^{m-1}
 \Big\{C^{k_{p}n}(d^+,d)- \sum_{n'=k_{p-1}}^{k_p-1}d^+_{n'k_p}d_{n'n}\Big\}\prod_{j=1}^pd_{k_{j-1}k_{j}}\bigg]\bigg)b_{nl}
 \nonumber\\
&& - \frac{1}{4}\sum\limits_{n=m+1}^{k} \Bigl[ d^+_{mn}-
\sum\limits_{n'=1}^{m-1} d^+_{n'n}d_{n'm} -
\sum_{n'=l+1}^{k}(1+\delta_{n'm})b^+_{n'n}b_{mn'}  \Bigr]{b}_{ln}\nonumber\\
&& + \frac{1}{4}\Bigl(\sum_{n=m}^kb^+_{ln}b_{ln} +  \sum_{n=
l+1}^k(1+\delta_{nm})b^+_{nm} b_{nm}  - \sum_{s>l}d_{ls}d_{ls} -
\sum_{s>m}d^+_{ms}d_{ms}\nonumber\\
&& +\sum_{r<l}d^+_{rl}d_{rl}
+\sum_{r<m}d^+_{rm}d_{rm} + h^{l'}+ h^{m'}\Bigr)b_{lm} \nonumber
\end{eqnarray}
 \begin{eqnarray}
&& -\frac{1}{4}  \sum\limits_{n=1}^{l-1} \Bigl[
\sum_{p=0}^{l-n-1}\bigg[\sum_{k_1=n+1}^{l-1}\ldots \sum_{k_p=n+p}^{l-1}
 \Big\{ C^{k_{p}n}(d^+,d)- \sum_{n'=k_{p-1}}^{k_p-1}d^+_{n'k_p}d_{n'n} \Big\}\prod_{j=1}^pd_{k_{j-1}k_{j}}\bigg]\nonumber\\
 && -\sum_{n'=n+1}^{k}(1+\delta_{n'l})b^+_{n'n}b_{n'l}
 \Bigr]{b}_{nm} - \frac{1}{4}\sum\limits_{n=l+1}^{k}(1+\delta_{nm})
\Bigl[ d^+_{ln} - \sum\limits_{n'=1}^{l-1}d^+_{n'n}d_{n'l}
\Bigr]{b}_{mn}\,.\nonumber
 \end{eqnarray}
In turn, for the ``mixed symmetry'' elements $t'_{rs}$, we have the representation
 \begin{eqnarray}
\label{t'lmFf} t^{\prime }_{rs} &=&
\sum_{p=0}^{s-r-1}\bigg[\sum_{k_1=r+1}^{s-1}\ldots \sum_{k_p=r+p}^{s-1}
 \Big\{C^{k_{p}s}(d^+,d)- \sum_{n'=k_{p-1}}^{k_p-1}d^+_{n'k_p}d_{n's}\Big\}\prod_{j=1}^pd_{k_{j-1}k_{j}}\bigg]
 \\
  && -\sum_{n=1}^{k}(1+\delta_{ns})b^+_{nr}
b_{ns}
  + \bigl[4\sum_{n=r+1}^{s-1}b^+_{rn}f_n +(2b^+_{rr}f_r-{f}^+_{r})\bigr]f_{s}
 \,, \quad k_0\equiv r, \nonumber
\end{eqnarray}
where the operators $C^{rs}(d,d^+)$ in Eqs. (\ref{t'iFf}), (\ref{l'lmf})--(\ref{t'lmFf})
were first derived in \cite{BRmixbos} for the symplectic $sp(2k)$ algebra, and
are determined, for $r<m$, as follows:
\begin{eqnarray}
 \label{Crsf}
C^{rs}(d^+,d)&\equiv &
\Bigl(h^{r}-h^{s}-\sum_{n=s+1}^{k}\bigl(d^+_{rn}d_{rn}+d^+_{sn}d_{sn}\bigl)+
\sum_{n=r+1}^{s-1}d^+_{ns}d_{ns}-d^+_{rs}d_{rs}\Bigr)d_{rs} \\
 &&   + \sum_{n=s+1}^{k}\Bigl\{d^+_{sn}  - \sum_{n'=r+1}^{s-1} d^+_{n'n} d_{n's}\Bigr\}d_{rn}.\nonumber.
  \end{eqnarray}
To construct the additional parts $o'_I$ in (\ref{t'+iFf})--(\ref{t'lmFf}),
we have introduced a new Fock superspace
$\mathcal{H}'$, generated by $2k$ fermionic, $f^+_i, f_i$, and $2k^2$ bosonic,
$b^{+}_{ij}, d^+_{rs}, b_{ij},
d_{rs}$, $i,j,r,s =1,\ldots, k; i\leq j; r<s$, creation and
annihilation operators, whose numbers are equal to those of the
second-class constraints $o'_a, o^{\prime +}_a$, with the standard
(only nonvanishing) commutation relations
\begin{equation}\label{commrelationsf}
\{f_i, f^+_j\}=\delta_{ij},,\ \qquad  [b_{i_1j_1}, b^+_{i_2j_2}] =
 \delta_{i_1i_2}\delta_{j_1j_2}\,, \   \qquad [d_{r_1s_1}\,,d^+_{r_2s_2}]
 =\delta_{r_1r_2}\delta_{s_1s_2}\,.
\end{equation}
 As usual, the additional parts $o^{\prime}_a(B,B^+), o^{\prime+}_a(B,B^+)$,
 regarded as polynomials in the oscillator variables $(B,B^+)\equiv ( f_i, b_{ij},
d_{rs}; f^+_i, b^{+}_{ij}, d^+_{rs})$, do not obey the standard properties
\begin{equation}
 \left(t^{\prime }_{i}\right)^+\neq t^{\prime +}_{i},\qquad \left(l^{\prime }_{ij}\right)^+\neq l^{\prime +}_{ij},\ i\leq j,  \qquad
\left(t^{\prime }_{rs}\right)^+\neq t^{\prime +}_{rs},\ r< s
\label{hermcong}
\end{equation}
if one should use the usual rules of Hermitian conjugation for
the new creation and annihilation operators,
\begin{equation}
(f_i)^+ = f^+_i,\qquad (b_{ij})^+=b_{ij}^+, \qquad (d_{rs})^+=d^+_{rs},
\end{equation}
with respect to the same definition of the odd scalar product (\ref{sproduct}),
however, residing in $\mathcal{H}'$.
To restore the proper Hermitian conjugation properties for the
additional parts, we define another odd scalar product in the Fock space
$\mathcal{H}'$, using the following relations:
\begin{eqnarray}
\langle\tilde{\Psi}_1|\Psi_2\rangle_{\mathrm{new}} =
\langle\tilde{\Psi}_1|K'|\Psi_2\rangle\,, \label{newsprod}
\end{eqnarray}
for any vectors $|\Psi_n\rangle$ (Dirac spinors), $n=1,2$, with a bosonic
operator $K'$, being nondegenerate in $\mathcal{H}'$ but yet unknown
otherwise. The operator is to be determined as a solution of the equations
\begin{align}\label{Kequation}
& \langle\tilde{\Psi}_1|K'E^{- \prime\alpha}|\Psi_2\rangle =
\langle\tilde{\Psi}_2|K'E^{\prime\alpha}|\Psi_1\rangle^* ,  &&
\langle\tilde{\Psi}_1|K'g_0^{\prime i}|\Psi_2\rangle =
\langle\tilde{\Psi}_2|K'g_0^{\prime i}|\Psi_1\rangle^*.
\end{align}
for all $(E^{- \prime\alpha}, E^{ \prime\alpha}) = (t^{\prime +}_{i}, l^{\prime +}_{ij}, t^{\prime +}_{rs}; t^{\prime }_{i}, l^{\prime }_{ij}, t^{\prime}_{rs})$.
A solution of Eqs.(\ref{Kequation}) exists in a form being Hermitian
with respect to the standard odd scalar product
in $\mathcal{H}'$, $\langle \ |\ \rangle$, similar to (\ref{sproduct}) in $\mathcal{H}$,
\begin{eqnarray}
\label{explicit K} && K'=Z^+Z, \quad
Z=\sum_{\vec{n}^0_{l}=\vec{0}^0_{l}}^{\vec{1}^0_{l}}\sum_{\vec{n}_{ij},\vec{p}_{rs})=(\vec{0},\vec{0})}^{\infty}
\left|\vec{N}^f\rangle_V\right.\frac{1}{(\vec{n}_{ij})!(\vec{p}_{rs})!}\langle
0|\prod_{r,s>r}^k{d}_{rs}^{p_{rs}}\prod_{i,j \geq
i}^k{b}_{ij}^{n_{ij}}\prod_{l=1}^kf_{k-l+1}^{n_{k-l+1}^0},\\
&& \phantom{K'=Z^+Z,}
Z^+= \sum_{\vec{n}^{\prime0}_{l}=\vec{0}^0_{l}}^{\vec{1}^0_{l}}\sum_{\vec{n}'_{ij},\vec{p}'_{rs})=(\vec{0},\vec{0})}^{\infty}
\frac{1}{(\vec{n}_{ij})!(\vec{p}_{rs})!}\prod_{l=1}^k(f^+_{l})^{n_{l}^{\prime 0}}\prod_{i,j \geq
i}^k({b}^+_{ij})^{n'_{ij}}\prod_{r,s>r}^k({d}^+_{rs})^{p'_{rs}}|0\rangle{}
{}_V\langle\left.\vec{N}^{\prime f}\right|,\nonumber
\end{eqnarray}
where the symbols $(\vec{n}_{ij})!$, $(\vec{p}_{rs})!$ imply the
products of factorials, $(\vec{n}_{ij})! =\prod^k_{i,j\geq
i}{n}_{ij}!$, $(\vec{p}_{rs})! = \prod^k_{r,s> r}{p}_{rs}!$; the
vector $\left|\vec{N}^f\rangle_V\right.$ is determined in
Appendix~\ref{addalgebra}, and
${}_V\langle\left.\vec{N}^{f}\right|$ is a dual vector. A detailed
calculation of the operator $K'$ is described in
Appendix~\ref{oscrealsp2kdet}.

Let now turn  to the case of massive fermionic HS fields whose
system of second-class constraints contains, in addition to the
elements of the $osp(1|2k)$ superalgebra, the constraints of the
isometry subalgebra of the  Minkowski space, $t_0,  l_0, l^i,
l^+_i$.

\subsection{On Auxiliary Representations of Superalgebra
$\mathcal{A}^f(Y(k),\mathbb{R}^{1,d-1})$ for Massive HS
Fields}\label{auxtheorem}

Analogous oscillator representations for the HS symmetry
superalgebra of massive fermionic HS fields with mass $m$, where
the massless Dirac equation, given by (\ref{Eq-0}), is to be
replaces by a massive equation, corresponding to the constraint
$t_0$ ($t_0=-\imath \tilde{\gamma}^\mu\partial_\mu
+\tilde{\gamma}m$), acting on the same string-vector (Dirac
spinor) $|\Psi\rangle$ (\ref{PhysState})
\begin{eqnarray}
\label{Eq-0m} && (\imath {\gamma}^\mu\partial_\mu-
m)\Psi_{(\mu^1)_{n_1},(\mu^2)_{n_2},...,(\mu^k)_{n_k}}
 =0 \Longleftrightarrow (\imath \tilde{\gamma}^\mu\partial_\mu-\tilde{\gamma}
m)\Psi_{(\mu^1)_{n_1},(\mu^2)_{n_2},...,(\mu^k)_{n_k}}
 =0,
\end{eqnarray}
can be constructed following the procedure described in
Section~\ref{oscrealosp2k} and  realized  in
Appendices~\ref{addalgebra},~\ref{oscrealsp2kdet} for the
$osp(1|2k)$ superalgebra (see the remarks in
Appendix~\ref{addalgebram} for massive spin-tensors). In addition,
because of the algebraic relation $(t_0)^2=-l_0$ from
Table~\ref{table inodd}, we should also replace in this case the
constraint $l_0$ by $l_0 =
\partial^\mu\partial_\mu +m^2$. Instead, following, in part, the
research of the integer-spin case \cite{BRmixbos}, we have used
the procedure of dimensional reduction of the initial superalgebra
$\mathcal{A}^f(Y(k),\mathbb{R}^{1,d})$ for massless fermionic HS
fields in a $(d+1)$-dimensional flat space-time to that of
dimension $d$, $\mathbb{R}^{1,d-1}$.

To this end, we, first, write down the rules of dimensional
reduction from the flat background $\mathbb{R}^{1,d}$ to
$\mathbb{R}^{1,d-1}$,
\begin{align} \label{reduction}
   &\partial_{M} = (\partial_{\mu}, \imath m)\,, &&a^{M}_i = (a^{\mu}_i, b_i)\,, &&
   a^{M{}+}_i = (a^{\mu{}+}_i, b_i^+)\,,  \\
    & \tilde{\gamma}^M = (\tilde{\gamma}^\mu, \tilde{\gamma})\,, &&M=(\mu,d)\,, && \eta^{MN} =
   diag (1,-1,\ldots,-1,-1)\,,\label{reduction1}
\end{align}
Second, we obtain, on a basis of the rules (\ref{reduction1})
for the set of the original elements $o_I$ from the massless
HS symmetry superalgebra $\mathcal{A}^f(Y(k),\mathbb{R}^{1,d})$,
a representation for $\tilde{o}_I$ in the massive HS symmetry
superalgebra $\mathcal{A}^f(Y(k),\mathbb{R}^{1,d-1})$, as follows:
\begin{align}
&\tilde{t}_0 = -\imath \tilde{\gamma}^{M}\partial_{M}= {t}_0 + \tilde{\gamma} m
, && \tilde{t}_i = \tilde{\gamma}^{M}a_M= {t}_i -\tilde{\gamma}b_i, \label{t0tilde}\\
    &\tilde{l}_0 = \partial^{M}\partial_{M}=
l_0+ m^2, && \tilde{t}^+_i = \tilde{\gamma}^{M}a^+_M= {t}^+_i -\tilde{\gamma}b^+_i,\label{l0tilde}
\end{align}
\vspace{-3ex}
\begin{align}
\label{litilde}& \tilde{l}_i = -ia^M_i\partial_{M}= {l}_i +mb_i,
&& \tilde{l}_i^+
= -ia^{+M}_i\partial_{M}= {l}^+_i + mb^+_i,\\
& \tilde{l}_{ij} = \frac{1}{2}a^M_ia_{Mj} = {l}_{ij} -
\frac{1}{2}b_ib_{j}, && \tilde{l}^+_{ij} =
\frac{1}{2}a^{M+}_ia^+_{Mj} = {l}^+_{ij} -
\frac{1}{2}b^+_ib^+_{j}, \\
& \tilde{t}_{ij} = a^{M+}_ia_{Mj}\theta^{ji} = {t}^+_{ij} -
b^+_{i}b_j\theta^{ji} , && \tilde{t}^+_{ij} =
a^{M}_ia^+_{Mj}\theta^{ji} = {t}^+_{ij} - b_{i}b^+_j\theta^{ji},
\label{exprnew}\\
& \tilde{g}^i_{0} = - a^+_{Mi}a^{M}_i+\frac{d+1}{2} =
g_0^i + b^+_ib_{i} +\frac{1}{2}. && \label{tildeg0i}
\end{align}
The set of odd $(\tilde{t}_0, {t}^+_i, {t}_i)$ and even  $(\tilde{l}_0, {l}^+_i, {l}_i), {l}_{ij},
{l}^+_{ij}, {t}_{ij}, {t}^+_{ij}, g_0^i)$ generators of the massive HS symmetry
superalgebra $\mathcal{A}^f(Y(k),\mathbb{R}^{1,d-1})$
satisfies the same algebraic relations as those in Table~\ref{table in}
and Table~\ref{table inodd} for the massless HS symmetry superalgebra,
except for the commutators
\begin{equation}\label{ll+}
   [t_i, l^+_j] = -\delta_{ij}(\tilde{t}_0 - \tilde{\gamma}m),\qquad [t^+_i, l_j] = \delta_{ij}(\tilde{t}_0 - \tilde{\gamma}m),\qquad  [l_i, l^+_j] = \delta_{ij}(\tilde{l}_0 - m^2).
\end{equation}
Definitions  (\ref{t0tilde}), (\ref{l0tilde}) and relations (\ref{ll+}) show
the presence of $2k$ additional second-class constraints, $l_i,
l_i^+$, with the corresponding oscillator operators $b_i, b_i^+$,
$[b_i, b_j^+] = \delta_{ij}$, in comparison with the massless case.

It is interesting to observe that the elements with a tilde in
Eqs.(\ref{l0tilde})--(\ref{exprnew}) satisfy the algebraic
relations for the massless HS symmetry superalgebra
$\mathcal{A}^f(Y(k),\mathbb{R}^{1,d-1})$ now without a central charge
(i.e., the quantities $\tilde{o}_I$ contain the same
second-class constraints as ${o}_I$ in the massless case),
however, in a wider (than $\mathcal{H}$) Fock space
$\mathcal{H}\otimes \mathcal{H}(b_i,b^+_i)$,
with a tensor co-multiplier  $\mathcal{H}(b_i,b^+_i)$
generated by the ``massive'' oscillator $b_i, b_i^+$.
Therefore, the converted constraints $O_I$, $O_I = o_I + o'_I$,
in the massive case are given by the relations
\begin{equation}\label{conv}
O_I = \tilde{o}_I + o'_I, \qquad M = m+{m'}=0,
\end{equation}
where the additional parts $o'_I = o'_I(f_i, f^+_i; b_{ij},b^+_{ij}, d_{i_1j_1},
d^+_{i_1j_1})$ are determined by relations (\ref{t'+iFf})--(\ref{t'lmFf}).

Thus, the auxiliary representation (generalized Verma module) for the $osp(1|2k)$
superalgebra completely determines, with the help of dimensional reduction,
an oscillator realization for the additional parts of the massive HS
symmetry superalgebra $\mathcal{A}^{\prime f}(Y(k),\mathbb{R}^{1,d-1})$.

In the following section, we determine the superalgebra of the extended
constraints and find the BRST operator corresponding to this superalgebra.

\section{BRST--BFV Operator}\label{BRSToperator}
\setcounter{equation}{0}

Now, we are in a position to find the BRST--BFV operator for the
Lie superalgebra of converted constraints $O_I$, following our
approach. Because the algebra under consideration is a Lie
superalgebra $\mathcal{A}^f(Y(k),\mathbb{R}^{1,d-1})$, this
operator can be constructed according to the standard prescription
\cite{BFV}. To this end, we introduce the set of ghost fields $C^I
= (q_0, q_i, q_i^+ ; \eta_0, \eta^i, \eta^+_i, \eta^{ij},
\eta^+_{ij}$, $\vartheta_{rs}$, $\vartheta^+_{rs}$, $\eta^i_{G})$
with the Grassmann parity opposite to that of the elements $O_I =
(T_0, T^+_i, T_i; L_0, L^+_i$, $L_i$, $L_{ij}$, $L^+_{ij}$,
$T_{ij}$, $T^+_{ij}$, $G_0^i)$\footnote{For massless HS fields,
the elements $T_0,
 L_0, L^+_i, L_i$ coincide with $t_0,  l_0,
l^+_i, l_i$, whereas in the massive case $T_0 = \tilde{t}_0$,
$L_0=\tilde{l}_0$, $L^+_i= {l}^+_i+l^{\prime+}_i,
L_i={l}_i+l^{\prime}_i$, with allowance for Eqs. (\ref{t0tilde}), (\ref{l0tilde}),
(\ref{litilde})}, subject to the properties
\begin{equation}\label{propgho}
    (\eta^{ij}, \eta^+_{ij})\ =\ (\eta^{ji} , \eta^+_{ji}),\qquad (\vartheta_{rs}, \vartheta^+_{rs})\ = \ (\vartheta_{rs},
 \vartheta^+_{rs})\theta^{sr},
\end{equation}
and their conjugated ghost
momenta $\mathcal{P}_I$ with the same properties as those for $C^I$
in (\ref{propgho}) with the only nonvanishing commutation
relations for bosonic ghosts
\begin{align}\label{fghosts}
 & [q_i, p^{+}_j] = [p_i, q^{+}_j] =\delta_{ij}\,, &&  [q_0, p_0]=\imath ;
\end{align}
and the anticommutation ones for fermionic ghosts
\begin{align}\label{bghosts}
& \{\vartheta_{rs},\lambda^+_{tu}\}= \{{\lambda_{tu}},
\vartheta_{rs}^+\}= \delta_{rt}\delta_{su}, &&
\{\eta_i,{\cal{}P}_j^+\}= \{{\cal{}P}_j, \eta_i^+\}=\delta_{ij}\,, \nonumber \\
& \{\eta_{lm},{\cal{}P}_{ij}^+\}= \{{\cal{}P}_{ij}, \eta_{lm}^+\}
=\delta_{li}\delta_{jm}\,,  &&  \{\eta_0,{\cal{}P}_0\}= \imath,\
\{\eta^i_{\mathcal{G}}, {\cal{}P}^j_{\mathcal{G}}\}
 = \imath\delta^{ij}.;
\end{align}
The ghost coordinates and momenta also possess the standard ghost
number distribution, $gh(\mathcal{C}^I)$ = $ - gh(\mathcal{P}_I)$
= $1$, providing the property $gh({Q}')$ = $1$, and posses the
Hermitian conjugation properties of zero-mode pairs,\footnote{By
means of the redefinition $\left( p_i,  {\cal{}P}_0,
{\cal{}P}^i_{G} \right) \mapsto \imath \left( p_i, {\cal{}P}_0,
{\cal{}P}^i_{G} \right)$, the BRST operator (\ref{Q'k}) and
relations (\ref{fghosts}), (\ref{bghosts}) are written in the
notation of \cite{symferm-flat}, \cite{symferm-ads}.}
\begin{eqnarray}\label{Hermnull}
 \left( q_0, \eta_0, \eta^i_{{G}}, p_0,  {\cal{}P}_0,
{\cal{}P}^i_{G} \right)^+ & = & \left(q_0, \eta_0, \eta^i_{G},
p_0,  - {\cal{}P}_0, -{\cal{}P}^i_{G}\right).
\end{eqnarray}
The BRST operator for the algebra of $O_I$, given by
Tables~\ref{table in},~\ref{table inodd}, can be found
in an exact form, with the use of the $(\mathcal{C} \mathcal{P})$-ordering
for the operators of ghost coordinates $\mathcal{C}^I$ and momenta $\mathcal{P}_I$,
as follows:
\begin{equation}\label{generalQ'}
    Q'  = {O}_I\mathcal{C}^I + \frac{1}{2}
    \mathcal{C}^I\mathcal{C}^Jf^K_{JI}\mathcal{P}_K (-1)^{\varepsilon({O}_K) + \varepsilon({O}_I)},
\end{equation}
with the constants $f^K_{IJ}$ (\ref{geninalg}) written in a
compact $x$-local representation. According to Tables~\ref{table
in},~\ref{table inodd} $Q'$, we finally have
\begin{eqnarray}
\label{Q'k} {Q}' \hspace{-0.4em} &=&\hspace{-0.4em}
\frac{1}{2}q_0T_0+q_i^+T^i + \frac{1}{2}\eta_0L_0+\eta_i^+L^i
+\sum\limits_{l\leq m}\eta_{lm}^+L^{lm} + \sum\limits_{l<
m}\vartheta^+_{lm}T^{lm} + \frac{1}{2}\eta^i_{{G}}{G}_i
\nonumber\\
&&  + \Bigl[\frac{1}{2}\sum_{l,m}(1+\delta_{lm})\eta^{lm}q_l^+
-\sum_{l<m}
q_l\vartheta^{lm}-\sum_{m<l}q_l\vartheta^{ml+}\Bigr]p_m^+ +
\frac{1}{2}
\sum_m\eta^m_{G}(q_mp_m^++q^+_mp_m)\nonumber\\
&& +
\imath\sum_l\Bigr[\frac{1}{2}\eta_l^+\eta^l{\cal{}P}_0
+\eta_l^+q^lp_0-  q^lq_l^+ {\cal{}P}^l_{G}\Bigr]- \frac{\imath}{2}q_0^2{\cal{}P}_0 \nonumber
\\\hspace{-0.4em}&&
{}\hspace{-0.4em} -\sum\limits_{i<l<j}
\vartheta^+_{lj}\vartheta^+_{i}{}^l\lambda^{ij} +
\frac{\imath}{2}
\sum\limits_{l<m}\vartheta_{lm}^+\vartheta^{lm}({\cal P}_G^m-{\cal
P}_G^l)-
\sum\limits_{l<n<m}\vartheta_{lm}^+\vartheta^{l}{}_n\lambda^{nm}
\nonumber\\&& +
\sum\limits_{n<l<m}\vartheta_{lm}^+\vartheta_{n}{}^m\lambda^{+nl}
- \sum_{n,l<m}(1+\delta_{ln})\vartheta_{lm}^+\eta^{l+}{}_{n}
\mathcal{P}^{mn}+
\sum_{n,l<m}(1+\delta_{mn})\vartheta_{lm}^+\eta^{m}{}_{n}
\mathcal{P}^{+ln}\nonumber\\
&&  + \frac{\imath}{8}\sum_{l\leq m}(1+\delta_{lm})
\eta_{lm}^+\eta^{lm}({\cal{}P}^l_{{G}}+{\cal{}P}^m_{{G}})+\frac{1}{2}\sum_{l\leq
m}(1+\delta_{lm})\eta^l_{{G}}\bigl(
\eta_{lm}^+{\cal{}P}^{lm}-\eta_{lm}{\cal{}P}^{lm+}\bigr)
\nonumber\\
\hspace{-0.4em} && \hspace{-0.4em} +
\frac{1}{2}\sum\limits_{l<m,n\le
m}\eta^+_{nm}\eta^{n}{}_l\lambda^{lm}
 -2 \sum_{l<m}q_lq_m^+\lambda^{lm}  +
\frac{1}{2}\sum\limits_{l<m}(\eta^m_{{G}}-
\eta^l_{{G}})\bigl(\vartheta^+_{lm}\lambda^{lm}- \vartheta_{lm}\lambda^{lm+}\bigr) \nonumber\\
&& - \Bigl[\frac{1}{2}\sum\limits_{l,
m}(1+\delta_{lm})\eta^m\eta_{lm}^+ +
\sum\limits_{l<m}\vartheta_{lm} \eta^{+m}
+\sum\limits_{m<l}\vartheta^+_{ml} \eta^{+m} +2\sum_lq_0q_l^+
\Bigr]\mathcal{P}^l \nonumber\\
&& -2\sum_{l,m}
q^+_lq^+_m\mathcal{P}^{lm} +
\frac{1}{2}\sum_l\eta^l_{{G}}\bigl(\eta_l^+\mathcal{P}^l-\eta_l\mathcal{P}^{l+}\bigr)+h.c.
\end{eqnarray}
In connection with the representation of the BRST operator
(\ref{Q'k}), note, first, that a nilpotent $Q'$ has a matrix-like
$2^{\left[\frac{d}{2}\right]}\times 2^{\left[\frac{d}{2}\right]}$
structure (providing its correct action on a Dirac spinor as the
string-vector $|\Psi\rangle$, (\ref{PhysState}), however, extended
to the total Hilbert space $\mathcal{H}_{tot}=\mathcal{H}\otimes
\mathcal{H}'\otimes \mathcal{H}_{gh}$). Second, it can be
presented as the sum of the BRST operator $Q'_b$, corresponding to
the symplectic $sp(2k)$ algebra, and an additional term, $Q'_f$,
vanishing for $(q_0, q_i, q_i^+, p_0, p_i, p_i^+) =0$:
\begin{equation}\label{Qrepr}
  Q'\ = \ Q'_b + Q'_f,\  Q'_f{}\big|_{\textstyle (q_0, q_i, q_i^+, p_0, p_i, p_i^+) =0}\ =\ 0,
\end{equation}
where the specific form of $Q'_b$ easily follows from Eq. (\ref{Q'k}),
with allowance for the boundary condition (\ref{Qrepr}),
and was already presented in \cite{BRmixbos},
however without the fermionic oscillators $f_i, f^+_i$, i.e.,
for $(f_i, f^+_i) = 0$. As in the case of bosonic
HS fields \cite{BRmixbos}, the property of the BRST operator
to be Hermitian is defined by the same rule
\begin{eqnarray}\label{HermQ}
  Q^{\prime +}K\  =\ K Q'\,, \qquad   K \ =\  \hat{1} \otimes K' \otimes \hat{1}_{gh}
  \end{eqnarray}
and is calculated with respect to the odd scalar product $\langle \ |\
\rangle$ in $\mathcal{H}_{tot}$ with the measure $d^dx$, which, in
its turn, is constructed as the direct product of odd scalar
products in $\mathcal{H}$ and even ones in $\mathcal{H}'$ and $\mathcal{H}_{gh}$.
The operator $K$ in (\ref{HermQ}) is the tensor product of the
operator $K'$ in $\mathcal{H}'$ (\ref{explicit K}) and the unit operators in
$\mathcal{H}$, $\mathcal{H}_{gh}$.

We have thus constructed a Hermitian BRST
operator for the superalgebra $\mathcal{A}^f_c(Y(k),
\mathbb{R}^{1,d-1})$ of converted operators $O_I$.
In the following section, we will use this operator
to construct a Lagrangian action for fermionic HS
fields of spin $(n_1+\frac{1}{2},..., n_k+\frac{1}{2})$ in a flat space-time.

%%%%%%%%%%%%%%%%%%%%%%%%%%%%%%%%%%%%%%%%%%%%%%%%%%%%%%%
\section{Unconstrained Gauge-invariant Lagrangians}\label{LagrFormulation}
\setcounter{equation}{0}

Let us develop a Lagrangian formulation for fermionic higher-spin
fields in a $d$-dimensional Minkowski space, partially following
the algorithm of \cite{mixfermiflat}, which is a particular case
of our construction, corresponding to a Young tableaux with 2
rows. In the beginning, one should extract the dependence of the
BRST operator $Q'$ (\ref{Q'k}) on the ``particle number'' of
ghosts $\eta^i_{G}, {\cal{}P}^i_{G}$, in order to obtain the BRST
operator $Q$ only for the system of converted first-class
constraints $\{O_I\} \setminus \{\mathcal{G}^i_0\}$ (to be
nilpotent after being restricted to the corresponding Hilbert
subspaces):
\begin{eqnarray} \label{Q'decomp}
{Q}'  &=& Q +
\eta^i_{G}(\sigma^i+h^i)+\mathcal{B}^i \mathcal{P}^i_{G}\,,
\end{eqnarray}
where the generalized spin operator $\vec{\sigma} =
(\sigma^1,\sigma^2,..., \sigma^k)$, extended by the ghost
Wick-pair variables is Hermitian, ${\sigma}^{i+}K = K{\sigma}^i$, and reads
\begin{eqnarray}
\label{sigmai}
  \sigma^i &=& G_0^i - h^i   - \eta_i \mathcal{P}^+_i +
   \eta_i^+ \mathcal{P}_i + \sum_{
m}(1+\delta_{im})(
\eta_{im}^+{\cal{}P}^{im}-\eta_{im}{\cal{}P}^+_{im})\nonumber\\
   &&  + \sum_{l<i}[\vartheta^+_{li}
\lambda^{li} - \vartheta^{li}\lambda^+_{li}]-
\sum_{i<l}[\vartheta^+_{il} \lambda^{il} -
\vartheta^{il}\lambda^+_{il}] +   q_ip_i^+ + q_i^+p_i\,.
\end{eqnarray}
The operator $Q$ in Eq. (\ref{Q'decomp}) (not yet nilpotent in
$\mathcal{H}_{tot}$) corresponds to a system of converted
first-class constraints and is unambiguously determined as
\begin{eqnarray}
\label{Q} {Q} \hspace{-0.4em} &=&\hspace{-0.4em}
 \frac{1}{2}q_0T_0+q_i^+T^i+\frac{1}{2}\eta_0L_0+\eta_i^+L^i
+\sum\limits_{l\leq m}\eta_{lm}^+L^{lm} + \sum\limits_{l<
m}\vartheta^+_{lm}T^{lm}
 + \frac{\imath}{2}\bigl(\sum_l\eta_l^+\eta^l-q_0^2\bigr){\cal{}P}_0
 \nonumber
\\
\hspace{-0.4em}
&&\hspace{-0.4em}  + \Bigl[\frac{1}{2}\sum_{l,m}(1+\delta_{lm})\eta^{lm}q_l^+
-\sum_{l<m}
q_l\vartheta^{lm}-\sum_{m<l}q_l\vartheta^{ml+}\Bigr]p_m^+  +
\imath\sum_l
\eta_l^+q^lp_0  -2 \sum_{l<m}q_lq_m^+\lambda^{lm} \nonumber
\\
\hspace{-0.4em}&& {}\hspace{-0.4em}  -2\sum_{l,m}
q^+_lq^+_m\mathcal{P}^{lm} -\sum\limits_{i<l<j}
\vartheta^+_{lj}\vartheta^+_{i}{}^l \lambda^{ij}-
\sum\limits_{l<n<m}\vartheta_{lm}^+\vartheta^{l}{}_n\lambda^{nm} +
\sum\limits_{n<l<m}\vartheta_{lm}^+\vartheta_{n}{}^m\lambda^{+nl}
\nonumber\\&& -
\sum_{n,l<m}(1+\delta_{ln})\vartheta_{lm}^+\eta^{l+}{}_{n}
\mathcal{P}^{mn}+
\sum_{n,l<m}(1+\delta_{mn})\vartheta_{lm}^+\eta^{m}{}_{n}
\mathcal{P}^{+ln}+ \textstyle\frac{1}{2}\sum\limits_{l<m,n\leq
m}\eta^+_{nm}\eta^{n}{}_l\lambda^{lm}
\nonumber\\
\hspace{-0.4em} && \hspace{-0.4em}  %-2 \sum_{l<m}q_lq_m^+
 - \Bigl[\frac{1}{2}\sum\limits_{l,
m}(1+\delta_{lm})\eta^m\eta_{lm}^+ +
\sum\limits_{l<m}\vartheta_{lm} \eta^{+m}
+\sum\limits_{m<l}\vartheta^+_{ml} \eta^{+m} +2\sum_lq_0q_l^+ \Bigr]\mathcal{P}^l
+h.c.
\end{eqnarray}
The operator $Q$ (modulo its spinor nature) contains
-- for vanishing fermionic oscillators $(f_i, f^+_i)$ and bosonic
ghosts $(q_0, q_i, q_i^+; p_0, p_i^+, p_i)$, for $i=1,\ldots,k$
--  the BRST operator $Q_b$ for the converted first-class constraints
corresponding to the HS symmetry algebra $\mathcal{A}_c(Y(k),
\mathbb{R}^{1,d-1})$ for bosonic HS fields, obtained earlier
\cite{BRmixbos}, in correspondence with the representation (\ref{Qrepr}) for $Q'$.
Finally, the  quantities $\mathcal{B}^i$ in (\ref{Q'decomp})
are uniquely determined by Eq. (\ref{Q'k}) as follows:
\begin{eqnarray}\label{Bi}
\mathcal{B}^i  = -2{\imath}p\sum_l  q^lq_l^+
   - {\imath}
\sum\limits_{l<m}\vartheta_{lm}^+\vartheta^{lm}(\delta^{mi}-\delta^{li})
+ \frac{\imath}{4}\sum_{l\leq m}(1+\delta_{lm})
\eta_{lm}^+\eta^{lm}(\delta^{il}+\delta^{mi}).
\end{eqnarray}

By construction, from the nilpotency of the BRST operator $Q'$,
considered in powers of the ghosts $(\eta^i_{G}, {\cal{}P}^i_{G})$,
the set of operators $Q$, $\sigma^i$, $\mathcal{B}^i$ supercommutes
with  each other,
\begin{eqnarray}\label{supercommQsB}
 {} [Q,\sigma^i\}\ = \ 0\,,\qquad [Q,\mathcal{B}^i\}\ = \ 0\,,\qquad [\sigma^i, \mathcal{B}^j\}\ = \ 0\,,\texttt{ for }i,j=1,\ldots,k\,,
\end{eqnarray}
providing an equation for $Q^2$:
\begin{equation}\label{Q2}
  Q^{\prime 2}\ =\ 0 \Longleftrightarrow Q^2 = -\imath \sum_i \mathcal{B}^i\sigma^i \,.
\end{equation}
We then choose the standard representation for the Hilbert space
$\mathcal{H}_{tot}$:
\begin{eqnarray}
(q_i, \eta_i, \eta_{ij},  \vartheta_{rs}, p_0, p_i, \mathcal{P}_0, \mathcal{P}_i,
\mathcal{P}_{ij}, \lambda_{rs},
\mathcal{P}^{i}_G)|0\rangle=0,\qquad |0\rangle\in
\mathcal{H}_{tot},
\end{eqnarray}
under the assumption that the field vectors $|\chi \rangle$, as well as the
gauge parameters $|\Lambda \rangle$ (Dirac spinors), do not depend on the ghosts
$\eta^{i}_G$ for particle number operators $G_0^i$,
\begin{eqnarray}
|\chi \rangle &=& \sum_n \prod_{c}^k ( f_c^+ )^{n^0_{c}}\prod_{l}^k ( b_l^+ )^{n'_{l}}\prod_{i\le
j, r<s}^k( b_{ij}^+ )^{n_{ij}}( d_{rs}^+ )^{p_{rs}}q_0^{n_{b{}0}}\eta_0
^{n_{f 0}}\nonumber
\\
&&{}\times  \prod_{e,g, i, j, l\le m, n\le o}(q_e^+)^{n_{a{}e}}(p_g^+)^{n_{b{}g}}( \eta_i^+ )^{n_{f i}} (
\mathcal{P}_j^+ )^{n_{p j}} ( \eta_{lm}^+ )^{n_{f lm}} (
\mathcal{P}_{no}^+ )^{n_{pno}} \prod_{r<s, t<u}(
\vartheta_{rs}^+)^{n_{f rs}} ( \lambda_{tu}^+ )^{n_{\lambda tu}}
\nonumber
\\
&&{}\times |\Psi(a^+_i)^{n_{b{}0} n_{f 0};  (n)_{a{}e} (n)_{b{}g} (n)_{f i}(n)_{p j}(n)_{f lm}
(n)_{pno}(n)_{f rs}(n)_{\lambda
tu}}_{(n^0)_{c};(n')_{l}(n)_{ij}(p)_{rs}}\rangle \,. \label{chif}
\end{eqnarray}
The brackets $(n^0)_{c}, (n)_{f i},(n)_{p j}, (n)_{ij}$ in the
definition of (\ref{chif}) imply, for instance, for $(n^0)_{c}$
and $(n)_{ij}$, the sets of indices $(n^0_{1},...,n^0_{k})$ and
$(n_{11},...,n_{1k},..., n_{k1},..., n_{kk})$. The above sum is
taken over $n_{b{}0}$, $n_{a{}e}$, $ n_{b{}g}$,  $h_{l}$,
$n_{ij}$, $p_{rs}$, running from $0$ to infinity, and over the
remaining $n$'s from $0$ to $1$, whereas for the massless basic HS
field $\Psi_{(\mu^1)_{n_1}...(\mu^k)_{n_k}}$ there are no
operators $b^+_l$ in the decomposition (\ref{chif}), i.e.,
$(n')_{l} = (0)_{l}$. We denote by $|\chi^k\rangle$ the state
(\ref{chif}) with the ghost number $-k$, i.e.,
$gh(|\chi^k\rangle)=-k$. Thus, the physical state having the zero
ghost number is $|\chi^0\rangle$; the gauge parameters $|\Lambda
\rangle$ having the ghost number $-1$ is $|\chi^1\rangle$, and so
on. Moreover, for the vanishing of all auxiliary creation
operators $f^+, b^+, d^+$ and ghost variables $q_0, q_i^+, \eta_0,
\eta^+_i, p_i^+, \mathcal{P}^+_i,...$, the vector $|\chi^0\rangle$
should contain only the physical string-like vector $|\Psi\rangle
= |\Psi(a^+_i)^{0_{b{}0} 0_{f 0};  (0)_{a{}e} (0)_{b{}g} (0)_{ f
i}(0)_{p j}(0)_{f lm} (0)_{pno}(0)_{f rs}(0)_{\lambda
tu}}_{(0^0)_{c}; (0)_l (0)_{ij}(0)_{rs}}\rangle$, so that
\begin{eqnarray}\label{decomptot}
|\chi^0\rangle&=&|\Psi\rangle+  |\Psi_A\rangle ,\qquad |\Psi_A\rangle\Big|_{\textstyle (f^+, b^+, d^+, q_0, q_i^+, \eta_0, \eta^+_i, p_i^+, \mathcal{P}^+_i,...)=0}=0,
\end{eqnarray}
with the vector $|\Psi_A\rangle$ containing only the set
of auxiliary spin-tensors as its components.
We will show in Appendix~\ref{reductionC} that the vector
$|\Psi_A\rangle$ can be completely gauged away, by using
a partial gauge fixing and solving some of the equations of motion.

Next, we derive, from the BRST-like equation, determining the
physical vector, $Q'|\chi\rangle$ = $0$, (for $|\chi\rangle =
|\chi^0\rangle$) and from the set of reducible gauge
transformations, $\delta|\chi\rangle$ = $Q'|\Lambda\rangle$,
$\delta|\Lambda\rangle = Q'|\Lambda^{(1)}\rangle$, $\ldots$,
$\delta|\Lambda^{(s-1)}\rangle = Q'|\Lambda^{(s)}\rangle$, a
sequence of relations underlying the $\eta^i_G$-independence of
all of the above vectors:
\begin{align}
\label{Qchi} & Q|\chi\rangle=0, && (\sigma^i+h^i)|\chi\rangle=0,
&& \left(\varepsilon, {gh}\right)(|\chi\rangle)=(1,0),
\\
& \delta|\chi\rangle=Q|\Lambda\rangle, &&
(\sigma^i+h^i)|\Lambda\rangle=0, && \left(\varepsilon,
{gh}\right)(|\Lambda\rangle)=(0,-1), \label{QLambda}
\\
& \delta|\Lambda\rangle=Q|\Lambda^{(1)}\rangle, &&
(\sigma^i+h^i)|\Lambda^{(1)}\rangle=0, && \left(\varepsilon,
{gh}\right)(|\Lambda^{(1)}\rangle)=(1,-2),\\
& \delta|\Lambda^{(s-1)}\rangle=Q|\Lambda^{(s)}\rangle, &&
(\sigma^i+h^i)|\Lambda^{(s)}\rangle=0, && \left(\varepsilon,
{gh}\right)(|\Lambda^{(s)}\rangle)= (s\mod{}2,-s-1). \label{QLambdai}
\end{align}
In the above equations, $s  = \sum_{l=1}^k n_l + k(k-1)/2 - 1$ is
the stage of reducibility for both massless and massive
fermionic HS field, since the only non-vanishing vector
(independent gauge parameter $|\Lambda^{s}\rangle$) has the lowest
negative ghost number, when, for instance, all powers of
commuting ghost momenta $(p^+_i)$ and all ``mixed-symmetry''
fermionic ghost momenta $\lambda^+_{rs}$ compose
$|\Lambda^{s}\rangle$, without the ghost coordinates
$\mathcal{C}^I$ in this vector. The solution of the spectral
problem given by Eqs. (\ref{Qchi})--(\ref{QLambdai}) is
compatible due to the validity of the second group in the set
of supercommutators (\ref{supercommQsB}), and is described
by the solution of the second column there. The middle set of
equations (\ref{Qchi})--(\ref{QLambdai}), with the generalized
spin operator $\sigma^i$, determines the set of proper
eigenvectors $|\chi^0\rangle_{(m)_k}$, $|\chi^1\rangle_{(m)_k}$,
$\ldots$, $|\chi^{s}\rangle_{(m)_k}$, $m_1 \geq m_2 \geq \ldots
m_k \geq 0$, and the set of the corresponding eigenvalues
for possible values of the parameters $h^i$,
\begin{eqnarray}
\label{hi} -h^i &=& m^i+\frac{d-4i}{2} \;, \quad
i=1,..,k\,,\quad m_1,...,m_{k-1} \in \mathbb{Z}, m_k \in
\mathbb{N}_0\,,
\end{eqnarray}
for massless,
\begin{eqnarray}
\label{him} -h^i_m &=& m^i+\frac{d+1-4i}{2} \;, \quad
i=1,..,k\,,\quad m_1,...,m_{k-1} \in \mathbb{Z}, m_k \in
\mathbb{N}_0\,,
\end{eqnarray}
and for massive half-integer HS fields. The values of $m_i$ are
related to the spin components $s_i=n_i+\frac{1}{2}$ of the
initial spin-tensor (\ref{Young k}), because the proper vector
$|\chi\rangle_{(n_1,...,n_k)}$, corresponding to $(h_1,...,h_k)$
has the leading term
$$|\Psi(a^+_i)^{0_{b{}0} 0_{f 0};  (0)_{a{}e} (0)_{b{}g} (0)_{ f i}(0)_{p j}(0)_{f
lm} (0)_{pno}(0)_{f rs}(0)_{\lambda tu}}_{(0^0)_{c}; (0)_l
(0)_{ij}(0)_{rs}}\rangle,$$ depending only on the operators
$a^+_i$, which corresponds to the spin-tensor
$\Psi_{(\mu^1)_{n_1},...,(\mu^k)_{n_k}}(x)$ with the initial value
of spin $\mathbf{s} = (s_1+\frac{1}{2},...,s_k+\frac{1}{2})$ in
the decomposition (\ref{chif}) and representation
(\ref{decomptot}). Let us denote by $|\chi\rangle_{(mp)_k}$ the
eigenvectors of $\sigma_i$ corresponding to the eigenvalues
$(m^i+\frac{d -4i}{2})$. Therefore, we conclude that
\begin{eqnarray}
\sigma_i | \chi \rangle_{(m)_k} = \left( m^i+\frac{d+\Theta(m)-4i}{2}
\right) | \chi \rangle_{(m)_k} \label{state} \,,
\end{eqnarray}
jointly for massless ($m=\theta(m)=0$) and massive ($m\ne 0
\Longrightarrow \theta(m)=1$) fermionic HS fields with the help of
the Heaviside $\theta$-function. We can show that, in order to
construct a Lagrangian for the field corresponding to a definite
Young tableau (\ref{Young k}), the numbers $m_i$ must be equal to
the numbers of the cells in the $i$-th row of the corresponding
Young tableau, i.e., $m_i=n_i$. Therefore, the state
$|\chi\rangle_{(n)_k}$ contains the physical field
(\ref{PhysState}) and all of its auxiliary fields. Let us fix some
values of $m_i=n_i$. Then, one has to substitute $h_i$
corresponding to the chosen $n_i$ (\ref{hi}) or (\ref{him}) into
(\ref{Q}), (\ref{Qchi})--(\ref{QLambdai}). Thus, the equation of
motion (\ref{Qchi}) corresponding to the field with a given spin
$(n_1+\frac{1}{2},...,n_k+\frac{1}{2})$ has the form
\begin{eqnarray}
Q_{(n)_k}|\chi^0\rangle_{(n)_k}=0, \label{Q12}
\end{eqnarray}
where the ordered value of spin $n_1\geq n_2 \geq \ldots \geq n_k$ for the vector
$|\chi^l\rangle_{(n)_k}$ should be composed
from the set of integers $( n_{b{}0}, n_{f 0},  n_{a{}e}, n_{b{}g}$, $n^0_{c}, n'_l, n_{ij}, p_{rs},
 n_{f{}i}$, $n_{p{}j},n_{f{}lm}, n_{p{}no}$, $n_{f{}rs},
n_{\lambda{}tu}, p_i)$, for $e,g,c,l,i,j,r,s,l,m,n,o, t,u
=1,...,k, \ i\leq j, r<s, l\leq m, n\leq o, t<u$; in (\ref{chif})
and (\ref{PhysState}) in the decomposition (\ref{chif}) the
coefficients\footnote{We replace the indices $n_i$ in
(\ref{PhysState}) for a vector of the initial Fock space
$\mathcal{H}$ by $p_i$ due to the use of $n_i$ in the value of
generalized spin for the basic HS field
$\Psi_{(\mu^1)_{n_1},\ldots,(\mu^k)_{n_k}}$} are to be restricted
for all the vectors $| \chi^l \rangle_{(n)_k}$, $l=0,\ldots,
\sum_{o=1}^k n_o + k(k-1)/2 - 1$, in view of the solution for the
spectral problem (\ref{state}), by the formulae
\begin{eqnarray}\label{nidecomposf}
n_i &= & p_i+ \Theta(m)n'_{i} + n_{a{}i}+ n_{b{}i}+ n^0_{i}+
\sum_{j=1}(1+\delta_{ij})(n_{ij}+n_{f{}ij}+n_{p{}ij} )
+n_{f{}i}+n_{p{}i} \nonumber \\
&{}& +\sum_{r<i} (p_{ri} + n_{f{}ri}+ n_{\lambda{}ri}) -\sum_{r>i}
(p_{ir}+ n_{f{}ir}+ n_{\lambda{}ir} )\,,\  i=1,\ldots,k.
\end{eqnarray}
In addition to the restrictions (\ref{nidecomposf}), valid
for a general case of HS fields subject to $Y(s_1,..., s_k)$,
the subset of ``ghost'' numbers
$(n_{f{}0}, n_{f{}i}$, $n_{b{}0},   n_{a{}e}, n_{b{}g} $,
$n_{p{}j},n_{f{}lm}, n_{p{}no}$, $n_{f{}rs}, n_{\lambda{}tu})$ in
(\ref{chif}) and (\ref{PhysState}) for fixed values of $n_i$,
satisfies the following equations for $|\chi^l\rangle_{(n)_k}$, $l
= 0,\ldots , \sum_{o=1}^k n_o + k(k-1)/2$
(with the identification  $|\chi^l\rangle_{(n)_k} = |\Lambda^{l-1}\rangle_{(n)_k}$ for $l>0$):
\begin{eqnarray}\label{ghnumf}
  \hspace{-0.7em} |\chi^l\rangle_{(n)_k}  \hspace{-0.7em}&:&\hspace{-0.7em} n_{b{}0}+ n_{f{}0}+\hspace{-0.2em}
     \sum_{i}\hspace{-0.2em}\bigl(n_{f{}i}- n_{p{}i} + n_{a{}i}- n_{b{}i}\bigr)+ \hspace{-0.2em}
\sum_{i\leq j}\hspace{-0.1em}\bigl(n_{f{}ij}-n_{p{}ij} \bigr) + \hspace{-0.2em}\sum_{r<s}\hspace{-0.1em} \bigl(
n_{f{}rs}- n_{\lambda{}rs}\bigr) = -l ,
\end{eqnarray}
Since the BRST--BFV operator ${Q}'$ is nilpotent (\ref{Q'k}) for
any values of $h_i$, and, due to the proportionality of $Q^2$
(\ref{Q2}) to the generalized spin operator, and because of a
joint solution of the spectral problem
(\ref{Qchi})--(\ref{QLambdai}), we have a sequence of reducible
gauge transformations:
\begin{align}
\label{dx0} &\delta|\chi^0 \rangle_{(n)_k}
=Q_{(n)_k}|\Lambda\rangle_{(n)_k} \,, &&
\delta|\Lambda\rangle_{(n)_k} =Q_{(n)_k}|\Lambda^{(1)}\rangle_{(n)_k}
\,,
\\
&\ldots &&\ldots
\nonumber \\
\label{dxs} & \delta|\Lambda^{(s-1)} \rangle_{(n)_k}
=Q_{(n)_k}|\Lambda^{(s)}\rangle_{(n)_k} \,, &&
\delta|\Lambda^{(s)} \rangle_{(n)_k} =0,\ s=\sum_{o=1}^k n_o +
k(k-1)/2-1\,,
\end{align}
with a nilpotent $Q_{(n)_k}$ in its action on proper
eigenfunctions of the operator $\sigma^i$, $|\chi\rangle_{(n)_k}$:
\begin{eqnarray}
Q_{(n)_k}^2\Bigl(|\chi \rangle_{(n)_k}, |\Lambda \rangle_{(n)_k}, \ldots, |\Lambda^{(\sum_{o=1}^k n_o + k(k-1)/2-1)} \rangle_{(n)_k}\Bigr) \equiv  0.
\end{eqnarray}
Summarizing the above, we have obtained the equations of motion
(\ref{Q12}) for an arbitrary half-integer-spin gauge theory
subject to $YT(s_1,...,s_k)$ with a mixed symmetry in any
space-time dimension, as well as the tower of reducible gauge
transformations (\ref{dx0})--(\ref{dxs}). An essential point is
that these equations are more than first order in the space-time
derivatives $\partial_{\mu}$, due to the presence of $L_0 \sim
\partial^2$ in the operator $Q$\footnote{Formally, we are able to
derive the equations of motion (\ref{Q12}), as in the case of
bosonic mixed-symmetric HS fields \cite{BRmixbos}, from the
Lagrangian action $\Pi\mathcal{S}_{(n)_k}$, $
\Pi\mathcal{S}_{(n)_k} = \int d \eta_0 \; {}_{(n)_k}\langle
\tilde{\chi}{}^0 |K_{(n)_k} Q_{(n)_k}| \chi^0 \rangle_{(n)_k}$,
with the operator $K$ (\ref{HermQ}), having in mind the
peculiarity found in the research on superfield Lagrangian BRST
quantization (e.g., \cite{AKSZ}, \cite{BV-BFV}, \cite{GMR}), that
the action $\Pi\mathcal{S}_{(n)_k}$ is an odd-valued quantity,
$\varepsilon(\Pi\mathcal{S}_{(n)_k}) =1$.}.

To obtain the Lagrangian formulation with first-order derivatives only,
we use the functional dependence of the operator $L_0$
on a fermionic operator $T_0$, $L_0=-T_0^2$ and attempt to gauged away
the dependence on $L_0, \eta_0$ from the BRST operator $Q$ (\ref{Q})
and from the whole set of the vectors $|\chi^l\rangle_{(n)_k}$.
To this end, we extract the zero-mode ghosts from the operator
$Q$ as follows:
\begin{eqnarray}
\label{strQ} Q &=& q_0\tilde{T}_0+\eta_0{L}_0 +
\imath(\eta_i^+q_i-\eta_iq_i^+)p_0 -
\imath(q_0^2-\eta_i^+\eta_i){\cal{}P}_0 +\Delta{}Q,
\end{eqnarray}
where
\begin{eqnarray}
\label{tildeT0} \tilde{T}_0 &=& T_0 -2q_i^+{\cal{}P}_i
-2q_i{\cal{}P}_i^+\,,
\\
\label{deltaQ} \Delta{}Q & = & q_i^+T^i+\eta_i^+L^i
+\sum\limits_{l\leq m}\eta_{lm}^+L^{lm} + \sum\limits_{l<
m}\vartheta^+_{lm}T^{lm}+ \Bigl[\frac{1}{2}\sum_{l, m}(1+\delta_{lm})\eta^{lm}q_l^+
 \nonumber
\\
\hspace{-0.4em}
&&\hspace{-0.4em}  -\sum_{l<m}
q_l\vartheta^{lm}-\sum_{m<l}q_l\vartheta^{ml+}\Bigr]p_m^+  -2 \sum_{l<m}q_lq_m^+\lambda^{lm}-2\sum_{l,m}
q^+_lq^+_m\mathcal{P}^{lm}-\sum\limits_{i<l<j}
\vartheta^+_{lj}\vartheta^+_{i}{}^l \lambda^{ij} \nonumber
\\
\hspace{-0.4em}&& {}\hspace{-0.4em}   -
\sum\limits_{l<n<m}\vartheta_{lm}^+\vartheta^{l}{}_n\lambda^{nm} +
\sum\limits_{n<l<m}\vartheta_{lm}^+\vartheta_{n}{}^m\lambda^{+nl}-
\sum_{n,l<m}(1+\delta_{ln})\vartheta_{lm}^+\eta^{l+}{}_{n}
\mathcal{P}^{mn}
\nonumber\\&& +
\sum_{n,l<m}(1+\delta_{mn})\vartheta_{lm}^+\eta^{m}{}_{n}
\mathcal{P}^{+ln}+ \frac{1}{2}\sum\limits_{l<m,n\leq
m}\eta^+_{nm}\eta^{n}{}_l\lambda^{lm}
\nonumber\\
\hspace{-0.4em} && \hspace{-0.4em}  %-2 \sum_{l<m}q_lq_m^+
 - \Bigl[\frac{1}{2}\sum\limits_{l,
m}(1+\delta_{lm})\eta^m\eta_{lm}^+ +
\sum\limits_{l<m}\vartheta_{lm} \eta^{+m}
+\sum\limits_{m<l}\vartheta^+_{ml} \eta^{+m}  \Bigr]\mathcal{P}^l
+h.c.
\,.
\end{eqnarray}
%
%
%ç
%
%
Here, $\tilde{T}_0$,  $\Delta{}Q$ are independent of $q_0$, $p_0$,
$\eta_0$, $\mathcal{P}_0$, and the relation ${\tilde{T}_0}^2 = - L_0$
holds true. We also expand the state vector and gauge parameters
in powers of the zero-mode ghosts, for $s=0, \ldots, \sum_{o=1}^k n_o + k(k-1)/2-1$:
\begin{align}
\label{0chi} |\chi\rangle &=\sum_{l=0}^{\sum_{o=1}^k n_o + k(k-1)/2-1}q_0^l(
|\chi_0^l\rangle +\eta_0|\chi_1^l\rangle), &
&gh(|\chi^{l}_{m}\rangle)=-(m+l), \ m=0,1
\\
\label{0L} |\Lambda^{(s)}\rangle
&=\sum_{l=0}^{\sum_{o=1}^k n_o + k(k-1)/2-1-s}q_0^k(|\Lambda^{(s)}{}^l_0\rangle
+\eta_0|\Lambda^{(s)}{}^l_1\rangle), &
&gh(|\Lambda^{(s)}{}^l_m\rangle)=-(s+l+m+1) .
\end{align}
Now, we can gauge away all the fields and gauge parameters by using
the equations of motion (\ref{Q12}) and the set of gauge transformations
(\ref{dx0})--(\ref{dxs}), except for two,
$|\chi^0_0\rangle$, $|\chi^1_0\rangle$
for the fields and $|\Lambda^{(s)}{}^l_0\rangle$, for
$l=0,1$ and $s=0, \ldots, \sum_{o=1}^k n_o + k(k-1)/2-1$,
for the gauge parameters. To do so, we follow, in part,
the procedure described in \cite{Sagnotti,symferm-flat}, \cite{mixfermiflat}.
Namely, after the extraction of zero-mode ghosts from the BRST
operator $Q$ (\ref{strQ}),  as well as from
 the state vector and the gauge parameter (\ref{0chi}), (\ref{0L}),
 the gauge transformation for the fields $|\chi_0^l\rangle$, $l \geq 2$
 has the form
\begin{equation}
    \delta|\chi_0^l\rangle = \Delta{}Q |\Lambda_0^l\rangle + \eta_i\eta_i^+
    |\Lambda_1^l\rangle +
    (l+1)(q_i\eta_i^+-\eta_iq_i^+)|\Lambda_0^{l+1}\rangle +
    \tilde{T}_0|\Lambda_0^{l-1}\rangle+
    |\Lambda_1^{l-2}\rangle\,,
\end{equation}
implying, by induction, that we can make all the fields
$|\chi_0^l\rangle$, $l \geq 2$ equal to zero by using the gauge
parameters $|\Lambda_1^{l-2}\rangle$. Then, considering  the
equations of motion for the powers $q_0^l, l\geq 3$ and taking
into account that $|\chi_0^l\rangle = 0$, $l \geq 2$, we can see
that these equations contain the subsystem
\begin{equation}\label{subsystem}
|\chi_1^{l-2}\rangle = \eta_i\eta_i^+ |\chi_1^l\rangle\,, \qquad l
\geq 3\,,
\end{equation}
which permits one to find, by induction, that all the fields
$|\chi_1^l\rangle$, $l \geq 1$ are equal to zero. Finally, we
examine the equations of motion for the power $q_0^2$:
\begin{equation}\label{lasteqs}
|\chi_1^{0}\rangle = - \tilde{T}_0 |\chi_0^1\rangle\,,
\end{equation}
  in order to express the vector $|\chi_1^0\rangle$ in terms of
  $|\chi_0^1\rangle$. Therefore, as in the mixed-symmetric case
  with two rows in YT \cite{mixfermiflat}, there remain
  only  two independent fields: $|\chi^l_0\rangle$, $l=0,1$.
  The above analysis is valid, of course, for the field vectors
  with a definite value of spin, i.e., for $|\chi^l_0\rangle_{(n)_k}$, $l=0,1$.

The first equation in (\ref{Qchi}), with the representation (\ref{strQ}), the
decomposition (\ref{0chi}), together with the above analysis, imply
(in view of the fact that the operators $Q$, $\tilde{T}_0$,
$\sum_l(\eta_l^+\eta_l)$ commute with  $\sigma^i$) that
independent equations of motion for these vectors have the form
\begin{eqnarray}
&& \Delta{}Q|\chi^{0}_{0}\rangle_{(n)_k}
+\frac{1}{2}\bigl\{\tilde{T}_0,\eta_i^+\eta_i\bigr\}
|\chi^{1}_{0}\rangle_{(n)_k} =0, \label{EofM1}
\\&&
\tilde{T}_0|\chi^{0}_{0}\rangle_{(n)_k} +
\Delta{}Q|\chi^{1}_{0}\rangle_{(n)_k} =0, \label{EofM2}
\end{eqnarray}
where $\{\ ,\ \}$ is the anticommutator and
the fields $|\chi_0^l\rangle_{(n)_k}$, $l=0,1$ are
assumed to obey the spin relations (\ref{state}) for $(m)_k = (n)_k$.

The equations of motion (\ref{EofM1}), (\ref{EofM2}) are
Lagrangian and can be deduced, in view of the invertibility of the
operator $K$ (\ref{explicit K}), (\ref{HermQ}), from the following
Lagrangian action for a fixed spin $(m)_k=(n)_k$ (defined, in the
standard manner, up to an overall factor),
\begin{eqnarray}
{\cal{}S}_{(n)_k} &=& {}_{(n)_k}\langle\tilde{\chi}^{0}_{0}|K_{(n)_k}\tilde{T}_0|\chi^{0}_{0}\rangle_{(n)_k} +
\frac{1}{2}\,{}_{(n)_k}\langle\tilde{\chi}^{1}_{0}|K_{(n)_k}\bigl\{
   \tilde{T}_0,\eta_i^+\eta_i\bigr\}|\chi^{1}_{0}\rangle_{(n)_k}\,,
\nonumber
\\&&
+ {}_{(n)_k}\langle\tilde{\chi}^{0}_{0}|K_{(n)_k}\Delta{}Q|\chi^{1}_{0}\rangle_{(n)_k} + {}_{(n)_k}\langle\tilde{\chi}^{1}_{0}|K_{(n)_k}\Delta{}Q|\chi^{0}_{0}\rangle_{(n)_k}\,, \label{L1}
\end{eqnarray}
where the standard odd scalar product for the creation and
annihilation operators is assumed, with the measure $d^dx$ over
the Minkowski space. The vectors (Dirac spinors)
$|\chi^{0}_{0}\rangle_{(n)_k}$, $|\chi^{1}_{0}\rangle_{(n)_k}$ (\ref{0chi}) as the solution
of the spin distribution relations (\ref{state})
are the respective vectors $|\chi^{l}_{0}\rangle$ in (\ref{chif})
for massless ($m=0$) and massive ($m\ne0$) HS fermionic field
$\Psi_{(\mu^1)_{n_1},...,(\mu^k)_{n_k}}(x)$ with
the ghost number $gh(|\chi^{l}_{0}\rangle_{(n)_k})=-l$, whereas
$K_{(n)_k}$ is obtained from $K$ (\ref{HermQ}) with the
substitution: $h^i\to-\bigl(n_i+(d-4i+\theta(m))/2\bigr)$.

The action (\ref{L1}) and the equations of motion (\ref{EofM1}), (\ref{EofM2})
are invariant with respect to gauge transformations, following
from the tower of relations (\ref{dx0})--(\ref{dxs}):
\begin{eqnarray}
\delta|\chi^{0}_{0}\rangle_{(n)_k} &=&
\Delta{}Q|\Lambda^{0}_{0}\rangle_{(n)_k}
 +
 \frac{1}{2}\bigl\{\tilde{T}_0,\eta_i^+\eta_i\bigr\}
 |\Lambda^{1}_{0}\rangle_{(n)_k}\,,
\label{GT1}
\\
\delta|\chi^{1}_{0}\rangle_{(n)_k} &=&
\tilde{T}_0|\Lambda^{0}_{0}\rangle_{(n)_k}
 +\Delta{}Q|\Lambda^{1}_{0}\rangle_{(n)_k}\,
 ,
\label{GT2}
\end{eqnarray}
which are reducible, with the gauge parameters
$|\Lambda^{(s)}{}^{j}_{0}\rangle_{(n)_k}$, $j=0,1$, subject to the
same conditions as those for $|\chi^j_0\rangle_{(n)_k}$ in
(\ref{state}),
\begin{align}
\delta|\Lambda^{(s)}{}^{0}_{0}\rangle_{(n)_k} &=
\Delta{}Q|\Lambda^{(s+1)}{}^{0}_{0}\rangle_{(n)_k}
 +
 \frac{1}{2}\bigl\{\tilde{T}_0,\eta_i^+\eta_i\bigr\}
 |\Lambda^{(s+1)}{}^{1}_{0}\rangle_{(n)_k},
& |\Lambda^{(0)}{}^0_0\rangle =|\Lambda^0_0\rangle\,, \label{GTi1}
\\
\delta|\Lambda^{(s)}{}^{1}_{0}\rangle_{(n)_k} &=
\tilde{T}_0|\Lambda^{(s+1)}{}^{0}_{0}\rangle_{(n)_k}
 +\Delta{}Q|\Lambda^{(s+1)}{}^{1}_{0}\rangle_{(n)_k},
& |\Lambda^{(0)}{}^1_0\rangle = |\Lambda^1_0\rangle\,,
\label{GTi2}
\end{align}
and with a finite number of reducibility stage
(the same as for the case of $q_0$-dependent vectors
$|\chi\rangle, |\Lambda^{(s)}\rangle$
in (\ref{Q12}), (\ref{dx0})--(\ref{dxs}))
to be equal to $s=\sum_{o=1}^k n_o + k(k-1)/2-1$.

A simultaneous construction of Lagrangian actions describing the
propagation of all massless (massive) fermionic fields with two
rows of the Young tableaux in the Minkowski space is similar to
the case of totally symmetric spin-tensors in the flat space
\cite{symferm-flat}. Note that a necessary condition for solving
this problem is to replace in $Q^{\prime}$, $Q$, $K$ the
parameters $-h^i$ by the operators $\sigma^i$ in an appropriate
manner and discard the condition (\ref{state}) for the fields and
gauge parameters.

In what follows, we consider some examples of the Lagrangian
formulation procedure.

\section{Examples}\label{examples}
\setcounter{equation}{0}

Here, we realize the general prescriptions of our Lagrangian formulation
in the case of mixed-symmetry fermionic fields of the lowest values
of rows and spins.

\subsection{Spin $(n_1+\frac{1}{2},n_2+\frac{1}{2})$ Mixed-symmetric Field}
As the first example, let us examine the mixed-symmetric
spin-tensor with two sets of indices corresponding to
spin-$(n_1+\frac{1}{2},n_2+\frac{1}{2})$. In this case, we expect
that our result will coincide with the one in the massless case
\cite{mixfermiflat}, and will be a new one for massive\footnote{In
\cite{mixfermiflat}, a Lagrangian formulation for massive
fermionic fields subject to a Young tableaux
$Y(n_1+\frac{1}{2},n_2+\frac{1}{2})$ on $\mathbb{R}^{1,d-1}$ was
realized by dimensional reduction from a massless theory in a
$(d+1)$-dimensional Minkowski space at the stage of a component
formulation only, see Eqs.(5.34)--(5.36) therein.}, where the
respective mixed-symmetric massless and massive spin-tensors,
subject to $Y(s_1,s_2)$, were considered. According to our
procedure, we have $(m_1,m_2)=(n_1,n_2)$, $m_i=0$, for $i=3,\ldots
, k$. We can show, given $m_i = 0$, that in (\ref{PhysState}) and
(\ref{chif}) all the components related to the rows $i \geq 3 $ in
the Young tableaux must vanish, i.e.,
\begin{eqnarray}
n'_{l}  &=& n_{a{}l} = n_{b{}l} = n^0_{l} = n_{1j}=n_{2m}= p_{1s}=p_{2t}=n_{f i}=n_{p j}= n_{f
1j}=n_{f 2m}=n_{p 1j}=n_{p 2o}\nonumber \\
&=& n_{f 1s}=n_{f 2t}=n_{\lambda 1s}=n_{\lambda 2t} =n_{s_m}
=0,\texttt{ for }l, j, m, s, t, i, o
>2.
\end{eqnarray}
As a result, the only surviving state vectors $|\chi_0^l \rangle$, $l=0,1$
from Eqs. (\ref{0chi}) are reduced to
\begin{eqnarray}
&& |\chi_0^l \rangle \ =\ \sum_{n} \prod_{c}^2 ( f_c^+ )^{n^0_{c}}\prod_{a}^2 ( b_{a}^+
)^{n'_{a}}\prod_{i\le j}^2( b_{ij}^+ )^{n_{ij}}( d_{12}^+
)^{p_{12}}\prod_{e,g}^2(q_e^+)^{n_{a{}e}}(p_g^+)^{n_{b{}g}}  \nonumber
\\
&&{}\qquad\quad \times\prod_{i, j, l\le m, n\le o}^2(
\eta_i^+ )^{n_{f i}} ( \mathcal{P}_j^+ )^{n_{p j}} ( \eta_{lm}^+
)^{n_{f lm}} ( \mathcal{P}_{no}^+ )^{n_{pno}}( \vartheta_{12}^+)^{n_{f 12}} (
\lambda_{12}^+ )^{n_{\lambda 12}} \nonumber
\\
&&{}\qquad\quad \times |\Psi(a^+_1,a^+_2)^{l 0_{f 0};(n)_{a{}e} (n)_{b{}g}
(n)_{f i}(n)_{p j}(n)_{f lm} (n)_{pno} n_{f 12}n_{\lambda
12}}_{(n^0)_{c};(n')_{l}(n)_{ij}p_{12}}\rangle\,, \label{x2}\\
&& {}|\Psi(a^+_1,a^+_2)^{l 0_{f 0};\ldots n_{\lambda
12}}_{(n^0)_{c};(n')_{l}(n)_{ij} p_{12}}\rangle \ = \
\sum_{p_1=0}^{\infty}\sum_{p_2=0}^{p_1}
\Psi^{l 0_{f 0};\ldots n_{\lambda
12}}_{(n^0)_{c};(n')_{l}(n)_{ij} p_{12}(\mu^1)_{p_1},(\mu^2)_{p_2}, (0)_{n_3}...,(0)_{n_k}}(x)\,\nonumber
\\
&&\phantom{{}|\Psi(a^+_1,a^+_2)^{l 0_{f 0};\ldots n_{\lambda
12}}_{(n^0)_{c};(n')_{l}(n)_{ij} p_{12}}\rangle \ } \times
\prod_{l_1=1}^{p_1}a^{+\mu^1_{l_1}}_1\prod_{l_2=1}^{p_2}
a^{+\mu^2_{l_2}}_2|0\rangle\label{x2cl}\,.
\end{eqnarray}
These vectors correspond to those in the massless case,
$(n')_{l}=(0)_{l}$, in \cite{mixfermiflat}, and are new in massive
cases. The representation (\ref{x2}), (\ref{x2cl}) is valid for
the sequence of gauge parameters
$|\Lambda^{(s)}{}^{l}_{0}\rangle_{(n)_2}$ from the tower of
reducible gauge transformations (\ref{GT1})--(\ref{GTi2}) with the
maximal value of reducibility stage $s_{max} = n_1+n_2$. The
operator $C^{rs}(d^+,d)$ in (\ref{Crsf}) has the same form as in
the integer-spin case \cite{BRmixbos}, and it is only
$C^{12}(d^+,d)$ that has a non-vanishing value:
\begin{eqnarray}
 \label{C122}
C^{12}(d^+,d) &\equiv &
\bigl(h^{1}-h^{2}-d^+_{12}d_{12}\bigr)d_{12},
 \end{eqnarray}
so that the expression for the $osp(1|4)$ algebra auxiliary
representation can be easily deduced from
Eqs.(\ref{t'+iFf})--(\ref{t'lmFf}).
 Then, one can easily show that equations (\ref{EofM1}), (\ref{EofM2}),
 relations (\ref{GT1})--(\ref{GTi2}), and action  (\ref{L1})
 with $|\chi^l_0\rangle$, $l=0,1$, as in
(\ref{x2}), (\ref{x2cl}), reproduce the same relations as those in
\cite{mixfermiflat} for the massless case, and new ones for the
massive case.

\subsection{Spin $(n_1+\frac{1}{2},n_2+\frac{1}{2},n_3+\frac{1}{2})$
General mixed-symmetric Field}

In this subsection, we examine a new, yet unknown, Lagrangian
formulation for the mixed-sym\-met\-ric HS field
$\Psi_{(\mu^1)_{n_1},(\mu^2)_{n_2},(\mu^3)_{n_3}}$ with three
families of symmetric indices subject to $Y(n_1,n_2,n_3)$. The
values of spin $(n_1+\frac{1}{2},n_2+\frac{1}{2},n_3+\frac{1}{2})$
for $n_1\geq n_2\geq n_3$ can be composed from the set of
coefficients $(l, n^0_{c}, n_{a{}e}, n_{b{}g},   n'_{l'},n_{ij},
p_{rs}, n_{f{}0}, n_{f{}i}$, $n_{p{}j},n_{f{}lm}, n_{p{}no}$,
$n_{f{}rs}, n_{\lambda{}tu}, p_i)$, for $l=0, 1$ and $c ,e, g,
l,i,j,r,s,l',m,n,o, t,u =1,2,3, \ i\leq j, r<s, l'\leq m, n\leq o,
t<u$, in (\ref{chif}), (\ref{0chi}) and
(\ref{PhysState})\footnote{We replace the indices $n_i$ in
(\ref{PhysState}) for the vector in the initial Fock space
$\mathcal{H}$ by $p_i$ due to the use of $n_i$ for the value of
generalized spin of the basic HS spin-tensor
$\Psi_{(\mu^1)_{n_1},(\mu^2)_{n_2},(\mu^3)_{n_3}}$.} to be
restricted for all vectors $| \chi^l_0 \rangle_{(n)_3}$,
$|\Lambda^{(s)}{}^{l}_{0}\rangle_{(n)_3}$ $l=0,1$; $s= 0,\ldots,
\sum_{o=1}^3 n_o + 2$, in view of the spectral problem solution
(\ref{state}) and the general decomposition (\ref{nidecomposf})
for spin [for the field subject to $Y(n_1,\ldots, n_k)$] by the
formulae
\begin{eqnarray}\label{nidecompos}
n_i &= &  p_i+ \Theta(m)n'_{i} + n_{a{}i}+ n_{b{}i}+ n^0_{i}+
\sum_{j=1}(1+\delta_{ij})(n_{ij}+n_{f{}ij}+n_{p{}ij} )
+n_{f{}i}+n_{p{}i} \nonumber \\
&{}& +\sum_{r<i} (p_{ri} + n_{f{}ri}+ n_{\lambda{}ri}) -\sum_{r>i}
(p_{ir}+ n_{f{}ir}+ n_{\lambda{}ir} )\,,\   i=1,2,3.
\end{eqnarray}
In addition to restrictions (\ref{nidecompos}), the subset of
``ghost'' numbers  $(l,  n_{f{}i}$,   $ n_{a{}e}, n_{b{}g} $,
$n_{p{}j}, n_{f{}lm}, n_{p{}no}$, $n_{f{}rs}, n_{\lambda{}tu})$ in
(\ref{chif}), without the $\eta_0$-number, should also obey such
relations as the general ones (\ref{ghnumf}) and (\ref{PhysState})
for fixed values of $n_i$; it satisfies the following equations
for the field vectors $|\chi^l_0\rangle_{(n)_3}$, and for the set
of gauge parameters $|\Lambda^{(s)}{}^{l}_{0}\rangle_{(n)_3}$, for
$l = n_{b{}0} = 0, 1$, $s=0 \ldots , \sum_{o=1}^3 n_o + 2$:
\begin{eqnarray}\label{ghnum}
     |\chi^l_0\rangle_{(n)_3}  \hspace{-0.7em}&:&\hspace{-0.7em}
     \sum_{i}\bigl(n_{f{}i}- n_{p{}i} + n_{a{}i}- n_{b{}i}\bigr)+
\sum_{i\leq j}\bigl(n_{f{}ij}-n_{p{}ij} \bigr) +\sum_{r<s} \bigl(
n_{f{}rs}- n_{\lambda{}rs}\bigr) = -l,\\
\label{ghnumg}
    |\Lambda^{(s)}{}^{l}_{0}\rangle_{(n)_3}  \hspace{-0.7em}&:&\hspace{-0.7em}
     \sum_{i}\bigl(n_{f{}i}- n_{p{}i} + n_{a{}i}- n_{b{}i}\bigr)+
\sum_{i\leq j}\bigl(n_{f{}ij}-n_{p{}ij} \bigr) +\sum_{r<s} \bigl(
n_{f{}rs}- n_{\lambda{}rs}\bigr) = -l-s-1,
\end{eqnarray}
which follows from the ghost number distributions
(\ref{Qchi})--(\ref{QLambdai}). Note that the above $2(k+3
+\sum_{o=1}^3 n_o)$ relations (\ref{nidecompos})--(\ref{ghnumg})
express both the fact of the general homogeneity of the vectors
$|\chi^l_0\rangle_{(n)_3}$,
$|\Lambda^{(s)}{}^{l}_{0}\rangle_{(n)_3}$ with respect to the spin
and ghost number distributions, and completely describe the
internal structure of these vectors in the powers of oscillators
in $\mathcal{H}_{tot}$.

The underlying part $\Delta Q$ (\ref{deltaQ}) of the BRST operator $Q$ (\ref{strQ})
for $7$ odd $(T_0, T_i, T_i^+)$ and $25$ even constraints
$(L_0, L_i, L_{ij}, T_{rs}, L^+_i, L^+_{ij}, T^+_{rs})$ reduces to the form
\begin{eqnarray}
\label{deltaQ3} \Delta {Q} \hspace{-0.4em} &=&\hspace{-0.4em}
 q^+_iT^i + \eta_i^+L^i
+\sum\limits_{l\leq m}\eta_{lm}^+L^{lm} + \sum\limits_{l<
m}\vartheta^+_{lm}T^{lm}+ \Bigl[\frac{1}{2}\sum_{l, m}(1+\delta_{lm})\eta^{lm}q_l^+
 -\sum_{l<m}
q_l\vartheta^{lm}
\\
\hspace{-0.4em}
&&\hspace{-0.4em}  -\sum_{m<l}q_l\vartheta^{ml+}\Bigr]p_m^+  -2 \sum_{l<m}q_lq_m^+\lambda^{lm}-2\sum_{l,m}
q^+_lq^+_m\mathcal{P}^{lm}
  -
\vartheta^+_{23}(\vartheta^+_{12}\lambda^{13}-
\vartheta_{13}\lambda^{+}_{12})\nonumber
\\ \hspace{-0.4em}
&& {}\hspace{-0.4em} - \vartheta_{13}^+\vartheta_{12}\lambda^{23}
- \sum_{n,l<m}(1+\delta_{ln})\vartheta_{lm}^+\eta^{l+}{}_{n}
\mathcal{P}^{mn}+
\sum_{n,l<m}(1+\delta_{mn})\vartheta_{lm}^+\eta^{m}{}_{n}
\mathcal{P}^{+ln}
\nonumber\\
\hspace{-0.4em} && \hspace{-0.4em} +
\textstyle\frac{1}{2}\hspace{-0.2em}\sum\limits_{l<m,n\leq
m}\hspace{-0.4em}\eta^+_{nm}\eta^{n}{}_l\lambda^{lm}
 - \Bigl[\frac{1}{2}\sum\limits_{l,
m}(1+\delta_{lm})\eta^m\eta_{lm}^+ +
\sum\limits_{l<m}\vartheta_{lm} \eta^{+m}
+\sum\limits_{m<l}\vartheta^+_{ml} \eta^{+m} \Bigr]\mathcal{P}^l
\hspace{-0.1em}+\hspace{-0.1em}h.c., \nonumber
\end{eqnarray}
whereas the extended (by the ghosts $\tilde{T}_0$) constraint
reads exactly as in (\ref{tildeT0}), however, for $k=3$. The BRST
operator $Q$ determined by decomposition (\ref{strQ}) is nilpotent
after the substitution $h_i \to -(p_i+\frac{d-4i+\theta(m)}{2})$,
for $i=1,2,3$, when restricted to the Hilbert subspace in
$\mathcal{H}_{tot}$, formed by the vectors
$|\chi^l_0\rangle_{(n)_3}$ (\ref{0chi}), and
$|\Lambda^{(s)}{}^{l}_{0}\rangle_{(n)_3}$ (\ref{0L}), proper for
the spin operator $(\sigma^1,\sigma^2,\sigma^3)$ (\ref{sigmai}).

The explicit form of the additional parts to the second-class
constraints $o'_a$ is determined by relations
(\ref{t'+iFf})--(\ref{t'lmFf}), however, for $k=3$ rows in YT, so
that the operators $t^{\prime +}_m$, $t^{\prime +}_{lm}$, for
$l,m=1,2,3$; $l<m$ in (\ref{t'+iFf}), (\ref{t'+lmf}) are written
as follows:
 \begin{eqnarray}
 \label{t'+1}
 t^{\prime  +}_1 & = & f^+_1 + 2b_{11}^+f_1
   \,,\\
 \label{t'+2}
 t^{\prime  +}_2 & = & f^+_2 + 2b_{22}^+f_2
 +4b_{12}^+f_1
  \,,\\
 \label{t'+3}
 t^{\prime  +}_3 & = & f^+_3 + 2b_{33}^+f_3
 +4(b_{13}^+f_1+b_{23}^+f_2)
  \,,\\
  t^{\prime+}_{12}   & = & d^+_{12}
   - \sum\nolimits_{n=1}^{3}(1+\delta_{1n})b^+_{n2}b_{1n}-({f}^+_{2}+2b^+_{22}f_2)f_{1}\,,
 \label{t'+12}
 \\
  t^{\prime+}_{13}   & = & d^+_{13}
   - \sum\nolimits_{n=1}^{3}(1+\delta_{1n})b^+_{n3}b_{1n}-\bigl[4b^+_{23}f_2 +{f}^+_{3}+2b^+_{33}f_3\bigr]f_{1}\,,
 \label{t'+13}
 \\
  t^{\prime+}_{23}   & = & d^+_{23} - d_{12}d^+_{13}
   - \sum\nolimits_{n=1}^{3}(1+\delta_{n2})b^+_{n3}b_{2n}-({f}^+_{3}+2b^+_{33}f_3)f_{2}\,,
 \label{t'+23}
  \end{eqnarray}
where one must take into account expressions for the operators
$C^{12}(d^+,d), C^{13}(d^+,d)$, $C^{23}(d^+,d)$ (defined for the
first time in \cite{BRmixbos} for a general bosonic field with
three groups of symmetric indices):
\begin{eqnarray}
  \label{C12}
C^{12}(d^+,d)&\equiv &
\bigl(h^{1}-h^{2}-d^+_{12}d_{12}-d^+_{13}d_{13}-d^+_{23}d_{23}\bigr)d_{12}
+ d^+_{23} d_{13},\\
  \label{C13}
C^{13}(d^+,d)&\equiv & \bigl(h^{1}-h^{3} -d^+_{13}d_{13} +
d^+_{23}d_{23}\bigr)d_{13}  ,\\
  \label{C23}
C^{23}(d^+,d)&\equiv &
\bigl(h^{2}-h^{3}-d^+_{23}d_{23}\bigr)d_{23}.
  \end{eqnarray}
Second, Eqs. (\ref{C12})--(\ref{C23})
permit one to present the expressions for the odd ``gamma-traceless''
elements $t^{\prime}_i$ as follows:
  \begin{eqnarray}
\label{t'1}% \nonumber to remove numbering (before each equation)
  t^{\prime }_1 &=&
2\sum_{n=2}^{3}\Bigl\{
d^+_{1n}  - \sum_{m
=1}^{3}(1+\delta_{1m})b^+_{nm}b_{1m}
  \Bigr\}f_n  \\
&& -2\left(\sum_{m=1}^3(1+\delta_{1m}) b^+_{1m}b_{1m}  -
d^+_{12}d_{12}-d^+_{13}d_{13}+ h^{1}\right)f_1 \nonumber \\
 && +  \sum\limits_{n=1}^3
(1+\delta_{1n}) \Bigl\{2\sum\limits_{m=n+1}^{3} b^+_{nm}f_m
-\frac{1}{2}\bigl(f_n^+ - 2b_{nn}^+f_n\bigr)
  \Bigr\}b_{1n}\,,  \nonumber\\
  %%%%%%%%%%%%%%%%%%%%%%%%%%%%%%%%%%%%%%
  \label{t'2}% \nonumber to remove numbering (before each equation)
  t^{\prime }_2 &=& -
2\Bigl\{  -
 C^{12}(d^+,d) +\sum_{m=1}^{3}(1+\delta_{m2})
b^+_{1m}b_{2m}+ f^+_1f_2
  \Bigr\}f_1\\
&&
  +2\Bigl\{
d^+_{23} - d^+_{13} d_{12}  - \sum_{m
=1}^{3}(1+\delta_{m2})b^+_{3m}b_{2m}
  \Bigr\}f_3 \nonumber \\
&& -2\left(\sum_{l=1}^3(1+\delta_{2l}) b^+_{2l}b_{2l}  -
d^+_{23}d_{23}+d^+_{12}d_{12} + h^{2}\right)f_2 \nonumber \\
 && + \sum\limits_{n=1}^3
(1+\delta_{n2}) \Bigl\{2\sum\limits_{m=n+1}^{3} b^+_{nm}f_m
-\frac{1}{2}\bigl(f_n^+ - 2b_{nn}^+f_n\bigr)
  \Bigr\}b_{n2}\,,  \nonumber\\
  %%%%%%%%%%%%%%%%%%%%%%%%%%%%%%%%%%%%%%%%%%
  \label{t'3}% \nonumber to remove numbering (before each equation)
  t^{\prime }_3 &=& -
2\Bigl\{ -C^{13}(d^+,d) -
 \big(C^{23}(d^+,d) - d^+_{12}d_{13}\big)d_{12}  +\sum_{m=1}^{3}(1+\delta_{m3})
b^+_{1m}b_{m3}  - \bigl[4 b^+_{12} f_2 -f^+_1
\bigr]f_3
  \Bigr\}f_1\\
&&-2\Bigl\{ d^+_{12}d_{13} -
 C^{23}(d^+,d) +\sum_{m=1}^{3}(1+\delta_{m3})
b^+_{m2}b_{m3}  + f^+_2
f_3
  \Bigr\}f_2\nonumber\\
&&
   -2\left(\sum_{l=1}^3(1+\delta_{3l}) b^+_{l3}b_{l3}  + d^+_{13}d_{13}+ d^+_{23}d_{23} + h^{3}\right)f_3 \nonumber \\
 && +  \sum\limits_{n=1}^3
(1+\delta_{n3}) \Bigl\{2\sum\limits_{m=n+1}^{3} b^+_{nm}f_m
-\frac{1}{2}\bigl(f_n^+ - 2b_{nn}^+f_n\bigr)
  \Bigr\}b_{n3}.  \nonumber
 \end{eqnarray}
In turn, for the even ``traceless'' elements
$l^{\prime}_{ll}$ we have
\begin{eqnarray}
\label{l'11text} l^{\prime }_{11} &=&
-\Bigl[2\Bigl\{
d^+_{12} -\sum_{n'=1}^3(1+\delta_{n'1})b^+_{n'2}b_{n'1}\Bigr\}f_2
+2\Bigl\{
d^+_{13} -\sum_{n'=1}^3(1+\delta_{n'1})b^+_{n'3}b_{n'1}\Bigr\}f_3\\
&&-  \sum\limits_{n=1}^3(1+\delta_{1n})
 \Bigl\{-2\sum\limits_{m=n+1}^{3}b^+_{nm}f_{m}
 +\frac{1}{2}\bigl[f^+_n -(1-\delta_{1n})2b^+_{nn}f_n\bigr]  \Bigr\}b_{1n}\Bigr]f_{1}
\nonumber\\
&&+
\frac{1}{4}\sum_{n=2}^{3}\Bigl[b^+_{nn}{b}_{1n} - 2d^+_{1n}+
2b^+_{n3}b_{13}
 \Bigr]{b}_{1n} +
\Bigl(\sum_{n= 1}^3
b^+_{1n}b_{1n}  - d^+_{12}d_{12}- d^+_{13}d_{13} + h^{1}\Bigr)b_{11}, \nonumber\\
\label{l'22text} l^{\prime }_{22} &=&
-\Bigl[2\Bigl\{
d^+_{23} -
d^+_{13}d_{12}-\sum_{n'=1}^3(1+\delta_{n'2})b^+_{n'3}b_{n'2}\Bigr\}f_3
\\
&&-  \sum\limits_{n=1}^3(1+\delta_{n2})
 \Bigl\{-2\sum\limits_{m=n+1}^{3}b^+_{nm}f_{m}
 +\frac{1}{2}\bigl[f^+_n -(1-\delta_{n2})2b^+_{nn}f_n\bigr]  \Bigr\}b_{2n}\Bigr]f_{2}\nonumber\\
&&+
\frac{1}{4}\Bigl(
b^+_{11}{b}^2_{12}+ b^+_{33}{b}^2_{23}\Bigr)+ \frac{1}{2}\Bigl[
2b^+_{12}b_{22}+ b^+_{13}b_{23}-
 C^{12}(d^+,d) \Bigr]{b}_{12} \nonumber
 \\
 && + \left(b^+_{22}b_{22}+b^+_{23}b_{23}  - d^+_{23}d_{23}+ d^+_{12}
 d_{12} + h^{2}\right)b_{22} - \frac{1}{2}\Bigl[d^+_{23} -
d^+_{13}d_{12} \Bigr]{b}_{23} , \nonumber\\
\label{l'33text} l^{\prime }_{33} &=& \sum\limits_{n=1}^3(1+\delta_{n3})
 \Bigl\{-2\sum\limits_{m=n+1}^{3}b^+_{nm}f_{m}
 +\frac{1}{2}\bigl[f^+_n -(1-\delta_{n3})2b^+_{nn}f_n\bigr]  \Bigr\}b_{3n}f_{3}\\
&&
\frac{1}{4}\Bigl(b^+_{11}{b}^2_{13}+b^+_{22}{b}^2_{23}\Bigr) +
\frac{1}{2}\Bigl[ d^+_{12}d_{13}
+2b^+_{23}b_{33} - C^{23}(d^+,d)\Bigr]{b}_{23} \nonumber\\
  && +
\frac{1}{2}\Bigl[b^+_{12}b_{23}+2b^+_{13}b_{33}
-C^{13}(d^+,d)-\big(C^{23}(d^+,d) - d^+_{12}d_{13}\big)d_{12} \Bigr]{b}_{13}
 \nonumber\\
 && + \left( b^+_{33}b_{33}  + d^+_{13}
 d_{13}+d^+_{23}
 d_{23} + h^{3}\right)b_{33}\nonumber ,
  \end{eqnarray}
and for the same elements $l^{\prime}_{lm}$, but for $l<m$,
\begin{eqnarray}
  \label{l'12bose}
 l^{\prime }_{12}&=&
 -\Biggl[  \frac{1}{2}\sum\limits_{n=1}^3(1+\delta_{n2})
 \Bigl\{2\sum\limits_{n'=n+1}^{3}b^+_{nn'}f_{n'}
 -\frac{1}{2}\bigl[f^+_{n} -(1-\delta_{n1})2b^+_{nn}f_{n}\bigr]  \Bigr\}b_{n2}\\
&& -\Bigl(\sum_{n=1}^3 (1+\delta_{n2})b^+_{2n}b_{2n}  -
d^+_{23}d_{23}+d^+_{12}d_{12} + h^{2}\Bigr)f_2\phantom{\qquad}
 \nonumber \\
 &&
+
  \Bigl\{d^+_{23}  -
d^+_{13}d_{12} -
\sum_{n'=1}^{3}(1+\delta_{n'2})b^+_{n'3}b_{2n'} \Bigr\}f_3\Biggr]f_{1} \nonumber\\
&& -\Biggl[  \frac{1}{2}\sum\limits_{n=1}^3(1+\delta_{n1})
 \Bigl\{2\sum\limits_{n'=n+1}^{3}b^+_{nn'}f_{n'}
 - \frac{1}{2}\bigl[f^+_{n} -(1-\delta_{n2})2b^+_{nn}f_{n}\bigr]  \Bigr\}b_{n1}
 \nonumber \\
&&+
 \Bigl\{d^+_{13} -
\sum_{n'=1}^{3}(1+\delta_{n'1})b^+_{3n'}b_{1n'}
  \Bigr\}f_{3}
\Biggr]f_{2}\nonumber\\
 %%%%%%%%%%%%%%%%%%%%%%%%%%%
 &&
  \frac{1}{4}\Bigl(\sum_{n=2}^3\bigl[b^+_{1n}b_{1n} +
(1+\delta_{n2})b^+_{2n} b_{2n}\bigr]   - d^+_{13}d_{13}-
d^+_{23}d_{23} + h^{1}+ h^{2}\Bigr)b_{12}\nonumber
\end{eqnarray}
\vspace{-3ex}
\begin{eqnarray}
&& - \frac{1}{2} \bigl[ C^{12}(d^+,d)
-\sum_{n=1}^{3}(1+\delta_{n2})
 b^+_{1n}b_{n2}
\bigr]b_{11}- \frac{1}{4} \bigl[
2d^+_{12}{b}_{22}+d^+_{13}{b}_{23} \bigr]
  \nonumber\\
 &&   - \frac{1}{4} \Bigl[ d^+_{23}-
 d^+_{13}d_{12} - \sum\limits_{n=2}^3
(1+\delta_{n2})b^+_{n3}b_{2n} \Bigr]{b}_{13}\nonumber, \\
  \label{l'13bose}
 l^{\prime }_{13}&=&
-\Biggl[ \hspace{-0.1em}\Bigl\{
 -
d^+_{12}d_{13}
  +
C^{23}(d^+,d)  -\sum_{n'=1}^{3}(1+\delta_{n'3})b^+_{n'2}b_{n'3}
 - f^+_{2}f_{3} \Bigr\}f_{2}\\
 &&
   -\Bigl(\sum_{n=1}^3 (1+\delta_{n3})b^+_{n3}b_{n3}  +d^+_{13}d_{13} +d^+_{23}d_{23} + h^{3}\Bigr)f_3
 \nonumber \\
 &&
+ \frac{1}{2} \sum\limits_{n=1}^3(1+\delta_{n3})
 \Bigl\{2\sum\limits_{n'=n+1}^{3}b^+_{nn'}f_{n'}
 -\frac{1}{2}\bigl[f^+_{n} -(1-\delta_{1n})2b^+_{nn}f_{n}\bigr]  \Bigr\}b_{n3}\Biggr]f_{1}
 \nonumber\\
&& - \frac{1}{2} \sum\limits_{n=1}^3(1+\delta_{1n})
 \Bigl\{2\sum\limits_{n'=n+1}^{k}b^+_{nn'}f_{n'}
 - \frac{1}{2}\bigl[f^+_{n} -(1-\delta_{n3})2b^+_{nn}f_{n}\bigr]  \Bigr\}b_{1n}
 f_{3}\nonumber\\
%%%%%%%%%%%%%%%%%%%%%%%%%%
&&
\frac{1}{4}\Bigl(b^+_{13}b_{13} +
\sum_{n=2}^3(1+\delta_{n3})b^+_{n3} b_{n3}  - d^+_{12}d_{12}+
d^+_{23}d_{23}+ h^{1}+ h^{3}\Bigr)b_{13}\nonumber  \\
 && +  \frac{1}{4} \Bigl[ \sum_{n=2}^{3}(1+\delta_{n3})
 b^+_{2n}b_{n3} + d^+_{12}d_{13} - C^{23}(d^+,d)\Bigr]b_{12 }  - \frac{1}{4} \Bigl[ d^+_{12}{b}_{23}+2d^+_{13}{b}_{33}
\Bigr] \nonumber \\
&& + \frac{1}{2} \Bigl[ \sum_{n=1}^{3}(1+\delta_{n3})
 b^+_{1n}b_{n3} - C^{13}(d^+,d)- \big(C^{23}(d^+,d) - d^+_{12}d_{13}\big)d_{12}\Bigr]b_{11 }\nonumber,
\\
  \label{l'23bose}
 l^{\prime }_{23}&=&
-\Biggl[  -\Bigl\{\sum_{n=1}^3 (1+\delta_{n3})b^+_{n3}b_{n3} +d^+_{13}d_{13}+d^+_{23}d_{23} + h^{3}\Bigr\}f_3
  \\
 &&
+ \frac{1}{2} \sum\limits_{n=1}^3(1+\delta_{n3})
 \Bigl\{2\sum\limits_{n'=n+1}^{3}b^+_{nn'}f_{n'}
 -\frac{1}{2}\bigl[f^+_{n} -(1-\delta_{n2})2b^+_{nn}f_{n}\bigr]  \Bigr\}b_{n3}\Biggr]f_{2}
 \nonumber\\
&& -\Bigl[  \frac{1}{2}\sum\limits_{n=1}^3(1+\delta_{n2})
 \Bigl\{2\sum\limits_{n'=n+1}^{3}b^+_{nn'}f_{n'}
 - \frac{1}{2}\bigl[f^+_{n} -(1-\delta_{n3})2b^+_{nn}f_{n}\bigr]  \Bigr\}b_{n2}
 \Bigr\}
\Bigr]f_{3}\nonumber\\
%%%%%%%%%%%%%%%%%%%%%%%%%%%%
&&
 \frac{1}{4} \Bigl[ \sum_{n=1}^{3}(1+\delta_{n3})
 b^+_{2n}b_{n3} - C^{13}(d^+,d)- \big(C^{23}(d^+,d) - d^+_{12}d_{13}\big)d_{12}\Bigr]b_{12
 }\nonumber\\
&& +  \frac{1}{2} \Bigl[ \sum_{n=2}^{3}(1+\delta_{n3})
 b^+_{2n}b_{n3} + d^+_{12}d_{13} - C^{23}(d^+,d)\Bigr]b_{22 }
\nonumber \\
&& + \frac{1}{4}\Bigl(b^+_{23}b_{23} + 2b^+_{33} b_{33}
+d^+_{12}d_{12} +d^+_{13}d_{13} +  h^{2}+
h^{3}\Bigr)b_{23} \nonumber\\
&& -\frac{1}{4} \Bigl[  C^{12}(d^+,d)
-\sum_{n=2}^{3}(1+\delta_{n2})b^+_{1n}b_{2n}
 \Bigr]{b}_{13} - \frac{1}{2}
\Bigl[ d^+_{23} - d^+_{13}d_{12} \Bigr]{b}_{33}\nonumber\,.
\end{eqnarray}
Finally,  for ``mixed-symmetry'' elements $t^{\prime}_{lm}$, we
have
 \begin{eqnarray}
t^{\prime }_{12} &=&
 C^{12}(d^+,d)
 -\sum\nolimits_{n=1}^{3}(1+\delta_{n2})b^+_{n1}
b_{n2} + (2b^+_{11}f_1-{f}^+_{1})f_{2}
 \,,  \label{t'12F}\\
t^{\prime }_{13} &=&
 C^{13}(d^+,d)+ \big(C^{23}(d^+,d) - d^+_{12}d_{13}\big)d_{12}
  -\sum\nolimits_{n=1}^{3}(1+\delta_{n3})b^+_{n1}
b_{n3} \label{t'13F}\\
&& + \bigl[4b^+_{12}f_2 +2b^+_{11}f_1-{f}^+_{1}\bigr]f_{3}
 \,, \nonumber\\
t^{\prime }_{23} &=& - d^+_{12}d_{13} +
 C^{23}(d^+,d)
  -\sum\nolimits_{n=1}^{3}(1+\delta_{n3})b^+_{n2}
b_{n3}+ (2b^+_{22}f_2-{f}^+_{2})f_{3}
 \,.\label{t'23F}
\end{eqnarray}
Relations (\ref{t'+1})--(\ref{t'23F}) together with Eqs.
(\ref{t'+iFf}), (\ref{g'0iFf})  for  $l^{\prime +}_{ij}$ and the
particle number operators $g^{\prime i}_0$ for the value of $k=3$,
compose the scalar oscillator realization of the $osp(1|6)$
superalgebra over the Heisenberg--Weyl superalgebra $A_{3|9}$ with
$6$ odd and $18$ even independent operators $(f^+_i, f_i)$,
$(b^+_{ij}, b_{ij}, d^+_{rs}, d_{rs})$, for $i\leq j, r<m$. The
above expressions for the vanishing of all operators $t^{\prime
+}_{i}, t^{\prime }_{i}, t^{\prime +}_{rs}, t^{\prime }_{rs},
l^{\prime +}_{lm}, l^{\prime }_{lm}, g^{\prime i}_0$ and $f^+_i,
f_i$, $b^+_{lm}, b_{lm}, d^+_{rs}, d_{rs}$, for $i,l,m, s=3$ are
reduced to the oscillator realization of the $osp(1|4)$
superalgebra over the Heisenberg--Weyl superalgebra $A_{2|6}$,
obtained for a massless mixed symmetry fermionic HS field in
Minkowski space with 2 rows in the YT \cite{mixfermiflat}.

It is now easy to present the Lagrangian equations of motion
(\ref{EofM1}), (\ref{EofM2}), the set of reducible Abelian gauge
transformations (\ref{GT1})--(\ref{GTi2}), and the unconstrained
action $\mathcal{S}_{(n)_3}$ (\ref{L1}), which have the final
(respective) form, for the HS field of spin
$(n_1+\frac{1}{2},n_2+\frac{1}{2},n_3+\frac{1}{2})$, with
$s=0,\ldots, n_1+n_2+n_3+ 2$,
\begin{eqnarray}
&& \hspace{-2em} \Delta{}Q|\chi^{0}_{0}\rangle_{(n)_3}
+\frac{1}{2}\bigl\{\tilde{T}_0,\eta_i^+\eta_i\bigr\}
|\chi^{1}_{0}\rangle_{(n)_3} =0, \qquad
\tilde{T}_0|\chi^{0}_{0}\rangle_{(n)_k} +
\Delta{}Q|\chi^{1}_{0}\rangle_{(n)_k} =0; \label{EofM123} \\
&& \hspace{-2em} \delta|\chi^{0}_{0}\rangle_{(n)_3} =
\Delta{}Q|\Lambda^{0}_{0}\rangle_{(n)_3}
 +
 \frac{1}{2}\bigl\{\tilde{T}_0,\eta_i^+\eta_i\bigr\}
 |\Lambda^{1}_{0}\rangle_{(n)_3}
\,, \quad \delta|\chi^{1}_{0}\rangle_{(n)_3} =
\tilde{T}_0|\Lambda^{0}_{0}\rangle_{(n)_3}
 +\Delta{}Q|\Lambda^{1}_{0}\rangle_{(n)_3}\,
 ;
\label{GT123}\\%
&& \hspace{-2em} \delta|\Lambda^{(s)}{}^{0}_{0}\rangle_{(n)_3} =
\Delta{}Q|\Lambda^{(s+1)}{}^{0}_{0}\rangle_{(n)_3}
 +
 \frac{1}{2}\bigl\{\tilde{T}_0,\eta_i^+\eta_i\bigr\}
 |\Lambda^{(s+1)}{}^{1}_{0}\rangle_{(n)_3}\,, \\
&& \hspace{-2em} \delta|\Lambda^{(s)}{}^{1}_{0}\rangle_{(n)_3} =
\tilde{T}_0|\Lambda^{(s+1)}{}^{0}_{0}\rangle_{(n)_3}
 +\Delta{}Q|\Lambda^{(s+1)}{}^{1}_{0}\rangle_{(n)_3}\,
\label{GTi123}
 ;\\
&&\hspace{-2em} {\cal{}S}_{(n)_3} = {}_{(n)_3}\langle\tilde{\chi}^{0}_{0}|K_{(n)_3}\tilde{T}_0|\chi^{0}_{0}\rangle_{(n)_3} +
\frac{1}{2}\,{}_{(n)_3}\langle\tilde{\chi}^{1}_{0}|K_{(n)_3}\bigl\{
   \tilde{T}_0,\eta_i^+\eta_i\bigr\}|\chi^{1}_{0}\rangle_{(n)_3}\,
\nonumber
\\&& \phantom{\hspace{-2em}{\cal{}S}_{(n)_3} =}
+ {}_{(n)_3}\langle\tilde{\chi}^{0}_{0}|K_{(n)_3}\Delta{}Q|\chi^{1}_{0}\rangle_{(n)_3} + {}_{(n)_3}\langle\tilde{\chi}^{1}_{0}|K_{(n)_3}\Delta{}Q|\chi^{0}_{0}\rangle_{(n)_3}\,, \label{L123}
\end{eqnarray}
where the operator $K_{(s)_3}$ is determined by relations
(\ref{explicit K}), (\ref{HermQ}), (\ref{Ka}) for $k=3$. The
corresponding Lagrangian formulation is an $(n_1+n_2+n_3+
2)$-th-stage reducible gauge theory for a free arbitrary HS
fermionnic field, subject to a Young tableaux $Y(n_1,n_2,n_3)$ in
a Minkowski space $\mathbb{R}^{1,d-1}$. To demonstrate an
application of the constructed Lagrangian formulations in the
above two examples, we will use our results to find a component
Lagrangian formulation for the spin-tensor $\Psi_{\mu \nu,\rho}$
with spin $\mathbf{s} =(2+\frac{1}{2},1+\frac{1}{2})$.

\subsection{Spin $(\frac{5}{2},\frac{3}{2})$ Mixed-symmetric Massless Spin-tensor} \label{ex5232}
In this section,  the general prescriptions of our Lagrangian
formulation will be applied to a rank-$3$ spin-tensor field,
$\Psi_{\mu \nu,\rho} \equiv \Psi_{\mu \nu,\rho{}A}$,
with a suppressed Dirac index
$A = 1,\ldots, 2^{\left[\frac{d}{2}\right]}$,
symmetric in the indices $\mu, \nu$, i.e. $\Psi_{\mu\nu,\rho} =
\Psi_{\nu\mu,\rho}$, starting from an analysis of the tower
of gauge transformations on a basis of the cohomological resolution complex.

\subsubsection{Reducible Gauge Transformations for Gauge
Parameters}\label{ggtrex}

For a spin-$(\frac{5}{2},\frac{3}{2})$ field, we have $(h^1,h^2) =
(\{-\frac{d}{2}\},\{3-\frac{d}{2}\}\})$.
Therefore, due to the analysis of the system of four spin
(\ref{nidecompos}) and two (\ref{ghnum}) ghost number equations
for all the indices of powers in the decomposition (\ref{chif}) and
(\ref{PhysState}) for the fields $|\chi^l_0\rangle_{(2,1)}$ and each
of the gauge parameters $|\Lambda^{(s)}{}^{l}_{0}\rangle_{(2,1)}$,
 $l = 0, 1$ and $s=0,1,2,3$, the gauge theory
 is that of an $(L = 3)$-th stage of reducibility.

As the initial step, the first lowest (independent) gauge parameter
$|\Lambda^{(3)}{}^{0}_{0}\rangle_{(2,1)}$ is determined only
by two-component spinor fields
$\psi^{(3)}{}^0_1(x), \psi^{(3)}{}^0_2(x)$ (\ref{x0-3}),
whereas the parameter $|\Lambda^{(3)}{}^{1}_{0}\rangle_{(2,1)}$
vanishes identically (see Appendix~\ref{example5232})
for all explicit expressions of the field
$|\chi^{l}_{0}\rangle_{(2,1)}$
and gauge $|\Lambda^{(s)}{}^{l}_{0}\rangle_{(2,1)}$
vectors obtained from the general representation
(\ref{chif}). For the reducible gauge parameters
of the second level $|\Lambda^{(2)}{}^{l}_{0}\rangle_{(2,1)}$,
given by Eqs. (\ref{x0-2}) --(\ref{x-2decompf}),
we have the gauge transformations from Eqs. (\ref{GTi1}), (\ref{GTi2}),
for $s=2$,
\begin{align}
\delta|\Lambda^{(2)}{}^{0}_{0}\rangle_{(2,1)} &=
\Delta{}Q|\Lambda^{(3)}{}^{0}_{0}\rangle_{(2,1)},
 &\delta|\Lambda^{(2)}{}^{1}_{0}\rangle_{(2,1)} &=
\tilde{T}_0|\Lambda^{(3)}{}^{0}_{0}\rangle_{(2,1)}
 .
\label{GTi3}
\end{align}
Due to Eqs. (\ref{GTi3}), the gauge transformations
for the component spin-tensors $\psi^{(2)}{}^{l}_{\ldots}$
have the form (omitting the coordinates $x$ ($x \in R^{1,d-1}$)
in the arguments)
\begin{align}\label{2level}
   & \delta\psi^{(2)}{}^{0}_{1} =  \psi^{(3)}{}^{0}_{1}, & \delta\psi^{(2)}{}^{0}_{2} = -3\psi^{(3)}{}^{0}_{1}
, && \delta\psi^{(2)}{}^{0}_{3} = \psi^{(3)}{}^{0}_{2},\\
& \delta\psi^{(2)}{}^{0}_{4} = -\psi^{(3)}{}^{0}_{2}, &  \delta\psi^{(2)}{}^{0}_{5} = -3\psi^{(3)}{}^{0}_{1} ,
&&\delta\psi^{(2)}{}^{0}_{6} = 2\psi^{(3)}{}^{0}_{2},\\
& \delta\psi^{(2)}{}^{0}_{7} =  0, & \delta\psi^{(2)}{}^{0}_{8} = - 2\psi^{(3)}{}^{0}_{2},&& \delta\psi^{(2)}{}^{0}_{5|\mu} = -3\gamma_{\mu}\psi^{(3)}{}^{0}_{1}+\imath\partial_\mu\psi^{(3)}{}^{0}_{2},\\
&  \delta\psi^{(2)}{}^{0}_{8|\mu} = -2\gamma_{\mu}\psi^{(3)}{}^{0}_{2},
& \delta\psi^{(2)}{}^{0}_{9} = 12\psi^{(3)}{}^{0}_{1} , &&
\delta\psi^{(2)}{}^{0}_{10} = 4\psi^{(3)}{}^{0}_{2} ,
\\
& \delta\psi^{(2)}{}^{1}_{2}= \imath \gamma^{\mu}\partial_\mu
\psi^{(3)}{}^{0}_{2}+ 6\psi^{(3)}{}^{0}_{1}, &
\delta\psi^{(2)}{}^{1}_{1}=-\imath \gamma^{\mu}\partial_\mu
\psi^{(3)}{}^{0}_{1} . && \label{2levelf}\end{align} In deducing
Eqs. (\ref{2level})--(\ref{2levelf}), we use the definition of odd
operators $\Delta{}Q$ (\ref{deltaQ3}) and $\tilde{T}_0$
(\ref{tildeT0}), for $i=1, 2$ and $k =2$ in the Young tableaux, as
well as
 the structure of additional parts for the constraints
 (\ref{t'+1}), (\ref{t'+2}), (\ref{t'+12}), (\ref{t'1}),
 (\ref{t'2}), (\ref{l'11text}), (\ref{l'22text}), (\ref{l'12bose}),
 (\ref{t'12F}), with the only restriction $k = 2$.

We then impose the gauge conditions (\ref{gLsmax-1}) for the first
lowest dependent gauge parameter
$|\Lambda^{(2)}{}^{0}_{0}\rangle_{(2,1)}$, so that the solution of
the equation $f_1|\Lambda^{(2)}{}^{0}_{0}\rangle_{(2,1)} =0 $ is
given by a gauged vector
$|\Lambda^{(2)}{}^{0}_{g0}\rangle_{(2,1)}$ with the vanishing
spinors $\psi^{(2)}{}^{0}_{p}$, $p= 5, 8$ and the remaining
(independent) component spin-tensors in
$|\Lambda^{(2)}{}^{0}_{g0}\rangle_{(2,1)}$,
$|\Lambda^{(2)}{}^{1}_{0}\rangle_{(2,1)}$. As a result, the theory
becomes a second-stage-reducible gauge theory, and the surviving
independent gauge parameter does not depend on the auxiliary
oscillator $f_1^+$.

In turn, the general gauge conditions (\ref{gLsmax-11}), (\ref{gLmax-2}),
when applied to the second lowest dependent gauge parameters
$|\Lambda^{(1)}{}^{l}_{0}\rangle_{(2,1)}$, given by
Eqs. (\ref{x1-0}) --(\ref{x-1decompff}), are as follows:
\begin{eqnarray}\label{g5232-1}
f_1\mathcal{P}_{11}^+|\Lambda^{(1)}{}^{0}_{0} \rangle =0,\ f_1|\Lambda^{(1)}{}^{1}_{0} \rangle =0, \  b_{11}\mathcal{P}_{11}^+|\Lambda^{(1)}{}^{0}_{0} \rangle =0,
\end{eqnarray}
and imply, first, the vanishing of the component functions
$\psi^{(1)}{}^{1}_{p}$, $p= 5, 8$, by means of all the degrees
of freedom in the vector
 $|\Lambda^{(2)}{}^{l}_{0}\rangle_{(2,1)}$,
 so that the gauged parameter
 $|\Lambda^{(1)}{}^{1}_{g0}\rangle_{(2,1)}$
 has the same structure as $|\Lambda^{(2)}{}^{0}_{g0}\rangle_{(2,1)}$,
 however, with the opposite Grassmann parity.
 Second, the requirements (\ref{g5232-1})
 lead to the vanishing of component functions:
  \begin{equation}\label{compfix1}
 \psi^{\prime(1)}{}^0_{n}, \psi^{(1)}{}^0_{r|\mu}, \psi^{(1)}{}^0_{r}, \psi^{(1)}{}^0_{t},\, \texttt{ for }n=1,6; r = 13, 18; t = 9, 10, 14, 15
 \end{equation}
 in the gauge vector $|\Lambda^{(1)}{}^{0}_{0}\rangle_{(2,1)}$.
 To find the degrees of freedom for which the component
 functions in the reduced independent parameter
 $|\Lambda^{(2)}{}^{0}_{g0}\rangle_{(2,1)}$ correspond
 to the vanishing of the above first-level components,
 we should examine the component form
 of gauge transformations for
 $|\Lambda^{(1)}{}^{0}_{0}\rangle_{(2,1)}$.
 They are given by the relations
 \begin{align}
\delta|\Lambda^{(1)}{}^{0}_{0}\rangle_{(2,1)} &=
\Delta{}Q|\Lambda^{(2)}{}^{0}_{g0}\rangle_{(2,1)},
 &\delta|\Lambda^{(2)}{}^{1}_{g0}\rangle_{(2,1)} &=
\tilde{T}_0|\Lambda^{(2)}{}^{0}_{g0}\rangle_{(2,1)}
 ,
\label{GTi4}
\end{align}
which follow from Eqs. (\ref{GTi1}), (\ref{GTi2}) for $s=1$. We do
not present here the sequence of component relations following
from (\ref{GTi4}), however, the restrictions on
$\psi^{(1)}{}^{0}_{\ldots}$ in (\ref{compfix1}) are due to the
degrees of freedom in the spinor
$|\Lambda^{(2)}{}^{0}_{g0}\rangle_{(2,1)}$, related, respectively,
to the component spinors $\psi^{(2)}{}^{0}_{1},
\psi^{(2)}{}^{0}_{3}$; $\psi^{(2)}{}^{0}_{5|\mu},
\psi^{(2)}{}^{0}_{8|\mu}$, $\psi^{(2)}{}^{0}_{9},
\psi^{(2)}{}^{0}_{10}$; $\psi^{(2)}{}^{0}_{t}$ for $t= 2, 4, 6, 7
$, which we must set equal to $0$ in (\ref{x0-1}) for
$|\Lambda^{(1)}{}^{l}_{0}\rangle_{(2,1)}$. As a result, all
degrees of freedom in the gauge parameters
$|\Lambda^{(2)}{}^{0}_{g0}\rangle_{(2,1)}$,
$|\Lambda^{(2)}{}^{1}_{0}\rangle_{(2,1)}$ are used, and the theory
becomes a first-stage-reducible gauge theory with the independent
parameter $|\Lambda^{(1)}{}^{l}_{g0}\rangle_{(2,1)}$, $l=0,1$, in
which only the component spin-tensors $ \psi^{(1)}{}^1_{r}$, for
$r = 1-4, 6, 7, 9$, $\psi^{(1)}{}^1_{t|\mu}$, for $t=5,8$ and
$\psi^{(1)}{}^0_{m|\mu\nu}, \psi^{\prime(1)}{}^0_{n|\mu},
\psi^{(1)}{}^0_{u|\mu}, \psi^{(1)}{}^0_{v}$, { for } $m = 13, 18$;
$n = 1,6$; $u = 1, 6, 9, 10, 14, 15, 21$;  $r = 13, 18$; $v = 1 -
8, 11, 12,  16, 17, 19-21$ survive. For the reducible gauge
parameters $|\Lambda^{l}_{0}\rangle_{(2,1)}$, $l=0,1$ determined
by Eqs. (\ref{x0-0})--(\ref{x-0decompf}) the general gauge
conditions (\ref{gLmax-3}), having the form (for $s_{max}= 3$)
\begin{eqnarray}\label{gLmax-31}
&& \bigl(f_1,\,b_{11}\bigr)\mathcal{P}_{11}^+|\Lambda^{l}_{0} \rangle =0,\qquad  \bigl(f_2\Pi^0_{p^+_1},\, b_{12} \bigr)\mathcal{P}_{11}^+\mathcal{P}_{12}^+ |\Lambda^{0}_{0} \rangle=0,
\end{eqnarray}
with $\Pi^0_{p^+_1}$ being the projector onto the
${p^+_1}$-independent part of $|\Lambda^{0}_{0} \rangle$,
introduced in Eqs. (\ref{gLmax-3})), lead to the vanishing of the
component spin-tensors
\begin{eqnarray}\label{compfix11}
&&  \psi^{\prime 1}_{n}, \psi^1_{r|\mu}, \psi^1_{r}, \psi^1_{t},\, \texttt{ for }n=1,6; r = 13, 18; t = 9, 10, 14, 15, \\
&& \psi^{\prime\prime 0}_{26|\mu}, \psi^{\prime 0}_{n|\mu}, \, \psi^0_{n|\mu},\, \psi^{\prime 0}_{n},\,
\psi^{\prime\prime0}_{26},\,
\psi^{0}_{n}  \,, \texttt{ for }n=1,26;  \label{compfix111}\\
&& \psi^{0}_{r}, \texttt{ for } r=6, 11, 17,  18, 20, 23, 28, 29, 34, 35;\ \psi^0_{t|\mu}, \texttt{ for } t= 20, 23, 35; \ \psi^0_{35|\mu\nu}.\label{compfix1111} \end{eqnarray}
The terms in Eqs. (\ref{compfix11}), due to the explicit
form of gauge transformations for the vectors $|\Lambda^{l}_{0} \rangle$,
 \begin{eqnarray}
 \delta|\Lambda^{0}_{0}\rangle_{(2,1)} &=&
\Delta{}Q|\Lambda^{(1)}{}^{0}_{g0}\rangle_{(2,1)}
 +
 \frac{1}{2}\bigl\{\tilde{T}_0,\eta_i^+\eta_i\bigr\}
 |\Lambda^{(1)}{}^{1}_{g0}\rangle_{(2,1)}, \nonumber\\
   \delta|\Lambda^{1}_{0}\rangle_{(2,1)} &= &
\tilde{T}_0|\Lambda^{(1)}{}^{0}_{g0}\rangle_{(2,1)}
 +\Delta{}Q|\Lambda^{(1)}{}^{1}_{g0}\rangle_{(2,1)}
\label{GTi5}
\end{eqnarray}
with the independent first-level gauge parameters
$|\Lambda^{(1)}{}^{l}_{g0}\rangle_{(2,1)}$, should be examined by
means of the final relation in (\ref{GTi5}), by analogy with the
components $ \psi^{(1)}{}^0_{\ldots}$ in Eq. (\ref{compfix1}), so
that all degrees of freedom of the reduced vector
$|\Lambda^{(1)}{}^{1}_{g0}\rangle_{(2,1)}$ are used. Next, the
vector $|\Lambda^{1}_{0}\rangle_{(2,1)}$ takes a form similar to
the vector $|\Lambda^{(1)}{}^0_{g0}\rangle_{(2,1)}$, however, with
the opposite Grassmann parities of its components and partially
used gauge transformations (\ref{GTi5}),
\begin{align}
&\delta|\Lambda^{0}_{0}\rangle_{(2,1)} =
\Delta{}Q|\Lambda^{(1)}{}^{0}_{g0}\rangle_{(2,1)},
   &\delta|\Lambda^{1}_{g0}\rangle_{(2,1)} =
\tilde{T}_0|\Lambda^{(1)}{}^{0}_{g0}\rangle_{(2,1)}
 .
\label{GTi6}
\end{align}
Without presenting explicitly the entire system
of linear equations for the component spin-tensors
(\ref{compfix111}), (\ref{compfix1111}) (which must vanish)
of $|\Lambda^{0}_{0}\rangle_{(2,1)}$, following from
Eqs. (\ref{GTi6}), we list the results of covariant
resolution in terms of the components of
$|\Lambda^{(1)}{}^{0}_{g0}\rangle_{(2,1)}$,
being the root of the mentioned system.
They are presented in the order corresponding
to the appearance of spin-tensors in (\ref{compfix111}), (\ref{compfix1111}) as follows:
\begin{eqnarray}
&& \psi^{(1)}{}^{0}_{14|\mu}, \,\psi^{\prime(1)}{}^{ 0}_{n|\mu}, \, \psi^{(1)}{}^{ 0}_{n|\mu},\, \psi^{(1)}{}^{ 0}_{11},\, \psi^{(1)}{}^{ 0}_{16},\, \psi^{(1)}{}^{ 0}_{17},\,
\psi^{(1)}{}^{0}_{n}  \,, \texttt{ for }n=1,6;  \label{compfix111r}\\
&& \psi^{(1)}{}^{0}_{r}, \texttt{ for } r=2-5, 19, 20, 7, 8, 12, 21;\ \psi^{(1)}{}^0_{t|\mu}, \texttt{ for } t= 9, 10, 21; \ \psi^{(1)}{}^0_{13|\mu\nu}.\label{compfix1111r} \end{eqnarray}
As a result, we have only two surviving first-level component
spin-tensors $\psi^{(1)}{}^0_{15|\mu}, \psi^{(1)}{}^0_{18|\mu\nu}$.
Then from the residual gauge transformation
for the component spin-tensor $\psi^0_{26|\mu\nu}$, remaining
in the restricted vector $|\psi^0_{26} \rangle_{(1,1)}$,
\begin{equation}\label{0level}
  \delta\psi^0_{26|\mu\nu} = -\psi^{(1)}{}^0_{18|\mu\nu},
\end{equation}
we can remove the second-rank spin-tensor $\psi^0_{26|\mu\nu}$ by
using the degree of freedom of the parameter
$\psi^{(1)}{}^0_{18|\mu\nu}$. The residual gauge transformations
for the remaining zero-level gauge parameter from the restricted
vector $|\Lambda^{0}_{g0}\rangle_{(2,1)}$ have the form (with a
unique first-level parameter spin-tensor
$\psi^{(1)}{}^0_{15|\mu}$)
\begin{align}\label{0levelr}
   & \delta\psi^{0}_{26|\mu,\nu} =  \imath \partial_\nu\psi^{(1)}{}^{0}_{15|\mu} , & \delta\psi^{0}_{23|\mu\nu} =  -\frac{\imath}{2} \partial_{\{\mu}\psi^{(1)}{}^{0}_{15|\nu\}} , && \delta\psi^{0}_{24} = \gamma^\mu\psi^{(1)}{}^{0}_{15|\mu},
   \\
& \delta\psi^{0}_{25}= - \imath \partial^\mu\psi^{(1)}{}^{0}_{15|\mu}, & \delta\psi^{1}_{15|\mu} = - \imath\gamma^\nu\partial_\nu\psi^{(1)}{}^{0}_{15|\mu}
, && \label{0levelr1} \\
 & \delta\psi^{0}_{n} = \delta\psi^{\prime0}_{31} = 0,\
 & \delta\psi^{0}_{p|\mu} = 0,\
&& \delta\psi^{0}_{r|\mu\nu} =  0,    \\
&  \delta\psi^{0}_{35|\mu\nu\rho} = 0,& \delta\psi^{0}_{1|\mu,\nu} =  0, && \delta\psi^{\prime\prime0}_{1} =  0,
 \\
&  \delta\psi^{\prime\prime0}_{1|\mu} =  0,& \delta\psi^{\prime 0}_{31|\mu} =   0, && \delta\psi^{1}_{\ldots} =  0, \label{0levelf}
\end{align}
for
$n = 2-5, 7-10, 12-16, 19, 21, 22, 24, 25, 27, 30-33$;
$p = 6, 11, 17, 18, 28, 29, 31, 34$; $r = 1, 20$,
where we have presented only those components in the last relation of
(\ref{0levelf}) which differ from the ones in Eq. (\ref{compfix11})
and $\psi^{1}_{15|\mu}$.

We are now able to examine the gauge transformation for the fields.

\subsubsection{Gauge Transformations for Fields}\label{gtrex}

For the gauge-dependent field vectors $|\chi^l_0\rangle_{(2,1)}$,
conditions (\ref{nidecompos}), (\ref{ghnum}) allow one, first, to
extract the dependence on the ghost variables, as done in
Appendix~\ref{example5232}, by Eqs. (\ref{x0-00}), (\ref{x0-01}),
whose components are determined by relations (\ref{x-00decompi})
and by Eqs. (\ref{x-2decompi}), (\ref{x-2decompf}),
(\ref{x-1decompf}), (\ref{x-1decompff}), (\ref{x-0decompi}),
(\ref{x-0decompf}), however, for the fields vector. The general
gauge conditions (\ref{G1}) with the operators
$\mathcal{C}^{k(k+1)}$, given by (\ref{Ck(k+1)}), when applied to
the fields in question, read as follows:
\begin{eqnarray}\label{gLmax-311}
&& \Bigl(\bigl(f_1,\,b_{11}\bigr)\mathcal{P}_{11}^+,\,\bigl(f_2\Pi^0_{p^+_1},\,b_{12}\bigr)\mathcal{P}_{11}^+\mathcal{P}_{12}^+ \Bigr)|\chi^{l}_{0} \rangle_{(2,1)} =0,\quad  \bigl(d_{12}\bigr)\lambda_{12}^+\mathcal{P}_{11}^+\mathcal{P}_{12}^+ |\chi^{0}_{0} \rangle_{(2,1)}=0.
\end{eqnarray}
Eqs. (\ref{gLmax-311}) lead, first, to the same elaboration of the
field vector $|\chi^{1}_{0} \rangle_{(2,1)}$ as was done for the
gauge parameter $|\Lambda^{0}_{g0}\rangle_{(2,1)}$ above, that
means, that the only gauge transformations (including trivial
ones) with the corresponding component spin-tensors as in Eqs.
(\ref{0levelr})--(\ref{0levelf}) hold true
\begin{align}\label{-1levelr}
   & \delta\varphi_{26|\mu,\nu} =  \imath \partial_\nu\psi^{1}_{15|\mu} + \imath\gamma^\rho\partial_\rho\psi^0_{26|\mu,\nu}  , & \delta\varphi_{23|\mu\nu} =  -\frac{\imath}{2} \partial_{\{\mu}\psi^{1}_{15|\nu\}}+ \imath\gamma^\rho\partial_\rho\psi^0_{23|\mu\nu} ,  \\&\delta\varphi_{24} = \gamma^\mu\psi^{1}_{15|\mu} - \imath\gamma^\rho\partial_\rho\psi^0_{24}-2\psi^0_{25} ,
& \delta\varphi_{25}=  -\imath \partial^\mu\psi^{1}_{15|\mu}+ \imath\gamma^\rho\partial_\rho\psi^0_{25},    \label{-1levelr1} \\
 & \delta\varphi_{n} = \delta\psi^{\prime 1}_{31} = 0,\
 & \delta\varphi_{p|\mu} = 0,\quad
 \delta\varphi_{r|\mu\nu} =  0,    \\
&  \delta\varphi_{35|\mu\nu\rho} = 0,& \delta\varphi_{1|\mu,\nu} =  0, \quad \delta\varphi^{\prime\prime}_{1} =  0,
 \\
&  \delta\varphi^{\prime\prime}_{1|\mu} =  0,& \delta\varphi^{\prime 1}_{31|\mu} =   0, \quad \delta\varphi_{\ldots} =  0, \label{-1levelf}
\end{align}
with the same values of indices $n, p, r$ as after Eqs.
(\ref{0levelf}). Second, the gauge (\ref{gLmax-311}) leads to the
vanishing of all component spin-tensors in the vector $|\Psi
\rangle_{(2,1)}$ (\ref{x-00decompi}), except for the original
spin-tensor $\Psi_{\mu\nu,\rho}$. To this end, the degrees of
freedom in the restricted gauge parameter
$|\Lambda^{0}_{g0}\rangle_{(2,1)}$ corresponding to
$\psi^{0}_{35|\mu\nu\rho}, \psi^{0}_{1|\mu\nu},
\psi^{0}_{20|\mu\nu}, \psi^{0}_{1|\mu,\nu}, \psi^{\prime\prime
0}_{1|\mu},\psi^{\prime 0}_{31|\mu}, \psi^{0}_{34|\mu},
\psi^{0}_{31|\mu}, \psi^{\prime 0}_{31}, \psi^{0}_{31},
\psi^{\prime\prime 0}_{1}$ are used, respectively, for the
auxiliary components in (\ref{x-00decompi}). Third, this leads to
the vanishing of the fields containing the auxiliary oscillators
$f_1^+, b_{11}^+, d_{12}^+$ in the vectors $|\psi_n
\rangle_{(0,1)}$ for $n=1,2,14,15$; $|\psi_m \rangle_{(1,0)}$ for
$m = 3,4, 16, 17, 23, 24, 26, 27,  38$; $|\psi_p \rangle_{(2,0)}$
for $p= 7, 20, 36, 37$; $|\psi_r \rangle_{(-1,1)}$ for $r = 5, 18$
and the oscillator $f_2^+$ in the vectors $|\psi_{14}
\rangle_{(0,1)}$, $|\psi_{15} \rangle_{(0,1)}$. As a result, there
remain only the component gauge spin-tensors
$\psi^{0}_{23|\mu\nu}, \psi^{0}_{26|\mu,\nu}, \psi^{0}_{24},
\psi^{0}_{25}$ and $\psi^{1}_{15|\mu}$.

The residual gauge transformations for the restricted field
vectors $|\chi^l_{g0}\rangle_{(2,1)}$
\begin{align}
&\delta|\chi^0_{g0}\rangle_{(2,1)} =
\Delta{}Q|\Lambda^{0}_{g0}\rangle_{(2,1)} +
 \frac{1}{2}\bigl\{\tilde{T}_0,\eta_i^+\eta_i\bigr\}
 |\Lambda^{1}_{g0}\rangle_{(2,1)},\nonumber\\
&\delta|\chi^1_{g0}\rangle_{(2,1)} =
\tilde{T}_0|\Lambda^{0}_{g0}\rangle_{(2,1)}+
\Delta{}Q|\Lambda^{1}_{g0}\rangle_{(2,1)}
\label{GTf}
\end{align}
 with allowance for the representation
\begin{eqnarray}\label{repren}
  && \frac{1}{2}\bigl\{\tilde{T}_0,\eta_i^+\eta_i\bigr\}
 |\Lambda^{1}_{g0}\rangle_{(2,1)} = \Bigl[\bigl(\eta_1^+ \mathcal{P}_2^+ - \eta_2^+ \mathcal{P}_1^+\bigr){T}_0 + \bigl(q_2^+ \mathcal{P}_1^+ - q_1^+\mathcal{P}_2^+\bigr)\Bigr]\tilde{\gamma}|\psi^1_{g|15} \rangle_{(1,0)}, \\
 && \tilde{T}_0|\Lambda^{0}_{g0}\rangle_{(2,1)} = {T}_0\Bigl[\mathcal{P}_2^+\Bigl( |\psi^0_{g|23} \rangle_{(2,0)} + \mathcal{P}_1^+\Bigl\{ q_1^+\tilde{\gamma}|\psi^0_{24} \rangle_{(0)} +\eta_1^+|\psi^0_{25} \rangle_{(0)}\Bigr\} \Bigr)
  +\mathcal{P}_1^+|\psi^0_{g|26} \rangle_{(1,1)} \Bigr]\nonumber\\
  && -2\mathcal{P}_2^+\mathcal{P}_1^+q_1^+|\psi^0_{25} \rangle_{(0)} \label{repren1}
\end{eqnarray}
 are written, in addition to (\ref{-1levelr})--(\ref{-1levelf})
 for the component functions with $l=0$, as follows:
 \begin{align}
    & \delta\Psi_{\mu\nu,\rho}  = - \imath\partial_\rho\psi^0_{23|\mu\nu}- \frac{\imath}{2}\partial_{\{\nu}\psi^0_{26|\mu\},\rho}, & \delta\psi_{14|\mu}  = - \imath\partial_\mu\psi^0_{24}+ \gamma^\nu\psi^0_{26|\nu,\mu},  \label{gtransfPsi}\\
    & \delta\psi_{15|\mu}  = - \imath\partial_\mu\psi^0_{25}- \imath\partial^\nu\psi^0_{26|\nu,\mu}, & \delta\psi_{16|\mu}  = \gamma^\nu\psi^0_{26|\mu,\nu}+ \psi^1_{15|\mu},  \\
    & \delta\psi_{17|\mu}  =  -\imath\partial^\nu\psi^0_{26|\mu,\nu}- \imath\gamma^\rho\partial_\rho\psi^1_{15|\mu}, & \delta\psi_{19}  =  - \frac{1}{2}\psi^0_{26|}{}^\mu{}_{,\mu} - \frac{1}{2}\psi^0_{25}, \\
    &   \delta\psi_{20|\mu\nu}  = \frac{1}{2}  \psi^0_{26|\{\nu,\mu\}} + \psi^0_{23|\mu\nu},
    &  \delta\psi_{26|\mu}   = 2\gamma^\nu\psi^0_{23|\mu\nu}  + \imath\partial_\mu\psi^0_{24} - \psi^1_{15|\mu},\\
        &\delta\psi_{27|\mu}  = -2\imath\partial^\nu\psi^0_{23|\nu\mu} + \imath\gamma^\rho\partial_\rho\psi^1_{15|\mu}+\imath\partial_\mu\psi^0_{25},
    & \delta\psi_{28}  = -\psi^0_{23|}{}^\mu{}_{\mu}+ \psi^0_{25} ,   \label{GTfields}\\
    & \delta\psi_{n}  = 0   &   \delta\psi_{p|\mu} = \delta\psi_{r|\mu\nu} = 0, \label{GTfieldstrr}
\end{align}
for $n = 1,2, 6, 8-13, 21, 22, 25,  29-35, 38$;  $p = 1-4, 23, 24, 33, 36-38$; $r = 7, 36, 37$.

Now, we can reduce the number of the zero-level gauge spin-tensors
in Eqs. (\ref{-1levelr})--(\ref{-1levelf}), (\ref{gtransfPsi})
--(\ref{GTfieldstr}), expressing the parameters
$\psi^0_{23|\mu\nu}, \psi^0_{25}, \psi^1_{15|\mu}$ in terms of the
field $\psi^0_{26|\nu,\mu}$ only. To do so, we completely remove
the field spin-tensors $\psi_{16|\mu}, \psi_{19},
\psi_{20|\mu\nu}$, and therefore obtain
\begin{equation}\label{expresfields }
  \psi^1_{15|\mu} = - \gamma^\nu\psi^0_{26|\mu,\nu}, \quad  \psi^0_{25}= - \psi^0_{26|}{}^\mu{}_{,\mu}, \quad \psi^0_{23|\mu\nu} = -\frac{1}{2}  \psi^0_{26|\{\nu,\mu\}}.
\end{equation}
As a result, the final non-trivial gauge transformations
(\ref{-1levelr})--(\ref{-1levelf}),  (\ref{gtransfPsi})
--(\ref{GTfieldstr}) take the form
 \begin{align}
    & \delta\Psi_{\mu\nu,\rho}  = \frac{\imath}{2}\partial_\rho\psi^0_{26|\{\nu,\mu\}}- \frac{\imath}{2}\partial_{\{\nu}\psi^0_{26|\mu\},\rho}, & \delta\psi_{14|\mu}  = - \imath\partial_\mu\psi^0_{24}+ \gamma^\nu\psi^0_{26|\nu,\mu},  \label{gtransfPsif}\\
    & \delta\psi_{17|\mu}  =  \imath\gamma^{\rho\nu}\partial_\rho\psi^0_{26|\mu,\nu},
    &  \delta\psi_{26|\mu}   = \imath\partial_\mu\psi^0_{24}- \gamma^\nu\psi^0_{26|\nu,\mu},\\
        &\delta\psi_{27|\mu}  = \imath\partial_{[\nu}\psi^0_{26|}{}^\nu{}_{,\mu]} - \imath\gamma^{\rho\nu}\partial_\rho\psi^0_{26|\mu,\nu},&
        \delta\psi_{15|\mu}  = \imath\partial_\mu \psi^0_{26|}{}^\nu{}_{,\nu}- \imath\partial^\nu\psi^0_{26|\nu,\mu},
  \label{GTfieldstr}\\
     & \delta\varphi_{26|\mu,\nu} =  \imath \gamma^\rho\partial_{[\rho}\psi^0_{26|\hat{\mu},\nu]}  , & \delta\varphi_{23|\mu\nu} =  \frac{\imath}{2}\gamma^\rho \partial_{\{\mu}\psi^0_{26|\nu\},\rho} -  \frac{\imath}{2}\gamma^\rho\partial_\rho\psi^0_{26|\{\mu,\nu\}} ,
          \label{gtransfPsif11} \\&\delta\varphi_{24} = \psi^0_{26|}{}^\mu{}_{,\mu} - \gamma^{\mu\nu}\psi^0_{26|\mu,\nu}  - \imath\gamma^\rho\partial_\rho\psi^0_{24} ,
& \delta\varphi_{25}=  \imath \partial^\mu \gamma^\nu\psi^0_{26|\mu,\nu} - \imath\gamma^\rho\partial_\rho\psi^0_{26|}{}^\mu{}_{,\mu},    \label{gtransfvarphi}
\end{align}
where we have introduced the matrix $\gamma^{\rho\nu} =
\frac{1}{2}(\gamma^{\rho}\gamma^{\nu}-
\gamma^{\nu}\gamma^{\rho})$, and the field $\psi_{28}$ becomes
 gauge-independent. Finally, due to the presence of the gauge parameter
 $\psi^0_{24}$ only with the derivative $\partial_\mu$, we can remove
 the spin-tensor $\psi_{14|\mu}$ by means of the degree of freedom
 corresponding to $\psi^0_{24}$. After that,
 we have an algebraic gamma-trace constraint for the gauge parameter
 $\psi^0_{26|\nu,\mu}$,
 \begin{equation}\label{algconstr}
 \gamma^\nu\psi^0_{26|\nu,\mu} = \imath\partial_\mu\psi^0_{24},
 \end{equation}
 whose substitution into the gauge transformations for the spin-tensors
 $\psi_{26|\mu}$, $\varphi_{24}$ makes them gauge-independent.
 Notice that the only remaining gauge $2$-nd rank spin-tensor is $\psi^0_{26|\mu,\nu}$,
 however, it is not arbitrary due to the gauge relation (\ref{0levelr}) and
 the algebraic constraint (\ref{algconstr}).

Let us now turn to the removal of the remaining auxiliary fields
from the field vectors $|\chi^l_{g0}\rangle_{(2,1)}$, by solving
some of the equations of motion.

\subsubsection{Gauge-invariant Unconstrained
Lagrangian}\label{glex}

We should find the action of the operator $\Delta Q $ on the
gauged field vectors
 $|\chi^l_{g0}\rangle_{(2,1)}$ , $l=0,1$, and of the operators
 $\tilde{T}_0$, $\{\tilde{T}_0, \eta_i^+\eta_i\}$,
 respectively, on $|\chi^0_{g0}\rangle_{(2,1)}$ and
 $|\chi^1_{g0}\rangle_{(2,1)}$,
 in order to solve the algebraic part of the equations
 of motion in the last general relation (\ref{EofM123})
 having, in this case, the form
 \begin{eqnarray}
&& \Delta{}Q|\chi^{0}_{g0}\rangle_{(2,1)}
+\frac{1}{2}\bigl\{\tilde{T}_0,\eta_i^+\eta_i\bigr\}
|\chi^{1}_{g0}\rangle_{(2,1)} =0, \qquad
\tilde{T}_0|\chi^{0}_{g0}\rangle_{(2,1)} +
\Delta{}Q|\chi^{1}_{g0}\rangle_{(2,1)} =0 \label{EofM123fin}
 .
 \end{eqnarray}
 Our starting point is to solve the final part of Eqs.
 (\ref{EofM123fin}),
 similar to the case of the gauged vector $|\Lambda^{0}_{g0}\rangle_{(2,1)}$
 in Section~\ref{gtrex}. In ding so, we have, due to the obvious consequence
 that all the fields $|\varphi_n\rangle_{\ldots}$ for $n= 16, 27, 31-34$
 with the multipliers $\mathcal{P}^+_{11}, \mathcal{P}^+_{12}$
 should vanish, except for the already gauged spinors
 $\varphi_{34}$, as well as with the linear (in $p_i^+$) spinors
 $|\varphi_m\rangle_{\ldots}$, $m=1, 20$
 (used earlier for the gauge-fixing of the vectors
 $|\chi^l_{0}\rangle_{(2,1)}$).
 Recalling that the only spinor $\psi_{33}$ in
 the gauged field vector $|\chi^{0}_{g0}\rangle_{(2,1)}$
 depends on the $f^+_1$ oscillator,
 we can choose for a solution of the last relation in
 (\ref{EofM123fin}) all the vectors $|\varphi_m\rangle_{\ldots}$
 for $m = 2, \ldots, 19$, equal to $0$.
 For the same reason (considering $f^+_2$, the oscillator arguments
 for the corresponding terms in $|\chi^{0}_{g0}\rangle_{(2,1)}$),
 the vectors $|\varphi_21\rangle_{0}$, $|\varphi_22\rangle_{0}$
 should vanish, as well.
 Then, the remaining terms, i.e.,
 $|\varphi_p\rangle_{\ldots}$, $p = 28 - 30, 35$,
 proportional to the momenta $\lambda^+_{12}$,
 should also vanish, because of the corresponding
 terms in $\tilde{T}_0|\chi^{0}_{g0}\rangle_{(2,1)}$
 (see Eqs. (\ref{EofM123fin})) do not depend on
 the $d_{12}^+$ oscillator.
 Therefore, only the four vectors $|\varphi_r\rangle_{\ldots}$, $r = 23, \ldots, 26 $,
 proportional to the momenta $\mathcal{P}_i^+$, survive at this stage of resolution
 of the equations of motion (\ref{EofM123fin}).

Turning to the first relation in (\ref{EofM123fin}), we can see
that, due to the absence of the auxiliary oscillators $f_i^+,
b_{ij}^+, d_{12}^+$ in the final field vector
$|\chi^{1}_{g0}\rangle_{(2,1)}$, all the vectors
$|\psi_n\rangle_{\ldots}$, with any of the multipliers
$\mathcal{P}_{ij}^+$, $\lambda^+_{12}$, $p_i^+$ should vanish.
Summarizing, the result of Eqs. (\ref{EofM123fin}) resolution
permits one to present the vectors
$|\chi^{l}_{g0}\rangle_{(2,1)}$, $l =0, 1$, in the form
\begin{eqnarray}\label{repren2}
  &&
 |\chi^{1}_{g0}\rangle_{(2,1)} = \mathcal{P}_2^+\Bigl( \tilde{\gamma}|\varphi_{g|23} \rangle_{(2,0)}  + \mathcal{P}_1^+\Bigl\{ q_1^+|
 \varphi_{24} \rangle_{(0)} +\eta_1^+\tilde{\gamma}|\varphi_{25} \rangle_{(0)}\Bigr\} \Bigr)
  +\mathcal{P}_1^+\tilde{\gamma}|\varphi_{g|26} \rangle_{(1,1)}, \\
 && |\chi^{0}_{g0}\rangle_{(2,1)} = |\Psi_g\rangle_{(2,1)} + \mathcal{P}_1^+\Bigl(\eta_1^+|\psi_{g|15} \rangle_{(0,1)} + \eta_2^+|\psi_{g|17} \rangle_{(1,0)}\Bigr)
  \nonumber\\
  && \phantom{|\chi^{0}_{g0}\rangle_{(2,1)}=}    + \mathcal{P}_2^+\Bigl( q_1^+\tilde{\gamma}|\psi_{g|26} \rangle_{(1,0)} +\eta_1^+|\psi_{g|27} \rangle_{(1,0)}
   + \eta_{11}^+|\psi_{28} \rangle_{(0)}\Bigr). \label{repren3}
\end{eqnarray}
The corresponding unconstrained gauge-invariant Lagrangian for the
spin-tensor $\Psi_{\mu\nu,\rho}$ with auxiliary $2$ second rank
spin-tensor fields, $\varphi_{23|\mu\nu}, \varphi_{26|\mu,\nu}$,
$4$ first rank spin-tensors $\psi_{15|\mu}, \psi_{17|\mu},
\psi_{26|\mu}, \psi_{27|\mu}$ and $3$ spinors  $\varphi_{24},
\varphi_{25}, \psi_{28}$ from Eqs. (\ref{repren2}),
(\ref{repren3}) in terms of the odd scalar product in
$\mathcal{H}\otimes \mathcal{H}_{gh}$ read as follows:
\begin{eqnarray}
&& {\cal{}S}_{(2,1)} = {}_{(2,1)}\langle\tilde{\chi}^{0}_{g0}|\tilde{T}_0|\chi^{0}_{g0}\rangle_{(2,1)} +
\frac{1}{2}\,{}_{(2,1)}\langle\tilde{\chi}^{1}_{g0}|\bigl\{
   \tilde{T}_0,\eta_i^+\eta_i\bigr\}|\chi^{1}_{g0}\rangle_{(2,1)}\,
\nonumber
\\&& \phantom{{\cal{}S}_{(n)_3} =}
+ {}_{(2,1)}\langle\tilde{\chi}^{0}_{g0}|\Delta{}Q|\chi^{1}_{g0}\rangle_{(2,1)} + {}_{(2,1)}\langle\tilde{\chi}^{1}_{g0}|
\Delta{}Q|\chi^{0}_{g0}\rangle_{(n)_3}\,. \label{L5232}
\end{eqnarray}
Now, we continue to eliminate the auxiliary fields from the
configuration space by solving the basic part of the remaining
equations of motion, following from the action (\ref{L5232}), with
allowance for the representation
\begin{eqnarray}\label{repren4}
  && \frac{1}{2}\bigl\{\tilde{T}_0,\eta_i^+\eta_i\bigr\}
 |\chi^{1}_{g0}\rangle_{(2,1)} = {T}_0\eta_1^+ \Bigl[\tilde{\gamma}|\varphi_{g|26} \rangle_{(1,1)} -
 \mathcal{P}_2^+ q_1^+|\varphi_{24} \rangle_{(0)}\Bigr] + {T}_0\eta_2^+ \Bigl[\tilde{\gamma}|\varphi_{g|23} \rangle_{(2,0)}\nonumber\\
  &&\quad +\mathcal{P}_1^+\Bigl\{ q_1^+|\varphi_{24} \rangle_{(0)} +\eta_1^+\tilde{\gamma}|\varphi_{25} \rangle_{(0)}\Bigr\}\Bigr]
 +  q_1^+\bigl(\eta_1^+ \mathcal{P}_2^+ - 2\eta_2^+ \mathcal{P}_1^+\bigr) \tilde{\gamma}|\varphi_{25} \rangle_{(0)}
  - q_1^+ \Bigl[\tilde{\gamma}|\varphi_{g|26} \rangle_{(1,1)}\nonumber\\
  && \quad   -
 \mathcal{P}_2^+ q_1^+|\varphi_{24} \rangle_{(0)} \Bigr] - q_2^+ \Bigl[\tilde{\gamma}|\varphi_{g|23} \rangle_{(2,0)}
 + \mathcal{P}_1^+\Bigl\{ q_1^+|\varphi_{24} \rangle_{(0)} +\eta_1^+\tilde{\gamma}|\varphi_{25} \rangle_{(0)}\Bigr\}\Bigr], \\
 && \tilde{T}_0|\chi^{0}_{g0}\rangle_{(2,1)} = {T}_0|\chi^{0}_{g0}\rangle_{(2,1)} + 2 q_1^+ \Bigl[\mathcal{P}_1^+ |\psi_{g|15} \rangle_{(0,1)} + \mathcal{P}_2^+|\psi_{g|27} \rangle_{(1,0)}\Bigr] + 2q_2^+
\mathcal{P}_1^+|\psi_{g|17} \rangle_{(1,0)}.  \label{repren5}
\end{eqnarray}
The equations of motion (except for ghost-independent terms) for
the second relation in Eqs. (\ref{EofM123fin}) have the form
 \begin{align}
     &   2\psi_{15|\mu}  - \imath\partial_\mu\varphi_{24}+ \gamma^\nu\varphi_{26|\nu,\mu} = 0, &
        - \varphi_{23|}{}^\mu{}_\mu - \varphi_{25}- \imath\gamma^\mu\partial_\mu\psi_{28} =0,  \label{EoMfin1}\\
    & \imath\gamma^\rho\partial_\rho\psi_{15|\mu} + \imath\partial_\mu\varphi_{25}+ \imath\partial^\nu\varphi_{26|\nu,\mu}= 0, &
    \gamma^\nu\varphi_{26|\mu,\nu}  +2 \psi_{17|\mu} = 0,  \\
    &   -\imath\partial^\nu\varphi_{26|\mu,\nu}- \imath\gamma^\rho\partial_\rho\psi_{17|\mu} = 0, &
     - \frac{1}{2}\varphi_{26|}{}^\mu{}_{,\mu} - \frac{1}{2}\varphi_{25}  = 0, \\
    &   \frac{1}{2}  \varphi_{26|\{\nu,\mu\}} + \varphi_{23|\mu\nu}   = 0,
    & 2\gamma^\nu\varphi_{23|\mu\nu}  + \imath\partial_\mu\varphi_{24}+2  \psi_{27|\mu} + \imath\gamma^\rho\partial_\rho\psi_{26|\mu} =0, \\
        &   -2\imath\partial^\nu\varphi_{23|\nu\mu} + \imath\partial_\mu\varphi_{25}-\imath\gamma^\rho\partial_\rho\psi_{27|\mu} =
        0.
 \label{EoMfin2}
  \end{align}
  For the first generating equation in (\ref{EofM123fin}), we have
   \begin{align}
 &-\imath\partial^\rho
\Psi_{\mu\nu, \rho} - \frac{\imath}{2}\partial_{\{\mu}\psi_{17|\nu\}} +  \imath\gamma^\rho\partial_\rho
\varphi_{26|\mu,\nu}  =0,  &
\Psi_{\{\mu\nu,\rho\}} =0 , \label{EoMfin3} \\
 %%%%%%%%%%
 %%%%%%%%%%
 & -2\gamma^\rho\Psi_{\mu\rho,\nu} - \imath\partial_{\nu}\psi_{26|\mu} -  \varphi_{26|\mu,\nu}  =0
, &
 -\gamma^\rho\Psi_{\mu\nu,\rho} -   \varphi_{23|\mu\nu}  =0,\label{EoMfin31} \\
 %%%%%%%%%%
%%%%%%%%%%
  &  2\imath\partial^\rho
\Psi_{\mu\rho,\nu} + \imath\partial_{\mu}\psi_{15|\nu} + \imath\partial_{\nu}\psi_{27|\mu} -  \imath\gamma^\rho\partial_\rho
\varphi_{23|\mu\nu}  =0,  & \Psi^\rho{}_{\rho, \mu} -  \imath\partial_{\mu}\psi_{28} - \psi_{15|\mu}=0,  \label{EoMfin32} \\
%%%%%%%%%%
  &  \frac{1}{2}
\Psi^\rho{}_{\mu,\rho}  - \frac{1}{2}\psi_{27|\mu}  - \frac{1}{2} \psi_{17|\mu} =0,  &
 \psi_{26|\mu} =0, \label{EoMfin321} \\
  & \imath\gamma^\rho\partial_\rho
 \varphi_{25}  + \imath \partial^\rho\psi_{15|\rho}  - \imath\partial^\rho\psi_{17|\rho} = 0\,,
%%%%%%%%%%%%%%%%%%%%%%%%%%%%%%%%%%%%%%%%%%%
 &
 -\psi_{15|\mu}   - \psi_{17|\mu}- \psi_{27|\mu}   =0, \label{EoMfin33} \\
 & 4 \psi_{28}  + \varphi_{24}+ \gamma^\rho\psi_{26|\rho} = 0 ,
&  - \varphi_{24}= 0 ,    \label{EoMfin4}\\
 & - \varphi_{25}  - \gamma^\nu\psi_{15|\nu} = 0 ,
&  2\varphi_{25}  +  \imath\gamma^\rho\partial_\rho
\varphi_{24}-  \gamma^\nu\psi_{17|\nu} =0 ,    \label{EoMfin42}
\\
&  -\varphi_{25}  -  \imath\gamma^\rho\partial_\rho
\varphi_{24}-  \gamma^\nu\psi_{27|\nu} -\imath\partial^\rho\psi_{26|\rho} =0 \,. &
    \label{EoMfin43}\end{align}
From the second equation in (\ref{EoMfin3}), which implies the
Young symmetry condition for the initial field
$\Psi_{\mu\nu,\rho}$, we can compose the spin-tensor
\begin{equation}\label{irreptens}
  \widetilde{\Psi}_{\mu\nu,\rho}= \Psi_{\mu\nu,\rho}-\frac{1}{2}\Psi_{\rho\mu,\nu}-\frac{1}{2}\Psi_{\nu\rho, \mu}
\end{equation}
identically satisfying  Eq. (\ref{EoMfin3}).  Therefore, the
solution of the system (\ref{EoMfin1})--(\ref{EoMfin43}) with
respect to the spin-tensor $\widetilde{\Psi}_{\mu\nu,\rho}$ has
the form
\begin{eqnarray}\label{finsol1}
    && \varphi_{23|\mu\nu}=- \gamma^\rho\Psi_{\mu\nu,\rho}, \  \varphi_{26|\mu,\nu}  = -2\gamma^\rho\Psi_{\mu\rho,\nu}, \ \varphi_{25}=
      2\gamma^\rho\Psi{}^\mu{}_{\rho,\mu},\  \psi_{17|\mu} = \gamma^\nu\gamma^\rho\Psi_{\mu\rho,\nu}, \\
     && \psi_{15|\mu} = \Psi^\nu{}_{\nu,\mu},\ \psi_{27|\mu} = \gamma^\nu\gamma^\rho\Psi_{\mu\nu,\rho},\ \varphi_{24} = \psi_{26|\mu} = \psi_{28}= 0  \label{finsol11}
\end{eqnarray}
From the Young symmetry relation, there follows the validity of
the algebraic consequences $\Psi^\nu{}_{\nu,\mu} = -
2\Psi{}^\nu{}_{\mu,\nu}$ for the spin-tensor $\Psi_{\mu\nu,\rho}$
components.

The final bra-vectors ${}_{(2,1)}\langle\tilde{\chi}^{l}_{g0}|$
read, due to the general formula (\ref{L5232}) for the action,
with allowance for the relations (\ref{finsol1}) and the Hermitian
conjugation rule, using the matrix $F^+ =
\tilde{\gamma}^0(F)^+\tilde{\gamma}^0$ and the spin-tensor $\psi^+
= (\psi)^+\tilde{\gamma}^0$, being compatible with the
conventional one \cite{symferm-flat}, \cite{symferm-ads}
\begin{eqnarray}
  {}_{(2,1)}\langle\tilde{\chi}_{g0}^0| &=&  {}_{(2,1)}\langle\tilde{\Psi_g}| +
  {}_{(0,1)}\langle\tilde{\psi}_{g|15}|\eta_1\mathcal{P}_1
+ {}_{(1,0)}\langle\tilde{\psi}_{g|17}|\eta_2\mathcal{P}_1
+ {}_{(1,0)}\langle\tilde{\psi}_{g|27}|\eta_1\mathcal{P}_2    \label{brachi00}
\\
%%%%%%%%%%%%%%%
  {}_{(2,1)}\langle\tilde{\chi}_{g0}^1| &=&
{}_{(2,0)}\langle\tilde{\varphi}_{g|23}| \tilde{\gamma}\mathcal{P}_2
   +
 {}_{(0,0)}\langle\tilde{\varphi}_{25}| \tilde{\gamma}\eta_1\mathcal{P}_1\mathcal{P}_2 + {}_{(1,1)}\langle\tilde{\varphi}_{g|26}|
 \tilde{\gamma}\mathcal{P}_1\,,
\end{eqnarray}
with the component bra-vectors
\begin{align}
& {}_{(2,1)}\langle\tilde{\Psi_g}|   \ = \  \langle0|
a^{\mu}_1a^{\nu}_1a^{\rho}_2 \Psi^+_{\mu\nu,\rho}(x)\tilde{\gamma}_0\,, &
{}_{(0,1)}\langle\tilde{\psi}_{g|15}|    \ = \ \langle0|
\psi^{+}_{15|\mu}(x)a^{\mu}_2
\tilde{\gamma}_0 \,, \label{braPsi}\\
 & {}_{(2,0)}\langle\tilde{\varphi}_{g|23}| \ =\ \langle 0|\varphi^{+}_{23|\mu\nu}(x)a^{\nu}_1a^{\mu}_1
\tilde{\gamma}_0\,, & {}_{(1,1)}\langle\tilde{\varphi}_{g|26}| \ =\ \langle 0|\varphi^{+}_{26|\mu,\nu}(x)a^{\nu}_2a^{\mu}_1
\tilde{\gamma}_0\,, \label{bravarphi}\\
 & {}_{(1,0)} \langle\tilde{\psi}_{g|27}|
   \ =  \langle 0|
   \psi^{+}_{27|\mu}(x)a^{\mu}_1
 \tilde{\gamma}_0, &  {}_{(0,0)} \langle\tilde{\psi}_{28}|
   \ =  \langle 0|
   \psi^{+}_{28}(x)
 \tilde{\gamma}_0\,.
  \label{chim}
  \end{align}
Explicitly, the action (\ref{L5232}) in terms of the spin-tensor
$\Psi_{\mu\nu,\rho}$ and auxiliary spin-tensors has the form
 \begin{eqnarray}
 {\cal{}S}_{(2,1)} &=& \int d^d x \Bigl\{\bar{\Psi}_{\mu\nu,\rho}
{\imath}\gamma^{\tau}\partial_{\tau} {\Psi}^{\mu\nu,\rho}- \bar{\psi}_{15|\mu}
{\imath}\gamma^{\tau}\partial_{\tau} {\psi}_{15|}{}^{\mu} - \bar{\psi}_{17|\mu}
{\imath}\gamma^{\tau}\partial_{\tau} {\psi}_{17|}{}^{\mu} -  \bar{\psi}_{27|\mu}
{\imath}\gamma^{\tau}\partial_{\tau} {\psi}_{27|}{}^{\mu} \nonumber \\ &&
  +2 \bar{\varphi}_{23|\mu\nu}
{\imath}\gamma^{\tau}\partial_{\tau} {\varphi}_{23|}{}^{\mu\nu}+ \bar{\varphi}_{26|\mu,\nu}
{\imath}\gamma^{\tau}\partial_{\tau} {\varphi}_{26|}{}^{\mu,\nu}  - \bar{\varphi}_{25}
{\imath}\gamma^{\tau}\partial_{\tau} {\varphi}_{25} +2\imath \bar{\Psi}_{\mu\nu,\rho}\partial^\rho {\varphi}_{23|}{}^{\mu\nu} \nonumber \\
&&  +  \imath
\bar{\Psi}_{\mu\nu,\rho}\partial^{\{\nu} {\varphi}_{26|}{}^{\mu\},\rho} -  \imath
\bar{\psi}_{15|\mu}\bigl(\partial_{\nu} {\varphi}_{26|}{}^{\nu,\mu} +  \partial^{\mu} {\varphi}_{25}\bigr)
- \imath \bar{\psi}_{17|\{\mu}\partial_{\nu\}} {\varphi}_{23|}{}^{\nu\mu} \nonumber \\
&&  +  \imath \bar{\psi}_{17|\mu}\partial^{\mu} {\varphi}_{25}
 - \imath \bar{\psi}_{27|\mu}\partial_{\nu} {\varphi}_{26|}{}^{\mu,\nu}
 +2\imath  \bar{\varphi}_{23|}{}^{\mu\nu}\partial^\rho{\Psi}_{\mu\nu,\rho}- \imath \bar{\varphi}_{23|}{}^{\nu\mu}\partial_{\{\nu}{\psi}_{17|\mu\}} \nonumber\\
 &&   -  \imath \bar{\varphi}_{25}
 \partial^{\mu}\bigl({\psi}_{15|\mu}-
 {\psi}_{17|\mu}\bigr)+ \imath\bar{\varphi}_{26|\{\mu,{\rho}}\partial_{\nu\}}{\Psi}^{\mu\nu,\rho}
  -  \imath
\bar{\varphi}_{26|}{}^{\mu,\nu}\bigl(\partial_{\mu}{\psi}_{15|\nu}+ \partial_{\nu}{\psi}_{27|\mu}\bigr)\Bigr\}. \label{S5232aux}
 \end{eqnarray}
 The above action determines a gauge-invariant Lagrangian description
 for a free massless fermionic particle of generalized spin
 $(\frac{5}{2},
 \frac{3}{2})$, described by a field ${\Psi}_{\mu\nu,\rho}$ and a set of auxiliary spin-tensors $\varphi_{23|\mu\nu}, \varphi_{25},
 \varphi_{26|\mu,\nu}$, ${\psi}_{15|\mu},
 {\psi}_{17|\mu}, {\psi}_{27|\mu}$.
 The reducible gauge transformations are given by Eqs. (\ref{gtransfPsif})-\-(\ref{gtransfPsif11})
 with the reducibility and constraint conditions in (\ref{0levelr}) and (\ref{algconstr}),
 respectively.

 The action (\ref{S5232aux}) is easily expressed through
 the spin-tensor $\Psi_{\mu\nu,\rho}$ components only,
 in view of Eqs. (\ref{finsol1}), (\ref{finsol11}):
 \begin{eqnarray}
  %%%%%%%%%%%%%%%%%%%%%%%%%%%%%%%%%%%%%
  {\cal{}S}_{(2,1)} &=& \int d^d x \bar{\Psi}_{\mu\nu,\rho}\Bigl\{
{\imath}\gamma^{\tau}\partial_{\tau} {\Psi}^{\mu\nu,\rho} - \eta^{\mu\nu}
{\imath}\gamma^{\tau}\partial_{\tau}\eta^{\sigma\lambda} {\Psi}_{{\sigma\lambda},}{}^{\rho} - \gamma^\nu\gamma^\rho
{\imath}\gamma^{\tau}\partial_{\tau} \gamma^\lambda\gamma^\sigma\Psi^{\mu}{}_{\sigma,\lambda} \nonumber \\ &&
   -
\gamma^\rho\gamma^\nu
{\imath}\gamma^{\tau}\partial_{\tau} \gamma^\sigma\gamma^\lambda\Psi^{\mu}{}_{\sigma,\lambda} + 2 \gamma^\rho
{\imath}\gamma^{\tau}\partial_{\tau} \gamma^\sigma {\Psi}^{\mu\nu}{,}_{\sigma} +
4\gamma^\nu
{\imath}\gamma^{\tau}\partial_{\tau}\gamma_\sigma\Psi^{\mu\sigma,\rho}-
4\imath \partial^\rho  \gamma^\sigma\Psi^{\mu\nu}{,}_{\sigma}
    \nonumber \\
&& - \eta^{\mu\nu}\gamma^\rho
{\imath}\gamma^{\tau}\partial_{\tau}\eta^{\sigma\lambda}\gamma^\alpha {\Psi}_{\sigma\lambda,\alpha}  -4  \imath
\partial^{\{\nu} \gamma^\sigma\Psi^{\mu\}}{}_{\sigma,}{}^{ \rho}  +  4\imath \eta^{\mu\nu}
\bigl(\partial_{\sigma} \gamma^\lambda\Psi_\lambda{}^{\sigma,\rho}  -   \partial^{\rho} \gamma^\sigma\eta^{\tau\alpha}
\Psi_{\tau\sigma,\alpha} \bigr) \nonumber \\
&&
+ 4\imath \gamma^\nu\gamma^\rho \partial_{\sigma} \gamma^\tau\Psi^{\mu\sigma}{}_{,\tau} +  4\imath \gamma^\nu\gamma^\rho \partial^{\mu}  \eta^{\sigma\lambda}\gamma^\alpha {\Psi}_{\sigma\alpha,\lambda}
 + 4\imath \gamma^\rho\gamma^\nu\partial_{\sigma} \gamma^\alpha {\Psi}_{\alpha}{}^{\mu,\sigma}
 \Bigr\}, \label{expL5/23/2fint}
 \end{eqnarray}
and is invariant with respect to the reducible gauge
transformations
\begin{align}
    & \delta\Psi_{\mu\nu,\rho}  = \frac{\imath}{2}\partial_\rho\xi_{\{\nu,\mu\}}- \frac{\imath}{2}\partial_{\{\nu}\xi_{\mu\},\rho},
    & \delta\xi_{\mu,\nu}  =  \imath \partial_\nu \xi^{(1)}_\mu, && \bigl(\xi_{\mu,\nu}, \xi^{(1)}_\mu\bigr) \equiv \bigl( \psi^0_{26|\mu,\nu},
    \psi^{(1)}{}^{0}_{15|\mu}\bigr),  \label{gtransffin}
\end{align}
subject to the relation $ \gamma^\mu\xi_{\mu,\nu} = \imath
\partial_\nu \xi  $, with an arbitrary spinor $\xi$.

We have thus obtained the gauge-invariant Lagrangian
(\ref{expL5/23/2fint}) only in terms of the initial free massless
mixed-symmetric spin-tensor field $\Psi_{\mu\nu,\rho}$. The
resulting theory is a first-stage reducible gauge theory. The
formulae (\ref{expL5/23/2fint}), (\ref{gtransffin}) present our
basic result of Section~\ref{ex5232}.

In view of the above result (see also the result for the
antisymmetric spin-tensor $\Psi_{[\mu,\nu]}$ in
\cite{mixfermiflat}), it should be noted that any gauge-invariant
unconstrained Lorentz-covariant Lagrangian formulation for the
fermionic mixed-symmetric spin-tensor
$\Psi_{(\mu^1)_{n_1},(\mu^2)_{n_2},...,(\mu^k)_{n_k}}$ in
Minkowski space realized only in terms of the initial spin-tensor
$\Psi_{(\mu^1)_{n_1},(\mu^2)_{n_2},...,(\mu^k)_{n_k}}$ (i.e.,
without any auxiliary fields) must possess a reducible gauge
symmetry transformation in such a way that the stage of
reducibility should be equal to $(k-1)$ for a field corresponding
to a Young tableaux $Y(s_1,\ldots, s_k)$. This obvious property is
in contradiction with the Lagrangian formulation suggested for
fermionic fields in the ``metric-like'' formalism \cite{Siegel} in
a flat space-time, where the resulting Lagrangian formulation for
any spin-tensor is an irreducible gauge theory\footnote{The author
is grateful to W.Siegel for an explanation of the peculiarities of
the formalism of \cite{Siegel}, which, unfortunately, was not
provided by any explicit example of Lagrangians for
mixed-symmetric spin-tensors, and, in addition, is based on the
hypothesis that all the algebraic gamma-traceless and
mixed-symmetry constraints for any initial spin-tensor be resolved
before the Lagrangian is derived.}. A similar conclusion can be
immediately enlarged for bosonic mixed-symmetric fields in
Minkowski space subject to $Y(s_1,\ldots, s_k)$, whose
unconstrained Lagrangian formulation was suggested in our previous
paper \cite{BRmixbos}. Indeed, the Lagrangian formulation for a
fourth-rank tensor $\Phi_{\mu\nu,\rho,\sigma}$ suggested therein
is a second-stage reducible gauge theory, whereas a general
unconstrained Lagrangian for any mixed-symmetric tensor in
\cite{Siegel1} has independent gauge parameters only; meanwhile,
no Lagrangian for a specific mixed-symmetric tensor was presented.

\subsection{Spin $(\frac{5}{2},\frac{3}{2})$ Mixed-symmetric Massive Spin-tensor} \label{ex5232m}

To obtain a Lagrangian description of a massive
 rank-3 mixed-symmetric spin-tensor $\Psi_{\mu\nu,\rho}$, having
 the Young tableaux $\begin{array}{|c|c|}\hline%\vphantom{\biggm|}
  \!\mu \!&\! \nu\!  \\
   \hline%\vphantom{\biggm|}
    \! \rho\!   \\
  \cline{1-1}
\end{array}\
$ and subject to conditions
 (\ref{Eq-1}), (\ref{Eq-2}) and equation (\ref{Eq-0m}),
 ($(\imath\gamma^{\mu}\partial_{\mu}-m)\Psi_{\mu\nu,\rho}(x)
 =0$, instead of (\ref{Eq-0}), we can follow, in part, the example
 of a massive second-rank antisymmetric spin-tensor from Ref.
 \cite{mixfermiflat},
 and apply the prescription
 (\ref{reduction}), (\ref{reduction1}), starting directly from the
 massless Lagrangian formulation with the action
 (\ref{expL5/23/2fint}) in a $(d+1)$-dimensional Minkowski space.

First, we have the following representation for the field and
gauge parameters:
\begin{eqnarray}
 \Psi^{MN,P}&=&  \bigl({\Psi}^{\mu\nu,\rho}, {\Psi}^{\mu\nu,d}, {\Psi}^{\mu{}d,\rho}, {\Psi}^{\mu{}d,d}, {\Psi}^{d{}d,\rho},
 {\Psi}^{d{}d,d}\bigr) \,, \label{red5232f}\\
 \xi_{M,N} &=& \bigl(\xi_{\mu,\nu}, \xi_{\mu,d}, \xi_{d,\nu}, \xi_{d,d}\bigr)   \,, \qquad  \xi^{(1)}_M \ =\
 \bigl(\xi^{(1)}_\mu, \xi^{(1)}_d\bigr) . \label{red5232gg}
\end{eqnarray}
 The Young symmetry condition for the spin-tensor $\Psi^{MN,P}$,
 i.e., $\Psi^{\{MN,P\}} = 0$, implies that after a projection onto
 $\mathbb{R}^{1,d-1}$ the $d$-dimensional spinor ${\Psi}^{d{}d,d}
 = \frac{1}{3}{\Psi}^{\{d{}d,d\}}$ vanishes, and there hold the following
 properties of the remaining projected spin-tensors (\ref{red5232f}):
\begin{equation}\label{projPsi}
  {\Psi}^{d{}d,\mu}\ = \ -2{\Psi}^{\mu{}d,d}\,, \qquad   {\Psi}^{\mu\nu,d} = - {\Psi}^{d{}\{\mu,\nu\}}\,, \qquad   {\Psi}^{\{\mu\nu,\rho\}}\ =\ 0.
\end{equation}
 Therefore, the total configuration space contains one third-rank
 massive mixed-symmetric spin-tensor ${\Psi}^{\mu\nu,\rho}$, two
 second-rank symmetric ${\varphi}_1^{\mu\nu}$,
 ${\varphi}_1^{\mu\nu}
 \equiv {\Psi}^{\mu\nu,d}$, and antisymmetric
 ${\varphi}_2^{[\mu\nu]}$, ${\varphi}_2^{[\mu\nu]} \equiv
 {\Psi}^{d{}[\mu,\nu]}$, spin-tensors and one first-rank
 spin-tensor ${\Psi}^{\mu}$, ${\Psi}^{\mu} \equiv
 {\Psi}^{d{}d,\mu}$. Three final spin-tensors play the role of
 Stueckelberg fields. The set of gauge parameters
 (\ref{red5232gg}) consists of one second-rank spin-tensor
 $\xi_{\mu,\nu}$, two first-rank spin-tensors $\xi_{\mu},
 \zeta_{\mu}$; $\xi_{\mu,d} \equiv  \xi_{\mu}; \xi_{d,\mu} \equiv
 \zeta_{\mu}$, and one spinor $\xi$, $\xi \equiv \xi_{d,d}$.

 Second, the corresponding action can be obtained from (\ref{expL5/23/2fint})
 by dimensional projection $\mathbb{R}^{1,d} \to
 \mathbb{R}^{1,d-1}$, and must be invariant with respect to the
 gauge transformations
\begin{align}
    & \delta{\Psi}^{\mu\nu,\rho}\hspace{1ex} =  \hspace{1ex}
    \frac{\imath}{2}\partial^\rho\xi^{\{\nu,\mu\}}- \frac{\imath}{2}\partial^{\{\nu}\xi^{\mu\},\rho}\,,
&&\delta{\varphi}_1^{\mu\nu}\hspace{1ex} = \hspace{1ex}\frac{m}{2}\xi^{\{\mu,\nu\}}+ \frac{\imath}{2}\partial^{\{\nu}\xi^{\mu\}}
    \,,
    \label{psi[munu]rhored1}\\
   &  \delta{\varphi}_2^{[\mu\nu]}\hspace{1ex} = \hspace{1ex}-\frac{\imath}{2}\partial^{[\nu}\bigl(\xi^{\mu]}- 2\zeta^{\mu]} \bigr)
   - \frac{m }{2}\xi^{[\mu,\nu]}\,, && \delta{\Psi}^{\mu} \hspace{1ex} = \hspace{1ex} \imath\partial^{\mu}\xi + m\zeta^\mu,   \label{psi[munu]rhored2}
\end{align}
which, in turn, are reducible:
\begin{align}
    & \delta\xi_{\mu,\nu}\hspace{1ex} = \hspace{1ex}\imath \partial_\nu \xi^{(1)}_\mu, && \delta \xi_{\mu}\hspace{1ex} = \hspace{1ex}-m
    \xi^{(1)}_\mu, \label{ximuxi1red1}\\
    & \delta\zeta_{\mu}\hspace{1ex} = \hspace{1ex}\imath \partial_\mu \xi^{(1)}_d
    \,,&& \delta\xi\hspace{1ex} =
\hspace{1ex} - m\xi^{(1)}_d
    \,.
    \label{ximuxi1red2}
\end{align}
Third, due to the specific character of the relation for the
quantity $\tilde{\gamma}T_0\tilde{\gamma}$, identical with $T_0$
for massless HS fields, and, when transformed, for massive fields,
as $\tilde{\gamma}T_0\tilde{\gamma}$ = $T_0^\ast$, $T_0^\ast =
-(\imath\tilde{\gamma}^\mu
\partial_\mu+\tilde{\gamma}m)$, with an odd $\tilde{\gamma}$-matrix
arising within $\tilde{\gamma}^\mu$-matrix. When calculating the
scalar products to obtain the final Lagrangian, we use the
identification
\begin{eqnarray}
   && i\gamma^M\partial_M{\Psi}^{NK,P} =
(i\gamma^\mu\partial_\mu-m){\Psi}^{NK,P},\quad
i\gamma^M\partial_M\gamma_N{\Psi}^{NK,P} =
(i\gamma^\mu\partial_\mu+m)\gamma_N{\Psi}^{NK,P}, \nonumber \\
&& i\gamma^M\partial_M\gamma_N\gamma_K{\Psi}^{NK,P} =
(i\gamma^\mu\partial_\mu-m)\gamma_N\gamma_K{\Psi}^{NK,P}, \ldots \label{reduc5/23/2}
\end{eqnarray}
being true, if instead  ${\Psi}^{NK,P}$ [$\gamma_N{\Psi}^{NK,P}$]
we substitute the quantities $(\gamma_N)^{2k}{\psi}^{NK...}$
[$(\gamma_N)^{2k+1}{\psi}^{NK...}$],
for a non-negative integer $k$.

Then, after removing the gauge parameters $\xi_{\mu}$, $\xi$, in
(\ref{ximuxi1red1}) (\ref{ximuxi1red2}) by shift transformations
with spinors $\xi^{(1)}_\mu$, $\xi^{(1)}_d$, respectively, we
obtain preliminarily an irreducible gauge theory with independent
gauge spin-tensors $\xi_{\mu,\nu}$, $\zeta_{\mu}$. Next, in the
same manner, we can gauge away the spin-tensors ${\Psi}_{\mu}$,
${\varphi}_1^{\mu\nu}$, ${\varphi}_2^{[\mu\nu]}$ in Eqs.
(\ref{psi[munu]rhored1}), (\ref{psi[munu]rhored2}), by means of
gauge transformations with the parameters $\zeta_\mu(x)$,
$\xi^{\{\mu,\nu\}}$ and $\xi^{[\mu,\nu]}$, respectively, so that
the theory becomes a non-gauge one only in terms of a massive
$3$-rd-rank massive mixed-symmetric spin-tensor
${\Psi}^{\mu\nu,\rho}$.

Finally, we obtain the Lagrangian for a massive spin-tensor of
generalized spin $(5/2,3/2)$ field in a $d$-dimensional flat
space:
 \begin{eqnarray}
  %%%%%%%%%%%%%%%%%%%%%%%%%%%%%%%%%%%%%
&&  {\cal{}L}^m(\Psi_{\mu\nu,\rho}) = \bar{\Psi}_{\mu\nu,\rho}\Bigl\{
\bigl({\imath}\gamma^{\tau}\partial_{\tau}-m\bigr) {\Psi}^{\mu\nu,\rho} - \eta^{\mu\nu}
\bigl({\imath}\gamma^{\tau}\partial_{\tau}-m\bigr)\eta^{\sigma\lambda} {\Psi}_{{\sigma\lambda},}{}^{\rho} -
\gamma^\nu\gamma^\rho
\bigl({\imath}\gamma^{\tau}\partial_{\tau} \nonumber \\ &&
\phantom{{\cal{}L}^m(\Psi^\mu{}{\nu,\rho})}  -m\bigr) \gamma^\lambda\gamma^\sigma\Psi^{\mu}{}_{\sigma,\lambda}  -
\gamma^\rho\gamma^\nu
\bigl({\imath}\gamma^{\tau}\partial_{\tau}-m\bigr) \gamma^\sigma\gamma^\lambda\Psi^{\mu}{}_{\sigma,\lambda} + 2 \gamma^\rho
\bigl({\imath}\gamma^{\tau}\partial_{\tau}+ m\bigr) \gamma^\sigma {\Psi}^{\mu\nu}{,}_{\sigma}
    \nonumber \\
&& \phantom{{\cal{}L}^m(\Psi^\mu{}_{\nu,\rho})} +
4\gamma^\nu
\bigl({\imath}\gamma^{\tau}\partial_{\tau}+ m\bigr)\gamma_\sigma\Psi^{\mu\sigma,\rho}-
4\imath \partial^\rho  \gamma^\sigma\Psi^{\mu\nu}{,}_{\sigma} - \eta^{\mu\nu}\gamma^\rho
\bigl({\imath}\gamma^{\tau}\partial_{\tau}+ m\bigr)\eta^{\sigma\lambda}\gamma^\alpha {\Psi}_{\sigma\lambda,\alpha}     \nonumber \\
&& \phantom{{\cal{}L}^m(\Psi_{\mu\nu,\rho})} -4  \imath
\partial^{\{\nu} \gamma^\sigma\Psi^{\mu\}}{}_{\sigma,}{}^{ \rho} +  4\imath \eta^{\mu\nu}
\bigl(\partial_{\sigma} \gamma^\lambda\Psi_\lambda{}^{\sigma,\rho}  -   \partial^{\rho} \gamma^\sigma\eta^{\tau\alpha}
\Psi_{\tau\sigma,\alpha} \bigr)
+ 4\imath \gamma^\nu\gamma^\rho \partial_{\sigma} \gamma^\tau\Psi^{\mu\sigma}{}_{,\tau}\nonumber \\
&& \phantom{{\cal{}L}^m(\Psi_{\mu\nu,\rho})} +  4\imath \gamma^\nu\gamma^\rho \partial^{\mu}  \eta^{\sigma\lambda}\gamma^\alpha {\Psi}_{\sigma\alpha,\lambda}
 + 4\imath \gamma^\rho\gamma^\nu\partial_{\sigma} \gamma^\alpha {\Psi}_{\alpha}{}^{\mu,\sigma}
 \Bigr\}. \label{mexpL5/23/2fint}
 \end{eqnarray}
Summarizing, we have obtained a Lagrangian formulation
(\ref{mexpL5/23/2fint}) only in terms of the initial free massive
mixed-symmetric spin-tensor field $\Psi_{\mu\nu,\rho}$. The
resulting theory is not a gauge theory, and formula
(\ref{mexpL5/23/2fint}) presents our basic result of
Section~\ref{ex5232m}.

%%%%%%%%%%%%%%%%%%%%%%%%%%%%%%%%%%%%%%%%%%%%%%%%%
%%%%%%%%%%%%%%%%%%%%%%%%%%%%%%%%%%%%%%%%%%%%%%%%%
%%%%%%%%%%%%%%%%%%%%%%%%%%%%%%%%%%%%%%%%%%%%%%%%%
%%%%%%%%%%%%%%%%%%%%%%%%%%%%%%%%%%%%%%%%%%%%%%%%%

\section{Conclusion}\label{summary}

In the present work, we have constructed a gauge-invariant
Lagrangian description of free half-integer HS fields, belonging
to an irreducible representation of the Poincare group
$ISO(1,d-1)$ with a corresponding Young tableaux having $k$ rows
in the ``metric-like" formulation. The results of this study are
the most general ones and can be applied in the unified way to
both massive and massless fermionic HS fields with a mixed
symmetry in a Minkowski space of any dimension.

In the standard manner, starting from an embedding of fermionic HS
fields into vectors (Dirac spinors) of an auxiliary Fock space, we
elaborate the fields as coordinates of Fock-space vectors and
reformulate the theory in terms of these objects. The conditions
that determine an irreducible Poincare-group representation with a
given mass and generalized half-integer spin are realized in terms
of differential operator constraints imposed on Fock-space
vectors. These constraints generate a closed Lie superalgebra of
HS symmetry, which contains, with the exception of $k$ basis
generators of its Cartan subalgebra, a system of first- and
second-class odd and even constraints. The above superalgebra
coincides, modulo the isometry group generators, with its
Howe-dual $osp(1|2k)$ orthosymplectic superalgebra.

We show that the construction of a correct Lagrangian description
requires a deformation of the initial symmetry superalgebra in
order to obtain from a system of mixed-class constraints a
converted system with the same number of first-class constraints
alone, whose structure provides the appearance of the necessary
number of auxiliary spin-tensor fields with lower generalized
spins within an opposite alphabetic ordering prescription. It is
demonstrated that this purpose can be achieved with the help of an
additional Fock space, by constructing an additive extension of a
symmetry subsuperalgebra, which consists of a subsystem of
second-class constraints alone and of the generators of the Cartan
subalgebra, which form an invertible even operator supermatrix,
composed of supercommutators of the second-class constraints.

The generalized Verma module construction \cite{Dixmier},
\cite{genVM} has been realized in order to obtain an auxiliary
representation in Fock space for the above superalgebra with
second-class constraints. As a consequence, the converted Lie
superalgebra of HS symmetry has the same algebraic relations as
the initial superalgebra with the only peculiarity that these
relations are realized in an enlarged Fock space. The generators
of the converted Cartan subalgebra contain linearly $k$ auxiliary
independent number parameters $h^i$, whose choice provides the
vanishing of these generators in the corresponding subspaces of
the total Hilbert space extended by the ghost operators in
accordance with the minimal BFV--BRST construction for the
converted HS symmetry superalgebra. Therefore, the above Cartan
generators, enlarged by the ghost contributions up to the
``particle number'' operators in the total Hilbert space,
covariantly determine Hilbert subspaces, in each of which the
converted  symmetry superalgebra consists of the first-class
constraints subsystem only. Each of the systems is labelled by the
values of the above parameters, and constructed from the initial
irreducible Poincare-group relations.

It is demonstrated that the Lagrangian description corresponding
to the BRST operator, which encodes the converted HS symmetry
superalgebra, yields a consistent Lagrangian dynamics for
fermionic fields of any generalized spin after a partial
gauge-fixing procedure, that permits one to gauge away the terms
with higher (second-order) derivatives from the consideration. The
resulting Lagrangian description, realized concisely in terms of
the total Fock space, presents a set of generating relations for
the action and the sequence of reducible gauge transformations for
given fermionic HS fields with a sufficient set of auxiliary
fields, and proves to be a reducible gauge theory with a finite
number of reducibility stages $s$, increasing with both the value
of generalized spin $(n_1+\frac{1}{2},\ldots,n_k+\frac{1}{2})$ and
the number of rows $k$ in the Young tableaux as $s=\sum_{o=1}^k
n_o + k(k-1)/2-1$. The basic results of the present work are given
by relations (\ref{L1}), where the action for a field with an
arbitrary generalized half-integer spin is constructed, as well as
by relations (\ref{GT1})--(\ref{GTi2}), where the gauge
transformations for the fields are presented, along with the
sequence of reducible gauge transformations and gauge parameters.

It has been proved that the solutions of the Lagrangian equations
of motion (\ref{EofM1}), (\ref{EofM2}), as a result of a new
partial gauge-fixing procedure and a resolution of some of the
equations of motion, correspond to the BRST cohomology space with
a vanishing ghost number, which is determined only by the
relations that extract the fields of an irreducible Poincare-group
representation with a given value of half-integer generalized
spin. One should notice that the case of totally antisymmetric
spin-tensors developed in Ref.\cite{brst1} is contained in the
general Lagrangian formulation for $s_1=s_2=...=s_k=\frac{3}{2}$,
$k=[(d-1)/2]$.

As examples demonstrating the applicability of the general scheme,
it is shown that it contains as a particular case the Lagrangian
formulation for a mixed-symmetric spin-tensors subject to a Young
tableaux with two rows, first developed in \cite{mixfermiflat}, as
well as the new unconstrained Lagrangian formulation in
(\ref{EofM123})--(\ref{L123}) for mixed-symmetry fermionic HS
fields with three groups of symmetric indices subject to a Young
tableaux with three rows, obtained in the literature for the first
time. We apply the above algorithm to obtain, first, a new
gauge-invariant Lagrangian (\ref{expL5/23/2fint}) and its
reducible gauge transformations (\ref{gtransffin}) for a massless
field of spin $(5/2,3/2)$, and, second, a new Lagrangian
(\ref{mexpL5/23/2fint}) for a massive field of spin $(5/2,3/2)$
only in terms of the corresponding initial spin-tensors of the
third rank. In principle, these results permit one to enlarge the
obtained Lagrangian formulations to those for an HS spin-tensor of
spin $(\frac{5}{2},\frac{3}{2},\ldots,\frac{3}{2})$ characterized
by $k$ rows in the corresponding Young tableaux.

Concluding, one should note that there are many ways to extend the
results obtained in this paper. We will outline only some of them.
First, the development of a Lagrangian construction for bosonic
and fermionic fields with an arbitrary index symmetry in AdS
space, along the lines of Ref.\cite{BRmixads}. Second, the
derivation of component Lagrangians for new simple cases. Third,
the development of the unconstrained formulation for fermionic
fields with an arbitrary Young tableaux similar to the component
formulation with the minimal number of auxiliary fields given in
\cite{quartmixbosemas} for totally symmetric spin-tensor fields
which (as shown \cite{quartmixbosemas} in the case of bosonic
fields) can also be derived from the obtained general Lagrangian
formulation by means of a partial gauge-fixing procedure. Fourth,
the derivation from an unconstrained formulation of a constrained
Lagrangian formulation for arbitrary fermionic fields in a flat
space-time (as well as those for bosonic mixed-symmetric HS fields
starting from the unconstrained formulation in \cite{BRmixbos}).
Fifth, the formulation of a diagrammatic technique within the
BRST--BFV approach, where the space-time variables $x^\mu$ should
be considered on equal footing with the total Fock space
variables, and all the field-antifield content has to be
determined in terms of Fock-space vectors. Finally, a consistent
deformation of the latter construction applied to bosonic and
fermionic mixed-symmetric HS fields will permit one to construct
an interacting theory with mixed-symmetry fermionic HS fields,
including the case of curved (AdS) backgrounds, following in part to the way suggested in \cite{BarnichHenneaux1}. We are going to
develop a research of these problems in our forthcoming work.

\section*{Acknowledgements}
The author is grateful to I.L. Buchbinder for numerous stimulating
discussions at all stages of the present research. He also thanks
Yu.M. Zinoviev and V.A. Krykhtin for discussions concerning the
examples, as well as P.M. Lavrov, K. Stepanyantz and V. Gershun
for their comments on the conversion procedure and the structure
of constraints. He is grateful to M.A. Vasiliev and V. Dobrev for
discussions on the difference of generalized Verma modules from
the true Verma module, as well as to V. Mazorchuk for remarks on
the generalized Verma module structure, to R. Rahman for
correspondence, to P. Moshin for careful reading of the text and to V, Tolstoy for the comments on classification of the HS symmetry superalgebra. The author is grateful to the RFBR grant, project
No. 12-02-000121, and the grant for LRSS, project No. 224.2012.2,
for partial support. The study was supported by The Ministry of education and science of Russian
Federation, project 14.B37.21.0774.
\appendix
\section*{Appendix}

\section{Construction of Additional Parts for $osp(1|2k)$ Superalgebra}\label{addalgebra}
\renewcommand{\theequation}{\Alph{section}.\arabic{equation}}
\setcounter{equation}{0}

Here, we describe the method of constructing an auxiliary
representation (known by mathematicians as the generalized Verma
module \cite{Dixmier}, \cite{genVM}, see Appendix~\ref{defVerma}
for definitions) for the orthosymplectic superalgebra $osp(1|2k)$
with the second-class constraints $\{o'_a, {o'}^+_a \} = \{t'_{i},
l'_{ij}, t^{\prime ij}, t^{\prime +}_{i}$, $l^{\prime +}_{ij},
t^{\prime +}_{ij}\}$ and the Cartan subalgebra elements
$g_0^{\prime i}$, having in mind the identification of $osp(1|2k)$
elements and those of the HS symmetry algebra $\mathcal{A}^f(Y(k),
\mathbb{R}^{1,d-1})$, given by Eqs.(\ref{osp2nhssa}).

Following the Poincare--Birkhoff--Witt theorem, we start with
constructing the generalized Verma module based on the Cartan
decomposition of $osp(1|2k)$ ($i\leq j$, $l<m$, $i,j, l, m =
1,...,k$)
\begin{equation}\label{Cartandecomp}
    osp(1|2k) =  \{t^{\prime +}_{i}, l^{\prime +}_{ij},
t^{\prime+}_{lm}\} \oplus \{g_0^{\prime i}\} \oplus \{t^{\prime
}_{i}, l^{\prime }_{ij}, t'_{lm}\} \equiv \mathcal{E}^-_k\oplus
H_k \oplus\mathcal{E}^+_k.\footnotemark
\end{equation}
\footnotetext{We may examine $osp(1|2k)$ in the Cartan--Weyl basis
for a unified description, however, without loss of generality,
the basis elements of the algebra under consideration will be
chosen as in Tables~\ref{table in},~\ref{table inodd}.}

We emphasize, first, that for the $sp(2k)$ subalgebra in
$osp(1|2k)$ the Verma module $V(sp(2k))$ was constructed in Ref.
\cite{BRmixbos}, second, in contrast to the case of
 totally-symmetric fermionic HS fields on $\mathbb{R}^{1,d-1}$,
 the negative root vectors in  $\mathcal{E}^-_k$ do not commute
 for $k \geq 2$ (see,
Refs. \cite{symferm-flat}, \cite{mixfermiflat}). However, we
examine the highest-weight representation of the orthosymplectic
algebra $osp(1|2k)$ with the highest-weight vector
$|0\rangle_V$,\footnote{Despite the fact that, in general, the
generalized Verma module can be generated by more than one vector
from the (non-diagonalizable by $H_k$-elements) representation
space, we will have a Verma module structure for the superalgebra
$osp(1|2k)$, see Appendix~\ref{defVerma}}, which should be
annihilated by the positive odd and even roots $(E^{\alpha^0_i},
E^{\alpha^1_i})\in \mathcal{E}^+_k$, and being a proper one for
the Cartan elements $g_0^i$,
\begin{align}\label{hwrep}
& E^{\alpha^0_i}|0\rangle_V =0,  && E^{\alpha^1_i}|0\rangle_V =0, && g_0^i |0\rangle_V =
h^i|0\rangle_V.
\end{align}
The general vector of the generalized Verma module $V(osp(1|2k))$,
written concisely as $|\vec{N}^f\rangle_V = |\vec{n}_l^0;
\vec{N}\rangle_V $, has -- in terms of occupation numbers with the
help of the general vector $|\vec{N}\rangle_V = |{\vec{n}}_{ij},
\vec{p}_{rs}\rangle_V$ -- the form of the Verma module $V(sp(2k))$
\cite{BRmixbos},
 \begin{equation} \label{GVMK}
 \left|\vec{n}_l^0; {\vec{N}} \rangle_V \right. \hspace{-0.2em}= \hspace{-0.1em} \left|
{n}_1^0,...,{n}_k^0;{{n}}_{11},...,{{n}}_{1k}, n_{22},...,{{n}}_{2k},...,{{n}}_{kk};
{p}_{12},\ldots, {p}_{1k}, {p}_{23},\ldots,
{p}_{2k},\ldots,p_{k-1k}\rangle_V \right.,
\end{equation}
where the non-negative integers ${n}_l^0 \in \mathbb{Z}_2$,
$(n_{ij}, p_{rs})\in \mathbb{N}_0$ mean the exponentials of the
corresponding negative root vectors $(E^{\alpha^0_i},
E^{\alpha^1_i})$, determined in a fixed ordering as
\begin{equation}\label{VM}
   |\vec{N}^f\rangle_V \equiv  \prod_{l=1}^k\bigl(t^{\prime
+}_{l}\bigr){}^{n^0_{l}}\prod_{i,j=1, i\leq j}^k\bigl(l^{\prime
+}_{ij}\bigr){}^{n_{ij}}\prod_{r=1}^{k-1}\Bigr[\prod_{s=r+1}^k \bigl(t^{\prime
+}_{rs}\bigr){}^{p_{rs}}\Bigl] |0\rangle_V.
\end{equation}
The action of the odd negative root vectors $t^{\prime  +}_i $ and
the Cartan generators $g_{0{}i}^{\prime }$ on
$|\vec{N}^f\rangle_V$ can be immediately found (for
$\left[\frac{n_{i'}^0+1}{2}\right] = 1(0)$ when $n_{i'}^0=1(0)$)
as follows:
\begin{eqnarray}\label{t'+i}
 t^{\prime  +}_{i'} |\vec{N}^f\rangle_V & = & \textstyle(-1)^{\sum\limits_{l'=1}^{i'-1}n_l'^0}\left(1+\left[\frac{n_{i'}^0+1}{2}\right]\right)
 \bigl|\vec{n}_l^0 + \delta_{i'l}{} mod{} 2; \vec{n}_{ij} + \delta_{i'i',ij}\left[\frac{n_{i'}^0+1}{2}\right]
 , {\vec{p}}_{rs}\bigr\rangle_V
 \\
 && +4\sum_{l'=1}^{i'-1}\textstyle (-1)^{\sum_{m=1}^{l'-1}n_{m}^0}n_{l'}^0
 \bigl|\vec{n}_l^0 - \delta_{l'l}; \vec{n}_{ij}
 + \delta_{l'i',ij},{\vec{p}}_{rs}\bigr\rangle_V \,,
 \nonumber \\
 g_{0{}i}^{\prime } |\vec{N}^f\rangle_V & = & \left(n_i^0 + 2n_{ii} +  \sum_{l\neq i} n_{il}  - \sum_{s>i}p_{is}+\sum_{r<i}p_{ri} + h^i\right)
 \left|\vec{N}^f\rangle_V
 \right.\,.\label{g'0i}
\end{eqnarray}
whereas the action of even negative root vectors,
$l^{\prime+}_{ij}, t^{\prime+}_{rs}$ on $|\vec{N}^f\rangle_V$ has
the form
 \begin{eqnarray}
l^{\prime+}_{i'j'}|\vec{N}^f\rangle_V & = & \left|\vec{N}^f +
\delta_{i'j',ij}\rangle_V
 \right. \,,
 \label{l'+ij} \\
 \label{t'+rs}
 t^{\prime+}_{r's'}  |\vec{N}^f\rangle_V & = & \left|\vec{n}_l^0;{\vec{n}}_{ij},
{\vec{p}}_{rs} + \delta_{r's',rs} \rangle_V \right. -
\sum_{k'=1}^{r'-1}p_{k'r'}\left|\vec{n}_l^0;{\vec{n}}_{ij},
{\vec{p}}_{rs} - \delta_{k'r',rs}+ \delta_{k's',rs} \rangle_V \right.  \\
 &&  -\delta_{lr'}{n}_{r'}^0\Biggl[4\sum_{n'=r'+1}^{s'-1}{n}_{n'}^0(-1)^{
 \sum\limits_{k'=r'+1}^{n'-1}n_{k'}^0}
 \left|\vec{n}_l^0-\delta_{lr'}-\delta_{ln'}; \vec{n}_{ij}+\delta_{n's',ij},
 {\vec{p}}_{rs}
 \right\rangle_V \nonumber\\
 && +(-1)^{\sum\limits_{k'=r'+1}^{s'-1}n_{k'}^0}
 \hspace{-0.15em}\textstyle\left(1+\left[\frac{n_{s'}^0+1}{2}\right]\hspace{-0.15em}\right)
\hspace{-0.15em}\left|\vec{n}_l^0-\delta_{lr'}+\delta_{ls'}mod{}2;
\vec{n}_{ij}+\delta_{s's',ij}\left[\frac{n_{s'}^0+1}{2}\right],
 {\vec{p}}_{rs}
 \right\rangle_V\hspace{-0.15em}\Biggr]
 \nonumber \\
  && - \sum_{k'=1}^{k}(1+\delta_{k'r'})n_{r'k'}\left|\vec{N}^f-\delta_{r'k',ij}  + \delta_{k's',ij}\rangle_V
 \right.\,.
\nonumber
 \end{eqnarray}
Notice that in Eqs.(\ref{t'+i})--(\ref{t'+rs}) we have used such
notation, e.g., for the vector $\left|\vec{N}^f +
\delta_{i'j',ij}\rangle_V\right.$ in the Eq.(\ref{l'+ij}), that is
subject to definition (\ref{GVMK}), increasing only the coordinate
$n_{ij}$ in the vector $|\vec{N}^f\rangle_V$, for $i=i', j=j'$, by
a unit with unchanged values of the remaining ones, whereas the
vector $\left|\vec{n}{}^0_{l};{\vec{n}}_{ij}, {\vec{p}}_{rs} -
\delta_{k'r',rs}+ \delta_{k's',rs} \rangle_V \right.$ implies
increasing the coordinate $p_{rs}$, for $r=k', s=s'$, by a unit
and decreasing by one unit the coordinate $p_{rs}$, for $r=k',
s=r'$, with unchanged values of the remaining coordinates in
$|\vec{N}^f\rangle_V$.

Derivation of relations (\ref{t'+i}), (\ref{g'0i}), (\ref{t'+rs})
is based on the the algebraic relations for $osp(1|2k)$ from
Tables~\ref{table in}, \ref{table inodd} and the formula for the
product of the graded operators $A$, $B$, $n\geq 0$,
\begin{eqnarray}
\label{product} &&    AB^n = \sum^{n}_{k=0} (-1)^{\varepsilon(A)\varepsilon(B)(n-k)}C^{(s)}{
}^n_k
B^{n-k}\mathrm{ad}^k_B{}A\,,   \   \mathrm{ad}^k_B{}A=
[[...[A,\stackrel{ k{\,} {\rm times}}{ \overbrace{B\},...\},B}\}},
\end{eqnarray}
with $s = \varepsilon(B)$, and the generalized coefficients for a
number of graded combinations, $C^{(s)}{ }^n_k$ (first introduced
in \cite{mixfermiflat}; for details, in particular, concerning the
odd Pascal triangle, see \cite{0905.2705}) coinciding with the
standard ones only for the bosonic operator $B$: ${C^{(0)}{}^n_k}
= C^n_k = \frac{n!}{k!(n-k)!}$. The coefficients are defined
recursively, by the relations
\begin{eqnarray}\label{combination}
&&    C^{(s)}{}^{n+1}_{k} = (-1)^{s(n+k+1)}C^{(s)}{ }^n_{k-1} +
C^{(s)}{}^n_k \,, \qquad
 n, k \geq 0\,,\\
&&  C^{(s)}{}^{n}_{0} = C^{(s)}{}^{n}_{n}=1\,, \qquad
C^{(s)}{}^{n}_{k}=0\,, \ n<k\,, s=0,1
\end{eqnarray}
and possess the properties $C^{(s)}{}^n_k=C^{(s)}{}^n_{n-k}$. The
corresponding values of $C^{(1)}{}^n_k$ are given, for $n\geq k$,
by the formulae
\begin{equation}\label{expressions}
C^{(1)}{}^n_k =
\sum^{n-k+1}_{i_k=1}\sum^{n-i_k-k+2}_{i_{k-1}=1}\ldots \sum^{n-
\sum^{k}_{j=3}i_j-1}_{i_2=1} \sum^{n-\sum^{k}_{j=2}i_j}_{i_1=1}
(-1)^{k(n+1) +
\sum\limits^{[(k+1)/2]}_{j=1}\left(i_{2j-1}+1\right)},
\end{equation}
which follow by induction. For our purposes, due to $n^0_k = 0,1$
in (\ref{GVMK}), (\ref{VM}), it is sufficient to know that
$C^{(1)}{}^0_0 = C^{(1)}{}^1_0 = 1$ and $C^{(1)}{}^{n^0_l}_1 =
n^0_l$. Second, as was shown in \cite{BRmixbos},
Eq.(\ref{product}) permits one to find both the identities
\begin{equation}\label{identllm}
t^{\prime }_{l'}\left|\vec{0}_l^0;{\vec{0}}_{ij},
{\vec{p}}_{rs}\rangle_V \right. = 0, \qquad
l^{\prime }_{i'j'} \left|\vec{0}{}^0_{l};{\vec{0}}_{ij}, {\vec{p}}_{rs}\rangle_V
\right. = 0
\end{equation}
and the equation in the action of the positive ``mixed-symmetry''
root vectors $ t^{'}_{r's'}$ on the vector
$|\vec{0}{}^0_{l};\vec{0}_{ij}, \vec{p}_{rs}\rangle_V$ (due to the
non-commutativity of the negative ``mixed-symmetry'' root vectors
$t^{\prime +}_{rs}$ among each other) in the form
\begin{eqnarray}
 \label{t'recurr}
  t^{\prime}_{r's'}
|\vec{0}{}^0_{l};\vec{0}_{ij},\vec{p}_{rs}\rangle_V &=&
\left|C^{r's'}_{\vec{p}_{rs}}\rangle_V
 \right.-
\sum_{n'=1}^{l'-1}p_{n's'}\left|\vec{0}{}^0_{l}; \vec{0}_{ij},
\vec{p}_{rs}-\delta_{n's',rs}+\delta_{n'r',rs}\rangle_V
 \right. \nonumber\\
  && + \sum_{k'=r'+1}^{s'-1}p_{r'k'}\Bigr[\prod_{l'< r', m'>l'}\prod_{l'=r', s'> m'>l'} \bigl(t^{\prime
+}_{l'm'}\bigr){}^{p_{l'm'}-\delta_{r'k',l'm'}}\Bigl]
t'_{k's'}\nonumber\\
&& \times\prod_{q'= r', t'\geq s'}\prod_{q'> r', t'>q'}
\bigl(t^{\prime +}_{q't'}\bigr){}^{p_{q't'}} \left|0\rangle_V
 \right.,
\end{eqnarray}
with the vector $ \left|C^{r's'}_{\vec{p}_{rs}}\rangle_V
 \right.$, $r'<s'$ determined as follows:
\begin{eqnarray}\label{Clmin}
 \left|C^{r's'}_{\vec{p}_{rs}}\rangle_V
 \right. &=& p_{r's'}\Big(h^{r'}-h^{s'}-\sum_{k'=s'+1}^{k}(p_{r'k'}+p_{s'k'})+\sum_{k'=r'+1}^{s'-1}p_{k's'}-p_{r's'}+1\Big)\times
 \nonumber
\\
&& \times \left|\vec{0}{}^0_{l};
\vec{0}_{ij}, \vec{p}_{rs}-\delta_{r's',rs}\rangle_V
 \right.  + \sum_{k'=s'+1}^{k}p_{r'k'}\Bigl\{\left|\vec{0}{}^0_{l}; \vec{0}_{ij},
\vec{p}_{rs}-\delta_{r'k',rs}+\delta_{s'k',rs}\rangle_V
 \right. \nonumber\\
  && - \sum_{n'=r'+1}^{s'-1} p_{n's'}\left|\vec{0}{}^0_{l};\vec{0}_{ij},
\vec{p}_{rs}-\delta_{r'k',rs}-\delta_{n's',rs}+\delta_{n'k',rs}\rangle_V
 \right. \Bigr\} .
 \end{eqnarray}
Since the  recurrent relation (\ref{t'recurr}) has exactly the
same form as  one in  case of  symplectic algebra
$sp(2k)$ in Ref. \cite{BRmixbos}, we have used the known solution
of (\ref{t'recurr}) in the form (for $k'_{-1}\equiv 1$)
\begin{eqnarray}
 \label{t'fin}
  t^{\prime}_{r's'}
|\vec{0}{}^0_{l};\vec{0}_{ij},\vec{p}_{rs}\rangle_V
&=&\sum_{p=0}^{s'-r'-1}\bigg[\sum_{k'_1=r'+1}^{s'-1}\ldots
\sum_{k'_p=r'+p}^{s'-1}\prod_{j=1}^{p}p_{k'_{j-1}k'_{j}}
\Big\{ \left|C^{k'_{p}s'}_{\vec{p}_{rs}-\sum_{j=1}^{p}\delta_{k'_{j-1}k'_j,rs}}\rangle_V
 \right.
   \nonumber\\
 &-&   \hspace{-1em}
\sum_{n'=k'_{p-1}}^{k'_p-1}\hspace{-0.25em}p_{n's'}\left| \vec{0}_{ij},
\vec{p}_{rs}-\hspace{-0.1em}\sum_{j=1}^{p}\delta_{k'_{j-1}k'_j,rs}-\delta_{n's',rs}+\delta_{n'k'_p,rs}\rangle_V
  \right.\Big\}\bigg] , \   k'_{0}\equiv  r'.
 \end{eqnarray}
Therefore, the final result for the action of $t^{\prime}_{r's'}$
on a vector $|\vec{N}^f\rangle_V$ can be written as follows:
   \begin{eqnarray}
\label{t'lm}
  t^{\prime}_{r's'}
|\vec{N}^f\rangle_V &=&
\sum_{p=0}^{s'-r'-1}\bigg[\sum_{k'_1=r'+1}^{s'-1}\ldots
\sum_{k'_p=r'+p}^{s'-1}\prod_{j=1}^{p}p_{k'_{j-1}k'_{j}}
\Big\{ \left|C^{k'_{p}s'}_{\vec{n}_l^0; \vec{n}_{ij},\vec{p}_{rs}-\sum_{j=1}^{p}\delta_{k'_{j-1}k'_j,rs}}\rangle_V
 \right.
   \\
 &&   -
\sum_{n'=k'_{p-1}}^{k'_p-1}\hspace{-0.25em}p_{n's'}\left| \vec{n}_l^0; \vec{n}_{ij},
\vec{p}_{rs}-\hspace{-0.1em}\sum_{j=1}^{p}\delta_{k'_{j-1}k'_j,rs}-\delta_{n's',rs}+\delta_{n'k'_p,rs}\rangle_V
  \right.\Big\}\bigg]
\nonumber\\
  && -\sum_{k'=1}^{k}(1+\delta_{k's'})n_{k's'}
\hspace{-0.2em}\left|\vec{n}_l^0;
\vec{n}_{ij}-\delta_{k's',ij}+\delta_{k'r',ij},
\vec{p}_{rs}\rangle_V
 \right.
 \nonumber\\
 && - n_{s'}^0\Biggl[4\sum_{k'=r'+1}^{s'-1}n^0_{k'} (-1)^{\sum\limits_{n'=k'+1}^{s'-1}n^0_{n'}}\left|\vec{n}_l^0 -  \delta_{lk'} - \delta_{ls'}; \vec{n}_{ij} + \delta_{k'r',ij},
 \vec{p}_{rs}\right\rangle_V\nonumber\\
 && + (-1)^{\sum\limits_{k'=r'+1}^{s'-1}n^0_{k'}}\textstyle \hspace{-0.15em}\left(1+\left[\frac{n_{r'}^0 +
1}{2}
 \right]\hspace{-0.15em}\right)\hspace{-0.15em}\left|\vec{n}_l^0 +  \delta_{lr'}{} mod{} 2 - \delta_{ls'}; \vec{n}_{ij} + \delta_{r'r',ij}\hspace{-0.15em}\left[\frac{n_{r'}^0+1}{2}\right]\hspace{-0.1em},
 \vec{p}_{rs}\right\rangle_V\hspace{-0.15em}\Biggr],
 \nonumber
\end{eqnarray}
where the vectors $\left|C^{k'_{p}s'}_{\vec{n}_l^0; \vec{n}_{ij},
\vec{p}_{rs}-\sum_{j=1}^{p+1}\delta_{k'_{j-1}k'_j,rs}}\rangle_V
 \right.$, have the same structure as in Eqs.(\ref{Clmin}),
 (\ref{t'fin}),
 where the only substitution
 $(\vec{0}_l^0; \vec{0}_{ij}) \to (\vec{n}_l^0; \vec{n}_{ij})$ should be made
 in $\left|C^{k'_{p}s'}_{\vec{p}_{rs}-\sum_{j=1}^{p+1}\delta_{k'_{j-1}k'_j,rs}}\rangle_V
 \right.$.

Then, it is not difficult to obtain the action of the odd
positive-root operators $E^{\prime\alpha^0}$ on the vector
$|\vec{N}^f\rangle_V$ in the form
\begin{eqnarray}
   \label{t'i}
 t^{\prime }_{i'} |\vec{N}^f\rangle_V &=&\hspace{-0.35em}
2\sum_{k'=1}^{i'-1}n_{k'}^0(-1)^{\sum\limits_{l'=1}^{k'-1}n_{l'}^0}\Biggl\{
   \sum_{p=0}^{i'-k'-1}\hspace{-0.2em}\bigg[\sum_{k'_1=k'+1}^{i'-1}\hspace{-0.2em}\ldots \hspace{-0.2em}\sum_{k'_p=k'+p}^{i'-1}
  \hspace{-0.2em} \Big(
 \left|C^{k'_{p}i'}_{\vec{N}{}^f-\delta_{k'l}-\sum_{j=1}^{p}\delta_{k'_{j-1}k'_j,rs}}\rangle_V
 \right.\\
 &&
 \hspace{-0.9em} -\hspace{-0.3em}
\sum_{n'=k'_{p-1}}^{k'_p-1}\hspace{-0.2em}p_{n'i'}\left|\vec{n}_l^0-\delta_{k'l};
\vec{n}_{ij},
\vec{p}_{rs}-\sum_{j=1}^{p}\delta_{k'_{j-1}k'_j,rs}-\delta_{n'i',rs}+\delta_{n'k'_p,rs}\rangle_V
 \right.\hspace{-0.2em}\Big)\prod_{j=1}^pp_{k'_{j-1}k'_{j}} \bigg]\nonumber\\
  && - \sum_{l'=1}^{k}(1+\delta_{l'i'})
n_{l'i'}\hspace{-0.2em}\left|\vec{n}_l^0-\delta_{k'l};
\vec{n}_{ij}-\delta_{l'i',ij}+\delta_{l'k',ij},
  \vec{p}_{rs}\rangle_V
 \right.
\nonumber\\
 && - n_{i'}^0\Bigl[4\sum_{n'=k'+1}^{i'-1}n^0_{n'} (-1)^{\sum\limits_{m'=n'+1}^{i'-1}n^0_{m'}}\left|\vec{n}_l^0 -  \delta_{ln'} - \delta_{li'}- \delta_{lk'}; \vec{n}_{ij} + \delta_{n'k',ij},
 \vec{p}_{rs}\right\rangle_V\nonumber\\
 && \qquad + (-1)^{\sum\limits_{m'=k'+1}^{i'-1}n^0_{m'}}\left|\vec{n}_l^0
  - \delta_{li'}; \vec{n}_{ij} ,
 \vec{p}_{rs}\right\rangle_V\Bigr]
  \Biggr\}\nonumber
  \end{eqnarray}
  \begin{eqnarray}
  \phantom{ t^{\prime }_{i'} |\vec{N}^f\rangle_V}
&& -2{n}_{i'}^0\Bigl(2n_{i'i'} +  \sum_{l\neq i'} n_{i'l}  -
\sum_{s>i'}p_{i's}+\sum_{r<i'}p_{ri'} +
h^{i'}\Bigr)(-1)^{\sum\limits_{l'=1}^{i'-1}n_{l'}^0}
 \left|\vec{N}^f-\delta_{i'l}\rangle_V
 \right. \nonumber \\
%%%%%%%%
 &&
  +2\sum_{k'=i'+1}^{k}n_{k'}^0(-1)^{\sum\limits_{l'=1}^{k'-1}n_{l'}^0}\Biggl\{-
\sum_{l'=1}^{i'-1}p_{l'i'}\left|\vec{n}_l^0-\delta_{k'l};{\vec{n}}_{ij},
{\vec{p}}_{rs} - \delta_{l'i',rs}+ \delta_{l'k',rs} \rangle_V \right.
 \nonumber\\
&&+\left|\vec{N}^f-\delta_{k'l} +
\delta_{i'k',rs} \rangle_V \right.   - \sum_{l'=1}^{k}(1+\delta_{l'i'})n_{i'l'}\left|\vec{N}^f-\delta_{k'l}-\delta_{i'l',ij}  + \delta_{k'l',ij}\rangle_V
 \right.
  \Biggr\} \nonumber \\
 && -\hspace{-0.15em}(-\hspace{-0.15em}1)^{\sum\limits_{l'=1}^{k}n_{l'}^0 } \textstyle  \hspace{-0.5em}\sum\limits_{k'=1}^k
(1\hspace{-0.15em}+\hspace{-0.15em}\delta_{k'i'})
n_{i'k'}\Biggl\{\hspace{-0.15em}2 \hspace{-0.5em}\sum\limits_{m'=k'+1}^{k} \hspace{-0.3em}n^0_{m'}(-1)^{\sum\limits_{n'=m'+1}^{k}n_{n'}^0
} \hspace{-0.3em}
\left|\vec{N}^f \hspace{-0.15em}-\hspace{-0.1em}\delta_{m'l}\hspace{-0.15em}-\hspace{-0.1em}\delta_{i'k',ij}\hspace{-0.15em}+\hspace{-0.1em}\delta_{k'm',ij}
\rangle_V
 \right. \nonumber\\
 && +\frac{1}{2}\textstyle
 \hspace{-0.2em}\left(1+
 \hspace{-0.2em}\left[\frac{n_{k'}^0+1}{2}
\hspace{-0.2em}\right]\hspace{-0.2em}\right)
(-1)^{\sum_{n'=k'+1}^{k}n_{n'}^0 }\left|\vec{N}^f+\delta_{k'l}
{}mod{}2
-\delta_{i'k',ij}+\delta_{k'k',ij}\hspace{-0.2em}\left[\frac{n_{k'}^0+1}{2}
\hspace{-0.2em}\right]\rangle_V
 \right.  \Biggr\},  \nonumber
   \end{eqnarray}
where we have taken into account the first relations in
(\ref{identllm}). In turn, for the even positive root operators
$l^{\prime}_{l'm'}$, for ${l'=m'}$, we have, with allowance for
the second relations in (\ref{identllm}),
   \begin{eqnarray}
  \label{l'll}
  l^{\prime}_{l'l'}
\left|\vec{N}^f \rangle_V
 \right.\hspace{-0.25em}&\hspace{-0.25em}=\hspace{-0.25em}&
 \hspace{-0.25em}   -n_{l'}^0\Biggl[2\sum_{k'=l'+1}^{k}n_{k'}^0(-1)^{\sum\limits_{n'=l'+1}^{k'-1}n_{n'}^0}\Biggl\{
\left|\vec{n}_l^0-\delta_{k'l}-\delta_{l'l};{\vec{n}}_{ij},
{\vec{p}}_{rs} + \delta_{l'k',rs} \rangle_V \right.  \\
&& -
\sum_{n'=1}^{l'-1}p_{n'l'}\left|\vec{n}_l^0-\delta_{l'l}-\delta_{k'l};{\vec{n}}_{ij},
{\vec{p}}_{rs} - \delta_{n'l',rs}+ \delta_{n'k',rs} \rangle_V \right. \nonumber \\
 &&  - \sum_{n'=1}^{k}(1+\delta_{l'n'})n_{l'n'}\left|\vec{N}^f-\delta_{k'l}-\delta_{l'l}-\delta_{l'n',ij}  + \delta_{k'n',ij}\rangle_V
 \right.\Biggr\} \nonumber
   \\
  && -(-1)^{\sum_{n'=l'+1}^{k}n_{n'}^0 } \textstyle
\sum\limits_{k'=1}^k(1+\delta_{k'l'})
 n_{l'k'}\Biggl\{2\sum\limits_{m'=k'+1,m'\neq l'}^{k}n^0_{m'}(-1)^{\sum\limits_{n'=m'+1}^{k}n_{n'}^0 }
 \nonumber\\
 && \times \left|\vec{N}^f-\delta_{m'l}-\delta_{l'l}-\delta_{l'k',ij}+\delta_{k'm',ij}\rangle_V
 \right. +\frac{1}{2}\textstyle
 \hspace{-0.2em}\left(1+
 \hspace{-0.2em}\left[\frac{n_{k'}^0-\delta_{l'k'}+1}{2}
\hspace{-0.2em}\right]\hspace{-0.2em}\right)
(-1)^{\sum\limits_{n'=k'+1}^{k}n_{n'}^0
} \nonumber\\
&& \textstyle\times \left|\vec{N}^f-\delta_{l'l}+\delta_{k'l}
{}mod{}2
-\delta_{l'k',ij}+\delta_{k'k',ij}\hspace{-0.2em}\left[\frac{n_{k'}^0-\delta_{l'k'}+1}{2}
\hspace{-0.2em}\right]\rangle_V
 \right.  \Biggr\}\Biggr]+l^{\prime{}}_{l'l'}
\left|\vec{0}_l^0;\vec{N}\rangle_V\right.\vert_{\vec{0}_l^0
\to \vec{n}_l^0}
 ,
 \nonumber
   \end{eqnarray}
 where the quantity $l^{\prime{}}_{l'l'}\left|\vec{0}_l^0;\vec{N}\rangle_V\right.$
 denotes the purely symplectic ($sp(2k)$) part of the action of $l^{\prime}_{l'l'}$
 on $\left|\vec{N}^f \rangle_V
 \right.$, first derived in Ref. \cite{BRmixbos}, with unchanged values of the ``odd''
 integers $\vec{n}{}^0_l$. Explicitly, the expression $l^{\prime{}}_{l'l'}
\left|\vec{0}_l^0;\vec{N}\rangle_V\right.$ reads
   \begin{eqnarray}
   \label{l'llb}
l^{\prime{}}_{l'l'}
\left|\vec{0}_l^0;\vec{N}\rangle_V\right.\vert_{\vec{0}_l^0
\to \vec{n}_l^0}
&\hspace{-0.25em}=\hspace{-0.25em}&
 \hspace{-0.25em}
 - \frac{1}{2}\hspace{-0.3em}\sum\limits_{k'=1}^{l'-1}{n}_{k'l'}\Biggl[\sum_{p=0}^{l'-k'-1}\bigg[\sum_{k'_1=k'+1}^{l'-1}\hspace{-0.3em}\ldots \hspace{-0.3em}\sum_{k'_p=k'+p}^{l'-1}\Big\{\left|C^{k'_{p}l'}_{\vec{N}^f-\delta_{k'l',ij}-\sum_{j=1}^{p}\delta_{k'_{j-1}k'_j,rs}}\rangle_V
 \right.\\
  &-&\hspace{-1em}
 \sum\limits_{n'=k'_{p-1}}^{k'_p-1}p_{n'l'}\left|\vec{N}^f-\delta_{k'l',ij}-\sum_{j=1}^{p}\delta_{k'_{j-1}k'_j,rs}-\delta_{n'l',rs}+\delta_{n'k'_p,rs}\rangle_V
 \right. \Big\}\prod_{j=1}^pp_{k'_{j-1}k'_{j}}\bigg]
\nonumber
 \end{eqnarray}
 \vspace{-3ex}
  \begin{eqnarray}
  && -\sum_{n'=k'+1}^{k}(1+\delta_{n'l'})n_{n'l'}\hspace{-0.2em}
\left|\vec{n}_l^0;
\vec{n}_{ij}-\delta_{k'l',ij}-\delta_{n'l',ij}+\delta_{k'n',ij},
\vec{p}_{rs}\rangle_V
 \right. \Biggr]
 \nonumber\\
 && + n_{l'l'}\left(n_{l'l'}-1 +  \sum_{k'> l'} n_{k'l'}  - \sum_{s>l'}p_{l's}+\sum_{r<l'}p_{rl'} + h^{l'}\right)\left|\vec{N}^f -
 \delta_{l'l',ij}\rangle_V
 \right.\nonumber\\
 && - \frac{1}{2}\sum\limits_{k'=l'+1}^{k}{n}_{l'k'}\Biggl[
\left|\vec{n}_l^0;{\vec{n}}_{ij}-\delta_{l'k',ij}, {\vec{p}}_{rs}
+ \delta_{l'k',rs} \rangle_V \right.\nonumber\\
&& -
\sum\limits_{n'=1}^{l'-1}p_{n'l'}\left|\vec{n}_l^0;{\vec{n}}_{ij}-\delta_{l'k',ij},
{\vec{p}}_{rs} - \delta_{n'l',rs}+ \delta_{n'k',rs} \rangle_V \right.  \nonumber\\
  && - \sum_{n'=k'+1}^{k}(1+\delta_{n'l'})n_{n'l'}
\left|\vec{N}^f-\delta_{l'k',ij}-\delta_{l'n',ij}  +
\delta_{k'n',ij}\rangle_V\mathcal{}
 \right.
 \Biggr]\nonumber\\
 && + \frac{1}{2}\sum_{k'=1,k'\neq l'}^{k}\frac{{n}_{l'k'}({n}_{l'k'}-1)}{2}
 \left|\vec{N}^f -2\delta_{l'k',ij} + \delta_{k'k',ij}\rangle_V
 \right..
 \nonumber
   \end{eqnarray}
Finally, for the operators $l^{\prime}_{l'm'}$, for ${l'< m'}$ we
have
   \begin{eqnarray}
    \hspace{-1.5em} l^{\prime}_{l'm'}
|\vec{N}^f\rangle_V \hspace{-0.7em}&=\hspace{-0.7em}&
\hspace{-0.3em}\frac{1}{2}n_{l'}^0\Biggl[
2\hspace{-0.5em}\sum_{k'=l'+1}^{m'-1}\hspace{-0.5em}n_{k'}^0(-1)^{
\sum\limits_{n'=l'+1}^{k'-1}n_{n'}^0}\hspace{-0.3em}\Biggl\{\sum_{p=0}^{m'-k'-1}\bigg[\sum_{k'_1=k'+1}^{m'-1}\hspace{-0.3em}\ldots \hspace{-0.3em}\sum_{k'_p=k'+p}^{m'-1}
 \hspace{-0.1em}\prod_{j=1}^pp_{k'_{j-1}k'_{j}}\Big\{\hspace{-0.2em} \sum_{n'=k'_{p-1}}^{k'_p-1}\hspace{-0.3em}p_{n'm'}\times \label{l'lm}
\\
 \hspace{-0.3em} &\hspace{-0.3em}\times & \hspace{-0.7em}\left|\vec{N}^f\hspace{-0.1em}-\delta_{l'l}-\delta_{k'l}-\sum_{j=1}^{p}\delta_{k'_{j-1}k'_j,rs}
-\delta_{n'm',rs}+\delta_{n'k'_p,rs}\rangle_V
 \right. - \left|C^{k'_{p}m'}_{\vec{N}{}^f-\delta_{l'l}-\delta_{k'l}-\sum\limits_{j=1}^{p}\delta_{k'_{j-1}k'_j,rs}}\rangle_V
 \right.
 \hspace{-0.3em}\Big\}\hspace{-0.2em}\bigg]\nonumber\\
 %%%%%%%%%%%%%%%%%%
  && +\sum_{n'=1}^{k}(1+\delta_{n'm'})n_{n'm'}\hspace{-0.2em}\left|\vec{n}_l^0-\delta_{l'l}-\delta_{k'l}; \vec{n}_{ij}-\delta_{n'm',ij}+\delta_{n'k',ij},
  \vec{p}_{rs}\rangle_V
 \right. \nonumber
  \\
&& + n_{m'}^0\Biggl(4\sum_{n'=k'+1}^{m'-1}n^0_{n'} (-1)^{\sum\limits_{n=n'+1}^{m'-1}n^0_{n}}\left|\vec{n}_l^0 -  \delta_{ln'} -  \delta_{ll'}- \delta_{lm'}-\delta_{lk'};
 \vec{n}_{ij} + \delta_{n'k',ij},
 \vec{p}_{rs}\right\rangle_V\nonumber\\
 && + (-1)^{\sum\limits_{n'=k'+1}^{m'-1}n^0_{n'}}\textstyle \left|\vec{n}_l^0
 -\delta_{lm'}-\delta_{ll'}; \vec{n}_{ij}
 ,
 \vec{p}_{rs}\right\rangle_V\Biggr) \Biggr\}\nonumber\\
 &&
  -2\sum_{k'=m'+1}^{k}n_{k'}^0(-1)^{\sum\limits_{n'=l'+1}^{k'-1}n_{n'}^0}\Biggl\{
\left|\vec{n}_l^0-\delta_{l'l}-\delta_{k'l};{\vec{n}}_{ij},
{\vec{p}}_{rs} + \delta_{m'k',rs} \rangle_V \right. \nonumber\\
&& -
\sum_{n'=1}^{m'-1}p_{n'm'}\left|\vec{n}_l^0-\delta_{l'l}-\delta_{k'l};{\vec{n}}_{ij},
{\vec{p}}_{rs} - \delta_{n'm',rs}+ \delta_{n'k',rs} \rangle_V \right. \nonumber \\
 &&  - \sum_{n'=1}^{k}(1+\delta_{n'm'})n_{m'n'}\left|\vec{N}^f-\delta_{l'l}-\delta_{k'l}-\delta_{m'n',ij}  + \delta_{k'n',ij}\rangle_V
 \right.
  \Biggr\} \nonumber \\
&& +2{n}_{m'}^0(-1)^{\sum\limits_{n'=l'+1}^{m'-1}n_{n'}^0}\hspace{-0.3em}\Bigl(
\sum_{l=1}^k \hspace{-0.1em}(1 \hspace{-0.1em}+ \hspace{-0.1em}\delta_{m'l})n_{m'l}  \hspace{-0.1em} - \hspace{-0.25em}
\sum_{s>m'} \hspace{-0.15em}p_{m's} \hspace{-0.1em}+ \hspace{-0.25em}\sum_{r<m'} \hspace{-0.15em}p_{rm'} \hspace{-0.1em} + h^{m'} \hspace{-0.2em}\Bigr) \hspace{-0.2em}
 \left|\vec{N}^f \hspace{-0.1em}- \hspace{-0.1em}\delta_{l'l} \hspace{-0.1em}- \hspace{-0.1em}\delta_{m'l}\rangle_V
 \right. \nonumber
 \end{eqnarray}
   \vspace{-3ex}
   \begin{eqnarray}
 && +(-1)^{\sum\limits_{n'=l'+1}^{k}n_{n'}^0 } \textstyle \sum\limits_{
k'=1}^k(1+\delta_{k'm'})
 n_{m'k'}\Biggl\{2\sum\limits_{n'=k'+1,n'\neq l'}^{k}n^0_{n'}(-1)^{\sum_{p'=n'+1}^{k}n_{p'}^0
}\nonumber\\
&&
\times\left|\vec{N}^f-\delta_{l'l}-\delta_{n'l}-\delta_{m'k',ij}+\delta_{k'n',ij}\rangle_V
 \right. +\frac{1}{2}\textstyle
 \hspace{-0.2em}\left(1+
 \hspace{-0.2em}\left[\frac{n_{k'}^0+1-\delta_{l'k'}}{2}
\hspace{-0.2em}\right]\hspace{-0.2em}\right)
(-1)^{\sum\limits_{n'=k'+1}^{k}n_{n'}^0
}\nonumber\\
 && \times \textstyle\left|\vec{N}^f-\delta_{l'l}+\delta_{k'l} {}mod{}2
-\delta_{m'k',ij}+\delta_{k'k',ij}\hspace{-0.2em}\left[\frac{n_{k'}^0-\delta_{l'k'}+1}{2}
\hspace{-0.2em}\right]\rangle_V
 \right.  \Biggr\}
 \Biggr]\nonumber \\
&& -\frac{1}{2}n_{m'}^0\Biggl[
2\sum_{k'=m'+1}^{k}n_{k'}^0(-1)^{\sum\limits_{n'=m'+1}^{k'-1}n_{n'}^0}\Biggl\{
\left|\vec{n}_l^0-\delta_{m'l}-\delta_{k'l};{\vec{n}}_{ij},
{\vec{p}}_{rs} + \delta_{l'k',rs} \rangle_V \right. \nonumber\\
&& -
\sum_{n'=1}^{l'-1}p_{n'l'}\left|\vec{n}_l^0-\delta_{m'l}-\delta_{k'l};{\vec{n}}_{ij},
{\vec{p}}_{rs} - \delta_{n'l',rs}+ \delta_{n'k',rs} \rangle_V \right. \nonumber \\
 &&  - \sum_{n'=1}^{k}(1+\delta_{n'l'})n_{l'n'}\left|\vec{N}^f-\delta_{m'l}-\delta_{k'l}-\delta_{l'n',ij}  + \delta_{k'n',ij}\rangle_V
 \right.
  \Biggr\} \nonumber\\
&&
 -(-1)^{\sum\limits_{n'=m'+1}^{k}n_{n'}^0 } \textstyle
\sum\limits_{k'=1}^k(1+\delta_{l'k'})
 n_{l'k'}\Biggl\{2\sum\limits_{n'=k'+1}^{k}n^0_{n'}(-1)^{\sum\limits_{p'=n'+1}^{k}n_{p'}^0 }
 \nonumber\\
 && \times\left|\vec{N}^f-\delta_{m'l}-\delta_{n'l}-\delta_{l'k',ij}+\delta_{k'n',ij}\rangle_V
 \right. +\frac{1}{2}\textstyle
 \hspace{-0.2em}\left(1+
 \hspace{-0.2em}\left[\frac{n_{k'}^0-\delta_{m'k'}+1}{2}
\hspace{-0.2em}\right]\hspace{-0.2em}\right)
(-1)^{\sum_{n'=k'+1}^{k}n_{n'}^0 }\nonumber \\
&& \textstyle\times \left|\vec{N}^f-\delta_{m'l}+\delta_{k'l}
{}mod{}2
-\delta_{l'k',ij}+\delta_{k'k',ij}\hspace{-0.2em}\left[\frac{n_{k'}^0-\delta_{m'k'}+1}{2}
\hspace{-0.2em}\right]\rangle_V
 \right.  \Biggr\}
\Biggr]\nonumber
\\
&& + l^{\prime{}}_{l'm'}
\left|\vec{0}_l^0;\vec{n}_{ij},\vec{p}_{rs}\rangle_V\right.\vert_{\vec{0}_l^0
\to \vec{n}_l^0}\nonumber ,   \end{eqnarray}
  where, in order to obtain Eq.(\ref{l'lm}),
  the formulae for the purely symplectic
  ($sp(2k)$) part of the action of
  $l^{\prime}_{l'l'}$ on $\left|\vec{N}^f \rangle_V
 \right.$, i.e., $l^{\prime{}}_{l'l'}\left|\vec{0}_l^0;\vec{N}\rangle_V\right.$,
  first derived in Ref. \cite{BRmixbos},
  with unchanged values of "odd" integers $\vec{n}{}^0_l$,
   have been written. Explicitly, we have
   \begin{eqnarray}
\label{l'lmaux}\hspace{-0.5em}l^{\prime }_{l'm'}
\bigl|\vec{0}_l^0;\vec{N}\bigr\rangle_V\vert_{\vec{0}_l^0
\to \vec{n}_l^0} \hspace{-0.5em}&=\hspace{-0.5em}& \hspace{-0.3em}
  \frac{1}{4}\hspace{-0.1em}\sum\limits_{k'=1}^{m'-1}(1+\delta_{k'l'}){n}_{k'l'}
\Biggl[ \hspace{-0.3em} \sum_{p=0}^{m'-k'-1}\bigg[\hspace{-0.3em}\sum_{k'_1=k'+1}^{m'-1}\ldots \hspace{-0.3em}\sum_{k'_p=k'+p}^{m'-1}\prod_{j=1}^pp_{k'_{j-1}k'_{j}}\bigg\{  \hspace{-0.25em}\sum\limits_{n'=k'_{p-1}}^{k'_p-1}\hspace{-0.2em}p_{n'm'} \times \\
 \hspace{-1.1em} &\hspace{-1.1em} & \hspace{-1.1em}\left|\vec{N}^f\hspace{-0.2em}
-\delta_{k'l',ij}\hspace{-0.2em}-\hspace{-0.2em}\sum_{j=1}^{p}\delta_{k'_{j-1}k'_j,rs} \hspace{-0.2em}-\delta_{n'm',rs}\hspace{-0.2em}+\delta_{n'k'_p,rs}\rangle_V
 \right. \hspace{-0.2em}-
 \left|C^{k'_{p}m'}_{\vec{N}{}^f\hspace{-0.2em}-\delta_{k'l',ij}\hspace{-0.2em}-\sum\limits_{j=1}^{p}\delta_{k'_{j-1}k'_j,rs}}\rangle_V
 \right.\hspace{-0.2em}\bigg\}\hspace{-0.25em}\bigg]\nonumber\\
  && +\sum_{n'=k'}^{k}(1+\delta_{n'm'})n_{n'm'}
\hspace{-0.2em}\left|\vec{n}_l^0; \vec{n}_{ij}-\delta_{k'l',ij}-
\delta_{n'm',ij}+\delta_{k'n',ij}, \vec{p}_{rs}\rangle_V
 \right. \Biggr]
 \nonumber\\
&& - \frac{1}{4}\sum\limits_{k'=m'+1}^{k}{n}_{l'k'}
\Biggl[ \left|\vec{n}_l^0;{\vec{n}}_{ij}-\delta_{l'k',ij},
{\vec{p}}_{rs} + \delta_{m'k',rs} \rangle_V \right. \nonumber\\
&&-
\sum\limits_{n'=1}^{m'-1}p_{n'm'}\left|\vec{n}_l^0;{\vec{n}}_{ij}-\delta_{l'k',ij},
{\vec{p}}_{rs} - \delta_{n'm',rs}+ \delta_{n'k',rs} \rangle_V \right. \nonumber \\
   && - \sum_{n'=l'+1}^{k}(1+\delta_{n'm'})n_{m'n'}\left|\vec{n}_l^0;
{\vec{n}}_{ij}-\delta_{l'k',ij}-\delta_{n'm',ij}  +
\delta_{k'n',ij}\rangle_V
 \right.  \Biggr]\nonumber
 \end{eqnarray}
   \vspace{-3ex}
   \begin{eqnarray}
&& + \frac{1}{4}n_{l'm'}\Bigl(n_{l'm'}-1 +  \sum_{k'>
l'} (1+\delta_{k'm'}) n_{k'm'}+\sum_{k'>
m'}n_{l'k'}  - \sum_{s>l'}p_{l's} \nonumber\\
&& - \sum_{s>m'}p_{m's}+\sum_{r<l'}p_{rl'}
+\sum_{r<m'}p_{rm'} + h^{l'}+ h^{m'}\Bigr)\left|\vec{n}_l^0;\vec{n}_{ij}-\delta_{l'm',ij},\vec{p}_{rs}\rangle_V\right. \nonumber\\
&& -\frac{1}{4}  \sum\limits_{k'=1}^{l'-1}{n}_{k'm'}
\Biggl[ \sum_{p=0}^{l'-k'-1}\bigg[\sum_{k'_1=k'+1}^{l'-1}\ldots \sum_{k'_p=k'+p}^{l'-1}\prod_{j=1}^pp_{k'_{j-1}k'_{j}} \Big\{ - \sum\limits_{n'=k'_{p-1}}^{k'_p-1}p_{n'l'} \times \nonumber\\
  &&\times  \left|\vec{N}^f-\delta_{k'm',ij}-\sum_{j=1}^{p}\delta_{k'_{j-1}k'_j,rs}-\delta_{n'l',rs}+\delta_{n'k'_p,rs}\rangle_V
 \right. +   
 \left|C^{k'_{p}l'}_{\vec{N}^f-\delta_{k'm',ij} -\sum\limits_{j=1}^{p}\delta_{k'_{j-1}k'_j,rs}}\rangle_V
 \right.\Big\}\bigg]\nonumber\\
  && -\sum_{n'=k'+1}^{k}(1+\delta_{n'l'})n_{n'l'}
\hspace{-0.2em}\left|\vec{n}_l^0; \vec{n}_{ij}-\delta_{k'm',ij}-
\delta_{n'l',ij}+\delta_{k'n',ij}, \vec{p}_{rs}\rangle_V
 \right. \Biggr]\nonumber\\
\hspace{-0.5em}&\hspace{-0.5em}&
 -
\frac{1}{4}\sum\limits_{k'=l'+1}^{k}(1+\delta_{k'm'}){n}_{m'k'}
\Biggl[ \left|\vec{n}_l^0;{\vec{n}}_{ij}-\delta_{m'k',ij},
{\vec{p}}_{rs} + \delta_{l'k',rs} \rangle_V \right
. \nonumber\\
&&- \sum\limits_{n'=1}^{l'-1}p\
_{n'l'}\left|\vec{n}_l^0;{\vec{n}}_{ij}-\delta_{m'k',ij},
{\vec{p}}_{rs} - \delta_{n'l',rs}+ \delta_{n'k',rs} \rangle_V
\right.  \Biggr].\nonumber
\end{eqnarray}

Thus, formulae (\ref{t'+i})-- (\ref{t'+rs}), (\ref{t'lm}) --
(\ref{l'lmaux}) completely solve the problem of an auxiliary
representation  (generalized Verma module) for the orthosymplectic
$osp(1|2k)$ algebra. Notice that the above result contains, as a
particular case, for $\vec{n}_l^0 = \vec{0}_l^0$, and without the
odd root vectors $t^{\prime }_{i}, t^{\prime +}_{i}$, the Verma
module for the symplectic $sp(2k)$ algebra constructed in
Ref.\cite{BRmixbos}.

\subsection{On Construction of Additional Parts for Massive Half-integer HS Fields}\label{addalgebram}}

The solution of a similar problem of constructing the auxiliary
representation for the HS symmetry massive superalgebra
$\mathcal{A}^f(Y(k),\mathbb{R}^{1,d-1})$ is provided by an
enlargement of the Cartan decomposition (\ref{Cartandecomp}) for
$osp(1|2k)$ up to the one for
$\mathcal{A}^f(Y(k),\mathbb{R}^{1,d-1})$. Then, we can make the
same steps again, adding, first, the ``divergence'', $l_i'$, and
``gradient'', $l^{\prime +}_i$, operators, respectively, to the
subsuperalgebras of positive $\mathcal{E}^+_k$ and negative
$\mathcal{E}^-_k$ root vectors in (\ref{Cartandecomp}), and,
second, with the peculiarity, that the Cartan-like subsuperalgebra
would now contain two elements $t_0';l_0'$\footnote{Despite the
fact that the anticommutators $\{t_0',t_i'\}, \{t_0',t^{\prime
+}_i\}$ are not proportional to $t_i'$, $t^{\prime +}_i$,
respectively.}. Simultaneously, the highest-weight vector
$|0\rangle_V$ and the basis vector $|\vec{N}^f_m\rangle_V$ of
$\mathcal{A}^f(Y(k),\mathbb{R}^{1,d-1})$, in addition to
definitions (\ref{hwrep})--(\ref{VM}), should be determined as
follows:
\begin{eqnarray}\label{massrep}
&& l'_i|0\rangle_V =0, \qquad \qquad \qquad \qquad t'_0 |0\rangle_V = \tilde{\gamma}m|0\rangle_V ,\\
&& |\vec{N}^f_m\rangle_V \sim
\prod_{i}^k\textstyle\bigl(\frac{l^{\prime
+}_{i}}{m_i}\bigr){}^{n_{i}}|\vec{N}^f\rangle_V,
\end{eqnarray}
for some parameters $m_i \in \mathbb{R}_+$ of the dimension of
mass, the odd matrix $\tilde{\gamma}$ is from the set of odd
gamma-like matrices (\ref{tgammas}), so that the central charge
$m$ in the initial algebra
$\mathcal{A}^f(Y(k),\mathbb{R}^{1,d-1})$ vanishes in the converted
algebra $\mathcal{A}^f_c(Y(k),\mathbb{R}^{1,d-1})$, because of the
additive composition law
\begin{align}\label{vancentcharge}
    & m \to M = m+{m'} =0, &&  \bigl(t'_0;l_0\bigr)  \to \bigl(T_0;L_0\bigr) = \bigl(t_0 + t'_0 = t_0+ \tilde{\gamma}m;\quad l_0 + l'_0 = l_0
+m^2\bigr),
\end{align}
for the central elements $m, {m'}$, we have the respective odd
Cartan-like operators $t_0, t'_0$, of the original superalgebra of
$o_I$ and the superalgebra of additional parts $o'_I$, with the
following relation to the corresponding Casimir operators $l_0,
l'_0$: $t_0^2=-l_0$ $t^{\prime 2}_0 = - l'_0$.

\subsection{on Verma modules}\label{defVerma}

The corresponding construction was suggested by the Indian
mathematician Daya-Nand Verma in his Ph.D. thesis \cite{Verma1} at
the 60-s of the last century. Let $\mathcal{F}$ be a field (i.g.,
real $\mathbb{R}$ or complex $\mathbb{C}$) and let us consider a
semisimple Lie algebra ${g}$ over $\mathcal{F}$ with the universal
enveloping algebra $\mathcal{U}(g)$ and with the Cartan
(triangular) decomposition
 \begin{equation}\label{LieCartdecomp}
   {g} \ =\ {g}^- \oplus \mathcal{H}\oplus  {g}^+,\texttt{ where } [\mathcal{H}, \mathcal{H}] = 0,\ [{g}^-, \mathcal{H}] \subset {g}^-, \ [{g}^+, \mathcal{H}] \subset {g}^+,
 \end{equation}
with the Cartan subalgebra $\mathcal{H}$ and the nilpotent
subalgebras ${g}^-$, ${g}^+$, which, within a matrix realization
of $g$, are associated with the vector spaces of diagonal matrices
and of upper- ${g}^-$ and lower- ${g}^+$ triangular matrices. The
direct sum of $\mathcal{H}$ and ${g}^+$ subalgebras,
$\mathcal{B}$, $ \mathcal{B} = \mathcal{H} \oplus {g}^+$ is called
the Borel subalgebra of $g$ and is the maximally solvable
subalgebra in it. Then, let $\lambda \in \mathcal{H}^*$ be a fixed
weight from the (dual to $g$) algebra $g^*$.

The definition of the Verma module implies the natural presence of
some other modules. Let $\mathcal{F}_\lambda$ be a one-dimensional
vector space over $\mathcal{F}$ together with a
$\mathcal{B}$-module structure, being such that
 \begin{equation}\label{defbsub}
   \mathcal{H}\mathcal{F}_\lambda =  \lambda  \mathcal{F}_\lambda, \qquad {g}^+ \mathcal{F}_\lambda = 0.
 \end{equation}
For any Lie algebra with a triangular decomposition
(\ref{LieCartdecomp}), the following decomposition for the
corresponding universal enveloping algebras
 $\mathcal{U}(g)$,  $\mathcal{U}(g^-)$,
 $\mathcal{U}(\mathcal{H})$, $\mathcal{U}(g^+)$,  $\mathcal{U}(\mathcal{B})$
 is valid:
 \begin{equation}\label{univdecomp}
   \mathcal{U}(g) \ =\ \mathcal{U}(g^-) \otimes \mathcal{U}(\mathcal{H})\otimes  \mathcal{U}(g^+)  =  \mathcal{U}(g^-) \otimes \mathcal{U}(\mathcal{B}).
 \end{equation}
Since the set $\mathcal{F}_\lambda$ is a left
$\mathcal{B}$-module, it can be presented as a left
$\mathcal{U}(\mathcal{B})$-module, as well.

An application of the Poincare--Birkhoff--Witt theorem concerning
the structure of the basis elements in $\mathcal{U}(g)$ provides a
natural right $\mathcal{U}(\mathcal{B})$-module structure on the
algebra $\mathcal{U}(g)$ by means of the right multiplication of
the Borel subalgebra $\mathcal{B}$. In addition, $\mathcal{U}(g)$
is a natural left $g$-module. Therefore, the universal enveloping
algebra  $\mathcal{U}(g)$
 is a $(g, \mathcal{U}(\mathcal{B})$-bimodule,
 \begin{equation}\label{bimodule}
   \forall a \in \mathcal{U}(g), b \in \mathcal{U}(\mathcal{B}), c \in g: \quad c\otimes a \otimes b \in   \mathcal{U}(g).
 \end{equation}
The \emph{Verma module} denoted as $M_\lambda$ (with respect to
the weight $\lambda$) is the induced $g$-module, determined by the
formula (see, for instance, \cite{VKac})
\begin{equation}\label{indmodule}
  M_\lambda \ = \ \mathcal{U}(g) \otimes{}_{\mathcal{U}(\mathcal{B})}\mathcal{F}_\lambda \ : = \ \bigl(\mathcal{U}(g) \otimes{}_{\mathcal{F}}\mathcal{F}_\lambda\bigr)\Big/ \sum_{a, b, v}   \mathcal{F}\bigl(a{}b \otimes v - a\otimes  b(v)\bigr),
\end{equation}
where the sum runs over all $a \in \mathcal{U}(g)$, $b \in
\mathcal{U}(\mathcal{B})$, $v \in \mathcal{F}_\lambda$ and the
elements $a{}b $, $b(v)$ belong to $\mathcal{U}(g)$ and
$\mathcal{F}_\lambda$, respectively.

The structure of the Verma module $M_\lambda$ for the algebra $g$
for a given weight $\lambda$, which is an infinite-dimensional
representation of $g$, is simplified due to the mentioned
Poincare--Birkhoff--Witt theorem (see, e.g., \cite{Verma1},
\cite{BGG}, \cite{Dixmier}). Indeed, the underlying vector space
(representation space) of the Verma module is isomorphic to
$\mathcal{U}(g^-) \otimes{}_{\mathcal{F}}\mathcal{F}_\lambda$,
\begin{equation}\label{indmodule1}
  M_\lambda \ = \ \mathcal{U}(g^-) \otimes{}_{\mathcal{F}}\mathcal{F}_\lambda,
\end{equation}
 with a nilpotent Lie subalgebra $g^-$ generated by the negative root spaces of $g$.

Verma modules can be equivalently determined (it can be regarded a
property if the definition (\ref{indmodule}) is a starting point
of Verma module introduction) via the notion of
$\mathcal{H}$-\emph{diagonalizable} $g$-module $V$. Recalling that
it possesses the last property if it has the decomposition
\begin{equation}\label{diaganV}
  V \ = \ \bigoplus_{\lambda \in \mathcal{H}^*} V_\lambda,\texttt{ where }V_\lambda  = \{v\in V| h(v) = \langle \lambda , h\rangle v,
  \texttt{ for } h \in \mathcal{H}\}
\end{equation}
 on the weight subspaces $V_\lambda$ with the non-vanishing vector
 $v$, $v\in V_\lambda$, called the \emph{weight-vector of the weight}
 $\lambda$. Then, let $P(V) = \{ \lambda \in \mathcal{H}^* | V_\lambda \ne 0 \}$
 denote the set of weights for the module $V$.
 For $\lambda \in \mathcal{H}^*$,
 we set $D(\lambda) = \{ \mu \in \mathcal{H}^* | \mu  \le \lambda \}$.
 Then, the \emph{category} $\mathcal{O}$,
 whose objects are $g$-modules $V$, being $\mathcal{H}$-\emph{diagonalizable} with
 the corresponding weight subspaces being finite-dimensional
 and possessing the property of the existence of finite weights
 $\lambda_1,\ldots, \lambda_s \in \mathcal{H}^*$, so that
 \begin{equation}\label{propVM}
   P(V) \subset \bigcup_{j=1}^s D(\lambda_j).
 \end{equation}
 The morphisms in $\mathcal{O}$ are homomorphisms of $g$-modules.

 Notice that any submodule, quotient module, sum or tensor product
 of a finite number of modules, from $\mathcal{O}$ belong
 to $\mathcal{O}$, as well. Let us examine the example
 of \emph{highest-weight modules} from the category $\mathcal{O}$.
 The module $V$ over a Lie algebra $g$ with the highest weight
 $\lambda \in \mathcal{H}^*$ is determined by non-zero vector
 $v_{\lambda}\in V$, such that
 \begin{equation}\label{defVermaot}
   g^+ (v_{\lambda}) = 0,\quad h (v_{\lambda}) = \lambda(h) v_{\lambda}\texttt{ for }h\in \mathcal{H}\texttt{ and }\mathcal{U}(g)(v_{\lambda}) = V.
  \end{equation}
 The vector $v_{\lambda}$ is called \emph{highest-weight vector}.
 The final condition due to the decomposition (\ref{univdecomp})
 can be replaced by $\mathcal{U}(g^-)(v_{\lambda}) = V$.

 From Eq. (\ref{defVermaot}), it follows that
 \begin{equation}\label{vectordecmp}
   V\ =\ \sum_{\mu \le \lambda}V_\mu, \quad V_\lambda = \mathcal{F}v_\lambda, \dim V_\lambda <
   \infty.
 \end{equation}
 Therefore, the highest-weight module belongs to $\mathcal{O}$,
 and any highest-weight vectors are proportional.

 Now, we can determine the Verma module in a way different than by (\ref{indmodule}).
 Namely, the highest-weight module $M_\lambda$ over a Lie algebra $g$ with
 the highest weight $\lambda$ is called \emph{Verma module}
 if any $g$-module $N_\lambda$ with the highest weight $\lambda$
 is a quotient module of the module $M_\lambda$, i.e., $N_\lambda = M_\lambda/ R_\lambda$
 for any submodule $R_\lambda \subset M_\lambda$.
 The following properties of the Verma module hold true \cite{Verma1}, \cite{Dixmier}:
\begin{description}
  \item[a)] for any highest weight $\lambda \in \mathcal{H}^*$ there exists
  a module $M_\lambda$, unique with accuracy up to a Verma isomorphism;
  \item[b)] $M_\lambda$ as a $\mathcal{U}(g^-)$-module is free of a rank-$1$ module
  generated by the highest-weight vector;
  \item[c)] $M_\lambda$ contains a unique proper maximal submodule $M'_\lambda$.
\end{description}
 Notice that the last property implies the existence of a unique irreducible highest-weight
 module $L_\lambda = M_\lambda/ M'_\lambda$.

 Now, we shortly describe an extension of the Verma module concept known
 as the generalized Verma module (GVM)  \cite{genVM}.
 For its introduction, with a given Lie algebra $g$, we consider,
 instead of the Borel subalgebra, some of its parabolic subalgebras
 $\mathfrak{p}$, where, by the definition of ``parabolicity'',
 $\mathfrak{p} \supset \mathcal{B}$.
 For any irreducible finite dimensional representation space
 $V$ of $\mathfrak{p}$ the \emph{generalized Verma module}
 is determined with the help of induced module terms such as
 \begin{equation}\label{genVM}
  M_{\mathfrak{p}}(V) : = \ \mathcal{U}(g) \otimes{}_{\mathcal{U}(\mathfrak{p})}V.
\end{equation}
 In case $\lambda$ is the highest weight of $V$, the GVM may be denoted
 as $M_{\mathfrak{p}}(\lambda)$, and it makes sense only for
 a so-called \emph{integral} and \emph{dominant} weight $\lambda$ in $\mathfrak{p}$.
 It is well known that a parabolic subalgebra
 $\mathfrak{p} \subset g$ determines a unique grading decomposition,
 $g = \bigoplus _{i=-m}^m g_i $, in such a way that
 $\mathfrak{p} = \bigoplus _{i\ge 0}^m g_i$.
 Denoting $g^- = \bigoplus _{i < 0}^m g_i$,
 we deduce from the Poincare--Birkhoff--Witt theorem
 a relation for the representation space of the algebra $g$ GVM:
\begin{equation}\label{genVM1}
  M_{\mathfrak{p}}(V) \simeq \ \mathcal{U}(g^-) \otimes{}_{\mathcal{F}}V.
\end{equation}
Among the properties of GVM we list the following: first, GVM is
the highest-weight module, because the highest weight $\lambda$ of
the representation space $V$ is the highest weight of
$M_{\mathfrak{p}}(\lambda)$; second, GVMs belong to the category
$\mathcal{O}$ of highest-weight modules, and therefore they are
quotients of the corresponding Verma module $M_{\lambda}$; third,
the kernel $K_{\lambda}$ of the projection $M_{\lambda} \to
M_{\mathfrak{p}}(\lambda)$ forms a (not direct) sum:
\begin{equation}\label{kern}
  K_{\lambda} : = \sum_{\alpha \in S} M_{s_\alpha \cdot \lambda} \subset M_{\lambda}.
\end{equation}
The set $S$ from the set $\Delta$ of all simple roots $\alpha$ of
the algebra $g$ is composed from such $\alpha \in S$ that the
negative root space $E^{-\alpha}$ of a root $(- \alpha)$ belongs
to a subalgebra $\mathfrak{p}$, i.e., $E^{-\alpha} \subset
\mathfrak{p}$, and, therefore, a basis of GVM is smaller than that
in the Verma module $M_{\lambda}$. Notice that the set $S$ is
uniquely determined by $\mathfrak{p}$, and $s_\alpha$ is the root
reflection with respect to the root $\alpha$ and $s_\alpha \cdot
\lambda$ represents the affine action of $s_\alpha$  on a
highest-weight $\lambda$.

In the case of a trivial set $S$, $S = \emptyset$, the parabolic
subalgebra $\mathfrak{p}$ coincides with the Borel subalgebra
$\mathcal{B}$ and GVM $M_{\mathfrak{p}}(\lambda) = M_{\lambda}$.
In the opposite case, when $S = \Delta$, and therefore
$\mathfrak{p}$ coincides with the semi-simple algebra Lie $g$, the
GVM is isomorphic to the induced representation $V$.

A consideration of the Verma module and generalized Verma module
concepts in the case of a Lie superalgebra $g$ is slightly
modified because of the existence of a natural
$\mathbb{Z}_2$-grading on $g$, but it can be formulated
straightforwardly (see, e.g., \cite{BurLeites} for algorithms of
constructing a first-order realization for a Lie (super)algebra),
as well as an extension of those concepts to the case
(super)algebras of more general than semi-simple ones.

To illustrate the applicability of the general constructions, let
us examine a semi-simple Lie algebra $g$ with a triangular
decomposition (\ref{LieCartdecomp}) and Cartan--Weyl basis
elements $E^{-\alpha_1}, \ldots, E^{-\alpha_k} \in g^-$, $H^i \in
\mathcal{H}$,  $E^{\alpha_1}, \ldots, E^{\alpha_k} \in g^+$, for
positive roots $\alpha_1, \ldots, \alpha_k$, and for $i=1,\ldots,
\mathrm{rank}g $. The independent commutation relations of the
basis elements have the form
\begin{equation}\label{indcommrel}
  [  H^i\,, E^{\alpha}]\ = \ \alpha_i E^{\alpha},\quad [E^{-\alpha}\,, E^{\alpha}]\ = \ \sum_i\alpha^i H^{i},\quad [E^{\alpha}\,, E^{\beta}]
  \ = \ N^{\alpha+\beta} E^{\alpha+\beta},
\end{equation}
 for $\alpha+\beta \ne 0$, some numbers $\alpha^i \in \mathcal{F}$.
 The corresponding Borel subalgebra of $g$ contains
 all positive root vectors $E^{\alpha}$ and basis of the Cartan subalgebra
 $H^i$, whereas the basis of the corresponding Verma module
 $M_\lambda$ for the highest weight $\lambda \in \mathcal{H}^*$
 in some $g$-module $V$ with the highest-weight vector
 $|0\rangle_V$, $|0\rangle_V \equiv v_{\lambda}$, determined
 in accordance with the representation (\ref{indmodule1}),
 (\ref{defVermaot}),
 has the form
 \begin{equation}\label{defVermaotex}
  M_\lambda \ =\ \{ \prod_{j=1}^k(E^{-\alpha_j})^{n_j}|0\rangle_V\},\quad    E^{\alpha_j}|0\rangle_V = 0, \quad   H^{i} |0\rangle_V = h^i |0\rangle_V ,
  \end{equation}
 for $j=1,\ldots,k$ and  $\lambda(H^{i}) = h^i$.
 Such a realization of the Verma module for a real simple Lie algebra
 $g$ and its oscillator realization was first examined in \cite{Burdik}.

 The corresponding representation was used (with a choice of a slightly different
 basis than the Cartan--Weyl basis) both in our previous paper \cite{BRmixbos}
 to construct the Verma module and an auxiliary oscillator representation
 for a semi-simple Lie algebra with second-class constraints $sp(2k)$ and
 in Appendix~\ref{addalgebra} for the case of $osp(1|2k)$ Lie
 superalgebra,
 as well as in the papers \cite{BurdikPashnev}, \cite{0001195}, \cite{symferm-flat},
 \cite{mixfermiflat}.
 The generalized Verma module structure can be realized for a Lie superalgebra
 which incorporates the odd isometry group element
 $t_0 = - \imath \tilde{\gamma}^\mu \partial_\mu$
 for massive half-integer HS fields in Minkowski  space,
 see Appendix~\ref{addalgebram} (with $t_0 = - \imath \tilde{\gamma}^\mu
 D_\mu$, for massive half-integer HS fields in AdS space it has been
 done in \cite{0905.2705}, \cite{symferm-ads})
 because its odd element from the Cartan subsuperalgebra
 of the HS symmetry superalgebra $\mathcal{A}^f(Y(k), \mathbb{R}^{1,d-1})$
 (for a quadratic HS symmetry superalgebra $\mathcal{A}^f(Y(1), AdS_d)$)
 will not diagonalize the representation space $V$.
 The same situation with GVM was realized in the case of massive
 bosonic HS fields in an AdS space both for totally-symmetric \cite{{symint-ads}}
 and mixed-symmetric with $Y(s_1,s_2)$ in \cite{BRmixads},
 where the bosonic element $l_0'$ from the corresponding
 Cartan subalgebras does not diagonalize the respective representation
 space $V$ due to non-trivial AdS-radius  presence.

 Another example of GVM, can be found if one adds to the Borel subsuperalgebra
 $\mathcal{B}= H_k \oplus\mathcal{E}^+_k$,
 from the decomposition of $osp(1|2k)$ given by Eq.
 (\ref{Cartandecomp}), the  negative root vectors from
 $\mathcal{E}^-_k$, which contain the index $k$, i.e., the nilpotent
 subsuperalgebra $[\mathcal{E}^-_k] =
 \{t'_{k}, l'_{ik},  t^{\prime rk}\}$ for
 $i = 1,\ldots, k; r=1,\ldots, k-1$.
 In this case, the parabolic subsuperalgebra $\mathfrak{p}$
 of $osp(1|2k)$ has the form $\mathfrak{p}$ = $[\mathcal{E}^-_k] \oplus H_k \oplus\mathcal{E}^+_k$.
 The basis $\{{|\vec{N}_g\rangle_V}\}$ of the representation space $V$
 with highest-weight vector $|0_g\rangle_V$ for GVM, $M_{\mathfrak{p}}(V)$,
 in accordance with the structure of the $g^-$ nilpotent subsuperalgebra
 in the decomposition
 $osp(1|2k) = g^- \oplus \mathfrak{p}$:
 $\mathcal{E}^+_k|0_g\rangle_V = [\mathcal{E}^-_k]|0_g\rangle_V = 0$,
 $H^{i}|0_g\rangle_V = h^i|0_g\rangle_V$,
 for $H^{i} \in H_k$, with allowance for the representation (\ref{genVM1}),
 can be presented as
 \begin{equation}\label{defgVermaotex}
|\vec{N}_g\rangle_V \equiv  \prod_{l=1}^{k-1}\bigl(t^{\prime
+}_{l}\bigr){}^{n^0_{l}}\prod_{i,j=1, i\leq j}^{k-1}\bigl(l^{\prime
+}_{ij}\bigr){}^{n_{ij}}\prod_{r=1}^{k-2}\Bigr[\prod_{s=r+1}^{k-1} \bigl(t^{\prime
+}_{rs}\bigr){}^{p_{rs}}\Bigl] |0_g\rangle_V,\texttt{ for
}n^0_{l}\in \mathbb{Z}_2;\ n_{ij}, p_{rs}\in \mathbb{N}_0,  \end{equation}
 following the conventions (\ref{VM})
 of Appendix~\ref{addalgebra}.
 Such a realization of GVM can be used if it is desirable
 to convert only a part of constraints from the whole system
 of second-class constraints
 (e.g., for the last example without
 conversion of constraints related to the $k$-th set of Lorentz
 indices $\mu^k_{1}\ldots \mu^k_{n_k}$
 in the initial spin-tensor
 $\Psi_{(\mu^1)_{n_1},(\mu^2)_{n_2},...,(\mu^k)_{n_k}}$).
 We finally note that the case of GVM construction
 and the study of its properties for infinite-dimensional
 super-Virasoro $N=1$ algebras were examined in \cite{Dobrev}.

\section{Oscillator Scalar Realization of Superalgerbra \\ $osp(1|2k)$
 in New Fock Space} \label{oscrealsp2kdet}
\setcounter{equation}{0}

Using the general results of Burdik \cite{Burdik}, initially
elaborated for simple Lie algebras and then enlarged to special
Lie superalgebras in Refs. \cite{symferm-flat},
\cite{mixfermiflat} (to nonlinear superalgebras for higher-spin
fields in AdS${}(d)$ spaces, in \cite{symferm-ads},
\cite{0905.2705}), and introducing a mapping between the basis of
a generalized Verma module for $osp(1|2k)$ given by the vector
$|\vec{N}^f\rangle_V$ \ref{GVMK} and the one a in new Fock space
$\mathcal{H}'$, we have
\begin{equation}\label{map}
    \left|\vec{n}_l^0, \vec{n}_{ij}, \vec{p}_{rs}\rangle_V \right.
    \leftrightarrow \left|\vec{n}_l^0, \vec{n}_{ij}, \vec{p}_{rs}\rangle \right.
 = \prod_{l=1}^k\bigl(f^+_{l}\bigr){}^{n^0_l}\prod_{i,j\geq i}^k\bigl(b^{+}_{ij}\bigr){}^{
 n_{ij}}\prod_{r,s,s>r}^k\bigl(d^{+}_{rs}\bigr){}^{p_{rs}}|0\rangle\,,
\end{equation}
with the vector $\left|\vec{n}_l^0, \vec{n}_{ij},
\vec{p}_{rs}\rangle\right.$ having the same structure as
$|\vec{N}^f\rangle_V$ in Eq. \ref{VM}, for ${n}_l^0 \in
\mathbb{Z}_2$, $n_{ij}, {p}_{rs} \in \mathbb{N}_0$. The set of
$\left|\vec{n}_l^0, \vec{n}_{ij}, \vec{p}_{rs}\rangle \right.$
presents the basis vectors of a Fock space $\mathcal{H}'$
generated by new fermionic, $f^+_{l}, f_{l}$, $l=1,\ldots , k $,
and bosonic, $b^{+}_{ij}, d^+_{rs}, b_{ij}, d_{rs}$, $i,j,r,s
=1,\ldots, k; i\leq j; r<s$, creation and annihilation operators
with the only nonvanishing supercommutation relations
\begin{equation}\label{commrelations}
 \{f_i\,, f^+_j\} = \delta_{ij}\,,\qquad [b_{i_1j_1}, b^+_{i_2j_2}] =
 \delta_{i_1i_2}\delta_{j_1j_2}\,, \   \qquad [d_{r_1s_1}\,,d^+_{r_2s_2}]
 =\delta_{r_1r_2}\delta_{s_1s_2}\,.
\end{equation}
Having the correspondence (\ref{map}), we can represent the action
of the elements $o'_I$ on the generalized Verma module vector
$|\vec{N}^f\rangle_V$ given by Eqs. (\ref{t'+i})-- (\ref{t'+rs}),
(\ref{t'lm})--(\ref{l'lmaux}) as polynomials in the creation and
annihilation operators of the Fock space $\mathcal{H}'$. In doing
so, we have to take into account the requirement of coincidence of
the numbers of the above fermionic $(f_l, f^+_l)$ and bosonic
$(b^{+}_{ij},  b_{ij}, d^+_{rs}, d_{rs})$ operators with the
numbers of second-class constraints, i.e., with the respective
numbers of odd and even root vectors in the Cartan decomposition
of $osp(1|2k)$ (\ref{Cartandecomp}).

As a result, the oscillator realization of the elements $o_I'$
over the Heisenberg superalgebra $A_{k,k^2}$ can be presented in a
unique way, first, for the Cartan elements $g^{\prime i}_0$ and
odd $t^{\prime +}_i$ and even $(l^{\prime +}_{ij}, t^{\prime
+}_{rs})$ negative root vectors as follows:
\begin{eqnarray}
   g_0^{\prime i}& = & f_i^+f_i + \sum_{l\leq m}
 b_{lm}^+b_{lm}(\delta^{il}+\delta^{im}) + \sum_{r< s}d^+_{rs}d_{rs}(\delta^{is}-
 \delta^{ir}) +h^i
 \,,\label{g'0iF} \\
\label{t'+iF}
 t^{\prime  +}_i & = & f^+_i + 2b_{ii}^+f_i
 +4\sum_{l=1}^{i-1}b_{li}^+f_l
  \,,
 \\
 l^{\prime+}_{ij} & = & b_{ij}^+\,,
 \label{l'+ijF}
\\
 t^{\prime+}_{rs}   & = & d^+_{rs} - \sum_{n=1}^{r-1}d_{nr}d^+_{ns}
  - \sum_{n=1}^{k}(1+\delta_{nr})b^+_{ns}b_{rn} -\bigl[4\sum_{n=r+1}^{s-1}b^+_{ns}f_n +({f}^+_{s}+2b^+_{ss}f_s)\bigr]f_{r}
 \,,
 \label{t'+lm}
 \end{eqnarray}
second, for the odd elements $t'_i$ of upper-triangular
subsuperalgebra $\mathcal{E}^+_k$,
\begin{eqnarray}
\label{t'iF}% \nonumber to remove numbering (before each equation)
 t^{\prime }_i &=&
2\sum_{n=1}^{i-1}\Bigl\{
\sum_{p=0}^{i-n-1} \bigg[\sum_{k_1=n+1}^{i-1}\ldots \sum_{k_p=n+p}^{i-1}\Big\{
 C^{k_{p}i}(d^+,d)- \sum_{n'=k_{p-1}}^{k_p-1}d^+_{n'k_p}d_{n'i} \Big\} \prod_{j=1}^pd_{k_{j-1}k_{j}}\bigg]\\
 &&  -\sum_{m=1}^{k}(1+\delta_{mi})
b^+_{mn}b_{mi}  + \bigl[4\sum_{m=n+1}^{i-1} b^+_{nm} f_m -f^+_n
\bigr]f_i
  \Bigr\}f_n\nonumber\\
  &&
  +2\sum_{n=i+1}^{k}\Bigl\{
d^+_{in} - \sum_{m=1}^{i-1}d^+_{mn} d_{mi}  - \sum_{m
=1}^{k}(1+\delta_{mi})b^+_{nm}b_{im}
  \Bigr\}f_n \nonumber\\
&& -2\left(\sum_{l=1}^k(1+\delta_{il}) b^+_{il}b_{il}  -
\sum_{s>i}d^+_{is}d_{is}+\sum_{r<i}d^+_{ri}d_{ri} + h^{i}\right)f_i \nonumber \\
 && + \textstyle \sum\limits_{n=1}^k
(1+\delta_{ni}) \Bigl\{2\sum\limits_{m=n+1}^{k} b^+_{nm}f_m
-\frac{1}{2}\bigl(f_n^+ - 2b_{nn}^+f_n\bigr)
  \Bigr\}b_{ni},  \nonumber
 \end{eqnarray}
and for even ones $l^{\prime }_{lm}$ of upper-triangular
subsuperalgebra $\mathcal{E}^+_k$ separately, for $l=m$ and for
$l<m$
 \begin{eqnarray}
\label{l'llosc}
l^{\prime }_{ll} &=&
 \hspace{-0.25em}
  -\Biggl[2\sum_{n=l+1}^{k}\Bigl\{
d^+_{ln} -
\sum_{n'=1}^{l-1}d^+_{n'n}d_{n'l}-\sum_{n'=1}^k(1+\delta_{n'l})b^+_{n'n}b_{n'l}\Bigr\}f_n
\\
&&-  \sum\limits_{n=1}^k(1+\delta_{nl})
 \Bigl\{-2\sum\limits_{m=n+1}^{k}b^+_{nm}f_{m}
 +\frac{1}{2}\bigl[f^+_n -(1-\delta_{nl})2b^+_{nn}f_n\bigr]  \Bigr\}b_{ln}\Biggr]f_{l}+ l^{\prime b}_{ll},
 \nonumber
 \end{eqnarray}
\vspace{-3ex}
\begin{eqnarray}\label{l'lmosc}
   l^{\prime }_{lm} &=&
\hspace{-0.3em}-\Biggl[ \hspace{-0.1em}\sum_{n=l+1}^{m-1}\Bigl\{
     \sum_{p=0}^{m-n-1}\bigg[\sum_{k_1=n+1}^{m-1}\ldots \sum_{k_p=n+p}^{m-1}
\Big\{C^{k_{p}m}(d^+,d)- \sum_{n'=k_{p-1}}^{k_p-1}d^+_{n'k_p}d_{n'm} \Big\} \prod_{j=1}^pd_{k_{j-1}k_{j}}\bigg]\\
 && -\sum_{n'=1}^{k}(1+\delta_{n'm})b^+_{n'n}b_{n'm}
 + \Bigl[4\sum_{n'=n+1}^{m-1}b^+_{n'n}f_{n'}
  - f^+_{n}\Bigr]f_{m} \Bigr\}f_{n}\nonumber\\
 &&
  +\sum_{n=m+1}^{k}\Bigl\{d^+_{mn}  -
\sum_{n'=1}^{m-1}d^+_{n'n}d_{n'm} -
\sum_{n'=1}^{k}(1+\delta_{n'm})b^+_{n'n}b_{mn'} \Bigr\}f_n \nonumber \\
&& -\Bigl(\sum_{n=1}^k (1+\delta_{nm})b^+_{mn}b_{mn}  -
\sum_{s>m}d^+_{ms}d_{ms}+\sum_{r<m}d^+_{rm}d_{rm} + h^{m}\Bigr)f_m
 \nonumber \\
 &&
+ \frac{1}{2}\textstyle \sum\limits_{n=1}^k(1+\delta_{nm})
 \Bigl\{2\sum\limits_{n'=n+1}^{k}b^+_{nn'}f_{n'}
 -\frac{1}{2}\bigl[f^+_{n} -(1-\delta_{nl})2b^+_{nn}f_{n}\bigr]  \Bigr\}b_{nm}\Biggr]f_{l}
 \nonumber\\
&& -\Biggl[ \sum_{n=m+1}^{k}\Bigl\{d^+_{ln} -
\sum_{n'=1}^{l-1}d^+_{n'n}d_{n'l}  -
\sum_{n'=1}^{k}(1+\delta_{n'l})b^+_{nn'}b_{ln'}
  \Bigr\}f_{n} \nonumber \\
&& + \frac{1}{2}\textstyle \sum\limits_{n=1}^k(1+\delta_{nl})
 \Bigl\{2\sum\limits_{n'=n+1}^{k}b^+_{nn'}f_{n'}
 - \frac{1}{2}\bigl[f^+_{n} -(1-\delta_{nm})2b^+_{nn}f_{n}\bigr]  \Bigr\}b_{nl}
 \Biggr]f_{m} + l^{\prime b}_{lm},  \nonumber
 \end{eqnarray}
where the purely bosonic operators $l^{\prime b}_{ll}$, $l^{\prime
b}_{lm}$, for $l<m$, correspond to those for the symplectic
algebra $sp(2k)$, derived from the actions of $l'_{l'l'}$,
$l'_{l'm'}$ on the bosonic part of the basis vector of a
generalized Verma module, $\left|\vec{0}_l^0;{\vec{N}}\rangle_V
\right.$, given by Eqs. (\ref{l'llb}),  (\ref{l'lmaux}) and first
found in \cite{BRmixbos},
 \begin{eqnarray}\label{l'llbose}
   l^{\prime b}_{ll}
  &=& \frac{1}{4}\sum_{n=1,n\neq l}^{k}b^+_{nn}{b}^2_{ln}
 + \frac{1}{2}\sum_{n=1}^{l-1}\bigg(
\sum_{n'=n+1}^{k}(1+\delta_{n'l})b^+_{nn'}b_{n'l}
\\
  && -
\sum_{p=0}^{l-n-1}\bigg[\sum_{k_1=n+1}^{l-1}\ldots \sum_{k_p=n+p}^{l-1}
\Big\{ C^{k_{p}l}(d^+,d)- \sum_{n'=k_{p-1}}^{k_p-1}d^+_{n'k_p}d_{n'l} \Big\}\prod_{j=1}^pd_{k_{j-1}k_{j}} \bigg]\bigg){b}_{nl}
 \nonumber\\
 && + \left(\sum_{n= l}^k b_{nl}  - \sum_{s>l}d^+_{ls}d_{ls}+\sum_{r<l}d^+_{rl}
 d_{rl} + h^{l}\right)b_{ll}\nonumber\\
 && - \frac{1}{2}\sum_{n=l+1}^{k}\Bigl[d^+_{ln} -
\sum\limits_{n'=1}^{l-1}d^+_{n'n}d_{n'l} -
\sum_{n'=n+1}^{k}(1+\delta_{n'l})b^+_{n'n}b_{n'l}
 \Bigr]{b}_{ln}\,, \nonumber
 \\
  \label{l'lmbose}
  l^{\prime b}_{lm}&=&
 -
\frac{1}{4}\sum\limits_{n=1}^{m-1}(1+\delta_{nl}) \bigg(
- \sum\limits_{n'=1}^{n-1}d^+_{n'n}d_{n'm}
-\sum_{n'=n}^{k}(1+\delta_{n'm})
 b^+_{n'n}b_{n'm}\label{l'lmbosef}\\
 && +
\sum_{p=0}^{m-n-1}\bigg[\sum_{k_1=n+1}^{m-1}\ldots \sum_{k_p=n+p}^{m-1}
 \Big\{C^{k_{p}n}(d^+,d)- \sum_{n'=k_{p-1}}^{k_p-1}d^+_{n'k_p}d_{n'n}\Big\}\prod_{j=1}^pd_{k_{j-1}k_{j}}\bigg]\bigg)b_{nl}
 \nonumber\\
&& - \frac{1}{4}\sum\limits_{n=m+1}^{k} \Bigl[ d^+_{mn}-
\sum\limits_{n'=1}^{m-1} d^+_{n'n}d_{n'm} -
\sum_{n'=l+1}^{k}(1+\delta_{n'm})b^+_{n'n}b_{mn'}  \Bigr]{b}_{ln}\nonumber\\
&& + \frac{1}{4}\Bigl(\sum_{n=m}^kb^+_{ln}b_{ln} +  \sum_{n=
l+1}^k(1+\delta_{nm})b^+_{nm} b_{nm}  - \sum_{s>l}d_{ls}d_{ls} -
\sum_{s>m}d^+_{ms}d_{ms}\nonumber
\end{eqnarray}
   \vspace{-3ex}
   \begin{eqnarray}
&& +\sum_{r<l}d^+_{rl}d_{rl}
+\sum_{r<m}d^+_{rm}d_{rm} + h^{l'}+ h^{m'}\Bigr)b_{lm} \nonumber\\
&&   -\frac{1}{4}  \sum\limits_{n=1}^{l-1} \Bigl[
\sum_{p=0}^{l-n-1}\bigg[\sum_{k_1=n+1}^{l-1}\ldots \sum_{k_p=n+p}^{l-1}
 \Big\{ C^{k_{p}n}(d^+,d)- \sum_{n'=k_{p-1}}^{k_p-1}d^+_{n'k_p}d_{n'n} \Big\}\prod_{j=1}^pd_{k_{j-1}k_{j}}\bigg]\nonumber\\
 && -\sum_{n'=n+1}^{k}(1+\delta_{n'l})b^+_{n'n}b_{n'l}
 \Bigr]{b}_{nm} - \frac{1}{4}\sum\limits_{n=l+1}^{k}(1+\delta_{nm})
\Bigl[ d^+_{ln} - \sum\limits_{n'=1}^{l-1}d^+_{n'n}d_{n'l}
\Bigr]{b}_{mn}\,.\nonumber
 \end{eqnarray}
Finally, for the ``mixed-symmetry'' operators $t^{\prime }_{rs}$
with purely fermionic input (proportional to $f_{l}, f^{+}_{l}$)
in the last row, we have
 \begin{eqnarray}
 \label{t'lmF}
 t^{\prime }_{rs} &=&
\sum_{p=0}^{s-r-1}\bigg[\sum_{k_1=r+1}^{s-1}\ldots \sum_{k_p=r+p}^{s-1}
 \Big\{C^{k_{p}s}(d^+,d)- \sum_{n'=k_{p-1}}^{k_p-1}d^+_{n'k_p}d_{n's}\Big\}\prod_{j=1}^pd_{k_{j-1}k_{j}}\bigg]
 \\
  && -\sum_{n=1}^{k}(1+\delta_{ns})b^+_{nr}
b_{ns}
  + \bigl[4\sum_{n=r+1}^{s-1}b^+_{rn}f_n +(2b^+_{rr}f_r-{f}^+_{r})\bigr]f_{s}
 \,, \quad k_0\equiv r,\nonumber
 \end{eqnarray}
where the operators $C^{rs}(d,d^+)$ are obtained from the vector
$\left|C^{rs}_{\vec{p}_{r's'}}\rangle_V \right.$, $r<m$,
determined in Eq. (\ref{Clmin}) by the rule (first derived in
\cite{BRmixbos}),
 \begin{eqnarray}
 \label{Crs}
C^{rs}(d^+,d)&\equiv &
\Bigl(h^{r}-h^{s}-\sum_{n=s+1}^{k}\bigl(d^+_{rn}d_{rn}+d^+_{sn}d_{sn}\bigl)+
\sum_{n=r+1}^{s-1}d^+_{ns}d_{ns}-d^+_{rs}d_{rs}\Bigr)d_{rs} \\
 &&   + \sum_{n=s+1}^{k}\Bigl\{d^+_{sn}  - \sum_{n'=r+1}^{s-1} d^+_{n'n} d_{n's}\Bigr\}d_{rn}\nonumber.
  \end{eqnarray}
To obtain an oscillator representation for $o'_I$, we have used,
for instance, following one-to-one correspondences inspired by
(\ref{map}), to find, for a given transformed  vector
$\left|\vec{N}^f \right\rangle_V$, a corresponding vector in
$\mathcal{H}'$,
\begin{eqnarray}\label{comp1}
&& -n_m\sum_{n=l+1}^{m-1}n^0_{n}
(-1)^{\sum\limits_{n'=n+1}^{m-1}n^0_{n'}}\left|\vec{N}^f -
\delta_{ln} - \delta_{lm} + \delta_{nl,ij}\right\rangle_V \longleftrightarrow  \sum_{n=l+1}^{m-1}b^+_{ln}f_n
 f_{m}\left|\vec{N}^f\rangle\right.,\\
&& -n_m(-1)^{\sum\limits_{n=l'+1}^{m-1}n^0_{n}}\textstyle
\left(1+\left[\frac{n_{l'}^0 + 1}{2}
 \right]\right)\left|\vec{N}^f +  \delta_{ll'}{} mod{} 2 - \delta_{lm}+ \delta_{l'l',ij}\left[\frac{n_{l'}^0+1}{2}\right]\right\rangle_V  \longleftrightarrow\\
 && \qquad \longleftrightarrow  (2b^+_{l'l'}f_{l'}-{f}^+_{l'})f_{m}\left|\vec{N}^f\rangle\right.,\nonumber
\end{eqnarray}

Let us find an explicit expression for the operator $K'$ used in
the definition of the scalar product (\ref{newsprod}), and given
in an exact form by (\ref{explicit K}).

One can show by direct calculation that the following relation
holds true:
\begin{equation}
{}_V\left\langle  \vec{p}'_{rs},\vec{n}'_{ij};\vec{n}^{\prime 0}_{l}\right.
\left|\vec{n}^{ 0}_{l};\vec{n}_{ij},\vec{p}_{rs}\rangle_V\right.\sim
\prod_{l}\delta^{\textstyle\sum_i n^0_{i}+\sum_i(1+\delta_{il})n_{il}-\sum_{i>l}
p_{li}+\sum_{i<l}
p_{il}}_{\textstyle\sum_i n^{\prime 0}_{i}+\sum_i(1+\delta_{il})n'_{il}-\sum_{i>l}
p'_{li}+\sum_{i<l} p'_{il}}.
\end{equation}
For practical calculations, with low pairs of numbers
\begin{eqnarray} && \Bigl(n^0_{1}+\sum_i(1+\delta_{i1})n_{1i}-\sum_{i>1}
p_{1i}, n^0_{2} + \sum_i(1+\delta_{i2})n_{2i}-\sum_{i>2} p_{2i} + p_{12},...
, \label{lowspin} \\
&& \qquad n^0_{k}+\sum_i(1+\delta_{ik})n_{ik} + \sum_{i<k} p_{ik}\Bigr),\nonumber
\end{eqnarray}
with the number of ``particles'' associated with ,$n^0_l, p_{rt},
n_{ij}$ being the numbers of Fermi and Bose ``particles''
associated with $f_l^+$ and $d^+_{rt}, b_{ij}^+$ for $i\leq j,
r<t$ (where $d^+_{rt}$ reduces the spin number $s_r$ by one unit
and increases the spin number $s_t$ by one unit simultaneously),
the operator $K'$ reads, with the use of the normalization
condition ${}_V\langle0|0\rangle_V = 1$,
\begin{eqnarray}\label{Ka}
K' &=& |0\rangle\langle0|  +
\sum_{r<s}(h^r-h^s)d^+_{rs}|0\rangle\langle0|d_{rs}
 +\sum_{i\le j}\Big(h^i(1+2\delta^{ij})+h^j\Big)
b_{ij}^+|0\rangle\langle 0|b_{ij}%\Big)
 \\
 && - 2\sum_ih^if^+_i |0\rangle\langle0|f_i + 2\sum_j\Big(f_j^+|0\rangle\langle0|\sum_{i<j}(h^i-h^j)
d_{ij}f_i\Big)
 + 2\sum_{i<j}(h^i-h^j)f^+_id^+_{ij}|0\rangle\langle0| f_j
   \nonumber \\
&&  - 2\sum_{i, r<s}f^+_id^+_{rs}|0\rangle\langle0| d_{rs}f_i(h^r-h^s)(h^i+ \delta^{is}-\delta^{ir})+2\sum_{ r<s}\sum_{ i=1}^{r-1}f^+_rd^+_{rs}|0\rangle\langle0| d_{is}f_i(h^r-h^s)\nonumber\\
&& -2\sum_{i, s=1}^{k, i-1}\sum_{ r=1}^{s-1}f^+_id^+_{rs}|0\rangle\langle0| d_{ri}f_s(h^r-h^s) +2\sum_{s,i, r }^{k,s-1,i-1}f^+_rd^+_{rs}|0\rangle\langle0| d_{is}f_i(h^i-h^s)\nonumber\\
&& -2\sum_{s, i, r}^{k,s-1,i-1}f^+_id^+_{rs}|0\rangle\langle0| d_{ri}f_s(h^r-h^i) + \frac{1}{2}\sum_{l<i}(h^i-h^l)\Big(
b^+_{ii}|0\rangle\langle0|b_{li}d_{li}+
b^+_{li}d^+_{li}0\rangle\langle0|b_{ii} \Big)\nonumber\\
&&
+\frac{1}{4} \sum_{i<j}b_{ij}^+|0\rangle\langle0|\Bigl(  4f_jf_i h^j + (1+\delta^{li})\sum_{l < j}(h^j-h^l)b_{il}d_{lj}
 \Bigr)\nonumber\\
&& + \frac{1}{4} \sum_{i<j}\Bigl(\sum_{l<i}b_{lj}^+d^+_{li}
|0\rangle\langle0|(h^i-h^l)
  + \sum_{l<j} b_{lj}^+d^+_{li}
|0\rangle\langle0|(1+\delta^{li})(h^j-h^l)
 \Bigr) b_{ij}
\nonumber \\
 % && +\sum_{l<i} b_{il}^+d^+_{lj}|0\rangle\langle0|\Bigl(b_{il}d_{lj}
%(h^i-h^j)(h^i-1)?
%  + \frac{1}{2}b_{ij} (h^j-h^i)? %+ 2f_jf_i (h^j-h^i)?
 %\Bigr) %\nonumber \\
 && + \sum_{i<j}f^+_if^+_j|0\rangle\langle0|\Bigl(4 f_jf_i (h^jh^i + h^j -
 h^i) + b_{ij} h^j %+ 2 \sum_{i\le l<j}b_{il}d_{lj} (h^l-h^i)?
 \Bigr) + \ldots \,.
\nonumber
\end{eqnarray}
The expression for the above operator $K'$ can be used to
construct LF for fermionic HS fields with low value of generalized
spin.

Summarizing, we state that the auxiliary scalar representation of
the orthosymplectic superalgebra $osp(1|2k)$ for the additional
parts of constraints $o'_I$ in the new Fock space $\mathcal{H}'$
is found. In addition, note that the result contains, as a
particular case, for vanishing fermionic oscillators,  $f_i, f^+_i
$, the auxiliary scalar representation of the symplectic $sp(2k)$
algebra constructed in Ref.\cite{BRmixbos}.

\section{Equivalence  to Initial Irreducible Relations}\label{reductionC}
\setcounter{equation}{0}

We examine here, for the most part, a massive case, and make
comments on massless HS spin-tensors. Our purpose to show that the
equations of motion (\ref{Eq-0m}), (\ref{Eq-1}) [or, equivalently,
(\ref{t0tilde}), for $\tilde{t}_0 = t_0+\tilde{\gamma}m$] can be
achieved by using the action (\ref{L1}) after gauge-fixing and
removing the auxiliary fields by using a part of the equations of
motion. Let us begin with gauge-fixing.

\subsection{Gauge-fixing Procedure}

Our starting point is the fields $|\chi^l_0 \rangle$ and the
sequence of $|\Lambda^{(s)}{}^l_0\rangle$, for $l=0,1$,
$s=0,\ldots,\sum_{o=1}^k n_o + k(k-1)/2-1$, at some fixed values
of spin $(n_1+\frac{1}{2},\ldots, n_k+\frac{1}{2})$. In this
section, we omit the subscripts associated with the eigenvalues of
the spin operators, $\sigma_i$, (\ref{state}). As a first step, we
examine the lowest-level gauge transformation, for $s_{max}=
\sum_{o=1}^k n_o + k(k-1)/2-1$,
\begin{eqnarray}\label{lowgtr}
\delta |\Lambda^{(s_{max}-1)}{}^0_0\rangle  = \Delta{}Q|\Lambda^{(s_{max})}{}^{0}_{0}\rangle,
&\qquad & \delta|\Lambda^{(s_{max}-1)}{}^{1}_{0}\rangle =
\tilde{T}_0|\Lambda^{(s_{max})}{}^{0}_{0}\rangle\,,
\end{eqnarray}
where, due to the ghost number (\ref{ghnumf}) and spin value
(\ref{nidecomposf}), we use the restrictions
$|\Lambda^{(s_{max})}{}^{1}_{0}\rangle \equiv 0$. Indeed, the
independent lowest gauge parameter
$|\Lambda^{(s_{max})}{}^{0}_{0}\rangle$ has the structure
\begin{eqnarray}\label{lowgv}
  |\Lambda^{(s_{max})}{}^{0}_{0}\rangle &=& \sum_n\prod_{i=1}^k\Bigl(\prod_{n_{p{}i}, n_{bi}}(\mathcal{P}^+_i)^{n_{p{}i}}(p^+_i)^{n_{b{}i}}\Bigr)\prod_{t<u}^{k-1,k}\lambda^+_{tu}  |\Lambda(d^+)^{(s_{max})(n)_{b{}i}(n)_{p{}i} (1)_{\lambda{}tu}}{}^{0}_{0}\rangle\,,
\end{eqnarray}
which does not contain any ghost coordinate operators from Wick
pairs $C^a$ ($\{C^a\} \subset \{C^I\}$) and creation  operators
$a^+_i, f_i^+, b_i^+,b_{ij}^+$. The corresponding sums of degrees
$n_{p{}i}, n_{bi}, n_{\lambda{}tu}$ satisfy the distributions
given by Eqs. (\ref{nidecomposf}), (\ref{ghnumf}. So, for the last
(ghost number) relation, we have $-[\sum_i(n_{p{}i}+ n_{bi}) +
\sum_{r<u}
 n_{\lambda{}ru}]= - s_{max}-1$, for $\sum_{r<u}
 n_{\lambda{}ru}=k(k-1)/2$,
 whereas the ghost momentum $p_1^+$ is always present
 in the decomposition (\ref{lowgv}) and the summands
 include the vector
 $$(p^+_i)^{\sum_{o=1}n_o}\prod_{t<u}^{k-1,k}(\lambda^+_{tu})|\Lambda(d^+)^{(s_{max})
 n_{b{}1}0_{b{}2}...0_{b{}k}(0)_{p{}i} (1)_{\lambda{}tu}}{}^{0}_{0}
 \rangle_{\bigl(-\sum_{i\geq 2}n_i +k-1, n_2 +(k-2)-1,...,n_k+(k-k)-k+1\bigr)}$$
 being independent, in addition,
 from the ghost momenta $\mathcal{P}^+_1, \ldots , \mathcal{P}^+_k$, ${p}^+_2, \ldots ,
 {p}^+_k$ and depends only on ``mixed-symmetry'' creation operators
 $d^+_{rs}$, e.g., as a multiplier $\prod_{s\geq 2}^{}(d^+_{1s})^{n_s-1}\prod_{r=2,r<p}^{}d^+_{rp}$.

Extracting an explicit dependence of the fields, of the gauge
parameters, and of the operator $\Delta{}Q$ (\ref{deltaQ}), on
$q_{1}$, ${p}_{1}^+$, bosonic ghost coordinate and momentum,
\begin{eqnarray}\label{Lambdal}
|\Lambda^{(s)}{}^{l}_{0}\rangle =|\Lambda^{(s)}{}^{l}_{00}\rangle+ p_{1}^+|\Lambda^{(s)}{}^{l}_{01}\rangle,\texttt{ for }l=0,1\,,
&&\hspace{-1em} \Delta Q=\Delta
Q_{11}+q_{1}(T_{11}^{0+} -2q_{1}\mathcal{P}_{11}^+) +U_{11}{p}_{1}^+,
\end{eqnarray}
where, first, for $s=-1$ we denote,
$|\Lambda^{(-1)}{}^{l}_{0}\rangle \equiv |\chi^{l}_{0}\rangle$,
second, the quantities $|\Lambda^{(s)}{}^{l}_{00}\rangle$,
$T_{11}^{0+}, T_{11}^{1+}$, $U_{11}$, $\Delta Q_{11}$ do not
depend on $q_{1}$, ${p}_{1}^+$ except for the vector
$|\Lambda^{(s)}{}^{l}_{01}\rangle$, we obtain the gauge
transformation of $|\Lambda^{(s_{max}-1)}{}^{l}_{00}\rangle$
\begin{eqnarray}
\delta |\Lambda^{(s_{max}-1)}{}^{0}_{01} \rangle & = & T_{11}^{0+} |\tilde{\Lambda}^{(s_{max})}{}^{0}_{01}
\rangle -2\mathcal{P}_{11}^+ |\widehat{\Lambda}^{(s_{max})}{}^{0}_{01}
\rangle. \label{dLsmax-1}
\end{eqnarray}
Here, we have used that
$|\Lambda^{(s_{max})}{}^{l}_{00}\rangle\equiv0$, due to the
decomposition (\ref{lowgv}), implying a ghost number restriction,
and the gauge parameter $|\tilde{\Lambda}^{(s_{max})}{}^{0}_{01}
\rangle$ (but not the vector
$|\widehat{\Lambda}^{(s_{max})}{}^{0}_{01} \rangle$) has the same
structure as $|\Lambda^{(s_{max}}{}^{0}_{0}\rangle$ in
(\ref{Lambdal}) lowered by 1 degree in $p_1^+$,
\begin{eqnarray}\label{trlowgv}
   |\tilde{\Lambda}^{(s_{max})}{}^{0}_{01}
\rangle  & = & \sum_nn_{b{}1}(\mathcal{P}^+_1)^{n_{p{}1}}(p^+_1)^{n_{b{}1}-1}\prod_{i=2}^k\Bigl(\prod_{n_{p{}i}, n_{bi}}(\mathcal{P}^+_i)^{n_{p{}i}}(p^+_i)^{n_{b{}i}}\Bigr)\prod_{t<u}^{k-1,k}\lambda^+_{tu} \\
 && \times |\Lambda(d^+)^{(s_{max})(n)_{b{}i}(n)_{p{}i} (1)_{\lambda{}tu}}{}^{0}_{0}\rangle\,.\nonumber
\end{eqnarray}
Since $T_{11}^{0+} = T_{1}^+ + O (\mathcal{C})= f_{1}^+ + \ldots$,
as follows from the structure of $\Delta Q$ in Eq. (\ref{deltaQ}),
we can remove the dependence of
$|\Lambda^{(s_{max}-1)}{}^{0}_{00}\rangle$ on the $f_{1}^+$
operator, by using all the degrees of freedom of
$|\Lambda^{(s_{max})}{}^{0}_{0}\rangle$. Therefore, after the
gauge-fixing at the lowest level of the gauge transformations, we
have conditions for $|\Lambda^{(s_{max}-1)}{}^{0}_{00}\rangle$:
\begin{eqnarray}
f_{1}|\Lambda^{(s_{max}-1)}{}^{0}_{01}\rangle=0 &\Longleftrightarrow&
f_{1}|\Lambda^{(s_{max}-1)}{}^{0}_{0}\rangle=0,
\label{gLsmax-1}
\end{eqnarray}
so that the theory becomes an $(s_{max}-1)$-reducible gauge
theory.

Let us turn to the next, $(s_{max}-2)$, level of gauge
transformation. Notice that the structure of the gauge parameter,
$ |\Lambda^{(s_{max}-1)}{}^{1}_{0}\rangle$, is the same as for
$|\Lambda^{(s_{max})}{}^{0}_{0}\rangle$ in (\ref{lowgv}) and the
gauge transformations for
$|\Lambda^{(s_{max}-2)}{}^{l}_{0}\rangle$, $l=0,1$, have the form
\begin{eqnarray}\label{nlowgtr}
\delta |\Lambda^{(s_{max}-2)}{}^0_0\rangle  &= &\Delta{}Q|\Lambda^{(s_{max}-1)}{}^{0}_{0}\rangle+
 \frac{1}{2}\bigl\{\tilde{T}_0,\eta_i^+\eta_i\bigr\}
 |\Lambda^{(s_{max}-1)}{}^{1}_{0}\rangle,\nonumber \\
 \delta|\Lambda^{(s_{max}-2)}{}^{1}_{0}\rangle &=&
\tilde{T}_0|\Lambda^{(s_{max}-1)}{}^{0}_{0}\rangle+ \Delta{}Q|\Lambda^{(s_{max}-1)}{}^{1}_{0}\rangle\,.
\end{eqnarray}
Obviously, first, the vector $|\Lambda^{(s_{max}-2)}{}^1_0\rangle$
has the same structure as
$|\Lambda^{(s_{max}-1)}{}^{0}_{0}\rangle$, and, second, the vector
$|\Lambda^{(s_{max}-1)}{}^{1}_{0}\rangle$ can be used to
gauge-away the dependence on the $f_1^+$ oscillator in
$|\Lambda^{(s_{max}-2)}{}^{1}_{0}\rangle$, as was done by
$|\Lambda^{(s_{max})}{}^{0}_{0}\rangle$ for
$|\Lambda^{(s_{max}-1)}{}^{0}_{0}\rangle $ in (\ref{gLsmax-1}).
Therefore, we have the gauge conditions
\begin{eqnarray}
f_{1}|\Lambda^{(s_{max}-2)}{}^{1}_{01}\rangle=0 &\Longleftrightarrow&
f_{1}|\Lambda^{(s_{max}-2)}{}^{1}_{0}\rangle=0,
\label{gLsmax-11}
\end{eqnarray}
Third, there arises a dependence on the $\mathcal{P}_{11}^+$ odd
ghost momentum in $|\Lambda^{(s_{max}-1)}{}^{0}_{0}\rangle$, as
follows from the decomposition (\ref{lowgv}), where one should
replace the multiplier $(p_1^+)^2$ by $\mathcal{P}_{11}^+$, thus
changing the Grassmann parity of the component vectors in it, and
so we extract the explicit dependence of the gauge parameters and
of the operator $\Delta Q$ on the ghosts $\eta_{11},
\mathcal{P}_{11}^+$, in addition to decomposition (\ref{Lambdal}),
\begin{eqnarray}\label{Lambdaladd}
|\Lambda^{(s)}{}^{l}_{0}\rangle =|\Lambda^{(s)}{}^{l}_{000}\rangle+ \mathcal{P}_{11}^+|\Lambda^{(s)}{}^{l}_{001}\rangle,\texttt{ for }l=0,1\,,
&&\hspace{-1em} \Delta Q=\Delta
Q^1_{11}+\eta_{11}T_{11}^{+} +U^1_{11}\mathcal{P}_{11}^+,
\end{eqnarray}
where the quantities $|\Lambda^{(s)}{}^{l}_{00p}\rangle, p=0,1$,
$\Delta Q^1_{11}, T_{11}^{+}, U_{11}^{1}$ do not depend on
$\eta_{11}, \mathcal{P}_{11}^+$. The remaining gauge
transformations in (\ref{nlowgtr}) for the parameter of $(s_{max}
-2)$ level $|\Lambda^{(s_{max}-2)}{}^0_0\rangle$ imply
\begin{eqnarray}
\delta |\Lambda^{(s_{max}-2)}{}^{0}_{000} \rangle & = & T_{11}^{+} |{\Lambda}^{(s_{max}-1p)}{}^{0}_{001}
\rangle + T_{11}^{0+} |\tilde{\Lambda}^{(s_{max}-1)}{}^{0}_{000}
\rangle . \label{dLsmax-2}
\end{eqnarray}
Since $T_{11}^{+}= L_{11}^++ q_1^+p_1^++O(\mathcal{C})=b_{11}^+  +
q_1^+p_1^+\ldots$, as follows from the structure $\Delta Q$ in Eq.
(\ref{deltaQ}), the dependence on the auxiliary oscillator
$b_{11}^+$ for $|\Lambda^{(s_{max}-2)}{}^{0}_{000} \rangle$
(however, not simultaneously for the product of ghosts
$q_1^+p_1^+$) can be gauged away by the residual, due to
restriction (\ref{gLsmax-1}), degrees of freedom of the vector
$|\Lambda^{(s_{max}-1)}{}^{0}_{001} \rangle$, whereas the
dependence on $f_1^+$ is removed by $\mathcal{P}_{11}^+$, the
independent vector $|\tilde{\Lambda}^{(s_{max}-1)}{}^{0}_{000}
\rangle$ having the same form as
$|{\Lambda}^{(s_{max}-1)}{}^{0}_{000} \rangle$ with allowance for
the representation (\ref{trlowgv}). Therefore, after gauge-fixing
at the $(s_{max}-2)$-level of gauge transformations, we have the
conditions for $|\Lambda^{(s_{max}-2)}{}^{0}_{000} \rangle$
\begin{eqnarray}\label{gLmax-2}
f_1|\Lambda^{(s_{max}-2)}{}^{0}_{000} \rangle =0,\ b_{11}\mathcal{P}_{11}^+|\Lambda^{(s_{max}-2)}{}^{0}_{000} \rangle =0, &\Longleftrightarrow& (f_1, b_{11})\mathcal{P}_{11}^+ |\Lambda^{(s_{max}-2)}{}^{0}_{0} \rangle=0,
\end{eqnarray}
so that the total set of gauge conditions for the parameters
$|\Lambda^{(s_{max}-2)}{}^{l}_{0} \rangle$ is listed in Eqs.
(\ref{gLsmax-11}), (\ref{gLmax-2}). Notice that the dependence on
the operators  $f_1^+, b_{11}^+$ in
$|\Lambda^{(s_{max}-2)}{}^{0}_{0}$ can only be in the
$\mathcal{P}_{11}^+$-dependent summands.

Let us now turn to the next, $(s_{max}-3)$, level of gauge
transformation. Extracting explicit dependence of the gauge
parameters and $\Delta{}Q$ on $q_1$, $p_1^+$, $\eta_{11}$,
$\mathcal{P}_{11}^+$, $q_2$, $p_2^+$, $\eta_{12}$,
$\mathcal{P}_{12}^+$, and using arguments similar to those at the
previous level of gauge transformation, one can show that one can
impose gauge on the vectors $|\Lambda^{(s_{max}-3)}{}^{l}_{0}
\rangle$:
\begin{eqnarray}\label{gLmax-3}
(f_1,\,b_{11}) \mathcal{P}_{11}^+|\Lambda^{(s_{max}-3)}{}^{l}_{0}
\rangle =0,\,  f_2 \mathcal{P}_{11}^+\mathcal{P}_{12}^+
\Pi^0_{p^+_1}|\Lambda^{(s_{max}-3)}{}^{0}_{0} \rangle=0, \
b_{12}\mathcal{P}_{11}^+\mathcal{P}_{12}^+|\Lambda^{(s_{max}-3)}{}^{0}_{0}
\rangle=0.
\end{eqnarray}
In (\ref{gLmax-3}), the quantity $\Pi^0_{p^+_1}$ is the projector
onto the $p^+_1$-independent monomials in the vector
$|\Lambda^{(s_{max}-3)}{}^{0}_{0} \rangle$, which compose the
system of projectors $\{\Pi^0_{p^+_1}, \Pi^1_{p^+_1}\}$, so that
$\sum_i\Pi^i_{p^+_1}=1$. To obtain these gauge conditions, one has
to use some of the degrees of freedom of the gauge parameters
$|\Lambda^{(s_{max}-2)}{}^{l}_{0} \rangle$, restricted by the
Eqs.(\ref{gLsmax-11}), (\ref{gLmax-2}).

Applying the above-described procedure, one can obtain,
step-by-step, first, for $s=s_{max}-4$,
\begin{eqnarray}\label{gLmax-4}
\Bigl[\bigl(f_1, b_{11}\bigr)\mathcal{P}_{11}^+,\, \bigl(f_2\Pi^0_{p^+_1}, b_{12}\bigr)\mathcal{P}_{11}^+\mathcal{P}_{12}^+ \Bigr] |\Lambda^{(s)}{}^{l}_{0} \rangle =0,\  \bigl(f_3\Pi^0_{p^+_1}\Pi^0_{p^+_2},\ b_{13}\bigr)\prod_{i}^3\mathcal{P}_{1i}^+|\Lambda^{(s)}{}^{0}_{0} \rangle =0,
\end{eqnarray}
(where $\Pi^0_{p^+_2}$ is the projector onto the
$p^+_2$-independent monomials in the vector
$|\Lambda^{(s}{}^{0}_{0} \rangle$), and, for $s=s_{max}-5$
\begin{eqnarray} \label{gLmax-5}
&&\left(\bigl(f_1 ,\,b_{11}\bigr)\mathcal{P}_{11}^+,\, \bigl(f_2\Pi^0_{p^+_1},\,b_{12}\bigr)\prod_i^2\mathcal{P}_{1i}^+,\,\bigl(f_3\prod_i^2\Pi^0_{p^+_i},\,
b_{13}\bigr)\prod_{i}^3\mathcal{P}_{1i}^+\right)|\Lambda^{(s)}{}^{l}_{0} \rangle =0, \nonumber \\
&&
\left(f_4\prod_i^3\Pi^0_{p^+_i},\, b_{14}\right)\prod_{i}^4\mathcal{P}_{1i}^+|\Lambda^{(s)}{}^{0}_{0} \rangle =0.
\end{eqnarray}
We then define the set of operators, used in
(\ref{gLsmax-1})--(\ref{gLmax-5}),
\begin{eqnarray} \label{As}
[\mathcal{A}^s]& = &\Bigl(\bigl(f_1,\,b_{11}\bigr)\mathcal{P}_{11}^+;..., \bigl(f_{k}\prod_i^{k-1}\Pi^0_{p^+_i},\,
b_{1k}\bigr)\prod_i^k\mathcal{P}_{1i}^+; b_{22}\mathcal{P}_{22}^+\prod_i^k\mathcal{P}_{1i}^+,\ldots,
 \\
&&  b_{k-1{}k}\prod_i^{k-1}\mathcal{P}_{ik}^+\prod_{i,j=1, i\leq
j}^{k-1} \mathcal{P}_{ij}^+, b_{k{}k}\prod_{i,j=1, i\leq j}^{k}
\mathcal{P}_{ij}^+\Bigr), s = 1,... , \frac{k(k+1)}{2},\texttt{ and } [\mathcal{A}^0]\equiv f_1.\nonumber
\end{eqnarray}
so that, for instance, the set $[\mathcal{A}^s]$, for $s>k$,
contains $1$-st, $k$-th and $(k+1)$-th components, equal,
respectively, to $(\mathcal{A}^{1}; \mathcal{A}^{k};
\mathcal{A}^{k+1}) = \bigl((f_1,\,b_{11})\mathcal{P}_{11}^+;
(f_k\prod_i^{k-1}\Pi^0_{p^+_i},b_{1k})\prod_i^k\mathcal{P}_{1i}^+;
b_{22}\mathcal{P}_{22}^+\prod_i^k\mathcal{P}_{1i}^+\bigr)$. With
the help of operators (\ref{As}), we can equivalently rewrite Eqs.
(\ref{gLsmax-1})--(\ref{gLmax-5}), and all subsequent gauge
conditions based on the decomposition of the gauge parameters in
all fermionic ghost momenta $P_{ij}^+, i\leq j$, as follows (for
$\mathcal{A}^{-1}\equiv 0$):
\begin{eqnarray} \label{Asgauge}
[\mathcal{A}^s]|\Lambda^{(s_{max}-s-1)}{}^{0}_{0}\rangle=0, \ \  [\mathcal{A}^{ s-1}]|\Lambda^{(s_{max}-s-1)}{}^{1}_{0}\rangle=0,\ \texttt{ for }s =
0,1,\ldots,  \frac{ k(k+1)}{2}.
\end{eqnarray}
At the next step, we apply the same procedure as above, however,
starting from the gauge parameters $|\Lambda^{(s_{max}-\frac{
k(k+1)}{2}-2)}{}^{l}_{0}\rangle, l=0,1$, and extract from it, as
well as from the operator $\Delta{}Q$ (\ref{deltaQ}), the ghost
coordinates and momenta $\eta_{ij}$, $\mathcal{P}_{ij}^+$, $i\leq
j$ and $\eta_1$, $\mathcal{P}_{1}^+$ (maintaining in the
coefficients of decomposition the bosonic ghosts $q_i, p^+_i$ as
parameters). As a result, we have obtained a set of gauge
conditions for the parameters $|\Lambda^{(s_{max}-\frac{
k(k+1)}{2}-2)}{}^{l}_{0}\rangle$:
\begin{equation}
\label{gLmax-k(k+1)2-1} \Bigl([\mathcal{A}^{\frac{1}{2}k(k+1)}],
b_1\mathcal{P}_1^+\prod_{i,j=1, i\leq j}^{k}
\mathcal{P}_{ij}^+\Bigr)|\Lambda^{(s_{max}-\frac{ k(k+1)}{2}-2)}{}^{0}_{0}\rangle =0,\ [\mathcal{A}^{\frac{1}{2}k(k+1)}]|\Lambda^{(s_{max}-\frac{ k(k+1)}{2}-2)}{}^{1}_{0}\rangle =0 .
\end{equation}
As we continue the process by extraction of the ghosts $\eta_1$,
$\eta_2$ $\mathcal{P}_{1}^+$, $\mathcal{P}_{2}^+$, and so on, we
find $k$ sets of gauge conditions for the parameters
$|\Lambda^{(s_{max}-\frac{ k(k+3)}{2}-m)}{}^{l}_{0}\rangle$,
$m=1,\ldots,k$:
\begin{eqnarray} \label{gk(k-1)2}
&& \Bigl([\mathcal{A}^{\frac{1}{2}k(k+1)}],
b_1\mathcal{P}_1^+\prod_{i,j=1, i\leq j}^{k} \mathcal{P}_{ij}^+,
b_2\prod_m^2\mathcal{P}_m^+\prod_{i,j=1, i\leq j}^{k}
\mathcal{P}_{ij}^+ \Bigr)|\Lambda^{(s_{max}-\frac{k(k+1)}{2}-3)}{}^{0}_{0}\rangle =0 , \nonumber\\
&& \Bigl([\mathcal{A}^{\frac{1}{2}k(k+1)}],
b_1\mathcal{P}_1^+\prod_{i,j=1, i\leq j}^{k} \mathcal{P}_{ij}^+,
\Bigr)|\Lambda^{(s_{max}-\frac{k(k+1)}{2}-3)}{}^{1}_{0}\rangle =0 ; \\
&& \hspace{3cm}\ldots \ldots \ldots \ldots \ldots \ldots\ldots \ldots \ldots\nonumber\\
 \label{gk(k-1)2f} &&
\Bigl([\mathcal{A}^{\frac{1}{2}k(k+1)}],
b_1\mathcal{P}_1^+\prod_{i,j=1, i\leq j}^{k} \mathcal{P}_{ij}^+,
 \ldots,
b_k\prod_m^k\mathcal{P}_m^+\prod_{i,j=1, i\leq j}^{k}
\mathcal{P}_{ij}^+ \Bigr)|\Lambda^{(s_{max}-\frac{k(k+3)}{2}-1)}{}^{0}_{0}\rangle =0,\nonumber\\
&&\Bigl([\mathcal{A}^{\frac{1}{2}k(k+1)}],
b_1\mathcal{P}_1^+\prod_{i,j=1, i\leq j}^{k} \mathcal{P}_{ij}^+,
 \ldots,
b_{k-1}\prod_m^{k-1}\mathcal{P}_m^+\prod_{i,j=1, i\leq j}^{k}
\mathcal{P}_{ij}^+ \Bigr)|\Lambda^{(s_{max}-\frac{k(k+3)}{2}-1)}{}^{1}_{0}\rangle =0.
\end{eqnarray}
Finally, realizing the same algorithm, however, starting from the
parameters
$|\Lambda^{(s_{max}-\frac{k(k+3)}{2}-2)}{}^{l}_{0}\rangle$, and
extracting from it, as well as from the operator $\Delta{}Q$
(\ref{deltaQ}), the ghost coordinates and momenta $\eta_m$,
$\mathcal{P}_{m}^+$, $\eta_{ij}$, $\mathcal{P}_{ij}^+$, $i\leq j$
and $\vartheta_{ps}$, $\lambda_{ps}^+$, for $p<s$, with a
parametric dependence on $q_i, p_i^+$, we obtain
$\frac{1}{2}k(k-3)$ sets of gauge conditions for the parameters
$|\Lambda^{(s_{max}-\frac{k(k+3)}{2}-1-m)}{}^{l}_{0}\rangle$, for
$m=1, \ldots, \frac{1}{2}k(k-1)$
\begin{eqnarray} \label{gk(k-1)2-1}
&& \Bigl([\mathcal{B}^{\frac{1}{2}k(k+3)}],
d_{12}\lambda_{12}^+\prod_{i,j=1, i\leq j}^{k} \mathcal{P}_{ij}^+,
b_2\prod_m^2\mathcal{P}_m^+\prod_{i,j=1, i\leq j}^{k}
\mathcal{P}_{ij}^+ \Bigr)|\Lambda^{(s_{max}-\frac{k(k+3)}{2}-2)}{}^{0}_{0}\rangle =0 ,\nonumber \\
&& [\mathcal{B}^{\frac{1}{2}k(k+3)}],
|\Lambda^{(s_{max}-\frac{k(k+3)}{2}-2)}{}^{1}_{0}\rangle =0 ; \\
&& \hspace{3cm}\ldots \ldots \ldots \ldots \ldots \ldots\ldots \ldots \ldots\nonumber\\
 \label{gk(k-1)2fin} &&
\Bigl([\mathcal{B}^{\frac{1}{2}k(k+3)}],
d_{12}\lambda_{12}^+\prod_{i,j=1, i\leq j}^{k} \mathcal{P}_{ij}^+,
 \ldots,
d_{k-1{}k}\prod_{p,s=1, p<
s}^{k}\lambda_{ps}^+\prod_{i,j=1, i\leq j}^{k}
\mathcal{P}_{ij}^+ \Bigr) |\Lambda^{(s_{max}-{k(k+1)}-1)}{}^{0}_{0}\rangle=0, \nonumber \\
&&
\Bigl([\mathcal{B}^{\frac{1}{2}k(k+3)}],
d_{12}\lambda_{12}^+\prod_{i,j=1, i\leq j}^{k} \mathcal{P}_{ij}^+,
 \ldots,
d_{k-2{}k}\prod_{r}^{k-2}\lambda_{rk}^+\prod_{p,s=1, p<
s}^{k-1}\lambda_{ps}^+\prod_{i,j=1, i\leq j}^{k}
\mathcal{P}_{ij}^+ \Bigr)\nonumber\\
&& \qquad  |\Lambda^{(s_{max}-{k(k+1)}-1)}{}^{1}_{0}\rangle=0.
\end{eqnarray}
The set of the
operators $[\mathcal{B}^{r}]$ is determined
in  Eqs. (\ref{gk(k-1)2-1}), (\ref{gk(k-1)2fin}) from  Eqs.
(\ref{gk(k-1)2f}) as
\begin{equation}
[\mathcal{B}^{\frac{1}{2}k(k+3)}] =
\Bigl([\mathcal{A}^{\frac{1}{2}k(k+1)}],
b_1\mathcal{P}_1^+\prod_{i,j=1, i\leq j}^{k} \mathcal{P}_{ij}^+,
 \ldots,
b_k\prod_m^k\mathcal{P}_m^+\prod_{i,j=1, i\leq j}^{k}
\mathcal{P}_{ij}^+ \Bigr).
\end{equation}
Because of absence of unused odd ghosts in the  rest set
of gauge parameters $|\Lambda^{(s_{max}-{k(k+1)}-1
-m)}{}^{l}_{0}\rangle$, for $m=1, \ldots  (s_{max}-{k(k+1)}-1)$,
we derive, as a result of the above procedure, the same gauge
conditions as in (\ref{gk(k-1)2fin}):
\begin{eqnarray}
\label{restgauge} &&
[\mathcal{C}^{k(k+1)}] |\Lambda^{(s_{max}-{k(k+1)}-1 -m)}{}^{l}_{0}\rangle=0,\texttt{ for }l=0,1\,, \\
&&
 [\mathcal{C}^{k(k+1)}] = \Bigl([\mathcal{B}^{\frac{1}{2}k(k+3)}],
d_{12}\lambda_{12}^+\prod_{i,j=1, i\leq j}^{k} \mathcal{P}_{ij}^+,
 \ldots,
d_{k-1{}k}\prod_{p,s=1, p<
s}^{k}\lambda_{ps}^+\prod_{i,j=1, i\leq j}^{k}
\mathcal{P}_{ij}^+ \Bigr)\label{Ck(k+1)}.
\end{eqnarray}
Finally, the gauge conditions for the fields $|\chi^l_0\rangle$
have the form
\begin{equation}
[\mathcal{C}^{k(k+1)}] |\chi^{l}_0\rangle =0, \label{G1}
\end{equation}
in terms of the operator-valued set $[\mathcal{C}^{k(k+1)}]$
introduced in Eqs. (\ref{Ck(k+1)}).

Let us now turn to removing the auxiliary fields, using the
equations of motion.

\subsection{Auxiliary Fields Removal by Solution of Equations of
Motion}

In the beginning, we decompose the fields
$\chi^{l}\equiv|\chi^{l}_0\rangle$ as follows:
\begin{align}
&|\chi^{l}\rangle = |\chi^{l}_0 \rangle + \mathcal{P}_{11}^+ |\chi^{l}_{1}
\rangle, \hspace{-2em}
&&\hspace{-2em}|\chi^{l}_{(0)_{\frac{1}{2}k(k+1)}} \rangle =
|\chi^{l}_{(0)_{\frac{1}{2}k(k+1)}0} \rangle + \mathcal{P}_{1}^+
|\chi^{l}_{(0)_{\frac{1}{2}k(k+1)}1}\rangle,
\\
&|\chi^{l}_{0}\rangle = |\chi^{l}_{00} \rangle + \mathcal{P}_{12}^+
|\chi^{l}_{01} \rangle, \hspace{-2em} &&\hspace{-2em}
|\chi^{l}_{(0)_{\frac{1}{2}k(k+1)}0} \rangle =
|\chi^{l}_{(0)_{\frac{1}{2}k(k+1)+2}} \rangle + \mathcal{P}_{2}^+
|\chi^{l}_{(0)_{\frac{1}{2}k(k+1)}01}\rangle,
\\
& \qquad\ldots , && \qquad\ldots ,
\nonumber
\end{align}
   \vspace{-3ex}
\begin{align}
&|\chi^{l}_{(0)_k} \rangle = |\chi^{l}_{(0)_k0} \rangle + \mathcal{P}_{1k}^+
|\chi^{l}_{(0)_k1} \rangle,\hspace{-1em}
&&\hspace{-2em}|S^0_{(0)_{\frac{1}{2}k(k+3)}} \rangle =
|\chi^{l}_{(0)_{\frac{1}{2}k(k+3)}0} \rangle + \mathcal{P}_{k}^+
|\chi^{l}_{(0)_{\frac{1}{2}k(k+3)}1} \rangle
\\
&|\chi^{l}_{(0)_{k+1}} \rangle = |\chi^{l}_{(0)_{k+2}} \rangle +
\mathcal{P}_{22}^+ |\chi^{l}_{(0)_{k+1}1} \rangle, \hspace{-1em}
&&\hspace{-2em} |\chi^{l}_{(0)_{\frac{1}{2}k(k+3)}0} \rangle =
|\chi^{l}_{(0)_{\frac{1}{2}k(k+3)}00} \rangle + \lambda_{12}^+
|\chi^{l}_{(0)_{\frac{1}{2}k(k+3)}01} \rangle
\\
&\qquad\ldots , \hspace{-2em} &&\hspace{-2em} \qquad \ldots ,
\nonumber\\
\hspace{-0.5em}&\hspace{-0.5em}|\chi^{l}_{(0)_{\frac{1}{2}k(k+1)-1}}\hspace{-0.1em}
\rangle \hspace{-0.2em}=\hspace{-0.2em}
|\chi^{l}_{(0)_{\frac{1}{2}k(k+1)}} \hspace{-0.2em}\rangle
\hspace{-0.1em}+ \hspace{-0.1em}\mathcal{P}_{kk}^+\hspace{-0.1em}
|\chi^{l}_{(0)_{\frac{1}{2}k(k+1)-1}1} \hspace{-0.2em}\rangle,
\hspace{-0.5em} &&|\chi^{l}_{(0)_{k(k+1)-1}} \hspace{-0.2em}\rangle
\hspace{-0.1em}= \hspace{-0.1em}|\chi^{l}_{(0)_{k(k+1)}}
\hspace{-0.1em}\rangle\hspace{-0.1em} +
\hspace{-0.2em}\lambda_{k-1{}k}^+ |\chi^{l}_{(0)_{k(k+1)-1}1}
\hspace{-0.1em}\rangle
\end{align}
It should be noted that, due to $gh(|\chi^{l}\rangle)=-l$ and the
spin value, first, the term independent on the ghost momenta is
absent from the vector (Dirac-spinor) $|\chi^{1}\rangle$, i.e.,
$|\chi^{1}_0\rangle = 0$, whereas the vector
$|\chi^{0}_{(0)_{k(k+1)}} \rangle$ contains the physical vector
$|\Psi\rangle$ (\ref{PhysState}) for the vanishing ghost momenta
$p_i^+$, for $i=1,\ldots, k$, in view of the representation
following from Eq. (\ref{chif}), and besides
\begin{equation}\label{decomppq}
  |\chi^{0}_{(0)_{k(k+1)}} \rangle_{(n)_k} = \sum_n \prod_{i=1}^k(q_i^+)^{n_{ai}}(p_i^+)^{n_{bi}}(\eta^+_{i})^{n_{f i}}|\Psi(a^+_i)^{(n)_{ai}(n)_{bi}(n)_{f i}}_{0}\rangle_{(n-n_{a}-n_{b}-n_{f})_k},
\end{equation}
under the requirement  that all the summands should obey the
relations $(n-n_{a}-n_{b}-n_{f})_k \geq (0)_k$, due to the absence
of the auxiliary even oscillators $f_i^+, b_i^+, b_{ij}^+,
d^+_{rs}$ [as a consequence of the gauge conditions (\ref{G1})] in
the vectors $|\Psi...\rangle$ in Eq. (\ref{decomppq}). Indeed, for
$(n_b)_{k} = (0)_k $, we have
$|\Psi(a^+_i)^{(n)_{ai}(0)_{bi}(n)_{f
i}}_{0}\rangle_{(n-n_{a}-0_{b}-n_{f})_k}$  =
$|\Psi(a^+_i)^{(0)_{ai}(0)_{bi}(0)_{f
i}}_{0}\rangle_{(n-0_{a}-0_{b}-0_{f})_k}$ = $|\Psi\rangle$.

After that, by analogy with the fields, we extract from
$\Delta{}Q$ (\ref{deltaQ}), first, the dependence on $\eta_{11}$,
$\mathcal{P}_{11}^+$,
 $\eta_{12}$, $\mathcal{P}_{12}^+$, \ldots ,
$\eta_{1k}$, $\mathcal{P}_{1k}^+$, next, the dependence on
$\eta_{l}$, $\mathcal{P}_{l}^+$, $l=1,\ldots , k$, and on
$\vartheta_{ps}$, $\lambda_{ps}^{+}$,  $p<s$, respectively.

Substituting these $k(k+1)$ decompositions into the equations of
motion
\begin{eqnarray}
 \Delta{}Q|\chi^{0}_{0}\rangle
+\frac{1}{2}\bigl\{\tilde{T}_0,\eta_i^+\eta_i\bigr\}
|\chi^{1}_{0}\rangle =0, \qquad
\tilde{T}_0|\chi^{0}_{0}\rangle +
\Delta{}Q|\chi^{1}_{0}\rangle =0, \label{EofM12a}
\end{eqnarray}
and using the gauge conditions (\ref{G1}), one can show that,
first, $|\chi^{1}_{(0)_{k(k+1)}} \rangle = 0$ in (\ref{EofM12a}),
second, $|\chi^{0}_{(0)_{k(k+1)}} \rangle = 0$ for the
$p_i^+$-dependent vector in (\ref{EofM12a}), so that only the
original $|\Psi\rangle$ vector survives in
$|\chi^{0}_{(0)_{k(k+1)}} \rangle $, and then we obtain that
$|\chi^{1}_{(0)_{k(k+1)-1}} \rangle = 0$ from the second equation
and $|\chi^{0}_{(0)_{k(k+1)1}} \rangle = 0$ from the first one in
(\ref{EofM12a}), and so on, until $|\chi^{1}_{0} \rangle = 0$ and
$|\chi^{0}_{1} \rangle = 0$, which implies
\begin{eqnarray}
{}&& \Delta{}Q|\chi^{0}_{0}\rangle=0,\qquad  \tilde{t}_0|\chi^0\rangle=0, \qquad |\chi^{1} \rangle = 0, \label{E1}\\
{}&& |\chi^{0}_{(0)_{k(k+1)-1}1}\rangle=|\chi^{0}_{(0)_{k(k+1)-2}1}\rangle=
\ldots = |\chi^{0}_{01}\rangle= |\chi^{0}_1\rangle=0. \label{E3}
\end{eqnarray}
Eqs. (\ref{E1}) and (\ref{E3}) imply that all the auxiliary fields
vanish and, as a result, we have
$|\chi^0\rangle_{(n)_k}=|\Psi\rangle$, and the equations of motion
(\ref{Eq-0m}), (\ref{Eq-1}), (\ref{Eq-2}) hold true. Thus, we have
proved that the space of BRST cohomologies of the operator $Q$
(\ref{Q}) with a vanishing ghost number is determined only by the
constraints (\ref{t0ti}),  (\ref{tij}), corresponding to an
irreducible Poincare-group representation with a given spin.

It should be noted that in the massless case the above proof of
one-to-one correspondence of the Lagrangian equations of motion
(\ref{EofM1}), (\ref{EofM2}) to Eqs. (\ref{Eq-0m}), (\ref{Eq-1}),
(\ref{Eq-2}) becomes slightly  corrected, because the $k$ gauge
fixing conditions (\ref{gLmax-k(k+1)2-1})--(\ref{gk(k-1)2f}) doe
not hold true, and in the remaining
Eqs.(\ref{gk(k-1)2-1})--(\ref{G1}) there are no operators $b_i
\prod_{s=1}^i\mathcal{P}^+_s, i=1,\ldots,k$. However, we can
straightforwardly prove the validity of the same conclusion as for
massive fermionic HS fields in the Lagrangian formulation for
massless fermionic HS fields.

\section{Decomposition of Fields and Gauge Fock Space Vectors  for
Spin $\mathbf{s}=(\frac{5}{2},\frac{3}{2})$
spin-tensor}\label{example5232} \setcounter{equation}{0}

We examine here only the structure of  corresponding Fock space
$\mathcal{H}_{tot}$ vectors $|\chi^l_0\rangle_{(2,1)},
|\Lambda^{(0)}{}^l_0\rangle_{(2,1)}$,
$|\Lambda^{(2)}{}^l_0\rangle_{(2,1)},
|\Lambda^{(3)}{}^l_0\rangle_{(2,1)}$, for $l=0,1$,
$|\Lambda^{(3)}{}^0_0\rangle_{(2,1)}$ in (\ref{GT1})--(\ref{GTi2})
for the example of Subsection~\ref{ex5232} to be used in a
Lagrangian formulation for a massless (and then in
Subsection~\ref{ex5232m} for massive) spin-tensor $\Psi_{\mu\nu,
\rho}$ in a $d$-dimensional flat space-time, which is
characterized by the Poincare-group irreducible coniditions
(\ref{Eq-0})--(\ref{Eq-2}) and a hook-like Young tableux
$\begin{array}{|c|c|}\hline%\vphantom{\biggm|}
  \!\mu \!&\! \nu\!  \\
   \hline%\vphantom{\biggm|}
    \! \rho\!   \\
  \cline{1-1}
\end{array}\
$. It should be noted that the maximal stage of reducibility, $L_k
= \sum_{o=1}^kn_o+ \frac{1}{2}k(k-1)-1$, for a spin-tensor with
$k$ group of symmetric indices is reached for any Young tableaux,
in contrast to the case of bosonic HS fields \cite{BRmixbos}.
Thus, in the case of $k=2$ rows, the value $L_2 = n_1 + n_2$ is
the stage of reducibility for any spin-tensor
$\Psi_{(\mu)_{n_1},(\nu)_{n_2}}$, and for $\Psi_{\mu\nu, \rho}$ we
have $L_2 =3$.

Therefore, $|\Lambda^{(3)}{}^1_0\rangle_{(2,1)} \equiv 0$
identically, whereas the lowest fermionic independent gauge
parameter $|\Lambda^{(3)}{}^0_0\rangle_{(2,1)}$ in the general
expression (\ref{chif}), subject to the spin (\ref{nidecompos})
and ghost number (\ref{ghnumg}) conditions for $i=1,2$ and $s=3$,
(for the minimal ghost number $gh_{\min}=-4$), has a
representation with $2$ summands:
\begin{eqnarray}
|\Lambda^{(3)}{}^0_0\rangle_{(2,1)} &=&
(p_1^+)^2\lambda^+_{12}\bigl(p_1^+ \tilde{\gamma}|\psi^{(3)}{}^0_1 \rangle_{(0)}+\mathcal{P}_1^+|\psi^{(3)}{}^0_2 \rangle_{(0)}\bigr)
, \quad  |\psi^{(3)}{}^0_m \rangle_{(0)} =
|0\rangle\psi^{(3)}{}^0_m(x) , \label{x0-3}
\end{eqnarray}
for the Dirac-spinors $\psi^{(3)}{}^0_m$, $m=1,2$, and where we
have used the notation $(0) \equiv (0,0)$, for $|\psi^{(3)}{}^0_m
\rangle_{(0)}\equiv |\psi^{(3)}{}^0_m \rangle_{(0,0)}$.

For the bosonic reducible gauge parameters
$|\Lambda^{(2)}{}^l_0\rangle_{(2,1)}$ of the second level for
$s=2$ in Eq.(\ref{ghnumg}), we have a decomposition in odd (for
$l=0$) and even (for $l=1$) powers of the ghosts, starting from
the third (for $l=0$) and fourth (for $l=1$) powers of the ghost
momenta, and consisting of $(10+2)$ summands, for $l=0, 1$,
respectively,
\begin{eqnarray}
%%%%%%%%%%
%%%%%%%%%%
|\Lambda^{(2)}{}^0_0\rangle_{(2,1)} &=&
p_1^+ \Bigl(p_1^+\Bigl\{p^+_{1}\tilde{\gamma}|\psi^{(2)}{}^0_1 \rangle_{(-1,1)}+ p^+_{2}\tilde{\gamma}|\psi^{(2)}{}^0_2 \rangle_{(0)}+ \mathcal{P}^+_{1}|\psi^{(2)}{}^0_3 \rangle_{(-1,1)}+ \mathcal{P}^+_{2}|\psi^{(2)}{}^0_4 \rangle_{(0)}\nonumber\\
&& + \lambda^+_{12}|\psi^{(2)}{}^0_5  \rangle_{(1,0)}\Bigr\} + \mathcal{P}_1^+\Bigl\{ p^+_{2}|\psi^{(2)}{}^0_6 \rangle_{(0)} + \mathcal{P}^+_{2}\tilde{\gamma}|\psi^{(2)}{}^0_7 \rangle_{(0)}+\lambda^+_{12}\tilde{\gamma}|\psi^{(2)}{}^0_8 \rangle_{(1,0)}\Bigr\}\nonumber\\
&& +\mathcal{P}_{11}^+\lambda^+_{12}\tilde{\gamma}|\psi^{(2)}{}^0_9 \rangle_{(0)}\Bigr)
 +\mathcal{P}_{1}^+\mathcal{P}_{11}^+\lambda^+_{12}|\psi^{(2)}{}^0_{10} \rangle_{(0)},
\label{x0-2}\\
|\Lambda^{(2)}{}^1_0\rangle_{(2,1)} &=&
(p_1^+)^2\lambda^+_{12}\bigl(p_1^+ |\psi^{(2)}{}^1_1 \rangle_{(0)}+\mathcal{P}_1^+\tilde{\gamma}|\psi^{(2)}{}^1_2 \rangle_{(0)}\bigr)
\,, \label{x1-2}
\end{eqnarray}
where the decomposition of ghost-independent vectors in powers of
the initial and auxiliary creation operators in
$\mathcal{H}\otimes \mathcal{H}'$ is written as
\begin{eqnarray}
&& |\psi^{(2)}{}^1_m \rangle_{(0)} =
|0\rangle\psi^{(2)}{}^1_m(x)  \,,
\qquad\qquad\qquad\qquad\qquad\qquad |\psi^{(2)}{}^0_n \rangle_{(0)} =|0\rangle\psi^{(2)}{}^0_n(x) \,,\label{x-2decompi}\\
&&|\psi^{(2)}{}^0_p \rangle_{(1,0)}  =  a_1^{+\mu} |0\rangle
\psi^{(2)}{}^0_{p|\mu} (x)+ f_1^+ \tilde{\gamma} |0\rangle
\psi^{(2)}{}^0_{p} (x)\,, \qquad |\psi^{(2)}{}^0_o \rangle_{(-1,1)} = d_{12}^+ |0\rangle \psi^{(2)}{}^0_o(x)\,,   \label{x-2decompf}
\end{eqnarray}
for $m=1,2,\,  n=2,4, 6,7, 9, 10, \,o = 1,3$ and $p=5,8$.

For reducible fermionic gauge parameters of the first level
$|\Lambda^{(1)}{}^l_0\rangle_{(2,1)}$ for $s=1$ in
Eq.(\ref{ghnumg}), and for the same spin value $(2,1)$ in the
Eqs.(\ref{nidecompos}), we obtain, using the general expression
(\ref{chif}), a decomposition in even (for $l=0$) and odd (for
$l=1$) powers of ghosts, consisting of $(21+10)$ summands, for
$l=0, 1$, and starting from the second order in $\mathcal{P}_I$
(from the third order in ghost momenta $\mathcal{P}_I$ for $l=1$),
\begin{eqnarray}
%%%%%%%%%%
%%%%%%%%%%%%%%%%%%%%%%%%%%%%
|\Lambda^{(1)}{}^0_0\rangle_{(2,1)} &=&
p_1^+\Biggl(p_1^+\Bigl\{ |\psi^{(1)}{}^0_1 \rangle_{(0,1)} + p_1^+\vartheta^+_{12}\tilde{\gamma}|\psi^{(1)}{}^0_2 \rangle_{(0)}  + \mathcal{P}_1^+\vartheta^+_{12}|\psi^{(1)}{}^0_3 \rangle_{(0)} + q_1^+\lambda^+_{12}\tilde{\gamma}|\psi^{(1)}{}^0_4 \rangle_{(0)}\nonumber
\\
  && + \eta_1^+\lambda^+_{12}|\psi^{(1)}{}^0_5 \rangle_{(0)}  \Bigr\} + \mathcal{P}_1^+\Bigl\{\tilde{\gamma}|\psi^{(1)}{}^0_6 \rangle_{(0,1)}
  + q_1^+\lambda^+_{12}|\psi^{(1)}{}^0_7 \rangle_{(0)} + \eta_1^+\lambda^+_{12}\tilde{\gamma}|\psi^{(1)}{}^0_8 \rangle_{(0)} \Bigr\}\nonumber
  \\
  && + p_2^+|\psi^{(1)}{}^0_9 \rangle_{(1,0)} + \mathcal{P}_2^+\tilde{\gamma}|\psi^{(1)}{}^0_{10} \rangle_{(1,0)}  + \mathcal{P}_{11}^+\tilde{\gamma}|\psi^{(1)}{}^0_{11} \rangle_{(-1,1)}+ \mathcal{P}_{12}^+\tilde{\gamma}|\psi^{(1)}{}^0_{12} \rangle_{(0)} \nonumber
  \end{eqnarray}
  \vspace{-3ex}
  \begin{eqnarray}
  && + \lambda_{12}^+\tilde{\gamma}|\psi^{(1)}{}^0_{13} \rangle_{(2,0)}\Biggr) + \mathcal{P}_1^+\Bigl(p_2^+\tilde{\gamma}|\psi^{(1)}{}^0_{14} \rangle_{(1,0)} + \mathcal{P}_2^+|\psi^{(1)}{}^0_{15} \rangle_{(1,0)}+ \mathcal{P}_{11}^+|\psi^{(1)}{}^0_{16} \rangle_{(-1,1)}\nonumber\\
  && +\mathcal{P}_{12}^+|\psi^{(1)}{}^0_{17} \rangle_{(0)} +  \lambda_{12}^+|\psi^{(1)}{}^0_{18} \rangle_{(2,0)}\Bigr) +  p_2^+ \mathcal{P}_{11}^+\tilde{\gamma}|\psi^{(1)}{}^0_{19} \rangle_{(0)} +  \mathcal{P}_2^+ \mathcal{P}_{11}^+|\psi^{(1)}{}^0_{20} \rangle_{(0)}
  \nonumber\\
  && + \mathcal{P}_{11}^+\lambda_{12}^+|\psi^{(1)}{}^0_{21} \rangle_{(1,0)}
\,, \label{x1-0}\\
%%%%%%%%%%
|\Lambda^{(1)}{}^1_0\rangle_{(2,1)} &=&
p_1^+ \Bigl(p_1^+\Bigl\{p^+_{1}|\psi^{(1)}{}^1_1 \rangle_{(-1,1)}+ p^+_{2}|\psi^{(1)}{}^1_2 \rangle_{(0)}+ \mathcal{P}^+_{1}\tilde{\gamma}|\psi^{(1)}{}^1_3 \rangle_{(-1,1)}+ \mathcal{P}^+_{2}\tilde{\gamma}|\psi^{(1)}{}^1_4 \rangle_{(0)}\nonumber\\
&& + \lambda^+_{12}\tilde{\gamma}|\psi^{(1)}{}^1_5  \rangle_{(1,0)}\Bigr\} + \mathcal{P}_1^+\Bigl\{ p^+_{2}\tilde{\gamma}|\psi^{(1)}{}^1_6 \rangle_{(0)} + \mathcal{P}^+_{2}|\psi^{(1)}{}^1_7 \rangle_{(0)}+\lambda^+_{12}|\psi^{(1)}{}^1_8 \rangle_{(1,0)}\Bigr\}\nonumber\\
&& +\mathcal{P}_{11}^+\lambda^+_{12}|\psi^{(1)}{}^1_9 \rangle_{(0)}\Bigr)
 +\mathcal{P}_{1}^+\mathcal{P}_{11}^+\lambda^+_{12}\tilde{\gamma}|\psi^{(1)}{}^1_{10} \rangle_{(0)},
\label{x1-1}
%%%%%%%%%%%%%%%%%%%%%%%%%%%%
\end{eqnarray}
where the ghost-independent vectors
$|\psi^{(1)}{}^1_n\rangle_{(0)}, |\psi^{(1)}{}^1_p
\rangle_{(1,0)}$ for $n=2, 4, 6, 7, 9, 10,\, o =1, 3 $ and $p=5,8$
in $\mathcal{H}\otimes \mathcal{H}'$ have the same decomposition
and the properties as those in Eqs. (\ref{x-2decompi}),
(\ref{x-2decompf}), whereas the decomposition of ghost-independent
vectors in Eq. (\ref{x1-0}), different from those in Eqs.
(\ref{x-2decompi}), (\ref{x-2decompf}), reads
\begin{eqnarray}
&& |\psi^{(1)}{}^0_n \rangle_{(0,1)} =a_1^{+\mu}d_{12}^+ |0\rangle
\psi^{\prime(1)}{}^0_{n|\mu} +a_2^{+\mu}|0\rangle
\psi^{(1)}{}^0_{n|\mu} + f_1^+ d_{12}^+\tilde{\gamma} |0\rangle
\psi^{\prime(1)}{}^0_{n}  + f_2^+\tilde{\gamma} |0\rangle
\psi^{(1)}{}^0_{n}  \,,    \label{x-1decompf}\\
&& |\psi^{(1)}{}^0_{r} \rangle_{(2,0)}  =  a_1^{+\mu}\bigl(a_1^{+\nu} |0\rangle
\psi^{(1)}{}^0_{r|\mu\nu} + f_1^+ \tilde{\gamma} |0\rangle
\psi^{(1)}{}^0_{r|\mu}\bigr)  + b_{11}^+|0\rangle
\psi^{(1)}{}^0_{r}\,,  \label{x-1decompff}
\end{eqnarray}
for $n = 1, 6,  \,r = 13, 18$.

Then, for the reducible bosonic gauge parameter of the zeroth
level (proper gauge parameters)
$|\Lambda^{(0)}{}^l_0\rangle_{(2,1)} \equiv
|\Lambda^l_0\rangle_{(2,1)}$ for $s=0$ in the Eq.(\ref{ghnumg})
and for the same spin value $(2,1)$ in Eqs.(\ref{nidecompos}), we
obtain, from the general expression (\ref{chif}), a decomposition
in odd (for $l=0$) and even( for $l=1$) powers of ghosts,
consisting of $(35+21)$ summands, respectively, for $l=0, 1$, and
starting from the first order in the ghost momenta $\mathcal{P}_I$
(from the second order in ghost momenta $\mathcal{P}_I$ for
$l=1$),
\begin{eqnarray}
%%%%%%%%%%
%%%%%%%%%%%%%%%%%%%%%%%%%%%%
|\Lambda^0_0\rangle_{(2,1)} &=&
p_1^+\Biggl(\tilde{\gamma}|\psi^0_1 \rangle_{(1,1)}+p_1^+\Bigl\{ q_1^+\tilde{\gamma}|\psi^0_2 \rangle_{(-1,1)} +\eta_1^+|\psi^0_3 \rangle_{(-1,1)}+q_2^+\tilde{\gamma}|\psi^0_4 \rangle_{(0)} +\eta_2^+|\psi^0_5 \rangle_{(0)} \nonumber\\
  && +\vartheta_{12}^+|\psi^0_6 \rangle_{(1,0)}\Bigr\}+\mathcal{P}_1^+\Bigl\{ q_1^+|\psi^0_7 \rangle_{(-1,1)} +\eta_1^+\tilde{\gamma}|\psi^0_8 \rangle_{(-1,1)}+q_2^+|\psi^0_9 \rangle_{(0)}+\eta_2^+\tilde{\gamma}|\psi^0_{10} \rangle_{(0)} \nonumber\\
  && +\vartheta_{12}^+\tilde{\gamma}|\psi^0_{11} \rangle_{(1,0)}\Bigr\} +p_2^+\Bigl\{ q_1^+\tilde{\gamma}|\psi^0_{12} \rangle_{(0)} +\eta_1^+|\psi^0_{13} \rangle_{(0)}\Bigr\}+\mathcal{P}_2^+\Bigl\{ q_1^+|\psi^0_{14} \rangle_{(0)}+\eta_1^+\tilde{\gamma}|\psi^0_{15} \rangle_{(0)}\Bigr\}\nonumber\\
  &&  +  \mathcal{P}_{11}^+\vartheta_{12}^+\tilde{\gamma}|\psi^0_{16} \rangle_{(0)} + \lambda_{12}^+ \Bigl\{q_1^+|\psi^0_{17} \rangle_{(1,0)}+\eta_1^+\tilde{\gamma}|\psi^0_{18} \rangle_{(1,0)} +\eta_{11}^+\tilde{\gamma}|\psi^0_{19} \rangle_{(0)} \Bigr\}\Biggr)\nonumber\\
  &&   +  p_2^+\Bigl( \tilde{\gamma}|\psi^0_{20} \rangle_{(2,0)} + \mathcal{P}_1^+\Bigl\{ q_1^+|\psi^0_{21} \rangle_{(0)} +\eta_1^+\tilde{\gamma}|\psi^0_{22} \rangle_{(0)}\Bigr\} \Bigr) +  \mathcal{P}_2^+\Bigl( |\psi^0_{23} \rangle_{(2,0)} \nonumber
  \\
  &&+ \mathcal{P}_1^+\Bigl\{ q_1^+\tilde{\gamma}|\psi^0_{24} \rangle_{(0)} +\eta_1^+|\psi^0_{25} \rangle_{(0)}\Bigr\} \Bigr)
  +\mathcal{P}_1^+\Biggl(|\psi^0_{26} \rangle_{(1,1)}+  \mathcal{P}_{11}^+\vartheta_{12}^+\tilde{\gamma}|\psi^0_{27} \rangle_{(0)}\nonumber
 \\
  &&   + \lambda_{12}^+ \Bigl\{q_1^+\tilde{\gamma}|\psi^0_{28} \rangle_{(1,0)}  +\eta_1^+|\psi^0_{29} \rangle_{(1,0)}+\eta_{11}^+|\psi^0_{30} \rangle_{(0)} \Bigr\}\Biggr)+  \mathcal{P}_{11}^+  \Bigl(|\psi^0_{31} \rangle_{(0,1)}\nonumber
  \\
  &&
  +  \lambda_{12}^+ \Bigl\{q_1^+\tilde{\gamma}|\psi^0_{32} \rangle_{(0)} +\eta_1^+|\psi^0_{33} \rangle_{(0)}\Bigr\} \Bigr)+\mathcal{P}_{12}^+  |\psi^0_{34} \rangle_{(1,0)}+  \lambda_{12}^+|\psi^0_{35} \rangle_{(3,0)}
\,, \label{x0-0}\end{eqnarray}
  %%%%%%%%%%%%%%%%%%%%%%%%%%%%
  %%%%%%%%%%%%%%%%%%%%%%%%%
    \begin{eqnarray}
%%%%%%%%%%
|\Lambda^1_0\rangle_{(2,1)} &=&
p_1^+\Biggl(p_1^+\Bigl\{ \tilde{\gamma}|\psi^1_1 \rangle_{(0,1)} + p_1^+\vartheta^+_{12}|\psi^1_2 \rangle_{(0)}  + \mathcal{P}_1^+\vartheta^+_{12}\tilde{\gamma}|\psi^1_3 \rangle_{(0)} + q_1^+\lambda^+_{12}|\psi^1_4 \rangle_{(0)}\nonumber\\
  && + \eta_1^+\lambda^+_{12}\tilde{\gamma}|\psi^1_5 \rangle_{(0)}  \Bigr\} + \mathcal{P}_1^+\Bigl\{|\psi^1_6 \rangle_{(0,1)}
  + q_1^+\lambda^+_{12}\tilde{\gamma}|\psi^1_7 \rangle_{(0)} + \eta_1^+\lambda^+_{12}|\psi^1_8 \rangle_{(0)} \Bigr\}\nonumber\\
  && + p_2^+\tilde{\gamma}|\psi^1_9 \rangle_{(1,0)} + \mathcal{P}_2^+|\psi^1_{10} \rangle_{(1,0)}  + \mathcal{P}_{11}^+|\psi^1_{11} \rangle_{(-1,1)}+ \mathcal{P}_{12}^+|\psi^1_{12} \rangle_{(0)} \nonumber\\
  && + \lambda_{12}^+|\psi^1_{13} \rangle_{(2,0)}\Biggr) + \mathcal{P}_1^+\Bigl(p_2^+|\psi^1_{14} \rangle_{(1,0)} + \mathcal{P}_2^+\tilde{\gamma}|\psi^1_{15} \rangle_{(1,0)}+ \mathcal{P}_{11}^+\tilde{\gamma}|\psi^1_{16} \rangle_{(-1,1)}\nonumber\\
  && +\mathcal{P}_{12}^+\tilde{\gamma}|\psi^1_{17} \rangle_{(0)} +  \lambda_{12}^+\tilde{\gamma}|\psi^1_{18} \rangle_{(2,0)}\Bigr) +  p_2^+ \mathcal{P}_{11}^+|\psi^1_{19} \rangle_{(0)} +  \mathcal{P}_2^+ \mathcal{P}_{11}^+\tilde{\gamma}|\psi^1_{20} \rangle_{(0)}
  \nonumber\\
  && + \mathcal{P}_{11}^+\lambda_{12}^+\tilde{\gamma}|\psi^1_{21} \rangle_{(1,0)}
\,, \label{x0-1}
%%%%%%%%%%%%%%%%%%%%%%%%%%%%
\end{eqnarray}
where the ghost-independent vectors $|\psi^1_n \rangle_{(0)},
|\psi^{(1)}{}^1_p \rangle_{(1,0)}, |\psi^{(1)}{}^1_r
\rangle_{(-1,1)}, |\psi^{(1)}{}^1_o \rangle_{(0,1)},
|\psi^{(1)}{}^1_t \rangle_{(2,0)}$  for $n=2-5,  7, 8, 12, 17, 19,
20,$ $p = 9, 10, 14, 15, 21$, $r= 11, 16, o= 1, 6$  and  $t = 13,
18$ in $\mathcal{H}\otimes \mathcal{H}'$ have the same
decomposition and the properties as those in Eqs.
(\ref{x-1decompf}), (\ref{x-1decompff}), whereas the
ghost-independent vectors in Eq. (\ref{x0-0}), different from
those in Eqs. (\ref{x-2decompi}), (\ref{x-2decompf}),
(\ref{x-1decompf}), (\ref{x-1decompff}) are decomposed as follows:
\begin{eqnarray}
&&\hspace{-2em} |\psi^0_{n} \rangle_{(1,1)}  =  a_1^{+\mu}\bigl(a_1^{+\nu}d^+_{12} |0\rangle
\psi^0_{n|\mu\nu} + f_1^+d^+_{12} \tilde{\gamma} |0\rangle
\psi^{\prime 0}_{n|\mu} + a_2^{+\nu} |0\rangle
\psi^0_{n|\mu,\nu} + f_2^+ \tilde{\gamma} |0\rangle
\psi^{\prime\prime 0}_{n|\mu}\bigr)\nonumber \\
&& \phantom{\hspace{-2em}|\psi^0_{n} \rangle_{(1,1)}} + f_1^+ a_2^{+\nu} \tilde{\gamma}|0\rangle
\psi^0_{n|\mu} + b_{11}^+d^+_{12}|0\rangle
\psi^{\prime 0}_{n} + b^+_{12}|0\rangle
\psi^{\prime\prime0}_{n} + f^+_{1}f^+_{2}|0\rangle
\psi^{0}_{n}  \,,  \label{x-0decompi}\\
&&\hspace{-2em} |\psi^0_{35} \rangle_{(3,0)}  =  a_1^{+\mu}\bigl(a_1^{+\nu}a_1^{+\rho} |0\rangle
\psi^0_{35|\mu\nu\rho} + a_1^{+\nu}f_1^+ \tilde{\gamma} |0\rangle
\psi^0_{35|\mu\nu} + b_{11}^+|0\rangle
\psi^0_{35|\mu} \bigr)  + f_1^+b_{11}^+ \tilde{\gamma}|0\rangle
\psi^0_{35}\,,  \label{x-0decompf}
\end{eqnarray}
for $n = 1, 26$.

Finally, conditions (\ref{nidecompos}), (\ref{ghnum}), as applied
to $(n)_2 = (2,1)$, permit one to  decompose the fermionic field
vectors $|\chi^l_0\rangle_{(s)_3}$ derived from the general
Eq.(\ref{chif}), in even (for $l=0$) and odd (for $l=1$) powers of
ghosts, consisting of $(39+35)$ summands, and starting from the
ghost-independent vector $|\Psi\rangle_{(2,1)}$ (from the first
order in ghost momenta $\mathcal{P}_I$ for $l=1$),
\begin{eqnarray}
%%%%%%%%%%
%%%%%%%%%%%%%%%%%%%%%%%%%%%%
|\chi^0_0\rangle_{(2,1)} &=& |\Psi\rangle_{(2,1)}+
p_1^+\Bigl(q_1^+|\psi_1 \rangle_{(0,1)} +\eta_1^+\tilde{\gamma}|\psi_2 \rangle_{(0,1)}+q_2^+|\psi_3 \rangle_{(1,0)}+\eta_2^+\tilde{\gamma}|\psi_4 \rangle_{(1,0)} \nonumber\\
  && +\eta_{11}^+\tilde{\gamma}|\psi_5 \rangle_{(-1,1)}+\eta_{12}^+\tilde{\gamma}|\psi_6 \rangle_{(0)}+\vartheta_{12}^+\tilde{\gamma}|\psi_7 \rangle_{(2,0)}+ p_1^+\Bigl\{q_1^+\vartheta_{12}^+\tilde{\gamma}|\psi_8 \rangle_{(0)}  \nonumber\\
  &&  + \eta_1^+\vartheta_{12}^+|\psi_9 \rangle_{(0)}  \Bigr\}+ \mathcal{P}_1^+\Bigl\{q_1^+\vartheta_{12}^+|\psi_{10} \rangle_{(0)} + \eta_1^+\vartheta_{12}^+\tilde{\gamma}|\psi_{11} \rangle_{(0)}  \Bigr\} + \lambda_{12}^+\Bigl\{(q_1^+)^2\tilde{\gamma}|\psi_{12} \rangle_{(0)} \nonumber\\
  &&  + q_1^+\eta_1^+|\psi_{13} \rangle_{(0)}  \Bigr\}
  \Bigr) +\mathcal{P}_1^+\Bigl(q_1^+\tilde{\gamma}|\psi_{14} \rangle_{(0,1)} +\eta_1^+|\psi_{15} \rangle_{(0,1)}+q_2^+\tilde{\gamma}|\psi_{16} \rangle_{(1,0)} \nonumber\\
  && +\eta_2^+|\psi_{17} \rangle_{(1,0)}+\eta_{11}^+|\psi_{18} \rangle_{(-1,1)}+\eta_{12}^+|\psi_{19} \rangle_{(0)}+\vartheta_{12}^+|\psi_{20} \rangle_{(2,0)}  \nonumber\\
  && + \lambda_{12}^+\Bigl\{(q_1^+)^2|\psi_{21} \rangle_{(0)} + q_1^+\eta_1^+\tilde{\gamma}|\psi_{22} \rangle_{(0)}  \Bigr\}
  \Bigr)
  +  p_2^+\Bigl( q_1^+|\psi_{23} \rangle_{(1,0)} +\eta_1^+\tilde{\gamma}|\psi_{24} \rangle_{(1,0)} \nonumber\\
  && +\eta_{11}^+\tilde{\gamma}|\psi_{25} \rangle_{(0)}\Bigr)+  \mathcal{P}_2^+\Bigl( q_1^+\tilde{\gamma}|\psi_{26} \rangle_{(1,0)} +\eta_1^+|\psi_{27} \rangle_{(1,0)}  +\eta_{11}^+|\psi_{28} \rangle_{(0)}\Bigr)\nonumber
  \\
  %%%%%%%%%%%%%%%
    &&
  +\mathcal{P}_{11}^+\Bigl(q_1^+\tilde{\gamma}|\psi_{29} \rangle_{(-1,1)} +\eta_1^+|\psi_{30} \rangle_{(-1,1)}+q_2^+\tilde{\gamma}|\psi_{31} \rangle_{(0)} +\eta_2^+|\psi_{32} \rangle_{(0)}\nonumber
  \\
  && +\vartheta_{12}^+|\psi_{33} \rangle_{(1,0)}  \Bigr)
  +\mathcal{P}_{12}^+\Bigl(q_1^+\tilde{\gamma}|\psi_{34} \rangle_{(0)} +\eta_1^+|\psi_{35} \rangle_{(0)} \Bigr)
    +  \lambda_{12}^+\Bigl(q_1^+\tilde{\gamma}|\psi_{36} \rangle_{(2,0)}\nonumber\\
     && +\eta_1^+|\psi_{37} \rangle_{(2,0)}+\eta_{11}^+|\psi_{38} \rangle_{(1,0)} \Bigr)
\,, \label{x0-00}
\end{eqnarray}
  \vspace{-3ex}
  \begin{eqnarray}
%%%%%%%%%%
|\chi^1_0\rangle_{(2,1)} &=&
p_1^+\Biggl(|\varphi_1 \rangle_{(1,1)}+p_1^+\Bigl\{ q_1^+|\varphi_2 \rangle_{(-1,1)} +\eta_1^+\tilde{\gamma}|\varphi_3 \rangle_{(-1,1)}+q_2^+|\varphi_4 \rangle_{(0)}+\eta_2^+\tilde{\gamma}|\varphi_5 \rangle_{(0)} \nonumber\\
  && +\vartheta_{12}^+\tilde{\gamma}|\varphi_6 \rangle_{(1,0)}\Bigr\}+\mathcal{P}_1^+\Bigl\{ q_1^+\tilde{\gamma}|\varphi_7 \rangle_{(-1,1)} +\eta_1^+|\varphi_8 \rangle_{(-1,1)}+q_2^+\tilde{\gamma}|\varphi_9 \rangle_{(0)} \nonumber\\
  && +\eta_2^+|\varphi_{10} \rangle_{(0)}+\vartheta_{12}^+|\varphi_{11} \rangle_{(1,0)}\Bigr\} +p_2^+\Bigl\{ q_1^+|\varphi_{12} \rangle_{(0)} +\eta_1^+\tilde{\gamma}|\varphi_{13} \rangle_{(0)}\Bigr\} +\mathcal{P}_2^+\Bigl\{ q_1^+\tilde{\gamma}|\varphi_{14} \rangle_{(0)}\nonumber\\
  &&  +\eta_1^+|\varphi_{15} \rangle_{(0)}\Bigr\}+  \mathcal{P}_{11}^+\vartheta_{12}^+|\varphi_{16} \rangle_{(0)} + \lambda_{12}^+ \Bigl\{q_1^+\tilde{\gamma}|\varphi_{17} \rangle_{(1,0)}+\eta_1^+|\varphi_{18} \rangle_{(1,0)} \nonumber\\
  && +\eta_{11}^+|\varphi_{19} \rangle_{(0)} \Bigr\}\Biggr)  +  p_2^+\Bigl( |\varphi_{20} \rangle_{(2,0)} + \mathcal{P}_1^+\Bigl\{ q_1^+\tilde{\gamma}|\varphi_{21} \rangle_{(0)} +\eta_1^+|\varphi_{22} \rangle_{(0)}\Bigr\} \Bigr)+  \mathcal{P}_2^+\Bigl( \tilde{\gamma}|\varphi_{23} \rangle_{(2,0)}\nonumber\\
  &&  + \mathcal{P}_1^+\Bigl\{ q_1^+|\varphi_{24} \rangle_{(0)} +\eta_1^+\tilde{\gamma}|\varphi_{25} \rangle_{(0)}\Bigr\} \Bigr)
  +\mathcal{P}_1^+\Biggl(\tilde{\gamma}|\varphi_{26} \rangle_{(1,1)} +  \mathcal{P}_{11}^+\vartheta_{12}^+|\varphi_{27} \rangle_{(0)} \nonumber\\
  &&  + \lambda_{12}^+ \Bigl\{q_1^+|\varphi_{28} \rangle_{(1,0)}  +\eta_1^+\tilde{\gamma}|\varphi_{29} \rangle_{(1,0)}+\eta_{11}^+\tilde{\gamma}|\varphi_{30} \rangle_{(0)} \Bigr\}\Biggr)+  \mathcal{P}_{11}^+  \Bigl(\tilde{\gamma}|\varphi_{31} \rangle_{(0,1)}\nonumber\\
  &&
  +  \lambda_{12}^+ \Bigl\{q_1^+|\varphi_{32} \rangle_{(0)} +\eta_1^+\tilde{\gamma}|\varphi_{33} \rangle_{(0)}\Bigr\} \Bigr)+\mathcal{P}_{12}^+  \tilde{\gamma}|\varphi_{34} \rangle_{(1,0)}+  \lambda_{12}^+\tilde{\gamma}|\varphi_{35} \rangle_{(3,0)}
\,, \label{x0-01}
%%%%%%%%%%%%%%%%%%%%%%%%%%%%
\end{eqnarray}
In Eqs. (\ref{x0-01}), the ghost-independent vectors $|\varphi_{n}
\rangle_{(...)}$, $n = 1,\ldots, 35$ have the same decomposition
and properties as $|\psi^0_{n} \rangle_{(...)}$ in Eqs.
(\ref{x0-0}) (\ref{x-0decompi}), (\ref{x-0decompf}), whereas the
ghost-independent vector in Eq. (\ref{x0-00}), different from the
remaining ones, is the vector $|\Psi \rangle_{(2,1)}$, which reads
\begin{eqnarray}
&& |\Psi \rangle_{(2,1)}  =  a_1^{+\mu}\Bigl(a_1^{+\nu}a_1^{+\rho}d^+_{12} |0\rangle
\psi_{\mu\nu\rho} +a_1^{+\nu}a_2^{+\rho} |0\rangle
\Psi_{\mu\nu,\rho} + a_1^{+\nu} f_1^+d^+_{12} \tilde{\gamma} |0\rangle
\psi_{\mu\nu} + a_1^{+\nu} f_2^+ \tilde{\gamma} |0\rangle
\psi^{\prime}_{\mu\nu} \nonumber \\
&& \phantom{|\psi^0_{n} \rangle_{(1,1)}}+ f_1^+a_2^{+\nu} \tilde{\gamma}|0\rangle
\psi^{\prime}_{\mu,\nu} + f_1^+f_2^+  |0\rangle
\psi^{}_{\mu} + b_{11}^+ d^+_{12}|0\rangle
\psi^{\prime}_{\mu} + b_{12}^+|0\rangle
\psi^{\prime \prime}_{\mu}\Bigr)\nonumber \\
&& \phantom{|\psi^0_{n} \rangle_{(1,1)}}  + b^+_{11}\Bigl( a_2^{+\mu}|0\rangle
\psi^{\prime\prime\prime}_{\mu} + f^+_{1}d^+_{12}\tilde{\gamma}|0\rangle
\psi^{}_{} + f^+_{2}\tilde{\gamma}|0\rangle
\psi^{\prime}\Bigr) + b^+_{12}f^+_{1}\tilde{\gamma}|0\rangle
\psi^{\prime\prime} \,,  \label{x-00decompi}
\end{eqnarray}
with the initial spin-tensor field $\Psi_{\mu\nu,\rho}$,
describing a massless particle with spin
$(\frac{5}{2},\frac{3}{2})$.


\begin{thebibliography}{9}

%\bibitem{HiggsLHC}ATLAS Collaboration et al, Observation of a new particle in the search for %the Standard Model Higgs boson with the ATLAS detector at the LHC, Phys. Lett. B 716 (2012) %1--29; CMS collaboration et al, Observation of a new boson at a mass of 125 GeV with the CMS %experiment at the LHC,
%    Phys. Lett. B 716 (2012) 30--61.

\bibitem{reviews}M.~Vasiliev, Higher spin gauge theories in various dimensions,
Fortsch. Phys. 52 (2004) 702--717, [arXiv:hep-th/0401177];
D.~Sorokin, Introduction to the classical theory of higher spins,
AIP Conf. Proc. 767 (2005) 172--202, [arXiv:hep-th/0405069];
N.~Bouatta, G.~Comp\`ere,  A.~Sagnotti, An introduction to free
higher-spin fields, [arXiv:hep-th/0409068]; A.~Sagnotti,
E.~Sezgin, P.~Sundell, On higher spins with a strong Sp(2,R)
sondition, [arXiv:hep-th/0501156]; X.~Bekaert, S.~Cnockaert,
C.~Iazeolla, M.A.~Vasiliev, Nonlinear higher spin theories in
various dimensions, [arXiv:hep-th/0503128]; A.~Fotopoulos,
M.~Tsulaia, Gauge Invariant Lagrangians for Free and Interacting
Higher Spin Fields. A review of BRST formulation, Int.J.Mod.Phys.
A24 (2008) 1--60, [arXiv:0805.1346[hep-th]].


\bibitem{flatin}M. Fierz, W. Pauli, On relativistic wave equations for particles
of arbitrary spin in an electromagnetic field, Proc. R. Soc. London,
Ser. A, 173 (1939) 211--232.

\bibitem{Singh}L.P.S.~Singh, C.R.~Hagen, Lagrangian formulation for arbitrary
spin. 1. The bosonic case, Phys. Rev. D9 (1974) 898--909;
Lagrangian formulation for arbitrary spin. 2. The fermionic case,
Phys. Rev. D9 (1974) 910--920.

\bibitem{Fronsdal}C.~Fronsdal, Massless fields with integer
spin, Phys. Rev. D18 (1978) 3624--3629; J.~Fang, C.~Fronsdal,
Massless fileds with half-integral spin, Phys. Rev. D18 (1978)
3630--3633;
C.~Fronsdal, Singletons and massless, integer-spin fileds on de
Sitter space, Phys. Rev. D20 (1979) 848--856; J.~Fang,
C.~Fronsdal, Massless, half-integer-spin fields in de Sitter
space, Phys. Rev. D22 (1980) 1361--1367; M.A.~Vasiliev,  'Gauge'
Form Of Description Of Massless Fields With Arbitrary Spin (in
Russian), Yad.Fiz. 32 (1980) 855--861.



\bibitem{Heslop}N.~Beisert, M.~Bianchi, J.F.~Morales, H.~Samtleben, Higher spin
symmetries and N=4 SYM, JHEP 0407 (2004) 058,
[arXiv:hep-th/0405057]; A.C.~Petkou, Holography, duality and
higher spin fields, [arXiv:hep-th/0410116];  P.J.~Heslop, F.~Riccioni, On the fermionic
Grande Bouffe: more on higher spin symmetry breaking in AdS/CFT,
JHEP 0510 (2005) 060, [arXiv:hep-th/0508086]; M.~Bianchi,
V.~Didenko, Massive higher spin multiplets and holography,
[arXiv:hep-th/0502220].

\bibitem{Deser}S.~Deser, A.~Waldron, Gauge invariances and phases of massive
higher spins in (A)dS, Phys. Rev. Lett. 87 (2001) 031601,
[arXiv:hep-th/0102166];  S.~Deser, A.~Waldron, Partial
Masslessness of Higher Spins in (A)dS, Nucl. Phys. B607 (2001)
577--604, [arXiv:hep-th/0103198];  K.~Hallowell, A.~Waldron, Constant
curvature algebras and higher spin action generating functions,
Nucl. Phys. B724  (2005) 453--486, [arXiv:hep-th/0505255];
E.D.~Skvortsov,  M.A.~Vasiliev, Geometric formulation for
partially massless fields, Nucl. Phys. B756  (2006) 117--147,
[arXiv:hep-th/0601095].

\bibitem{Bonelli1}G.~Bonelli,
On the boundary gauge dual of closed tensionless free strings in
AdS, JHEP 0411 (2004) 059, [arXiv:hep-th/0407144]; On the
tensionless limit of bosonic strings, infinite symmetries and
higher spins, Nucl. Phys. B669 (2003) 159--172,
[arXiv:hep-th/0305155].


\bibitem{mg}
G.~Barnich, M.~Grigoriev, A.~Semikhatov, I.~Tipunin, Parent field
theory and unfolding in BRST first-quantized terms, Commun. Math.
Phys. 260 (2005) 141; G.~Barnich, M.~Grigoriev, Parent form for
higher spin fields in anti-de Sitter space, JHEP 0608 (2006) 013;
M.~Grigoriev, Off-shell gauge fields from BRST quantization,
[arXiv:hep-th/0605089];
O.A.~Gelfond, M.A.~Vasiliev, Unfolding versus BRST and
currents in $Sp(2M)$ invariant higher-spin theory,
[arXiv:1001.2585[hep-th]].

\bibitem{alkalaev}K.B.~Alkalaev, M.~Grigoriev, I.Y.~Tipunin,
Massless Poincare modules and gauge invariant equations, Nucl.
Phys. B823 (2009) 509, [arXiv:0811.3999 [hep-th]];
K.B.~Alkalaev, M. Grigoriev, Unified BRST description of AdS gauge
fields, Nucl. Phys. B835 (2010) 197, [arXiv:0910.2690[hep-th]]; K.~Alkalaev,
Mixed-symmetry tensor conserved currents and AdS/CFT correspondence, [arXiv:1207.1079 [hep-th]].

\bibitem{Vasiliev_inter}M.A.~Vasiliev, Cubic
interactions of bosonic higher spin gauge fields in AdS(5), Nucl.
Phys. B616 (2001) 106--162 [Erratum-ibid. B 652 (2003) 407],
[arXiv:hep-th/0106200];   Higher Spin Superalgebras in any
Dimension and their Representations, JHEP 12 (2004) 046,
[hep-th/0404124]; Holography, Unfolding and Higher-Spin Theory, [arXiv:1203.5554[hep-th]].

\bibitem{skvortsov}V.E. Didenko, E.D. Skvortsov, Towards higher-spin holography in ambient space of any dimension,
[arXiv:1207.6786[hep-th]]; Exact higher-spin symmetry in CFT: all correlators in unbroken Vasiliev theory, [arXiv:1210.7963[hep-th]].

\bibitem{consist1}
M.~Porrati, Universal limits of massless higher-spin particles,
Phys.Rev. D78 (2008) 065016; M.~Porrati, R.~Rahman,   A model independent
ultraviolet cutoff for theories with charged massive higher spin
fileds, Nucl.Phys. B814 (2009) 370;  M.~Porrati, R.~Rahman, Causal propagation of a charged
spin 3/2 field in an external electromagnetic background, Phys.Rev.
D80 (2009) 025009 [arXiv:0906.1432[hep-th]];  M.~Porrati, R.~Rahman, A.~Sagnotti, String
Theory and The Velo-Zwanziger Problem, [arXiv:1011.6411[hep-th]].

\bibitem{henneauxrahmanfermiint}M.~Henneaux, G.L.~Gomez, R.~Rahman, Higher-Spin Fermionic Gauge Fields and Their Electromagnetic Coupling, JHEP 1208 (2012) 093, [arXiv:1206.1048[hep-th]].

\bibitem{JoungTaronna}A.~Sagnotti and M.~Taronna, String
Lessons for Higher-Spin Interactions, Nucl. Phys. B842 (2011) 299,
[arXiv:1006.5242[hep-th]]; E.~Joung, M.~Taronna, Cubic interactions
of massless higher spins in (A)dS: metric-like approach,
[arXiv:1110.5918[hep-th]].

\bibitem{Metsaev-recent}R.R.~Metsaev, Shadows, currents and AdS,
Phys.Rev. D 78 (2008) 106010, [arXiv:0805. 3472[hep-th]]; CFT
adapted gauge invariant formulation of arbitrary spin fields in
AdS and modified de Donder gauge, Phys.Lett. B671 (2009) 128--134,
[arXiv:0808.3945[hep-th]]; CFT adapted gauge invariant formulation
of massive arbitrary spin fields in AdS, Phys.Lett. B682 (2010)
455--461, [arXiv:0907.2207[hep-th]]; Gauge invariant approach to
low-spin anomalous conformal currents and shadow fields, Phys.Rev.
D83 (2011) 106004, [arXiv:1011.4261[hep-th]]; Extended Hamiltonian Action for Arbitrary Spin Fields in Flat And AdS Spaces,
[arXiv:1112.0976[hep-th]].

\bibitem{Zinovievint}Yu.A.~Zinoviev, Frame-like gauge invariant
formulation for massive high spin particles, Nucl. Phys. B808
(2009) 185, [arXiv:0808.1778[hep-th]]; Yu.M.~Zinoviev,
Gravitational cubic interactions for a massive mixed symmetry
gauge field, [arXiv:1107.3222[hep-th]]; On electromagnetic
interactions for massive mixed symmetry field, JHEP 1103 (2011)
082, arXiv:1012.2706 [hep-th]]; N.~Boulanger, E.D.~Skvortsov,
Yu.M.~Zinoviev, Gravitational cubic interactions for a simple
mixed-symmetry gauge field in AdS and flat backgrounds, J.Phys.A
A44 (2011) 415403, [arXiv:1107.1872[hep-th]];  I.L. Buchbinder, T.V.
Snegirev, Yu.M. Zinoviev, Cubic interaction vertex of higher-spin fields with external electromagnetic field,
Nucl.Phys. B 864 (2012) 694-721,
[arXiv:1204.2341[hep-th]]; Gauge invariant Lagrangian formulation of massive higher spin fields in (A)dS${}_3$ space,
Phys.Lett. B716 (2012) 243-248,
[arXiv:1207.1215[hep-th]].


\bibitem{quartmixbosemas}I.L.~Buchbinder, A.V.~Galajinsky, V.A.~Krykhtin,
Quartet unconstrained formulation for
massless higher spin fields, Nucl. Phys. B779 (2007) 155, [arXiv:
hep-th/0702161]; I.L.~Buchbinder, A.V.~Galajinsky, Quartet
unconstrained formulation for massive higher spin fields, JHEP
0811 (2008) 081 [arXiv:0810.2852[hep-th]].


\bibitem{sorvas}D.P.~Sorokin, M.A.~Vasiliev, Reducible higher-spin
multiplets in flat and AdS spaces and their geometric frame-like
formulation, Nucl.Phys. B809 (2009) 110--157,
[arXiv:0807.0206[hep-th]].

\bibitem{Francia1}D.~Francia, Geometric Lagrangians for massive higher-spin fields,
Nucl. Phys. B {796} (2008)   77, [arXiv:0710.5378 [hep-th]].

\bibitem{Boulanger}N.~Boulanger, P.~Sundell, An action principle for Vasiliev's
four-dimensional higher-spin gravity, [arXiv:1102.2219[hep-th]];
N.~Doroud, L.~Smolin, An action for higher spin gauge theory in
four dimensions, [arXiv:1102.3297[hep-th]].

\bibitem{massless AdS}M.A.~Vasiliev, Free massless fermionic fields of arbitrary spin in
D-dimensional anti-de~Sitter space, Nucl. Phys. B301 (1988)
26--51; V.E.~Lopatin, M.A.~Vasiliev, Free massless bosonic fields
of arbitrary spin in D-dimensional de~Sitter space, Mod. Phys.
Lett. A3 (1998) 257--265.


\bibitem{massive AdS}Yu.M.~Zinoviev, On massive high spin particles in
AdS, [arXiv:hep-th/0108192]; R.R.~Metsaev, Massive totally
symmetric fields in AdS(d), Phys.Lett. B590 (2004) 95--104,
[arXiv:hep-th/0312297]; Fermionic fields in the d-dimensional
anti-de Sitter spacetime, Phys. Lett. B419 (1998) 49--56,
[arXiv:hep-th/9802097]; Light-cone form of field dynamics in
anti-de Sitter space-time and AdS/CFT correspondence, Nucl. Phys.
B563 (1999) 295--348, [arXiv:hep-th/9906217]; Massless arbitrary
spin fields in AdS(5) Phys. Lett. B531 (2002) 152--160,
[arXiv:hep-th/0201226];  P.~de~Medeiros, Massive gauge-invariant
field theories on space of constant curvature, Class. Quant. Grav.
21 (2004) 2571--2593, [arXiv:hep-th/0311254].

\bibitem{Labastida}J.M.F. Labastida, T.R. Morris, Massless mixed symmetry
bosonic free fields, Phys. Lett. B180 (1986) 101--106; J.M.F.
Labastida, Massless fermionic free fields, Phys. Lett. B186 (1987)
365--369; Massless bosonic free fields,
 Phys. Rev. Lett. 58 (1987) 531--534; Massless
particles in arbitrary representations of the Lorentz group, Nucl.
Phys. B322 (1989) 185--209.

\bibitem{Vasilievmix}
L.~Brink, R.R.~Metsaev, M.A.~Vasiliev, How massless are massless
fields in $AdS_d$, Nucl. Phys. B586 (2000) 183--205,
[arXiv:hep-th/0005136].

\bibitem{metsaevmixirrep}R.R.~Metsaev,
Massless mixed symmetry bosonic free fields in d-dimensional
anti-de Sitter space-time, Phys. Lett. B354 (1995) 78--84;
Mixed-symmetry massive fields in AdS(5), Class. Quant. Grav. 22
(2005) 2777--2796, [arXiv:hep-th/0412311].


\bibitem{Curtright}T. Curtright, Massless field supermultiplets with arbitrary
spin, Phys. Lett. B85 (1979) 219--224; Generalized gauge fields
 Phys. Lett. B165 (1985) 304--308.

\bibitem{franciamixfermi}A. Campoleoni,  D. Francia,  J. Mourad, A. Sagnotti,
Unconstrained Higher Spins of Mixed Symmetry. II. Fermi Fields.
Nucl.Phys. B828 (2010) 405--514, [arXiv:0904.4447[hep-th]].

\bibitem{franciamixbos}A. Campoleoni,  D. Francia,  J. Mourad, A. Sagnotti,
Unconstrained Higher Spins of Mixed Symmetry. I. Bose Fields.
Nucl.Phys. B815 (2009) 289--357, [arXiv:0810.4350[hep-th]]

\bibitem{framefermimix}E.D.~Skvortsov, Yu.M.~Zinoviev,
Frame-like Actions for Massless Mixed-Symmetry Fields in Minkowski
space. Fermions, Nucl.Phys. B843 (2011) 559--569,
[arXiv:1007.4944[hep-th]].

\bibitem{Zinovievfermi}Yu.M.~Zinoviev,
Frame-like gauge invariant formulation for mixed symmetry fermionic fields, Nucl.Phys. B821 (2009) 21--47,
[arXiv:0904.0549[hep-th]].

\bibitem{BRmixbos}I.L. Buchbinder and  A.
Reshetnyak, General Lagrangian Formulation for Higher Spin Fields
with Arbitrary Index Symmetry. I. Bosonic fields, Nucl. Phys. B
862 (2012)  270--327, [arXiv:1110.5044[hep-th]].

\bibitem{mixboseResh} A.A. Reshetnyak, On Lagrangian formulations for arbitrary
bosonic HS fields on
Minkowski backgrounds, Phys. of Particles and Nuclei  43 (2012) 689-693, [arXiv:1202.4710[hep-th]].

\bibitem{Pashnev1}A. Pashnev,  M.M. Tsulaia, Description of the higher massless
irreducible integer spins in the BRST approach, Mod. Phys. Lett.
A13 (1998) 1853--1864, [arXiv:hep-th/9803207].

\bibitem{Bekaert}X. Bekaert, N. Boulanger, Tensor gauge fields in arbitrary
representations of GL(D,R): duality and Poincare lemma, Commun.
Math. Phys. 245 (2004) 27--67, [arXiv:hep-th/0208058]; On
geometric equations and duality for free higher spins, Phys. Lett.
B561 (2003) 183--190, [arXiv:hep-th/0301243]; Mixed symmetry gauge
fields in a flat background, [arXiv:hep-th/0310209]; Tensor gauge
fields in arbitrary representations of GL(D,R). II. Quadratic
actions, Commun. Math. Phys. 271 (2007) 723--773,
[arXiv:hep-th/0606198]; X. Bekaert, N. Boulanger, S. Cnockaert, No
self-interaction for two-column massless fields, J. Math. Phys. 46
(2005) 012303, [arXiv:hep-th/0407102].

\bibitem{Medeiros}P. de Medeiros,  C. Hull, Geometric second order field
equations for general tensor gauge fields, JHEP 0305 (2003) 019,
[arXiv:hep-th/0303036].

\bibitem{BFV}E.S. Fradkin,  G.A. Vilkovisky, Quantization of relativistic
systems with constraints, Phys. Lett. B55 (1975) 224--226; I.A.
Batalin, G.A. Vilkovisky, Relativistic S-matrix of dynamical
systems with boson and fermion constraints, Phys. Lett. B69 (1977)
309--312; I.A. Batalin, E.S. Fradkin, Operator quantization of
relativistic dynamical systems subject to first class constraints,
Phys. Lett. B128 (1983) 303.

\bibitem{bf}I.A. Batalin, E.S. Fradkin, Operator quantization method and
abelization of dynamical systems subject to first class
constraints, Riv. Nuovo Cimento, 9, No.~10 (1986) 1; I.A. Batalin,
E.S. Fradkin, Operator quantization of dynamical systems subject
to constraints.  A further study of the construction, Ann. Inst.
H. Poincare, A49 (1988) 145--214.

\bibitem{Henneaux}M. Henneaux, Hamiltonian form of the path integral for theories
with a gauge freedom, Phys. Rept. 126 (1985) 1--66; M. Henneaux,
C. Teitelboim, Quantization of gauge systems, Princeton Univ.
Press, 1992.

\bibitem{AKSZ} M. Alexandrov, M. Kontsevich, A.Schwarz, O. Zaboronsky, The geometry of the master equation and topological quantum field theory, Int. I. Mod. Phys. A12 (1997) 1405--1430, [arXiv:hep-th/9502010].

\bibitem{BV-BFV}M. Grigoriev,  P.H. Damgaard,
Superfield BRST charge and the master action, Phys. Lett. B474
(2000) 323--330, [arXiv:hep-th/9911092]; G. Barnich,  M.
Grigoriev, Hamiltonian BRST and Batalin-Vilkovisky formalisms for
second quantization of gauge theories, Commun. Math. Phys. 254
(2005) 581--601, [arXiv:hep-th/0310083].

\bibitem{GMR} D.M. Gitman, P.Yu. Moshin, A.A. Reshetnyak, Local
superfield Lagrangian BRST quantization, J. Math. Phys. {46}
(2005) 072302-01--072302-24, [arXiv:hep-th/0507160]; An embedding
of the BV quantization into an N=1 local superfield formalism,
Phys. Lett. B 621 (2005) 295--308, [arXiv:hep-th/0507049].

\bibitem{Francia}D.~Francia, A.~Sagnotti, Free geometric equations for higher
spins,  Phys. Lett. B543 (2002) 303--310, [arXiv:hep-th/0207002];
A.~Sagnotti, M.~Tsulaia, On higher spins and the tensionless limit
of String Theory, Nucl. Phys.  B682 (2004) 83--116,
[arXiv:hep-th/0311257]; D.~Francia, J.~Mourad, A.~Sagnotti,
Current exchanges and unconstrained higher spins, Nucl. Phys. B773
(2007) 203 [arXiv:hep-th/0701163].


\bibitem{Zinoviev_m}Y.M.~Zinoviev, On massive mixed symmetry tensor
fields in Minkowski space and (A)dS, [arXiv:hep-th/0211233]; First
order formalism for mixed symmetry tensor fields,
[arXiv:hep-th/0304067]; First order formalism for massive mixed
symmetry tensor fields in Minkowski and (A)dS spaces,
[arXiv:hep-th/0306292].

\bibitem{Ouvry}C.S. Aulakh, I.G. Koh, S. Ouvry, Higher spin fields with
mixed symmetry, Phys. Lett. B173 (1986) 284--288; S. Ouvry, J.
Stern, Gauge fields of any spin and symmetry, Phys. Lett. B177
(1986) 335--340; A.K.H. Bengtsson, A unified action for higher
spin gauge bosons from covariant string theory, Phys. Lett. B182
(1986) 321--325.

\bibitem{Metsaev-1}R.R.Metsaev, Free totally (anti)symmetric massless fermionic
fields in d-dimensional anti-de Sitter space, Class. Quant. Grav. 14
(1997) L115--L121, [arXiv:hep-th/9707066]; Fermionic fields in the
d-dimensional anti-de Sitter spacetime, Phys. Lett. B419 (1998)
49--56, [arXiv:hep-th/9802097]; Arbitrary spin massless bosonic
fields in d-dimensional anti-de~Sitter space,
[arXiv:hep-th/9810231]; Light-cone form of field dynamics in
anti-de Sitter space-time and AdS/CFT correspondence, Nucl. Phys.
B563 (1999) 295--348, [arXiv:hep-th/9906217]; Massless arbitrary
spin fields in AdS(5) Phys. Lett. B531 (2002) 152--160,
[arXiv:hep-th/0201226]; Massive totally symmetric fields in
AdS(d), Phys. Lett. B590 (2004) 95--104, [arXiv:hep-th/0312297];
R.R. Metsaev, Mixed-symmetry massive fields in AdS(5),
Class. Quant. Grav. 22 (2005) 2777--2796, [arXiv:hep-th/0412311];
Cubic interaction vertices of massive and massless higher spin
fields, Nucl. Phys. B759  (2006) 147--201, [arXiv:hep-th/0512342];
R.R. Metsaev, Gauge invariant formulation of massive totally
symmetric fermionic fileds in (A)dS space, Phys. Lett. B643 (2006)
205--212, [arXiv:hep-th/0609029].

\bibitem{Mets-amb}R.R. Metsaev, Massless mixed symmetry bosonic free fields
in d-dimensional anti-de Sitter space-time, Phys. Lett. B354
(1995) 78--84; Fermionic fields in the d-dimensional anti-de
Sitter spacetime, Phys. Lett. B419 (1998) 49--56,
[arXiv:hep-th/9802097].


\bibitem{BurdikPashnev}C. Burdik, A.
Pashnev, M. Tsulaia, On the mixed symmetry irreducible
representations of the Poincare group in the BRST approach, Mod.
Phys. Lett. A16 (2001) 731--746, [arXiv:hep-th/0101201]; The
Lagrangian description of representations of the Poincare group,
Nucl. Phys. Proc. Suppl. 102 (2001) 285--292, [arXiv:hep-th/0103143].


\bibitem{Metsaev-0}R.R.~Metsaev,   Cubic interaction vertices of massive and
massless higher spin fields, Nucl. Phys. B759  (2006) 147--201,
[arXiv:hep-th/0512342].


\bibitem{Tsulaiaint}A.~Fotopoulos, K.L.~Panigrahi, M.~Tsulaia,
Lagrangian formulation of higher spin theories on AdS, Phys. Rev.
D74 (2006) 085029, [arXiv:hep-th/0607248];
A.~Fotopoulos and
M.~Tsulaia, On the Tensionless Limit of String theory, Off - Shell
Higher Spin Interaction Vertices and BCFW Recursion Relations,
JHEP 1011 (2010) 086, [arXiv:1009.0727[hep-th]]; P.~Dempster and
M.~Tsulaia, On the Structure of Quartic Vertices for Massless Higher Spin Fields on
Minkowski Background, Nucl.Phys. B 865 (2012) 353--375,
[arXiv:1203.5597[hep-th]].

\bibitem{Manvelyan}R.~{Manvelyan}, K.~Mkrtchyan, W.~Ruehl, Off-shell construction of some
trilinear higher spin gauge field interactions, Nucl.Phys. B826
(2010) 1--17, [arXiv:0903.0243[hep-th]]; Direct construction of a
cubic selfinteraction for higher spin gauge fields, Nucl.Phys.
B844 (2011) 348--364, [arXiv:1002.1358[hep-th]]; General trilinear
interaction for arbitrary even higher spin gauge fields,
Phys.Lett. B696 (2011) 410--415, [arXiv:1009.1054[hep-th]];
Radial Reduction and Cubic Interaction for Higher Spins in (A)dS space,
[arXiv:1210.7227[hep-th]];
R.~Manvelyan, K.~Mkrtchyan, W.~Ruehl, M.~Tovmasyan, On Nonlinear
Higher Spin Curvature,  Phys.Lett. B699 (2011) 187--191,
[arXiv:1102.0306[hep-th]].

\bibitem{AlkalaevVasiliev}
K.B. Alkalaev,
Free fermionic higher spin fields in AdS(5), Phys. Lett. B519
(2001) 121--128, [arXiv: hep-th/0107040];
K.B. Alkalaev, M.A. Vasiliev,
N=1 supersymmetric theory of higher spin gauge fields in AdS(5)
at the cubic level, Nucl. Phys. B655 (2003) 57--92, [arXiv:hep-th/0206068];
K.B. Alkalaev,
Two column higher spin massless fields in AdS(d), Theor. Math. Phys.
140 (2004) 1253--1263, [arXiv:hep-th/0311212];
K.B. Alkalaev, O.V. Shaynkman, M.A. Vasiliev,
On the frame-like formulation of mixed-symmetry massless fields in (A)dS(d),
Nucl. Phys. B692 (2004) 363--393, [arXiv:hep-th/0311164];  Lagrangian formulation
for free mixed-symmetry bosonic gauge fields in (A)dS(d), JHEP
0508 (2005) 069, [arXiv:hep-th/0501108]; Frame-like formulation for free
mixed-symmetry bosonic massless higher-spin fields in AdS(d),
[arXiv:hep-th/0601225];
K.B. Alkalaev,
Mixed-symmetry massless gauge fields
in AdS(5), Theor. Math. Phys. 149 (2006) 1338--1348.

\bibitem{Zinoviev}Yu.M. Zinoviev, Massive N=1 supermultiplets with arbitrary
superspins, [arXiv:0704.1535]; Massive supermultiplets with spin
3/2, [arXiv:hep-th/0703118]





\bibitem{Tsulaia}A. Fotopoulos, M. Tsulaia, Interacting higher spins and the high energy
limit of the bosonic string, [arXiv:0705.2939].


\bibitem{Sagnotti}A. Sagnotti, M.  Tsulaia, On higher spins
and the tensionless limit of string theory, Nucl. Phys.  B682
(2004) 83--116, [arXiv:hep-th/0311257].


%\bibitem{0505092}I.L. Buchbinder, V.A. Krykhtin, Gauge invariant Lagrangian
%construction for massive bosonic higher spin fields in D
%dimensions, Nucl. Phys. B727 (2005) 536--563,
%[arXiv:hep-th/0505092].

\bibitem{symferm-flat}I.L. Buchbinder, V.A. Krykhtin, A. Pashnev, BRST approach to
Lagrangian construction for fermionic massless higher spin fields,
Nucl. Phys. B711 (2005) 367--391, [arXiv:hep-th/0410215]; I.L.
Buchbinder, V.A. Krykhtin, L.L. Ryskina, H. Takata, Gauge
invariant Lagrangian construction for massive higher spin
fermionic fields,  Phys. Lett. B641 (2006) 386--392,
[arXiv:hep-th/0603212].

\bibitem{symferm-ads}I.L. Buchbinder, V.A. Krykhtin, A.A.
Reshetnyak, BRST approach to Lagrangian construction for fermionic
higher spin fields in AdS space, Nucl. Phys. B787 (2007) 211,
[arXiv:hep-th/0703049].

\bibitem{symint-ads}I.L. Buchbinder, A. Pashnev, M. Tsulaia, Lagrangian formulation of
the massless higher integer spin fields in the AdS background,
Phys. Lett. B523 (2001) 338--346, [arXiv:hep-th/0109067];
 X. Bekaert, I.L. Buchbinder, A.
Pashnev, M. Tsulaia, On higher spin theory: strings, BRST,
dimensional reductions, Class. Quant. Grav. 21 (2004) 1457--1464,
[arXiv:hep-th/0312252]; I.L.~Buchbinder, V.A.~Krykhtin, P.M.~Lavrov, Gauge invariant
Lagrangian formulation of higher massive bosonic field theory in
AdS space, Nucl. Phys. B762 (2007) 344--376,
[arXiv:hep-th/0608005].


\bibitem{BRmixads}C. Burdik, A. Reshetnyak, On representations of Higher Spin symmetry  algebras for
mixed-symmetry HS fields on AdS-spaces. Lagrangian formulation, J. Phys. Conf. Ser. 343 (2012) 012102,  [arXiv:1111.5516[hep-th]]

\bibitem{mixfermiflat}{Moshin P.Yu. and Reshetnyak A.A.}, BRST approach to Lagrangian formulation for mixed-symmetry
fermionic higher-spin fields,  JHEP. 10 (2007) 040,
[arXiv:0707.0386[hep-th]].

\bibitem{brst1}
I.L.~Buchbinder,
V.A.~Krykhtin, L.L.~Ryskina, Lagrangian formulation of massive
fermionic totally antisymmetric tensor field theory in AdS${}_d$
space, Nucl. Phys. B819 (2009) 453--477, [arXiv:0902.1471[hep-th]].



\bibitem{Howe1}R. Howe, Transcending classical invariant theory, J. Amer.
Math. Soc. 3 (1989) 2; Remarks on classical invariant theory,
Trans. Amer. Math. Soc. 2 (1989) 313.

\bibitem{conversion}L.D. Faddeev, S.L. Shatashvili, Realization of
the Schwinger term in the Gauss low and the possibility of correct
quantization of a theory with anomalies, Phys.Lett. B167 (1986)
225--238; I.A. Batalin, E.S. Fradkin, T.E. Fradkina, Another
version for operatorial quantization of dynamical systems with
iireducible constraints, Nucl. Phys. B314 (1989) 158--174; I.A.
Batalin, I.V. Tyutin, Existence theorerm for the effective gauge
algebra in the generalized canonical formalism and Abelian
conversion of second class constraints, Int. J. Mod. Phys. A6
(1991) 3255--3282; E. Egorian, R. Manvelyan, Quantization of
dynamical systems with first and second class constraints, Theor.
Math. Phys. 94 (1993) 241--252.



\bibitem{0905.2705}{A. Kuleshov  and A. Reshetnyak}
Programming realization of symbolic computations for non-linear
commutator superalgebras over the Heisenberg-Weyl superalgebra: data structures
and processing methods,
[arXiv:0905.2705[hep-th]].

\bibitem{0001195}C. Burdik, A.
Pashnev, M. Tsulaia, Auxiliary representations of Lie algebras and
the BRST constructions, Mod. Phys. Lett. A15 (2000) 281--291,
[arXiv:hep-th/0001195].

\bibitem{Siegel}W.~Siegel, Gauging Ramond
string fields via Osp(1,1/2), Nucl.Phys. B284 (1987) 632--644.

\bibitem{Siegel1}W.~Siegel, B.~Zwiebach, Gauge string fields from the light
cone, Nucl.Phys. B 282 (1987) 125--143.

\bibitem{BarnichHenneaux1}G. Barnich, M. Henneaux,  Consistent couplings
between fields with a gauge freedom and deformations of the master
equation, Phys. Lett. B311 (1993) 123--129,
[arXiv:hep-th/9304057]; M. Henneaux, Consistent interactions
between gauge fields: the cohomological approach, Contemp. Math.
219 (1998) 93--105, [arXiv:hep-th/9712226], M. Dubois-Violette, M.
Henneaux, Generalized cohomology for irreducible tensor fields of
mixed Young symmetry type, Lett. Math. Phys. 49 (1999) 245--252,
[arXiv:math.qa/9907135].

\bibitem{Verma1} D.-N. Verma,   Structure of certain induced representations of complex semisimple Lie algebras,
Yale Univ.,dissertation, 1966, Bull. Amer. Math. Soc. 74, No. 1 (1968)  160--166.

\bibitem{BGG} I.N. Bernstein, I.M. Gelfand, S.I. Gelfand, The structure of representations by vectors of highest weight, Funct. Anal. i Prilozhen. 5 (1971) 1--9.

\bibitem {Dixmier}J. Dixmier, Algebres enveloppantes,
Gauthier-Villars, Paris (1974), English transl., Enveloping algebras, North-Holland, Amsterdam, 1977.

\bibitem{genVM}A. Rocha-Caridi, Splitting criteria for $\mathfrak{p}$-modules induced from a parabolic and a Bernstein-Gelfand-Gelfand resolution of a finite-dimensional, irreducible $\mathfrak{p}$-module, Trans. Amer. Math. Soc., 262 (1980) 335--366;\\
    A.J. Coleman, V.M. Futorny, Stratified L-modules, J. Algebra, 163 (1994) 219--234;\\
     V. Futorny, V. Mazorchuk,  Structure of $\alpha$-Stratified L-modules for finite-dimensional Lie algebras, J. Algebra, I 183 (1996) 456--482;\\
      V. Mazorchuk, S. Ovsienko, Submodule  structure of Generalized Verma modules induced from generic Gelfand-Zeitlin modules, Alg. and Repr. Theory 1 (1998) 3--26.

\bibitem{Burdik}C. Burdik, Realizations of the real simple Lie algebras:
the method of construction, J. Phys. A: Math. Gen. 18 (1985)
3101--3112.



\bibitem{VKac} V. G. Kac Infinite dimensional Lie algebras, Cambridge University Press, Cambridge (1990).

\bibitem{BurLeites} C. Burdik, P. Grozman, D. Leites, A.  Sergeev,
Realization of Lie Algebras and Superalgebras in Terms of
Creation and Annihilation Operators: I, Theor. Math.
Phys. {124} (2000) 1048--58.

\bibitem{Dobrev} V.K. Dobrev, Representations and characters of the
Virasoro algebra and N=1 super-Virasoro algebras, [arXiv:0709.0105[hep-th]].

\end{thebibliography}
\end{document}